\documentclass[twocolumn]{aastex631} %

\accepted{for publication in The Astrophysical Journal, January 7, 2024}
\begin{document}

\title{The Cosmos in its Infancy: JADES Galaxy Candidates at $z > 8$ in GOODS-S and GOODS-N}

\author[0000-0003-4565-8239] {Kevin N.\ Hainline}
\affiliation{Steward Observatory, University of Arizona, 933 N. Cherry Ave, Tucson, AZ 85721, USA}

\author[0000-0002-9280-7594] {Benjamin D.\ Johnson}
\affiliation{Center for Astrophysics $|$ Harvard \& Smithsonian, 60 Garden St., Cambridge MA 02138 USA}

\author[0000-0002-4271-0364] {Brant Robertson}
\affiliation{Department of Astronomy and Astrophysics, University of California, Santa Cruz, 1156 High Street, Santa Cruz CA 96054, USA}

\author[0000-0002-8224-4505] {Sandro Tacchella}
\affiliation{Kavli Institute for Cosmology, University of Cambridge, Madingley Road, Cambridge CB3 0HA, UK}
\affiliation{Cavendish Laboratory, University of Cambridge, 19 JJ Thomson Avenue, Cambridge CB3 0HE, UK}

\author[0000-0003-4337-6211] {Jakob M.\ Helton}
\affiliation{Steward Observatory, University of Arizona, 933 N. Cherry Ave, Tucson, AZ 85721, USA}

\author[0000-0002-4622-6617] {Fengwu Sun}
\affiliation{Steward Observatory, University of Arizona, 933 N. Cherry Ave, Tucson, AZ 85721, USA}

\author[0000-0002-2929-3121] {Daniel J.\ Eisenstein}
\affiliation{Center for Astrophysics $|$ Harvard \& Smithsonian, 60 Garden St., Cambridge MA 02138 USA}

\author[0000-0003-4770-7516] {Charlotte Simmonds}
\affiliation{Kavli Institute for Cosmology, University of Cambridge, Madingley Road, Cambridge CB3 0HA, UK}
\affiliation{Cavendish Laboratory, University of Cambridge, 19 JJ Thomson Avenue, Cambridge CB3 0HE, UK}

\author[0000-0001-8426-1141] {Michael W.\ Topping}
\affiliation{Steward Observatory, University of Arizona, 933 N. Cherry Ave, Tucson, AZ 85721, USA}

\author[0000-0003-1432-7744] {Lily Whitler}
\affiliation{Steward Observatory, University of Arizona, 933 N. Cherry Ave, Tucson, AZ 85721, USA}

\author[0000-0001-9262-9997] {Christopher N.\ A.\ Willmer}
\affiliation{Steward Observatory, University of Arizona, 933 N. Cherry Ave, Tucson, AZ 85721, USA}

\author[0000-0002-7893-6170] {Marcia Rieke}
\affiliation{Steward Observatory, University of Arizona, 933 N. Cherry Ave, Tucson, AZ 85721, USA}

\author[0000-0002-1714-1905] {Katherine A.\ Suess}
\affiliation{Department of Astronomy and Astrophysics, University of California, Santa Cruz, 1156 High Street, Santa Cruz, CA 95064 USA}
\affiliation{Kavli Institute for Particle Astrophysics and Cosmology and Department of Physics, Stanford University, Stanford, CA 94305, USA}

\author[0000-0002-4684-9005] {Raphael E.\ Hviding}
\affiliation{Steward Observatory, University of Arizona, 933 N. Cherry Ave, Tucson, AZ 85721, USA}

\author[0000-0002-0450-7306] {Alex J.\ Cameron}
\affiliation{Department of Physics, University of Oxford, Denys Wilkinson Building, Keble Road, Oxford OX1 3RH, UK}

\author[0000-0002-8909-8782] {Stacey Alberts}
\affiliation{Steward Observatory, University of Arizona, 933 N. Cherry Ave, Tucson, AZ 85721, USA}

\author[0000-0003-0215-1104] {William M. Baker }
\affiliation{Kavli Institute for Cosmology, University of Cambridge, Madingley Road, Cambridge CB3 0HA, UK}
\affiliation{Cavendish Laboratory, University of Cambridge, 19 JJ Thomson Avenue, Cambridge CB3 0HE, UK}

\author[0000-0002-4735-8224]{Stefi Baum}
\affiliation{Department of Physics and Astronomy, University of Manitoba, Winnipeg, MB R3T 2N2, Canada}

\author[0000-0003-0883-2226] {Rachana Bhatawdekar}
\affiliation{European Space Agency (ESA), European Space Astronomy Centre (ESAC), Camino Bajo del Castillo s/n, 28692 Villanueva de la Cañada, Madrid, Spain; European Space Agency, ESA/ESTEC, Keplerlaan 1, 2201 AZ Noordwijk, NL}

\author[0000-0001-8470-7094]{Nina Bonaventura}
\affiliation{Cosmic Dawn Center (DAWN), Copenhagen, Denmark}
\affiliation{Niels Bohr Institute, University of Copenhagen, Jagtvej 128, DK-2200, Copenhagen, Denmark}
\affiliation{Steward Observatory, University of Arizona, 933 N. Cherry Ave, Tucson, AZ 85721, USA}

\author[0000-0003-4109-304X]{Kristan Boyett}
\affiliation{School of Physics, University of Melbourne, Parkville 3010, VIC, Australia}
\affiliation{ARC Centre of Excellence for All Sky Astrophysics in 3 Dimensions (ASTRO 3D), Australia}

\author[0000-0002-8651-9879] {Andrew J.\ Bunker}
\affiliation{Department of Physics, University of Oxford, Denys Wilkinson Building, Keble Road, Oxford OX1 3RH, UK}

\author[0000-0002-6719-380X] {Stefano Carniani }
\affiliation{Scuola Normale Superiore, Piazza dei Cavalieri 7, I-56126 Pisa, Italy}

\author[0000-0003-3458-2275] {Stephane Charlot}
\affiliation{Sorbonne Universit\'e, CNRS, UMR 7095, Institut d'Astrophysique de Paris, 98 bis bd Arago, 75014 Paris, France}

\author[0000-0002-7636-0534]{Jacopo Chevallard}
\affiliation{Department of Physics, University of Oxford, Denys Wilkinson Building, Keble Road, Oxford OX1 3RH, UK}

\author[0000-0002-2178-5471]{Zuyi Chen}
\affiliation{Steward Observatory University of Arizona 933 N. Cherry Avenue Tucson AZ 85721, USA}

\author[0000-0002-2678-2560]{Mirko Curti}
\affiliation{European Southern Observatory, Karl-Schwarzschild-Strasse 2, 85748 Garching, Germany}
\affiliation{Kavli Institute for Cosmology, University of Cambridge, Madingley Road, Cambridge CB3 0HA, UK}
\affiliation{Cavendish Laboratory, University of Cambridge, 19 JJ Thomson Avenue, Cambridge CB3 0HE, UK}

\author[0000-0002-9551-0534] {Emma Curtis-Lake}
\affiliation{Centre for Astrophysics Research, Department of Physics, Astronomy and Mathematics, University of Hertfordshire, Hatfield AL10 9AB, UK}

\author[0000-0003-2388-8172] {Francesco D'Eugenio}
\affiliation{Kavli Institute for Cosmology, University of Cambridge, Madingley Road, Cambridge CB3 0HA, UK}
\affiliation{Cavendish Laboratory, University of Cambridge, 19 JJ Thomson Avenue, Cambridge CB3 0HE, UK}

\author[0000-0003-1344-9475] {Eiichi Egami}
\affiliation{Steward Observatory, University of Arizona, 933 N. Cherry Ave, Tucson, AZ 85721, USA}

\author[0000-0003-4564-2771]{Ryan Endsley}
\affiliation{Department of Astronomy, University of Texas, Austin, TX 78712, USA}

\author[0000-0002-8543-761X] {Ryan Hausen}
\affiliation{Department of Physics and Astronomy, The Johns Hopkins University, 3400 N. Charles St. Baltimore, MD 21218}

\author[0000-0001-7673-2257] {Zhiyuan Ji}
\affiliation{Steward Observatory, University of Arizona, 933 N. Cherry Ave, Tucson, AZ 85721, USA}

\author[0000-0002-3642-2446] {Tobias J. Looser}
\affiliation{Kavli Institute for Cosmology, University of Cambridge, Madingley Road, Cambridge CB3 0HA, UK}
\affiliation{Cavendish Laboratory, University of Cambridge, 19 JJ Thomson Avenue, Cambridge CB3 0HE, UK}

\author[0000-0002-6221-1829] {Jianwei Lyu}
\affiliation{Steward Observatory, University of Arizona, 933 N. Cherry Ave, Tucson, AZ 85721, USA}

\author[0000-0002-4985-3819] {Roberto Maiolino}
\affiliation{Kavli Institute for Cosmology, University of Cambridge, Madingley Road, Cambridge CB3 0HA, UK}
\affiliation{Cavendish Laboratory, University of Cambridge, 19 JJ Thomson Avenue, Cambridge CB3 0HE, UK}
\affiliation{Department of Physics and Astronomy, University College London, Gower Street, London WC1E 6BT, UK}

\author[0000-0002-7524-374X] {Erica Nelson}
\affiliation{Department for Astrophysical and Planetary Science, University of Colorado, Boulder, CO 80309, USA}

\author[0000-0001-8630-2031]{D\'{a}vid Pusk\'{a}s}
\affiliation{Kavli Institute for Cosmology, University of Cambridge, Madingley Road, Cambridge CB3 0HA, UK}
\affiliation{Cavendish Laboratory, University of Cambridge, 19 JJ Thomson Avenue, Cambridge CB3 0HE, UK}

\author[0000-0002-7028-5588]{Tim Rawle}
\affiliation{European Space Agency (ESA), European Space Astronomy Centre (ESAC), Camino Bajo del Castillo s/n, 28692 Villafranca del Castillo, Madrid, Spain}

\author[0000-0001-9276-7062]{Lester Sandles}
\affiliation{Kavli Institute for Cosmology, University of Cambridge, Madingley Road, Cambridge CB3 0HA, UK}
\affiliation{Cavendish Laboratory, University of Cambridge, 19 JJ Thomson Avenue, Cambridge CB3 0HE, UK}

\author[0000-0001-5333-9970] {Aayush Saxena}
\affiliation{Department of Physics, University of Oxford, Denys Wilkinson Building, Keble Road, Oxford OX1 3RH, UK}
\affiliation{Department of Physics and Astronomy, University College London, Gower Street, London WC1E 6BT, UK}

\author[0000-0001-8034-7802]{Renske Smit}
\affiliation{Astrophysics Research Institute, Liverpool John Moores University, 146 Brownlow Hill, Liverpool L3 5RF, UK}

\author[0000-0001-6106-5172]{Daniel P. Stark}
\affiliation{Steward Observatory, University of Arizona, 933 N. Cherry Ave, Tucson, AZ 85721, USA}

\author[0000-0003-2919-7495] {Christina C. Williams}
\affiliation{NSF’s National Optical-Infrared Astronomy Research Laboratory, 950 North Cherry Avenue, Tucson, AZ 85719, USA}

\author[0000-0002-4201-7367]{Chris Willott}
\affiliation{NRC Herzberg, 5071 West Saanich Rd, Victoria, BC V9E 2E7, Canada}

\author[0000-0002-7595-121X] {Joris Witstok}
\affiliation{Kavli Institute for Cosmology, University of Cambridge, Madingley Road, Cambridge CB3 0HA, UK}
\affiliation{Cavendish Laboratory, University of Cambridge, 19 JJ Thomson Avenue, Cambridge CB3 0HE, UK}

\begin{abstract}

We present a catalog of 717 candidate galaxies at $z > 8$ selected from 125 square arcminutes of NIRCam imaging as part of the JWST Advanced Deep Extragalactic Survey (JADES). We combine the full JADES imaging dataset with data from the JEMS and FRESCO JWST surveys along with extremely deep existing observations from HST/ACS for a final filter set that includes fifteen JWST/NIRCam filters and five HST/ACS filters. The high-redshift galaxy candidates were selected from their estimated photometric redshifts calculated using a template fitting approach, followed by visual inspection from seven independent reviewers. We explore these candidates in detail, highlighting interesting resolved or extended sources, sources with very red long-wavelength slopes, and our highest redshift candidates, which extend to $z_{phot} \sim 18$. Over 93\% of the sources are newly identified from our deep JADES imaging, including 31 new galaxy candidates at $z_{phot} > 12$. We also investigate potential contamination by stellar objects, and do not find strong evidence from SED fitting that these faint high-redshift galaxy candidates are low-mass stars. Using 42 sources in our sample with measured spectroscopic redshifts from NIRSpec and FRESCO, we find excellent agreement to our photometric redshift estimates, with no catastrophic outliers and an average difference of $\langle \Delta z = z_{phot}- z_{spec} \rangle= 0.26$. These sources comprise one of the most robust samples for probing the early buildup of galaxies within the first few hundred million years of the Universe's history.

\end{abstract}


\section{Introduction} \label{sec:intro}

The earliest galaxies that appeared from the Cosmic Dark Ages fundamentally changed the Universe. For hundreds of millions of years after recombination, the decoupling of matter and radiation, the Universe's baryon content consisted of predominantly neutral hydrogen that was gravitationally pooling and collecting, pulled by early dark matter halos. Eventually these massive clouds collapsed and formed the first stars which gave off energetic ultraviolet (UV) radiation, ionizing the neutral hydrogen medium throughout the universe. Reionization is thought to have taken place across the the first billion years after the Big Bang, but exactly how this process occurred, and more specifically, what types of galaxies are responsible for this phase transition, has been an active area of research for decades \citep{barkana2001, stark2016, dayal2018, finkelstein2019, ouchi2020, robertson2022}. Observations of early galaxies offer us a vital insight into the first stages of galaxy formation and evolution, and help us understand emergence of the elements heavier than helium. To aid in understanding these distant sources, in this paper we present a sample of 717 galaxies and candidate galaxies with spectroscopic and photometric redshifts corresponding to the first 200 to 600 Myr after the Big Bang and describe their selection and properties. 

To explore the very early universe, researchers search for galaxies at increasingly high redshifts using deep observations from space. One of the pioneering early universe surveys was the \textit{Hubble Space Telescope} (HST) Deep Field project \citep[HDF, ][]{williams1996}, a set of observations at wavelengths spanning the near-ultraviolet to near-infrared (IR). These data provided an opportunity to explore galaxy evolution out to $z = 4 - 5$ \citep{madau1996}. Following the success of the HDF, the next decades were spent observing multiple deep fields down to unprecedented observational depths of 30 mag (AB) at optical and near-IR wavelengths. These surveys included the Hubble Ultra-Deep Field \citep[HUDF, ][]{beckwith2006}, HUDF09 \citep{bouwens2011b}, HUDF12 \citep{ellis2013, koekemoer2013}, the UVUDF \citep{teplitz2013}, the HST Great Observatories Origins Deep Survey \citep[GOODS, ][]{giavalisco2004}, the Cosmological Evolution Survey \citep[COSMOS, ][]{scoville2007}, and the Cosmic Assembly Near-infrared Deep Extragalactic Legacy Survey \citep[CANDELS, ][]{grogin2011, koekemoer2011}. The Brightest of Reionizing Galaxy survey \citep[BoRG, ][]{trenti2011} was a campaign to search for bright high-redshift galaxies across a wide (274 square arcminutes), but relatively shallow area. Researchers hoping to target fainter galaxies also focused on lensing clusters, leading to the Cluster Lensing and Supernova Survey with Hubble \citep[CLASH, ][]{postman2012}, the Hubble Frontier Fields \citep[HFF, ][]{lotz2017}, and the Reionization Lensing Cluster Survey \citep[RELICS, ][]{coe2019}. 

It has therefore been exciting to see the fruits of these observations: the discovery of many thousands of galaxies at $z > 4$ \citep{bunker2004, bunker2010, bouwens2011, lorenzoni2011, ellis2013, mclure2013, oesch2013, schenker2013, oesch2014, bouwens2015, finkelstein2015, ishigaki2015, harikane2016, mcleod2016, oesch2018,  morishita2018, bridge2019, rojasruiz2020, bouwens2022, bagley2022, finkelstein2023}. While these sources have been found through multiple methods, the primary method of high-redshift galaxy selection relies on photometry alone. Neutral hydrogen within, surrounding, and between distant galaxies serves to absorb ultraviolet radiation, leading to what is commonly referred to as the ``Lyman break'' in the spectral energy distribution (SED) at 912 \AA. At redshifts above $z > 5$, the increasingly neutral hydrogen in the universe results in Lyman-$\alpha$ forest absorption between 912\AA\, and 1216\AA, and sources at these redshifts are more commonly known as ``Lyman-$\alpha$ break'' or ``Lyman-$\alpha$ dropout'' galaxies. By identifying galaxies where the Lyman break and Lyman-$\alpha$ break fell between two adjacent filters at a given redshift, these sources could be selected in large quantities, as done initially in \citet{guhathakurta1990} and \citet{steidel1992}. A similar approach involves fitting galaxy photometry to simulated or observed galaxy SEDs, a method that utilizes more data than pure color selection \citep{koo1985, lanzetta1996, gwyn1996, pello1996, koo1999, bolzonella2000}. These results require accurate template sets that span the full color space of the photometric data, and include the effects of both dust extinction and intergalactic medium (IGM) absorption. This template-fitting procedure is uncertain at high-redshifts given the current lack of UV and optical SEDs for galaxies in the early universe \citep[for a review of selection methods for finding high-redshift galaxies, see][]{stern1999, giavalisco2002, dunlop2013, stark2016}. 

While both galaxy selection techniques have been used to find galaxies out to $z \sim 10$ with HST, the reddest filter on the telescope's Wide-Field Camera 3 (WFC3/IR) is at 1.6$\mu$m, such that potential galaxies at higher redshifts would have their Lyman-$\alpha$ break shifted out of the wavelength range of the instrument. Exploring the evolution of galaxies at earlier times was limited by the availability of deep, high-resolution near- and mid-IR observations. This changed with the launch of the James Webb Space Telescope (JWST) in late 2021, an observatory carrying a suite of sensitive infrared instruments behind a 6.5m primary mirror. The instruments include NIRCam \citep{rieke2005}, a high-resolution camera operating at $0.7 - 5.0$ $\mu$m across a 9.7 square arcminute field of view, and NIRSpec \citep{jakobsen2022}, a spectrograph operating at similar wavelengths with a unique multi-object shutter array capable of obtaining spectra at multiple resolutions. 

In the first year of JWST science, researchers have identified scores of candidate high-redshift galaxies at $z > 8$ \citep{castellano2022, naidu2022, harikane2023, finkelstein2023, leethochawalit2023, morishita2023, whitler2023, yan2023, donnan2023, adams2023, austin2023, harikane2023, perezgonzalez2023, atek2023}. Some of these sources have been spectroscopically confirmed at $z > 8$ \citep{haro2023, haro2023b}, demonstrating the efficacy of using NIRCam for early universe observations. It should be noted, however, that this is an imperfect science - \citet{haro2023} describe how the early bright $z \sim 16$ candidate CEERS-93316 was spectroscopically found to be at $z_{spec} = 4.9$ with strong line emission and dust obscuration simulating the colors of a distant galaxy, a possibility discussed in \citet{naidu2022} and \citet{zavala2023}. 

One of the largest JWST Cycle 1 extragalactic surveys by time allocation is the JWST Advanced Deep Extragalactic Survey \citep[JADES; ][]{rieke2023}, a GTO program that will eventually encompass 770 hours of observations from three of the telescope's instruments: NIRCam, NIRSpec, and the mid-infrared instrument MIRI. These data, which focus on the GOODS-S and GOODS-N regions of the sky, are ideal for finding and understanding the most distant galaxies through imaging and follow-up spectroscopy. Because the JADES target regions have been observed by multiple telescopes and instruments across the electromagnetic spectrum, there is a rich quantity of ancillary data for comparing with JWST images and spectroscopy. 

Early JADES observations resulted in the discovery of the highest-redshift spectroscopically-confirmed galaxy thus far, JADES-GS-z13-0 ($z_{spec} = 13.20^{+0.04}_{-0.07}$ )\citep{robertson2022, curtislake2022}. Because of NIRCam's wavelength range and dichroic offering simultaneous short wavelength (0.7 - 2.3 $\mu$m) and long wavelength (2.4 - 5.0 $\mu$m) images, these and other high-redshift candidates are detected in multiple bands at wavelengths longward of the Lyman-$\alpha$ break. The high-redshift galaxies that can be observed thanks to the wavelength coverage of JWST are vital for exploring the potential downturn in the number density of ultra-high-redshift ($z \gtrsim 10$) galaxies previously predicted by HST observations alone \citep{oesch2018}.

In this study, we present the results of a search through the first year of JADES NIRCam imaging of the GOODS-S and GOODS-N regions for galaxy candidates at $z > 8$, where we combine the deepest HST optical and near-IR observations with JADES NIRCam data taken across ten filters. These data are supplemented by medium-band JWST imaging in five additional filters from both the publicly available JWST Extragalactic Medium Survey (JEMS) \citep{williams2023} and First Reionization Epoch Spectroscopic COmplete Survey (FRESCO) \citep{oesch2023} programs. We perform template fitting in order to select candidate high-redshift candidates, capitalizing on the large number of filters at wavelengths longer than $2\mu$m. Because of both the unparalleled HST coverage and the mixture of medium and wide NIRCam filters present in the JADES data, these data currently represent the best opportunity for uncovering galaxies at $z > 8$ with minimal low-redshift interlopers. The deepest portions of the JADES dataset probe down to $5\sigma$ depths of 2.17 nJy (30.6 mag AB) at 2.7$\mu$m, currently deeper than the other similar JWST extragalactic fields studied in the literature. In addition, because of the FRESCO grism spectra and the JADES NIRSpec spectroscopy, we also have a number of spectroscopic redshifts for these sources confirming their selection, providing constraints on the accuracy of photometric redshifts for galaxies in the early universe. 

The structure of this paper is as follows. We begin by introducing the JADES dataset used in this study, and we discuss our data reduction and photometric and spectroscopic measurements in Section \ref{sec:jadessurvey}. In Section \ref{sec:galaxyselection} we describe how we estimate photometric redshifts and, from these results, select candidate galaxies at $z > 8$. We then spend the bulk of this study exploring the resulting sample in Section \ref{sec:results}, separating the objects into three bins: $z = 8 - 10$ (\S \ref{sec:z_8_10}), $z = 10 - 12$ (\S \ref{sec:z_10_12}), and $z > 12$ (\S \ref{sec:z_gt_12}). We then consider candidate galaxies that fall out of our primary selection either because of their template fits (\S \ref{sec:deltachisqlt4}) or their proximity to brighter sources (\S \ref{sec:flag_bn_2_objects}). We also discuss the possibility of these sources being low-mass stars (\S \ref{sec:browndwarfs}), describe which candidates have been included in samples previous to this study (\S \ref{sec:literature}), and explore the impact of different galaxy template sets for photometric redshifts (\S \ref{sec:alternatetemplates}). Finally, we examine the selection and further properties of these sources in Section \ref{sec:discussion} and conclude in Section \ref{sec:conclusions}. Throughout this paper we assume the \citet{planck2018} cosmology with $H_0 = 67.4$ km s$^{-1}$ Mpc$^{-1}$, $\Omega_{\mathrm{M}} = 0.315$ and $\Omega_\Lambda = 0.685$. All magnitudes are provided using the AB magnitude system \citep{oke1974, oke1983}.

\section{JADES Imaging and Photometry} \label{sec:jadessurvey}

JADES is a joint Guaranteed Time Observations (GTO) program between the NIRCam and NIRSpec extragalactic GTO teams that consists of NIRCam imaging, NIRSpec spectroscopy, and MIRI imaging across the GOODS-S (RA = 53.126 deg, DEC = -27.802 deg) and GOODS-N (RA = 189.229, DEC = +62.238 deg) \citep{giavalisco2004} fields. In this section we describe the Cycle 1 JADES observations taken as of February 8 2023, the data reduction, and the measurement of fluxes and spectroscopic redshifts. The full description of these observations is provided in \citet{eisenstein2023r}. 

\subsection{Observations}\label{sec:observations}
In this paper we will discuss galaxy candidates selected from the NIRCam imaging in both GOODS-S, with observations taken on UT 2022-09-29 through 2022-10-10 (Program 1180, PI:Eisenstein), and GOODS-N, with observations taken on UT 2023-02-03 through 2023-02-07 (Program 1181, PI:Eisenstein). In addition, a set of NIRCam parallels (9.8 square arcmin each) were observed during NIRSpec observation PID 1210 (PI:Ferruit) on UT 2022-10-20 to 2022-10-24 within and southwest of the JADES Medium footprint in GOODS-S. Another set of NIRCam observations (9.8 square arcmin) parallel to NIRSpec PID 1286 (PI:Ferruit) were observed on UT 2023-01-12 to 2023-01-13 to the northwest of the JADES Deep footprint in GOODS-S. These data were partly presented in both \citet{robertson2022} as well as \citet{tacchella2023}, although here we combine the full suite of JADES data observed as of February 8 2023. 

The total current survey area of the JADES GOODS-S is 67 square arcminutes, with 27 square arcminutes for the JADES Deep program, and 40 square arcminutes for the JADES Medium program. The filters used for JADES Deep are NIRCam F090W, F115W, F150W, F200W, F277W, F335M, F356W, F410M, and F444W ($\lambda = 0.8 - 5.0 \mu m$), while JADES Medium uses the same filters without F335M. For the 1286 parallel, the JADES observations include the F070W filter. 

The total current area of the NIRCam GOODS-N program is 58 square arcminutes. The NIRCam filters observed for GOODS-N are F090W, F115W, F150W, F200W, F277W, F335M, F356W, F410M, and F444W ($\lambda = 0.8 - 5.0 \mu m$). The GOODS-N observations are separated into two portions: the northwest (NW) portion, which covers 30.4 square arcminutes, and a southeast (SE) portion, which covers 27.6 square arcminutes. The NW portion was taken under PID 1181 (PI:Eisenstein) with NIRCam as the prime instrument and MIRI in parallel, while the SW portion was taken as part of the same program with NIRSpec as prime and NIRCam in parallel. 

We also include observations taken for the JWST Extragalactic Medium-band Survey \citep[JEMS, ][]{williams2023}. These data, which are part of program PID 1963 (PIs C. Williams, S. Tacchella, M. Maseda) were taken on UT 2022-10-12. For this study, we use the NIRCam data from JEMS which covers the Ultra Deep Field \citep[UDF, ][]{beckwith2006} by the NIRCam A module, with the NIRCam B module to the southwest, spanning the JADES Deep and Medium portions, for a total area of 10.1 square arcminutes. The NIRCam observations in the JEMS survey were taken with the F182M, F210M, F430M, F460M, and F480M filters \citep{williams2023}.

We also supplement our observations with NIRCam data from the The First Reionization Epoch Spectroscopic COmplete Survey (FRESCO, PID 1895, PI P. Oesch). While nominally a NIRCam grism survey across GOODS-S and GOODS-N, we use the FRESCO F182M, F210M, and F444W imaging of GOODS-S and GOODS-N to supplement the filters available in JADES. The FRESCO area extends beyond the JADES Deep and Medium region, and we do not select galaxies in this region due to the lack of NIRCam filter coverage afforded by the JADES observations. We use the FRESCO grism data as well as the NIRSpec observations from PID 1210 and 1286 to measure spectroscopic redshifts for sources within our sample. 

The GOODS-S and GOODS-N regions have been the target of deep HST observations, and we utilize existing HST/ACS and WFC3 mosaics. We use the HST/ACS mosiacs from the Hubble Legacy Fields (HLF) v2.0 for GOODS-S and v2.5 for GOODS-N \citep[$25' \times 25'$ for GOODS-S, and  $20.5' \times 20.5'$ for GOODS-N,][]{illingworth2013, whitaker2019}. We use data in the HST/ACS F435W, F606W, F775W, F814W, and F850LP filters. 

\subsection{Data Reduction}\label{sec:datareduction}

\subsubsection{JADES NIRCam}

The data reduction techniques used in this present study will be fully described in a future paper (Tacchella et al. in prep), but they follow the methods outlined in \citet{robertson2022} and \citet{tacchella2023}, which we briefly summarize here. For both the JADES GOODS-S and GOODS-N observations, the data were first reduced using the JWST calibration pipeline v1.9.2, with the JWST Calibration Reference Data System (CRDS) context map 1039. The raw images ({\tt uncal} frames) are processed using the default JWST Stage 1 pipeline, which performs the detector-level corrections and results in count-rate images ({\tt rate} frames). 

The JWST pipeline Stage 2 involves flat fielding and flux calibration, and was run largely with the default values. We convert from counts/s to MJy/sr following \citet{boyer2022}. During the data reduction, we discovered that the current long wavelength flats used in the JWST pipeline result in non-astrophysical artifacts in the final mosaics. To mitigate this effect, we developed our own sky-flats, stacking in each filter 80 - 200 source-masked raw {\tt uncal} frames from across PID 1180, 1210, 1286, and JEMS. For F335M and F410M, where we did not have enough exposures to properly perform this stacking procedure, we instead constructed these sky flats via interpolation using the other wide-band LW sky flats. 

After Stage 2, we used custom corrections for common features seen in JWST/NIRCam data \citep{rigby2023}. We fit and subtracted the 1/f noise \citep{schlawin2020} assuming a parametric model. To fit for the scattered-light ``wisps'' in the NIRCam SW channel we constructed templates by stacking our images from the JADES program (PID 1180, 1210, 1286) as well as other publicly available programs (PIDs 1063, 1345, 1837, and 2738), and then subtracted these scaled templates for the SW channel detectors A3, A4, B3, and B4 (Tacchella et al. in prep). The background was removed using the {\tt photutils} Background2D class \citep{photutils2023}.

We created our final mosaics using the JWST Pipeline Stage 3, after performing an astrometric alignment using a custom version of the JWST {\tt TweakReg} software. In both GOODS-S and GOODS-N, we calculated the relative and absolute astrometric corrections for the individual images grouped by visit and by photometric band. We matched to sources in a reference catalog created from HST F814W and HST F160W mosaics with astrometry tied to Gaia-EDR3 \citep[][private communication from G. Brammer]{gaia2021}. Following this alignment, we performed the default steps of Stage 3 of the JWST pipeline for each filter and visit. For our final mosaics we chose a pixel scale of 0.03 arcsec/pixel and drizzle parameter of \texttt{pixfrac=1} for both the SW and LW images. 

\subsubsection{FRESCO}

The FRESCO \citep{oesch2023} NIRCam grism spectroscopic data in the F444W filter ($\lambda = 3.9-5.0$\,\micron) were reduced and analyzed following the routines in \citet{sun2022} and \citet{helton2023}. Here we briefly summarize the main steps of the process. Because we aim to conduct a targeted emission line search of [O\,III] and H$\beta$ lines for our $z>8$ galaxy candidates, and we do not expect any of them to have strong continuum emission that can be detected with grism data, we used a median-filtering technique to subtract out the remaining continuum or background on a row-by-row basis, following the methods outlined by \citet{kashino2022}. We extracted 2D grism spectra using the continuum-subtracted emission-line maps for all objects that are brighter than 28.5\,AB mag in the F444W band and within the FRESCO survey area. The emission lines from sources fainter than 28.5\,AB mag are not expected to be detected with FRESCO. The FRESCO short-wavelength parallel imaging observations were used for both astrometric and wavelength calibration of the F444W grism spectroscopic data. 

We extracted 1D spectra from the 2D grism spectra using the optimal extraction algorithm \citep{horne1986} using the light profiles of sources in the F444W filter. We then performed automatic identifications of $>3\sigma$ peaks in 1D spectra \citep[see][]{helton2023}, and fit these detected peaks with Gaussian profiles. We tentatively assigned spectroscopic redshifts for $>3\sigma$ peaks which minimize the difference from the estimated photometric redshifts (Section \ref{sec:speczs}). Visual inspection was performed on these tentative spectroscopic redshift solutions and spurious detections caused by either noise or contamination were removed. The final grism spectroscopic redshift sample of JADES sources will be presented in a forthcoming paper from the JADES collaboration. 

\subsubsection{JADES NIRSpec}

In addition to FRESCO data, we discuss NIRSpec spectroscopic redshifts in Section \ref{sec:speczs}, and these were reduced following the same procedure as outlined in \citet{curtislake2022}, \citet{cameron2023}, \citet{bunker2023}, and \citet{bunker23r}. For the present study, we are only using the derived spectroscopic redshifts from these data. 

\subsection{Photometry}\label{sec:photometry}

\begin{deluxetable*}{cccccccc}
\tabletypesize{\footnotesize}
\tablecolumns{8}
\tablewidth{0pt}
\tablecaption{ 5$\sigma$ Photometric Depth In JADES Areas Measured in $0.2^{\prime\prime}$ Apertures (nJy) \label{tab:five_sigma_limits}}
\tablehead{
\multicolumn{2}{c}{} & \multicolumn{4}{c}{GOODS-S} & \multicolumn{2}{c}{GOODS-N}\\
\colhead{Instrument} &  \colhead{Filter}  &   \colhead{JADES Deep}  &  \colhead{JADES Medium} & \colhead{1210 Parallel}& \colhead{1286 Parallel} & \colhead{SE} & \colhead{NW}}
\startdata
	HST/ACS & F435W & 2.33 & 10.77 & 10.9 & 10.43 & 7.75 & 9.40\\ 
	HST/ACS & F606W & 3.61 & 6.96 & 6.68 & 8.72 & 9.81 & 8.55\\
	HST/ACS & F775W & 2.22 & 15.79 & 16.46 & 15.03 & 13.61 & 9.21\\
	HST/ACS & F814W & 8.20 & 7.2 & 7.02 & 11.03 & 8.21 & 7.7\\
	HST/ACS & F850LP & 4.28 & 17.53 & 18.58 & 19.55 & 15.23 & 17.55\\    \hline
	JWST/NIRCam & F070W & - & - & - & 8.29 & - & - \\
	JWST/NIRCam & F090W & 3.55 & 6.26 & 2.40 & 5.92 & 6.04 & 11.03\\
	JWST/NIRCam & F115W & 2.93 & 5.44 & 2.27 & 5.26 & 4.51 & 8.08\\
	JWST/NIRCam & F150W & 2.89 & 5.53 & 2.15 & 5.28 & 5.16 & 8.44\\
	JWST/NIRCam & F182M & 8.04 & 9.53 & 10.37 & 11.29 & - & - \\
	JWST/NIRCam & F200W & 3.01 & 5.27 & 2.37 & 4.63 & 4.66 & 7.78\\
	JWST/NIRCam & F210M & 5.83 & 12.11 & 13.71 & 13.53 & - & - \\
	JWST/NIRCam & F277W & 2.17 & 4.24 & 1.64 & 3.69 & 3.92 & 6.17\\
	JWST/NIRCam & F335M & 3.64 & 3.81 & 2.86 & 6.08 & 5.7 & 9.12\\
	JWST/NIRCam & F356W & 2.46 & 4.07 & 1.62 & 3.60 & 3.81 & 5.74\\
	JWST/NIRCam & F410M & 3.23 & 6.43 & 2.39 & 5.60 & 6.33 & 9.524\\
	JWST/NIRCam & F430M & 7.84 & 7.31 & - & - & - & - \\
	JWST/NIRCam & F444W & 2.79 & 5.11 & 2.00 & 4.62 & 5.07 & 7.31\\
	JWST/NIRCam & F460M & 10.71 & 9.61 & - & - & - & - \\
	JWST/NIRCam & F480M & 7.98 & 6.51 & - & - & - & - \\    \hline
\enddata
\end{deluxetable*}

To compute the photometry from both the GOODS-S and GOODS-N mosaics in each filter, we used the software package {\tt jades-pipeline} developed by authors BR, BDJ, and ST. We began by creating an inverse-variance-weighted stack of the NIRCam F277W, F335M, F356W, F410M, and F444W images as an ultra-deep signal-to-noise ratio (SNR) image. From this SNR image, {\tt jades-pipeline} utilizes software from the {\tt Photutils} package to define a catalog of objects with five contiguous pixels above a SNR of 3 \citep{photutils2023}, creating a segmentation map in the process. 

From this catalog, we calculated circular and Kron aperture photometry on both the JWST NIRCam mosaics as well as the 30mas pixel scale HST Legacy Fields mosaics \citep{illingworth2016, whitaker2019} for ACS F435W, F606W, F775W, F814W, and F850LP filters. Forced photometry was performed using a range of aperture sizes. The uncertainties we report were measured by combining in quadrature both the Poisson noise from the source and the noise estimated from random apertures placed throughout the image \citep[e.g.][]{labbe2005, quadri2007, whitaker2011}. Elliptical Kron aperture fluxes were measured using {\tt Photutils} with a Kron parameter of $K = 2.5$ and the default circularized radius six times larger than the Gaussian-equivalent elliptical sizes while masking segmentation regions of any neighboring source. We created empirical HST/ACS and JWST/NIRCam point spread functions to estimate and apply aperture corrections assuming point source morphologies (Z. Chen, private communication). 

For this present study we will fit to the JADES ``CIRC1'' ($0.2^{\prime\prime}$ diameter aperture) fluxes, which reduces the background noise associated with the use of larger apertures, and is appropriate given the typically small sizes found for high-redshift galaxies \citep{shibuya2015, curtislake2016, robertson2022, tacchella2023}. We note that in Section \ref{sec:z_8_10} we discuss a sample of morphologically extended sources with photometric redshifts $z_a = 8 - 9$, although these sources consist of multiple smaller clumps, supporting the use of the smaller aperture photometry for their selection. We also use Kron aperture fluxes to calculate some derived parameters, such as the UV magnitude $M_{UV}$ to better encompass the full flux from more extended sources. We estimated the 5$\sigma$ limiting flux across both GOODS-S and GOODS-N from the $0.2^{\prime\prime}$ diameter fluxes and uncertainties. In Table \ref{tab:five_sigma_limits}, we report these 5$\sigma$ limiting fluxes in nJy for these portions of GOODS-S: JADES Deep, JADES Medium, the 1210 Parallel, and the 1286 Parallel. In addition, we report the limiting fluxes for both the shallower NW portion of the GOODS-N field and the SE portion. Understanding these depths is important for exploring the recovery of high-redshift galaxies across the JADES data.  

\section{Galaxy Selection At \lowercase{\textit{z}} $ > 8$} \label{sec:galaxyselection}

Our final photometric catalogs span 20 optical and near-IR filters, including both HST/ACS and JWST/NIRCam observations. Because of the multiple datasets included in these catalogs, however, objects will only have coverage in a subset of these filters, with the maximum number being in the area of the JEMS survey in the GOODS-S region, where there is coverage in 19 filters (F070W was only observed in the 1286 parallel, no portion of which overlaps with JEMS). In this section we describe how we identified $z > 8$ sources from the measured line-flux catalog. Throughout this study, we will identify sources using ``JADES-GS-'' or ``JADES-GN-'' followed by the right ascension (RA) and declination (DEC) values in decimal degrees corresponding to the source. 

As discussed in the introduction, we choose to employ template fitting in this study due to the large quantity of available data in the JADES data set, especially longward of the potential Lyman-$\alpha$ break for objects at $z > 8$. The rest-frame UV and optical continuum can be fit with the templates as well, better constraining the exact redshift than with color selection alone. In addition, potential strong optical emission lines such as [OIII]$\lambda$5007 observed in high-redshift galaxies can boost the flux in photometric filters, and can be modeled with template fitting. The JADES data set includes multiple medium-band filters longward of 3$\mu$m, where these effects can be more significant.

\subsection{{\tt EAZY} Photometric Redshifts} \label{sec:eazy}

\begin{figure*}
  \centering
  \includegraphics[width=\textwidth]{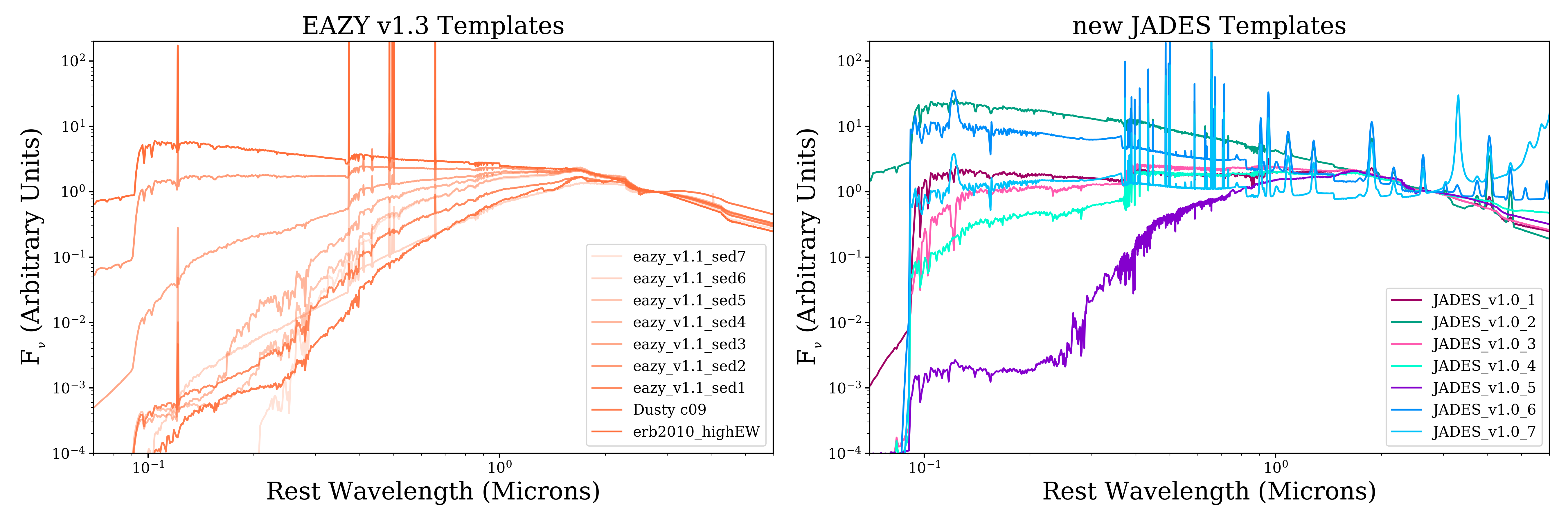}
  \caption{{\tt EAZY} templates used in this work. In the left panel, we show the {\tt EAZY} ``v1.3'' templates, and in the right panel, the new {\tt EAZY} templates created for fitting to JADES galaxies. The templates are presented in $F_{\nu}$ units normalized at ${2.8\mu}$m. These templates include both UV-faint quiescent and UV-bright star-forming populations, and were both compiled from the literature and created using Flexible Stellar Population Synthesis \citep[\texttt{fsps}, ][]{conroy2010}.}
  \label{fig:EAZY_templates}
\end{figure*}

In order to estimate the redshifts of the GOODS-S galaxies, we used the photometric redshift code {\tt EAZY} \citep{brammer2008}. {\tt EAZY} combines galaxy templates and performs a grid-search as a function of redshift. We used the EAZY photometric redshift $z_a$, corresponding to the minimum $\chi^2$ of the template fits, to identify high-redshift galaxies. For the fits, we started with the {\tt EAZY} ``v1.3'' templates, which we plot in the left panel in Figure \ref{fig:EAZY_templates}. These templates include the original seven templates modified from \citet{brammer2008} to include line emission, the dusty template ``{\tt c09\_del\_8.6\_z\_0.019\_chab\_age09.40\_av2.0.dat},'' and the high-equivalent-width template taken from \citet{erb2010} (the equivalent width of [OIII]$\lambda$5007 measured for the galaxy this template is derived from, Q2343-BX418, is 285 \AA). We supplemented these with seven additional templates that were designed to optimize photometric redshift estimates for mock galaxy observations from the JAGUAR simulations \citep{williams2018}. These templates were created to better span the observed color space of the JAGUAR galaxies, including both red, dusty and blue, UV-bright populations. Similar to what has been demonstrated by other authors \citep[e.g.][]{larson2022}, we found that young galaxies with very high specific star-formation rates can have very blue observed UV continuum slopes, which is made more complex due to strong nebular continuum and line emission \citep{topping2022}. To aid in fitting these galaxies, we generated additional templates using Flexible Stellar Population Synthesis \citep[\texttt{fsps}, ][]{conroy2010}, and added these to the ``5Myr'' and ``25Myr'' simple stellar population models introduced in \citet{coe2006} for fitting blue galaxies in HUDF. We show our additional templates in the right panel of Figure \ref{fig:EAZY_templates}, and we provide these templates online hosted on Zenodo: \url{https://doi.org/10.5281/zenodo.7996500}. Multiple templates from the full set contain nebular continuum and line emission, including that from Lyman-$\alpha$.

In each redshift bin considered, {\tt EAZY} combines all of the available templates together and applies an IGM absorption consistent with the redshift \citep{madau1995}. The best fit in that redshift bin, measured using the minimum $\chi^2$, is recorded in a $\chi^2(z)$ surface that is output from the program. We explored the redshift range $z = 0.01 - 22$, with a redshift step size $\Delta z = 0.01$. We did not adopt any apparent magnitude priors, as the exact relationship between galaxy apparent magnitude and redshift at $z > 8$ is currently not well constrained, so any attempt to impose a prior would serve to only remove faint objects from the sample. To prevent bright fluxes from overly constraining the fits and to account for any photometric calibration uncertainties not captured by the offset procedure described below \citep[e.g. due to detector specific offsets as observed in][]{bagley2023}, we set an error floor on the photometry of 5\%, and additionally, we used the {\tt EAZY} template error file ``{\tt template error.v2.0.zfourge}'' to account for any uncertainties in the templates as a function of wavelength. We also explored the use of the {\tt EAZY} templates discussed in \citet{larson2022}, which were used in finding high-redshift galaxies in the JWST Cosmic Evolution Early Release Science (CEERS) observations, and we describe how using these photometric redshifts affects our final sample in Section \ref{sec:alternatetemplates}. 

To match the {\tt EAZY} template set to the observed fluxes in our catalog, we estimated photometric offsets with {\tt EAZY}. We calculated the offsets for GOODS-S and GOODS-N data separately, where we first fit the observed photometry for a sample of galaxies with an SNR in F200W between 5 and 20, and calculated the offsets from the observed photometry to the template photometry. We then applied these offsets to the photometry and re-fit, iterating on this procedure. We list the final photometric offsets that we used for GOODS-S and GOODS-N, normalized to F200W, in Table \ref{tab:photometric_offsets}. These offsets are within 10\% of unity for all of the filters, with the exception of a large offset used for the F850LP observations in GOODS-N. We find that the F850LP depths are among the shallowest in our dataset (Table \ref{tab:five_sigma_limits}) which is likely contributing to the large offset. While we observe differences between the GOODS-S and GOODS-N offsets, this is primarily driven by the comparison of the HST/ACS photometry to the NIRCam photometry. To demonstrate this, we recalculated the photometric redshifts but used identical filter sets between the sources in the two fields excluding the HST/ACS bands and the NIRCam medium bands F182M, F210M, F430M, F460M, and F480M where we have limited coverage in these filters in GOODS-S and GOODS-N. The median difference in the photometric offsets between the GOODS-S and GOODS-N fits for the remaining filters as provided in Table \ref{tab:photometric_offsets} is 0.006 with a standard deviation of 0.012. However, when we calculate the offsets without the HST/ACS bands or the NIRCam medium bands, the median difference goes down to 0.001 and a standard deviation of only 0.004, consistent with no difference. 

\begin{deluxetable}{l c c c}
\tabletypesize{\footnotesize}
\tablecolumns{8}
\tablewidth{0pt}
\tablecaption{{\tt EAZY}-derived photometric offsets, normalized to F200W. \label{tab:photometric_offsets}}
\tablehead{
\colhead{} & \colhead{} & \colhead{GOODS-S} & \colhead{GOODS-N} \\
\colhead{Instrument} &  \colhead{Filter}  &   \colhead{Offset}  &  \colhead{Offset}}
\startdata
	HST & F435W & 1.021 & 1.072\\ 
	HST & F606W & 1.002 & 0.976\\
	HST & F775W & 1.009 & 0.996\\
	HST & F814W & 0.962 & 0.998\\
	HST & F850LP & 0.919 & 0.774\\    \hline
	NIRCam & F070W & 0.981 & -\\
	NIRCam & F090W & 0.987 & 1.012\\
	NIRCam & F115W & 1.008 & 1.020\\
	NIRCam & F150W & 0.994 & 0.989\\
	NIRCam & F182M & 1.001 & 0.991\\
	NIRCam & F200W & 1.000 & 1.000\\
	NIRCam & F210M & 1.014 & 1.006\\
	NIRCam & F277W & 0.998 & 0.990\\
	NIRCam & F335M & 1.035 & 1.024\\
	NIRCam & F356W & 1.057 & 1.047\\
	NIRCam & F410M & 1.071 & 1.057\\
	NIRCam & F430M & 1.014 & -\\
	NIRCam & F444W & 1.015 & 1.009\\
	NIRCam & F460M & 0.956 & -\\
	NIRCam & F480M & 1.017 & -\\ 
 \enddata
\end{deluxetable}

We used the $\chi^2(z)$ values output from {\tt EAZY} to calculate a probability $P(z)$ assuming a uniform redshift prior: $P(z) = \exp{[-\chi^2(z) / 2]}$, where we normalize such that $\int P(z) dz = 1.0$. The $P(z)$ and $\chi^2(z)$ values allowed us to calculate $P(z > 7)$, the summed probability from {\tt EAZY} that the galaxy is at $z > 7$, as well as the $\chi^2$ minimum for {\tt EAZY} fits restricted to $z < 7$. These statistics, and others, are helpful for identifying and removing interlopers from our sample. 

\begin{figure}
  \centering
  \includegraphics[width=1\linewidth]{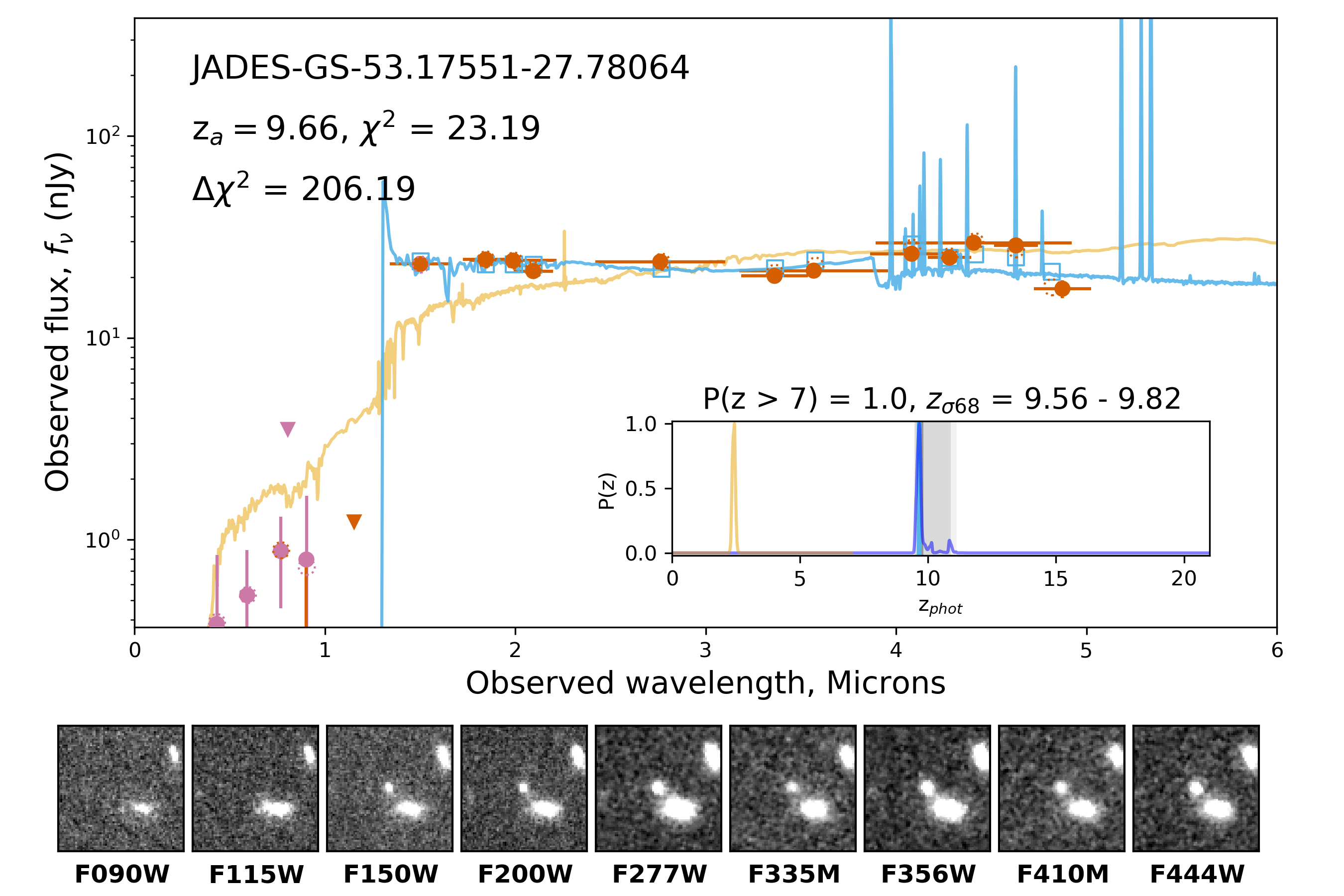}
  \caption{Example best-fit SED for object JADES-GS-53.17551-27.78064, at the best-fit redshift $z_a = 9.66$. We plot the NIRCam photometry with red points, and the HST photometry with light purple points. The error bars represent the 1$\sigma$ uncertainties on the fluxes. We plot $2\sigma$ upper limits with downward pointing triangles. The blue line represents the fit corresponding to $z_a$ and the gold line shows the best fit at $z < 7$. We show the minimum $\chi^2$ value for the $z_a$ fit, as well as the $\Delta\chi^2$ value corresponding to the difference between the minimum $\chi^2$ and the $\chi^2$ for the fit at $z < 7$. In the inset, we plot $P(z)$, with the 1$\sigma$, 2$\sigma$, and 3$\sigma$ uncertainty regions derived from the $P(z)$ surface with shades of grey, and $z_a$ with a blue line, along with the $P(z)$ distribution for the $z < 7$ best fit in gold (also normalized to 1). Above the inset we provide the summed probability of the source being at $z > 7$, along with the {\tt EAZY} redshifts corresponding to $\sigma_{68,low}$ and $\sigma_{68,high}$. Below the SED we plot $2^{\prime\prime} \times 2^{\prime\prime}$ thumbnails for the JADES NIRCam filters.}
  \label{example_SED_fit}
\end{figure}

In Figure \ref{example_SED_fit} we show the {\tt EAZY} fit to an object in GOODS-S, JADES-GS-53.17551-27.78064, along with the $P(z)$ surface, and the JADES NIRCam thumbnails. The source is an F115W dropout, with no visible flux at shorter wavelengths. The fit constrained at $z < 7$ produces significantly more F115W flux than is observed, lending evidence of this galaxy being at $z > 9$. 

\subsection{High-Redshift Galaxy Selection and Catalogs} \label{sec:highzselection}

Because of the extensive deep photometric data for the GOODS-S and GOODS-N fields, we chose to use the {\tt EAZY} photometric redshifts for finding $z > 8$ candidates, as template fitting utilizes more photometric data points in the fit than color selection by itself. Following work done in the literature \citep{finkelstein2023}, we selected galaxies at $z > 8$ by imposing these rules on the {\tt EAZY} fits:

\begin{enumerate}
    \item The redshift of the fit corresponding to the minimum $\chi^2$, $z_a$, must be greater than 8.
    \item The SNR in at least two photometric bands must be above 5. For this study, we chose NIRCam F115W, F150W, F200W, F277W, F335M, F356W, F410M, or F444W, as these filters are longward of the Lyman-$\alpha$ break at $z > 8$. We used the photometry derived using $0.2^{\prime\prime}$ diameter apertures for measuring this SNR. 
    \item The summed probability of the galaxy being above $z > 7$ must be greater than 70\%, or ${\int_{7}^{22} P(z) dz > 0.7}$.
    \item The difference between the overall minimum $\chi^2$ and the minimum $\chi^2$ at $z < 7$, $\Delta \chi^2$, must be greater than 4. 
    \item There should be no object within $0.3^{\prime\prime}$ (10 pixels in the final JADES mosaics), or within the object's bounding box, that is 10 times brighter that the object. 
\end{enumerate}

For this study, we targeted galaxies at $z_a > 8$ as galaxies above this redshift should have no observed flux in the JWST/NIRCam F090W filter. This allows us to use the deep JADES F090W observations to aid in visually rejecting lower-redshift contaminants. The second requirement, that the source be detected in multiple bands, was chosen to ensure that the sources we selected were not artifacts found in individual exposures such as cosmic rays or bad pixels. We imposed the {\tt EAZY} $\int_{7}^{22} P(z) > 0.7$ (which we will shorten to ``$P(z > 7)$'') and $\Delta \chi^2$ limits in order to help remove objects where {\tt EAZY} could fit the observed SED at low redshift with high probability. In \citet{harikane2023}, the authors recommend the use of a more strict cut, $\Delta \chi^2 > 9$, and we consider this cut in Section \ref{sec:discussion}. We also, in Section \ref{sec:deltachisqlt4}, discuss those objects where $\Delta \chi^2 < 4$ in our sample, as these sources, though faint, may contain true high-redshift galaxies that should be considered. Finally, we remove objects with close proximity to bright sources because of the possibility of selecting tidal features or stellar clusters near to the edges of relatively nearby galaxies.  We list those objects that satisfied our other requirements but were close to a brighter source, along with discussion of these targets, in Section \ref{sec:flag_bn_2_objects}. We chose not to implement a direct cut on $\chi^2$ as this metric is dependent on the flux uncertainties, which vary across the field in such a way as to make a comparison of the value between objects difficult and potentially non-meaningful. We still report the resulting $\chi^2$ values, however. In comparison, the $\Delta\chi^2$ value is calculated from two fits to the same photometry and uncertainties, and is helpful in exploring the relative goodness of fits at different redshifts.
 
These cuts resulted in 1078 objects in GOODS-S and 636 objects in GOODS-N. From here, we began the process of visual inspection, first to remove obvious non-astrophysical data artifacts, including extended diffraction spikes from stars, and hot pixels caused by cosmic rays. We also removed extended, resolved low-redshift, dusty sources, many of which were not visible in HST imaging. These sources were identified by very red slopes between 1 and 5$\mu$m, and have half-light radii $r_{\mathrm{half}} \gtrsim 1^{\prime\prime}$ in the filters where they are observed. These sources comprise only $\sim 0.4$\% of the objects that satisfy our cuts. After removing these sources, we were left with 580 possible objects in GOODS-S, and 212 objects in GOODS-N. There is a much larger fraction of spurious sources in GOODS-N as compared to GOODS-S as these data had a larger number of bright pixels and cosmic rays, which primarily affected the NIRCam long-wavelength channels. 

After this initial inspection, authors KH, JH, DE, MWT, CNW, LW, and CS independently graded each target with a grade of Accept, Reject, or Review. For those objects where $50\%$ or more of the reviewers accepted the candidate, it was then added to the final candidate list. In cases where greater than $50\%$ of the reviewers chose to reject the candidate, this candidate was removed entirely from the candidate list. In all other cases (57 objects in GOODS-N and 102 objects in GOODS-S), the reviewers did one more round of visual inspection with only the grades Accept or Reject, with a larger discussion occurring for objects where necessary. Again, a 50\% of Accept grades was required for these galaxies under review to be listed as part of the final sample. 

\section{Results} \label{sec:results}

\begin{deluxetable*}{l l}
\tabletypesize{\footnotesize}
\tablecolumns{2}
\tablewidth{0pt}
\tablecaption{Overview of Columns in the $z > 8$ Source Catalog \label{tab:z_gt_8_table_columns}}
\tablehead{
\colhead{Column} &  \colhead{Description}}
\startdata
	1 & JADES ID \\
	2, 3 & Right ascension and declination, in decimal degrees, of the source \\
	4 & $m_{\mathrm{F277W, Kron}}$ (AB) \\
	5 - 11 & {\tt EAZY} $z_a$, $\sigma_{68, low}$, $\sigma_{68, high}$, $\sigma_{95, low}$, $\sigma_{95, high}$, $\sigma_{99, low}$, $\sigma_{99, high}$  \\
	12 & {\tt EAZY} $\int_7^{22} P(z) dz$ \\
	13 & {\tt EAZY} minimum $\chi^2$ \\
	14 & {\tt EAZY} minimum $z_a$ $(z < 7)$ \\
	15 & {\tt EAZY} $\chi^2$ $(z < 7)$ \\
	16 & {\tt EAZY} $\Delta\chi^2$ \\
	17, 18 & Spectroscopic Redshift, Source (FRESCO or NIRSpec) \\
	19 & $M_{UV}$ \\
	20 & Flag indicating the source is fit by a brown dwarf model within $\Delta\chi^2 < 4$ \\
	21 & Flag indicating whether or not the source is unresolved ($r_{\mathrm{eff,F444W}} < 0.063^{\prime\prime}$) \\
	22, 23 & HST/ACS F435W Flux, 1$\sigma$ uncertainty (nJy) \\
	24, 25 & HST/ACS F606W Flux, 1$\sigma$ uncertainty (nJy) \\
	26, 27 & HST/ACS F775W Flux, 1$\sigma$ uncertainty (nJy) \\
	28, 29 & HST/ACS F814W Flux, 1$\sigma$ uncertainty (nJy) \\
	30, 31 & HST/ACS F850LP Flux, 1$\sigma$ uncertainty (nJy) \\
	32, 33 & JWST/NIRCam F070W Flux, 1$\sigma$ uncertainty (nJy) \\
	34, 35 & JWST/NIRCam F090W Flux, 1$\sigma$ uncertainty (nJy) \\
	36, 37 & JWST/NIRCam F115W Flux, 1$\sigma$ uncertainty (nJy) \\
	38, 39 & JWST/NIRCam F150W Flux, 1$\sigma$ uncertainty (nJy) \\
	40, 41 & JWST/NIRCam F182M Flux, 1$\sigma$ uncertainty (nJy) \\
	42, 43 & JWST/NIRCam F200W Flux, 1$\sigma$ uncertainty (nJy) \\
	44, 45 & JWST/NIRCam F210M Flux, 1$\sigma$ uncertainty (nJy) \\
	46, 47 & JWST/NIRCam F277M Flux, 1$\sigma$ uncertainty (nJy) \\
	48, 49 & JWST/NIRCam F335M Flux, 1$\sigma$ uncertainty (nJy) \\
	50, 51 & JWST/NIRCam F356W Flux, 1$\sigma$ uncertainty (nJy) \\
	52, 53 & JWST/NIRCam F410M Flux, 1$\sigma$ uncertainty (nJy) \\
	54, 55 & JWST/NIRCam F430M Flux, 1$\sigma$ uncertainty (nJy) \\
	56, 57 & JWST/NIRCam F444W Flux, 1$\sigma$ uncertainty (nJy) \\
	58, 59 & JWST/NIRCam F460M Flux, 1$\sigma$ uncertainty (nJy) \\
	60, 61 & JWST/NIRCam F480M Flux, 1$\sigma$ uncertainty (nJy) \\
	62 & JADES Footprint Region \\
\enddata
\tablecomments{We provide these values for the primary $z > 8$ sample with $\Delta\chi^2 > 4$, as well as the subsamples outlined in the text: $\Delta\chi^2 < 4$ and those proximate to brighter sources.}
\end{deluxetable*}

Our final $z > 8$ samples consist of 535 objects in GOODS-S and 182 objects in GOODS-N. In Table \ref{tab:z_gt_8_table_columns} we provide the descriptions of the columns in our final catalog; the catalog itself is provided as an online table on Zenodo: \url{https://doi.org/10.5281/zenodo.7996500}. We include $0.2^{\prime\prime}$ diameter aperture photometry in each of the observed photometric bands, as well as the {\tt EAZY} $z_a$, $\chi^2$, $P(z > 7)$, and $\Delta\chi^2$ values used in selecting the galaxies. We also provide the $\sigma_{68}$, $\sigma_{95}$, and $\sigma_{99}$ confidence intervals estimated from the $P(z)$ distribution. In this table we also list the $z > 8$ candidates that have {\tt EAZY} $\Delta\chi^2 < 4$, and we will discuss these sources in Section \ref{sec:deltachisqlt4}. Similarly, in our output table, we list those $z > 8$ candidates that were either within $0.3^{\prime\prime}$ or within the bounding box of a target 10 times brighter than the candidate, which we discuss in Section \ref{sec:flag_bn_2_objects}.

\begin{figure*}[t!]
  \centering
  \includegraphics[width=0.49\linewidth]{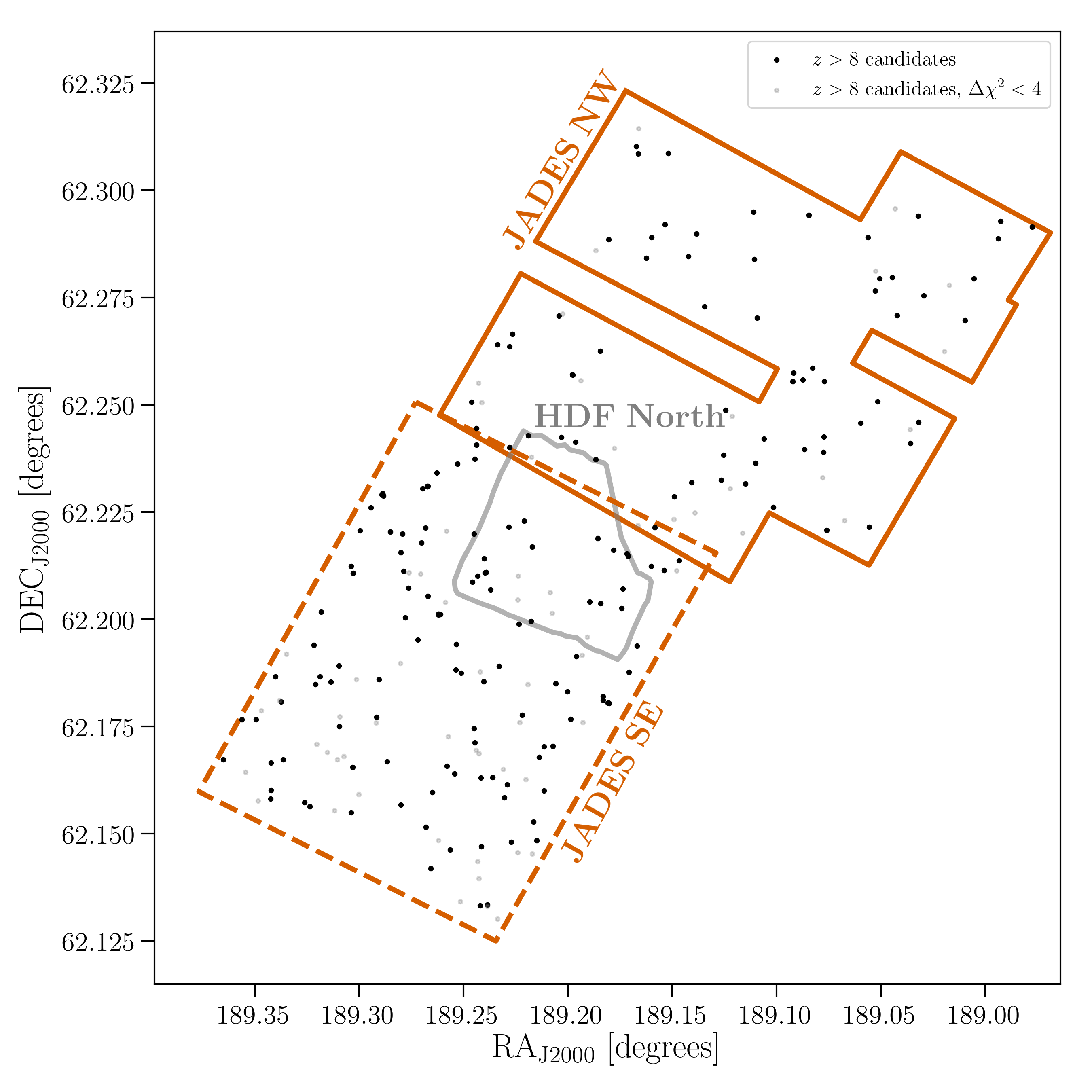}\
  \includegraphics[width=0.49\linewidth]{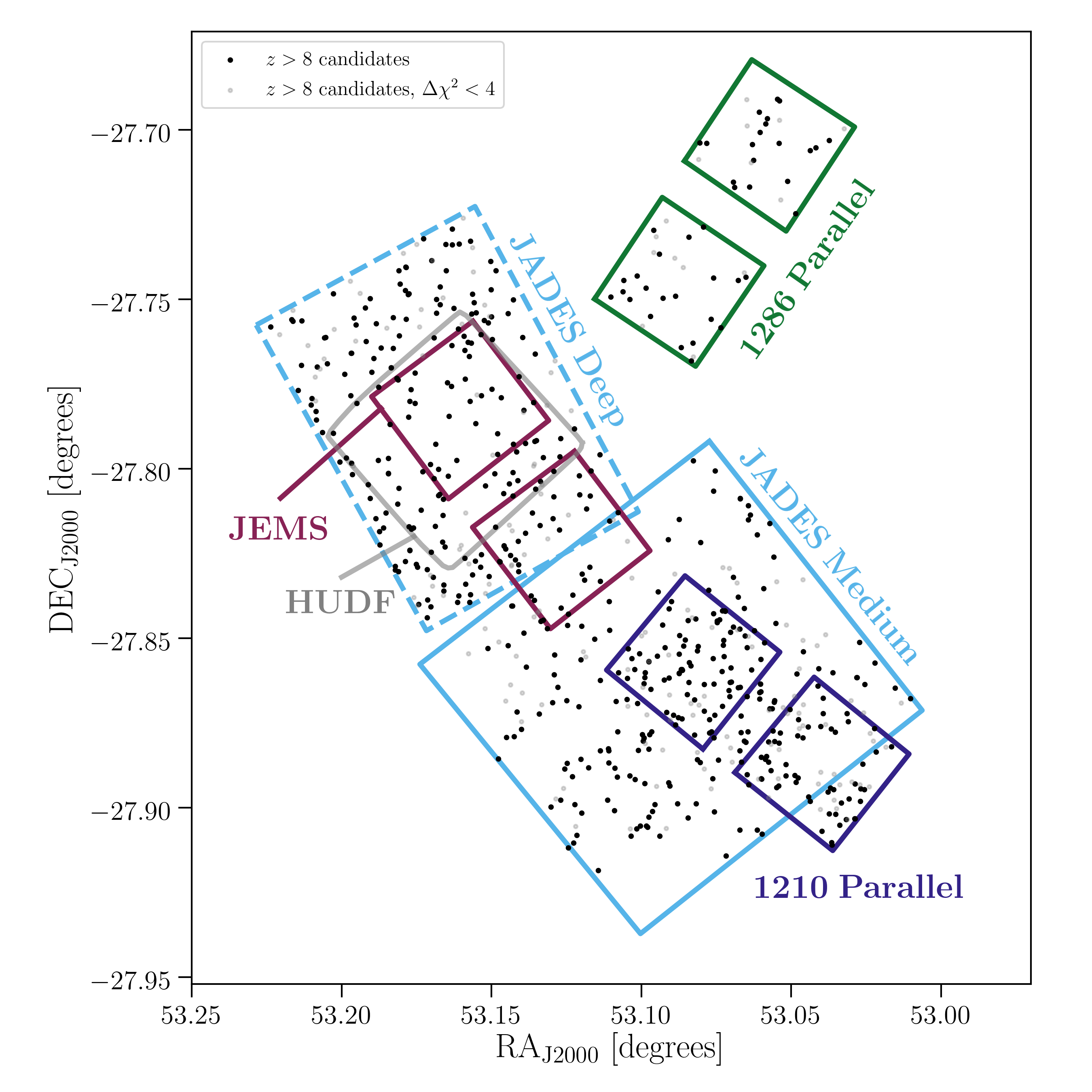}
  \caption{(Left) GOODS-N footprint showing the positions of the $z > 8$ candidates. The southeastern portion (dashed) of the GOODS-N area is deeper than the northwestern portion (solid), resulting in a larger density of candidate high-redshift galaxies. In grey, we plot the outline of the Hubble Deep Field (North) \citep{williams1996} as a comparison to the JADES survey area. (Right) GOODS-S footprint. The dashed blue outline highlights the JADES Deep GOODS-S area, and the solid blue outline highlights the JADES Medium GOODS-S region. The colored squares denote additional NIRCam pointings from the JEMS survey (burgundy), the 1210 parallels (purple), and the 1286 parallels (green). In grey, we plot the outline of the Hubble Ultra Deep Field \citep{beckwith2006}. There is a noticeable increase in the density of sources in the JADES Deep and the ultra-deep 1210 parallel footprint.}
  \label{fig:GS_GN_footprint}
\end{figure*}

We show the positions of the GOODS-N sources in the left panel and the GOODS-S sources in the right panel of Figure \ref{fig:GS_GN_footprint}. On these figures, we include both those with {\tt EAZY} $\Delta\chi^2 > 4$ (dark points) and {\tt EAZY} $\Delta\chi^2 < 4$ (lighter points). The relatively higher density of sources in the southern portion of the GOODS-N observations compared to the northern portion is a result of the increased observational depth in that region. In GOODS-N, we find 2.1 objects in our $z > 8$ sample per square arcminute in the NE footprint, and 4.3 objects per square arcminute in the SW footprint. Similarly, the deepest portions of the JADES GOODS-S coverage are the large rectangular JADES Deep region, and the smaller 1210 parallels, where a significantly higher density of objects are detected. In GOODS-S, we find 7.8 objects in our $z > 8$ sample per square arcminute in JADES Deep, 4.5 objects per square arcminute in JADES Medium, 4.1 objects per square arcminute in the 1286 parallel, and 13.9 objects per square arcminute in the 1210 parallel. 

\begin{figure*}
  \centering
  \includegraphics[width=\textwidth]{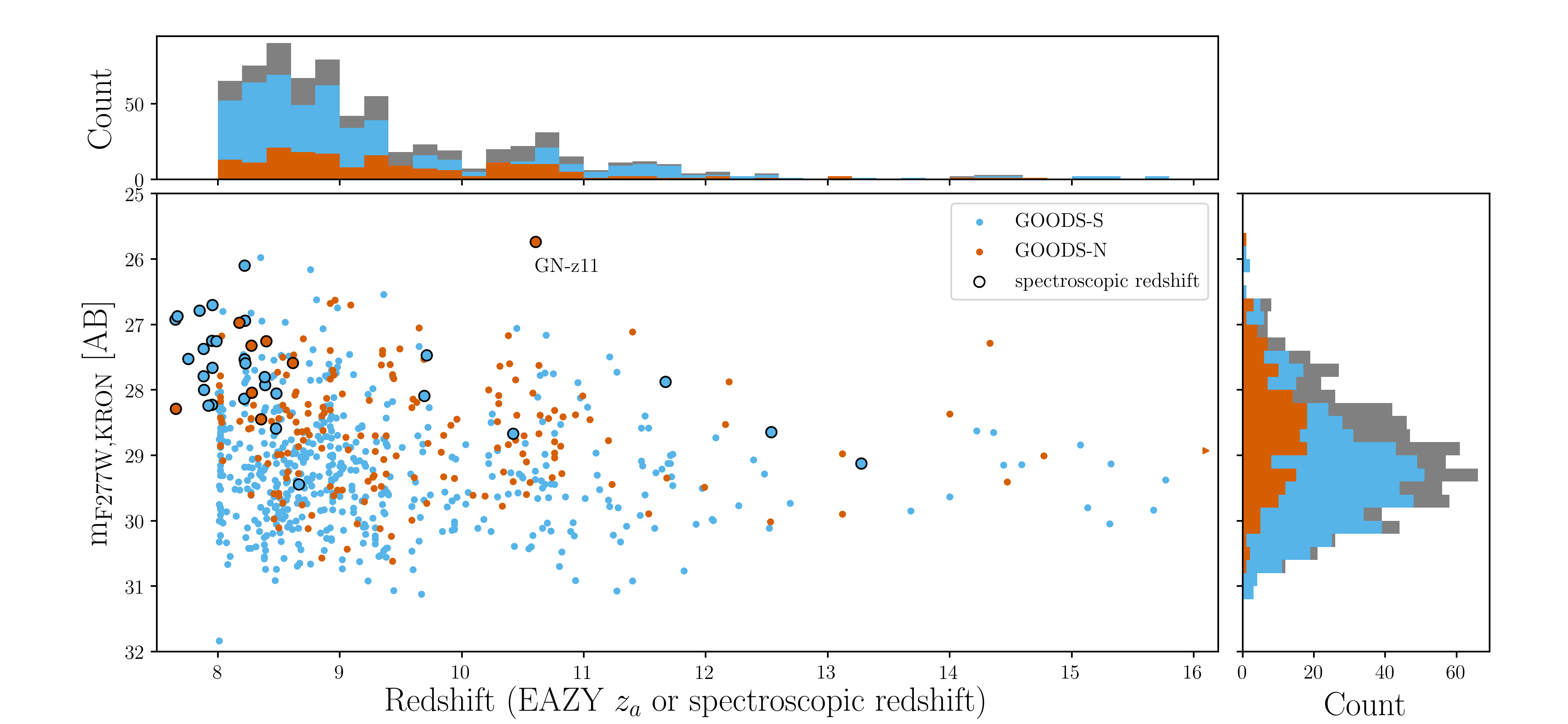}
  \caption{F277W AB Kron Magnitude plotted against the best-fitting {\tt EAZY} $z_a$ photometric redshift for the 717 galaxies and candidate galaxies in the GOODS-S (blue) and GOODS-N (red) $z > 8$ samples. Along the top we show the redshift distribution for the full sample (grey) as well as the GOODS-S and GOODS-N samples. On the right we show the magnitude distribution in a similar manner. The points colored with dark circles are plotted with the available spectroscopic redshifts for those sources, which, in many cases, extends to $z < 8$. We discuss these sources in Figure \ref{sec:speczs}. There is a lack of sources at $z \sim 10$ because of how the Lyman-$\alpha$ break falls between the NIRCam F115W and F150W at this redshift, making exact photometric redshift estimates difficult. The brightest source in the sample is the spectroscopically-confirmed galaxy GN-z11 at $z_{spec} = 10.6$, and the highest-redshift spectroscopically-confirmed source is JADES-GS-z13-0 at $z_{spec} = 13.2$. There is one source JADES-GN-189.15981+62.28898, at $z_a = 18.79$, which we plot as a right-facing arrow in the plot and discuss in Section \ref{sec:z_gt_12}.}
  \label{mag_vs_za}
\end{figure*}

In Figure \ref{mag_vs_za}, following similar work done in the literature \citep{finkelstein2023, perezgonzalez2023, austin2023}, we show the F277W observed AB magnitude measured using a Kron aperture against the {\tt EAZY} photometric redshift for each candidate $z > 8$ galaxy in GOODS-S and GOODS-N. Across the top we show the distribution of the photometric redshifts, and on the right side we show the F277W magnitude distribution for the photometric redshift sample as well as the GOODS-N and GOODS-S sample independently. For those objects where we have spectroscopic redshifts from either NIRSpec or FRESCO, we plot this value instead of the photometric redshift, and indicate those galaxies with larger points with black outlines. The GOODS-S sample, by virtue of the deeper coverage, extends to much fainter F277W magnitudes. On this diagram, the galaxy GN-z11 \citep{oesch2016, bunker2023, tacchella2023} is the brightest source, as one of two galaxies at $m_{\mathrm{F277W,Kron}} < 26$ (the other is JADES-GS-53.10394-27.89058, at $z_a = 8.35$). The redshifts seen in the main panel are discrete because of how {\tt EAZY} fits galaxies at specific redshift steps. 

In Figure \ref{mag_vs_za} we can see how the usage of wide filters for estimating photometric redshift leads to relative dearths of objects at $z \sim 10$ and $z \sim 13$, as these redshifts are where the Lyman-$\alpha$ break is between the F090W, F115W, and F150W filters. This is an artificial effect - for Lyman-$\alpha$-break galaxies, estimations of precise redshifts are highly predicated on the flux in the band that probes the break, and when the break sits between filters, the resulting redshifts are more uncertain. For example, faint galaxies at $z \sim 9 - 11$ can have similar SEDs where there is flux measured in the F150W filter and none measured in the F115W filter. This degeneracy results in photometric redshifts of these galaxies of $z_a \sim 9.5$, with broad $\chi^2$ minima reflecting larger redshift uncertainties. At slightly higher redshifts, however, the Lyman-$\alpha$ break moves into the F150W filter and this results in red F150W - F200W colors, leading to more precise photometric redshifts. This same effect is seen between the F150W and F200W filters at $z \sim 13$. The usage of medium-band filters, like NIRCam F140M, F162M, F182M, and F210M would help mitigate this effect somewhat for galaxies at these redshifts. We further explore these gaps by simulating galaxies across a uniform redshift range in Appendix \ref{sec:PhotoZGaps}. 

In this section we discuss the candidates in three subcategories: $z_{a} = 8 - 10$ (Section \ref{sec:z_8_10}), $z_{a} = 10 - 12$ (Section \ref{sec:z_10_12}), and $z_{a} > 12$ (Section \ref{sec:z_gt_12}). For each subcategory we describe the properties of the sample, plot example SEDs for galaxies spanning the magnitude and redshift range, and discuss notable examples. 

\subsection{$z_{phot} = 8 - 10$ Candidates}\label{sec:z_8_10}

\begin{figure*}
\centering
Example $z = 8 - 10$ Candidates\par\medskip
{\includegraphics[width=0.49\textwidth]{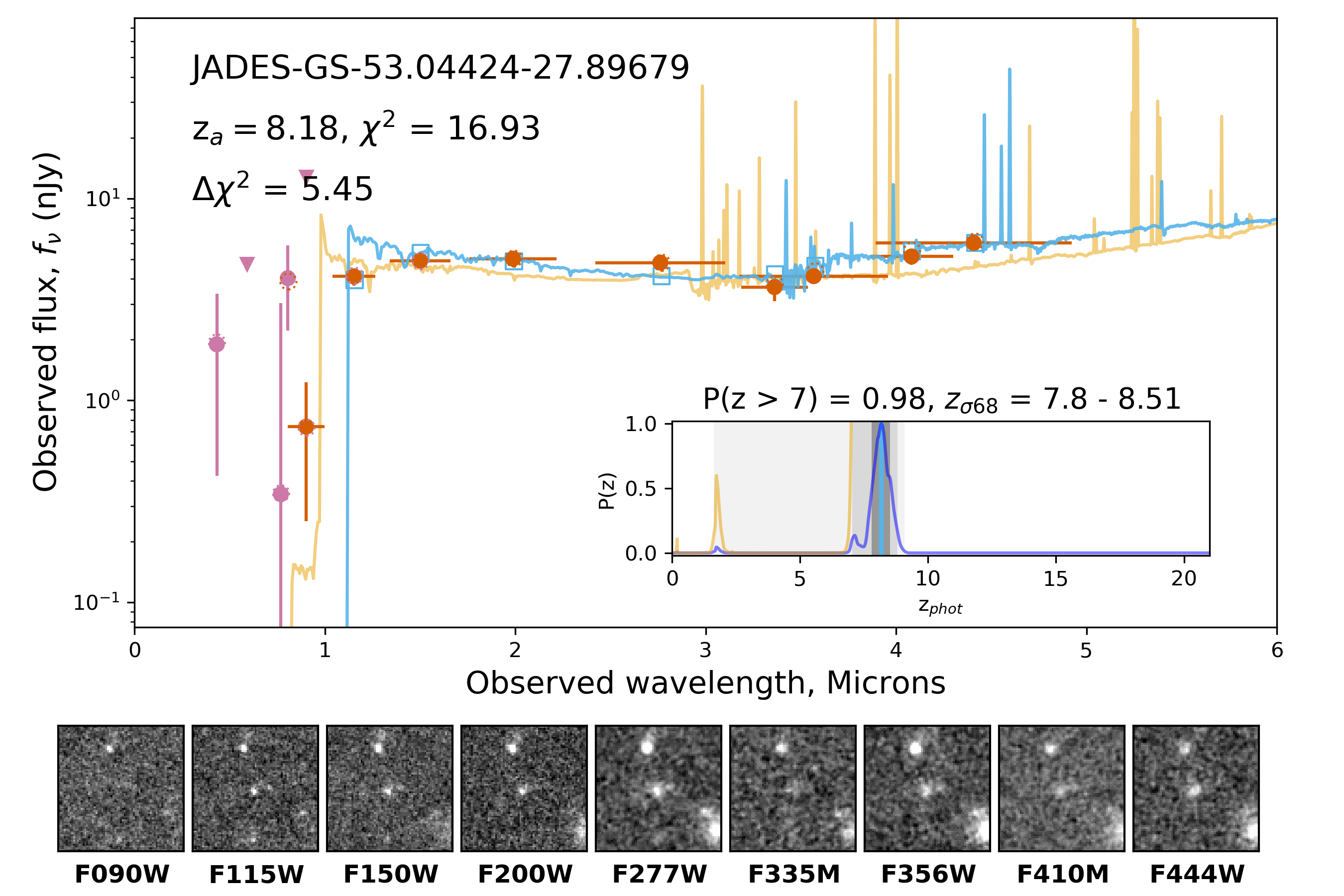}}\ 
{\includegraphics[width=0.49\textwidth]{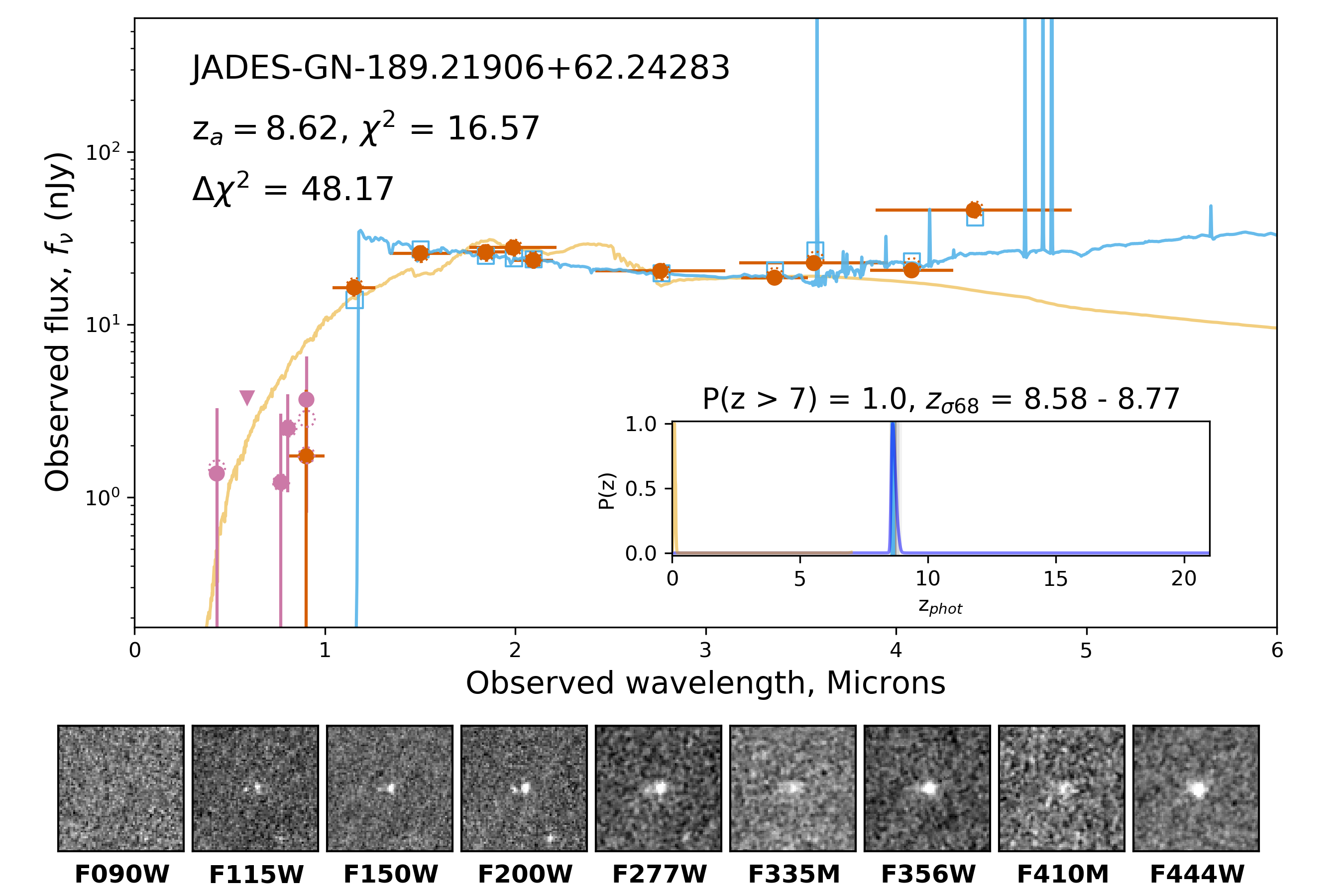}}\ 
{\includegraphics[width=0.49\textwidth]{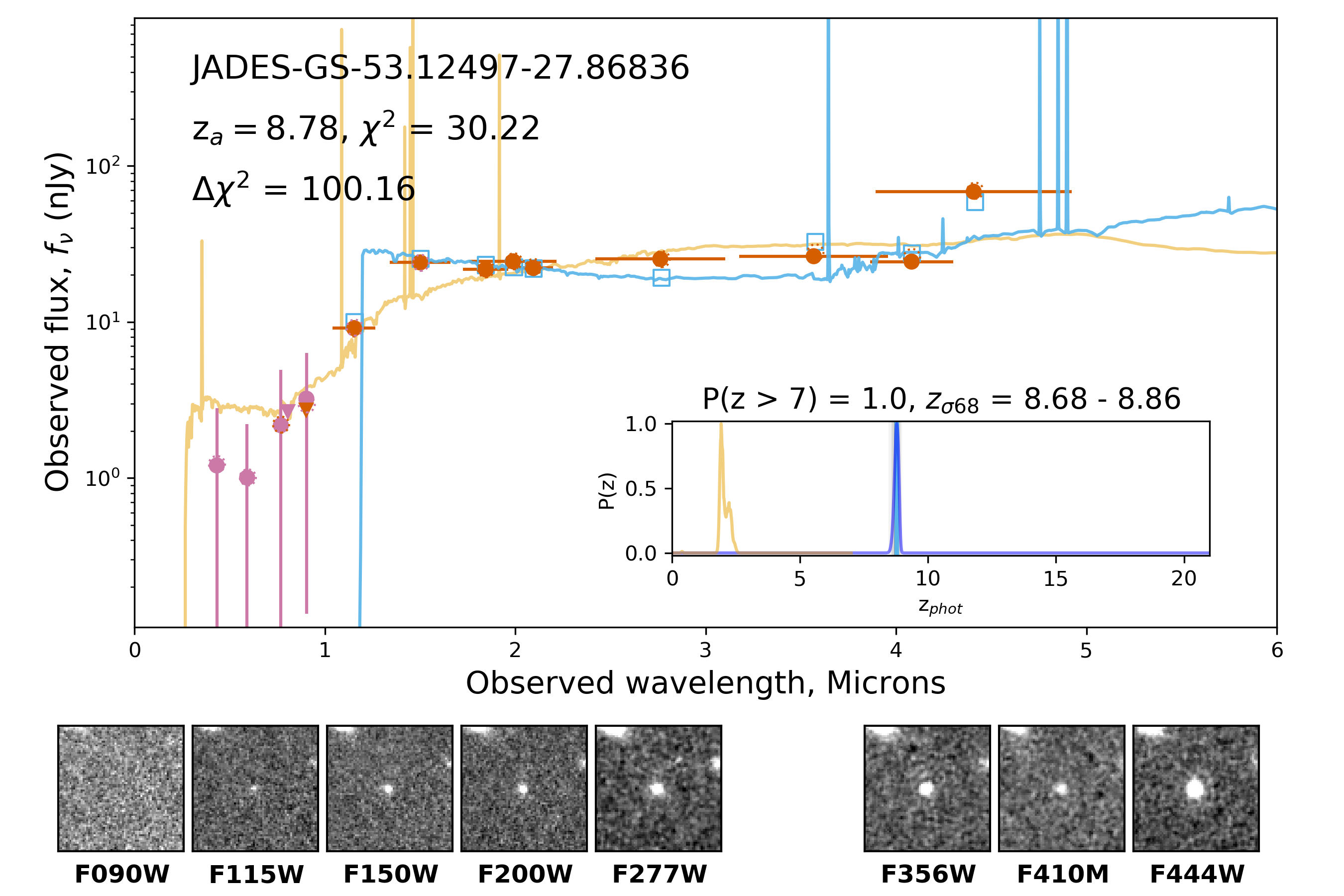}}\ 
{\includegraphics[width=0.49\textwidth]{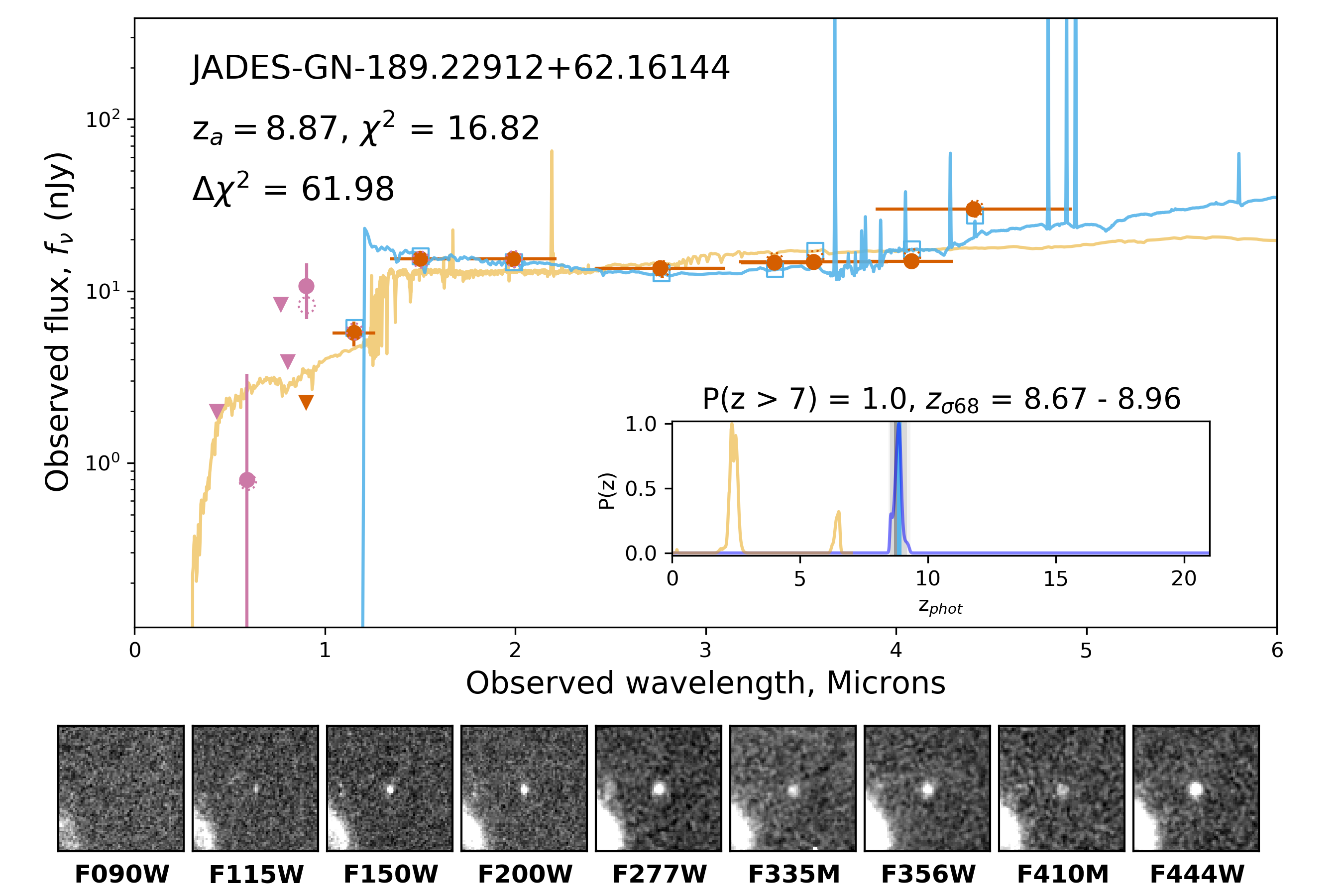}}\ 
{\includegraphics[width=0.49\textwidth]{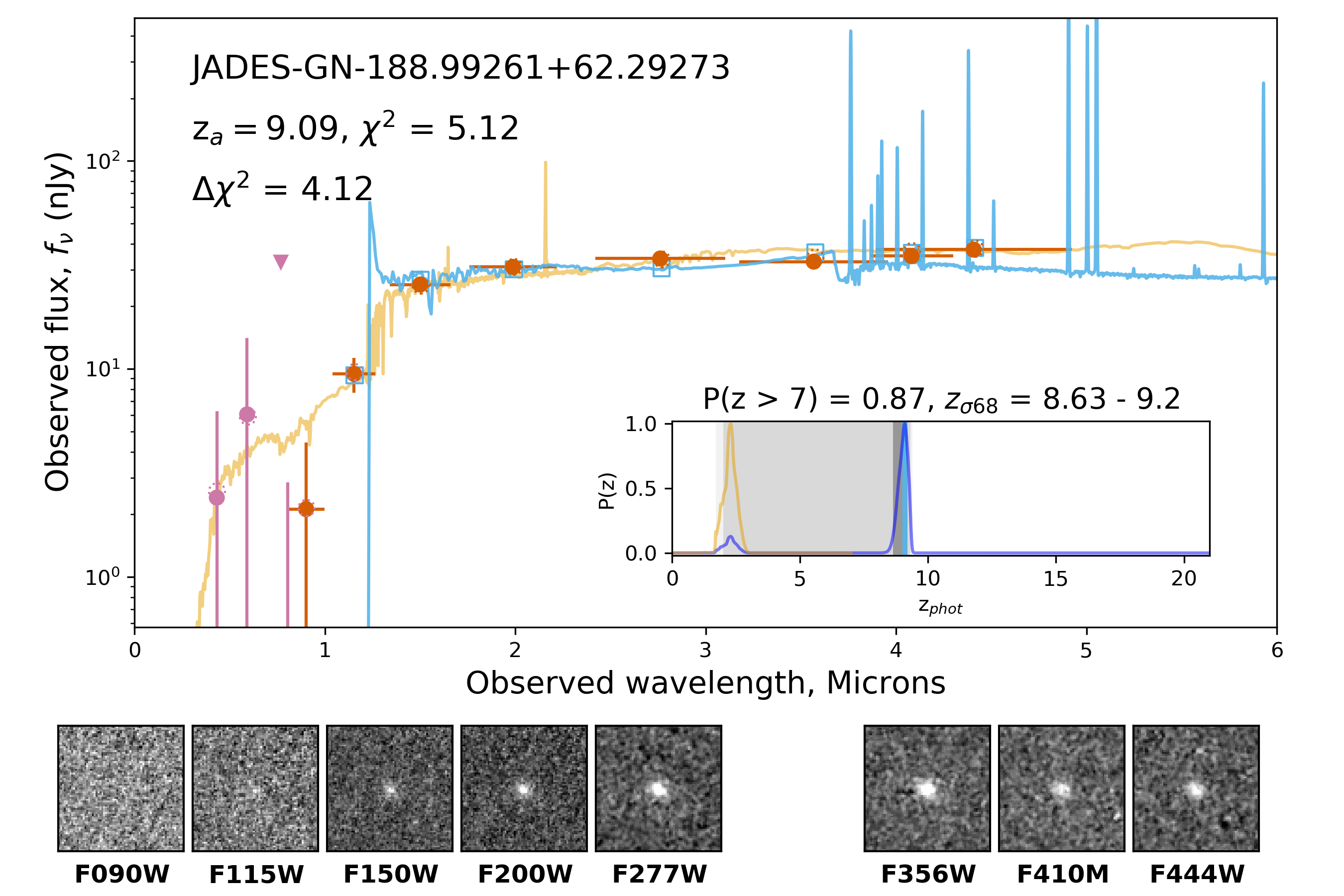}}\ 
{\includegraphics[width=0.49\textwidth]{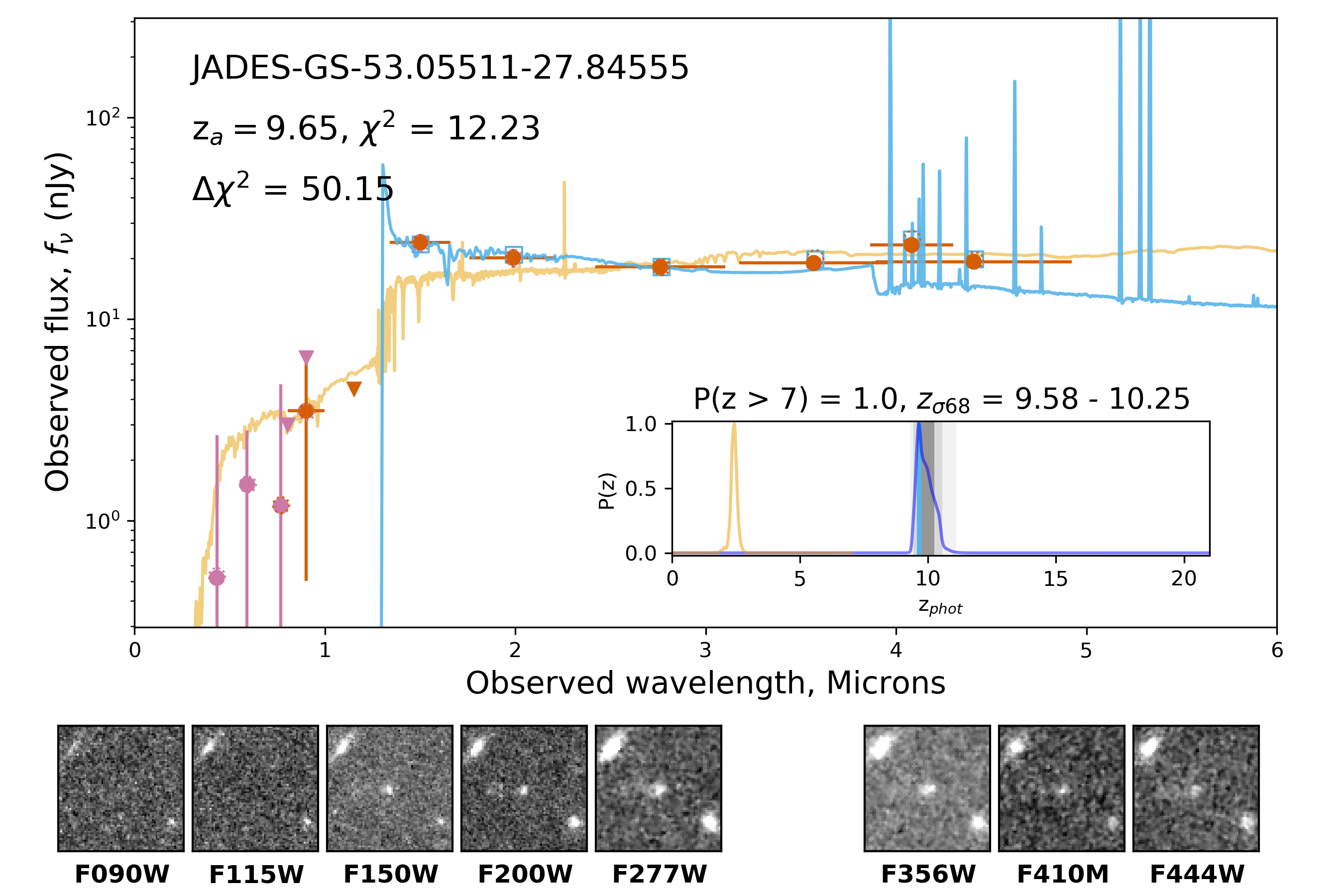}}\ 
{\includegraphics[width=0.49\textwidth]{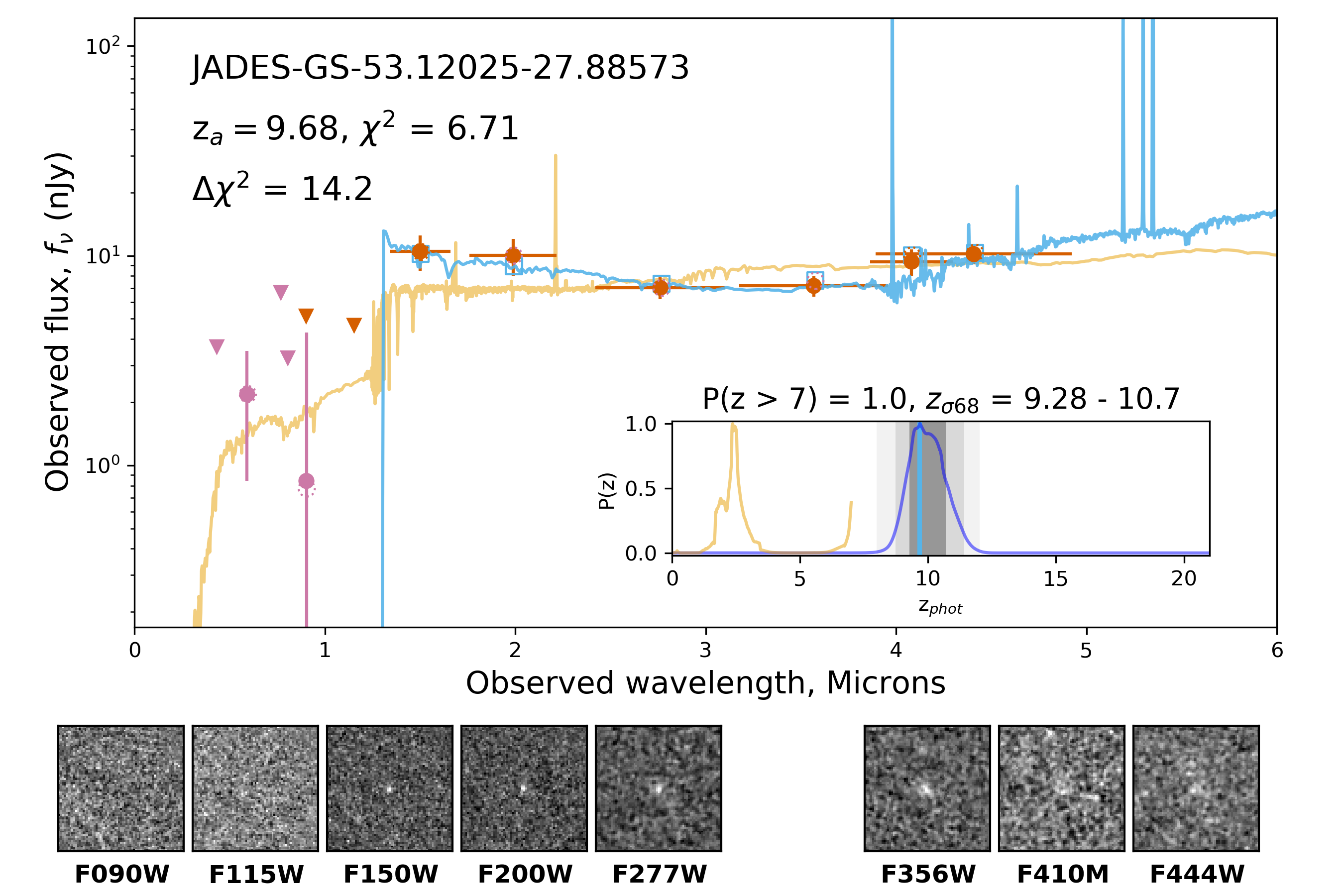}}\ 
{\includegraphics[width=0.49\textwidth]{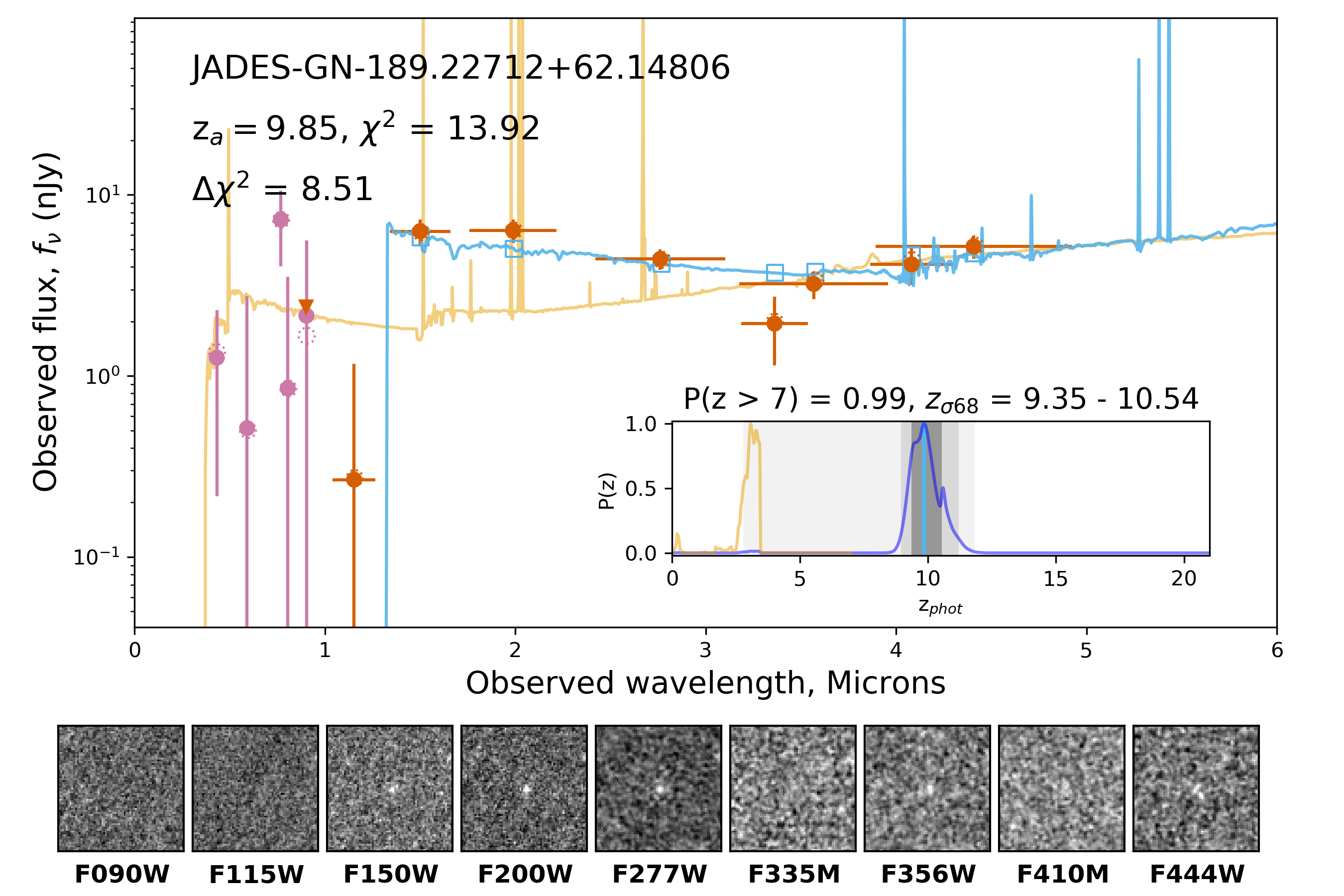}} 

\caption{
Example SEDs for eight $\Delta\chi^2 > 4$ candidate GOODS-S and GOODS-N galaxies at $z_a = 8 - 10$. In each panel, the colors, lines, and symbols are as in Figure \ref{example_SED_fit}.}
\label{fig:z_8_10_example_SEDs}
\end{figure*}

We find 547 total galaxies and galaxy candidates combined across the JADES GOODS-S (420 sources) and GOODS-N (127 sources) areas at $z_{a} = 8 - 10$. We show a subsample of the {\tt EAZY} SED fits and the JADES thumbnails for eight example candidate high-redshift galaxies in this photometric redshift range in Figure \ref{fig:z_8_10_example_SEDs}. In each plot, we show both the minimum $\chi^2$ fit, as well as the fit constrained to be at $z < 7$. We chose these objects from the full sample to span a range of F277W Kron magnitudes as well as photometric redshifts. 

Because of the availability of both NIRSpec and FRESCO spectroscopy for our sample, there are 34 (27 in GOODS-S and 7 in GOODS-N) galaxies in this photometric redshift range where a spectroscopic redshift has been measured. For 14 of these sources (13 in GOODS-S and 1 in GOODS-N), the resulting spectroscopic redshift $z_{spec} = 7.65 - 8.0$. Because these objects satisfied our photometric redshift selection criteria, we choose to include them in our sample, and discuss their spectroscopic redshifts in Section \ref{sec:speczs}.  

\begin{figure*}
\centering
Example $z = 8 - 9$ Resolved Candidates \par\medskip
{\includegraphics[width=0.3\textwidth]{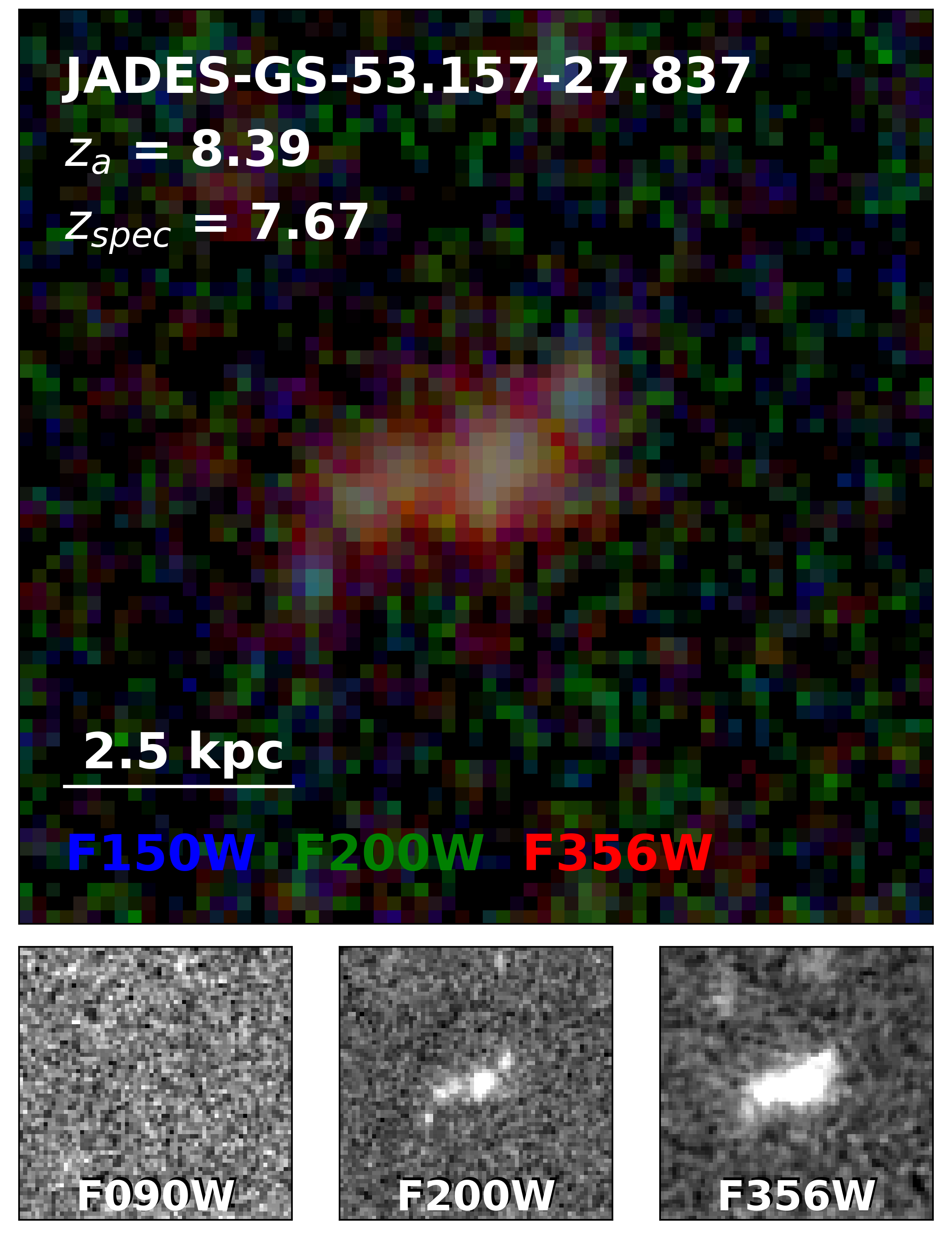}}\ 
{\includegraphics[width=0.3\textwidth]{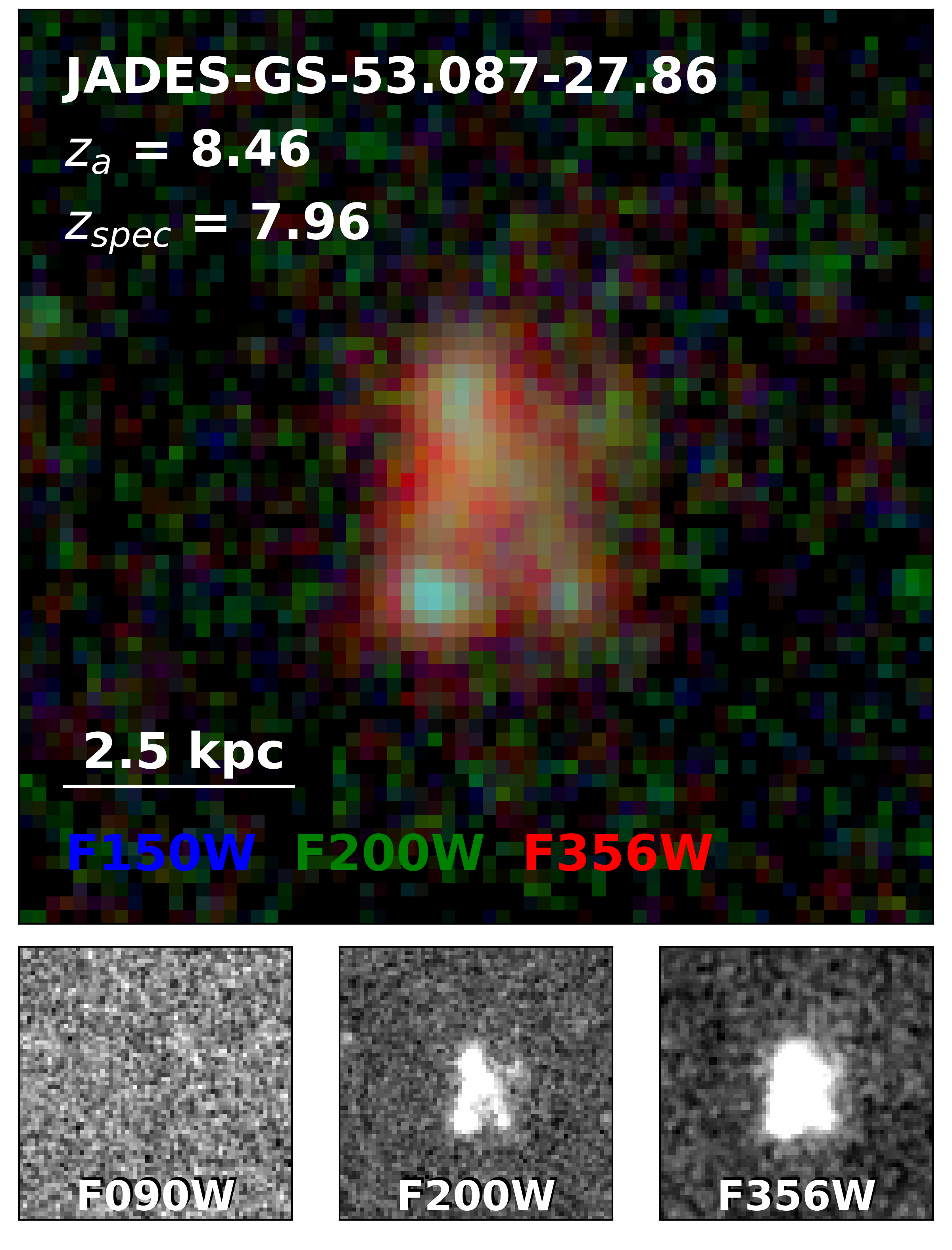}}\ 
{\includegraphics[width=0.3\textwidth]{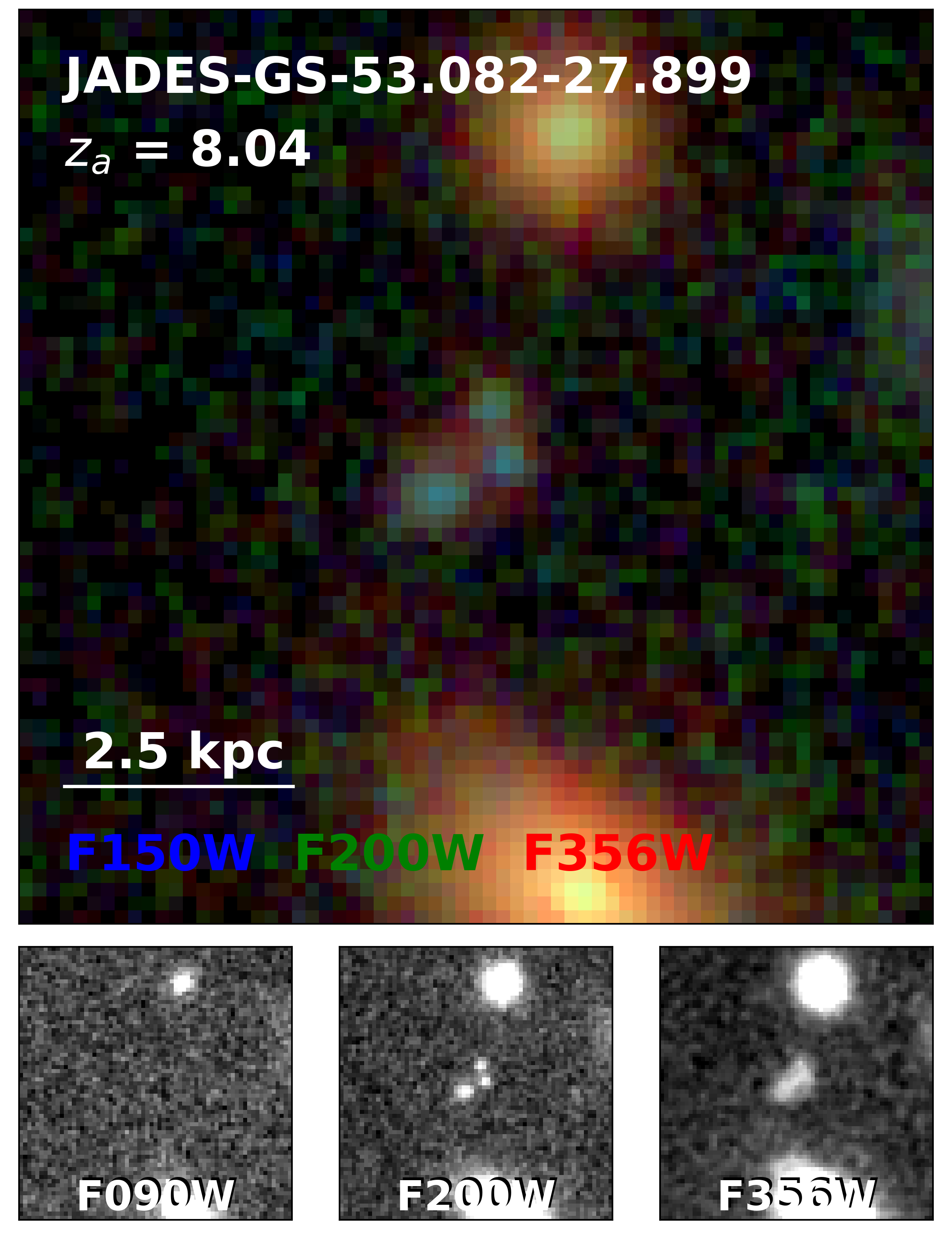}}\  
{\includegraphics[width=0.3\textwidth]{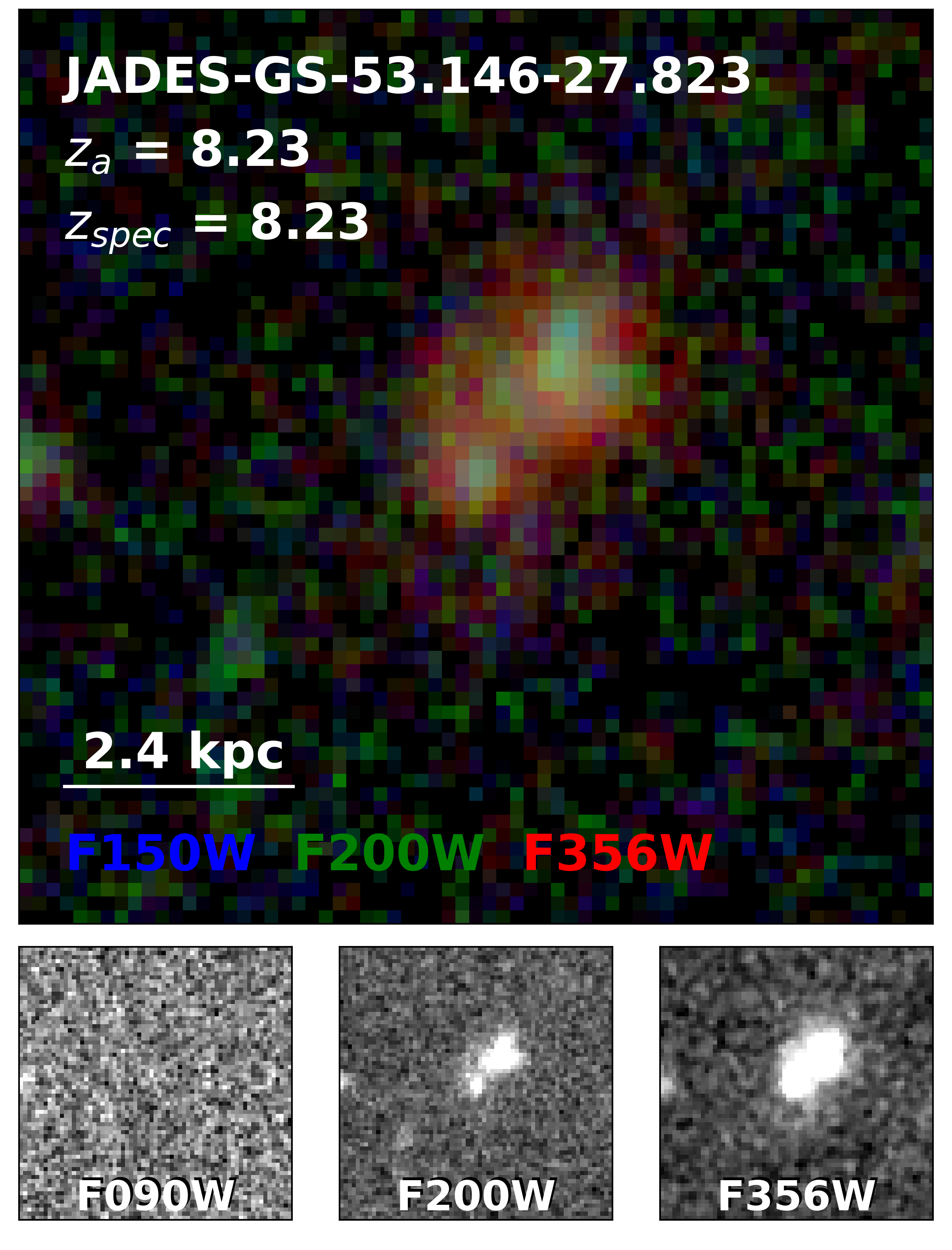}}\
{\includegraphics[width=0.3\textwidth]{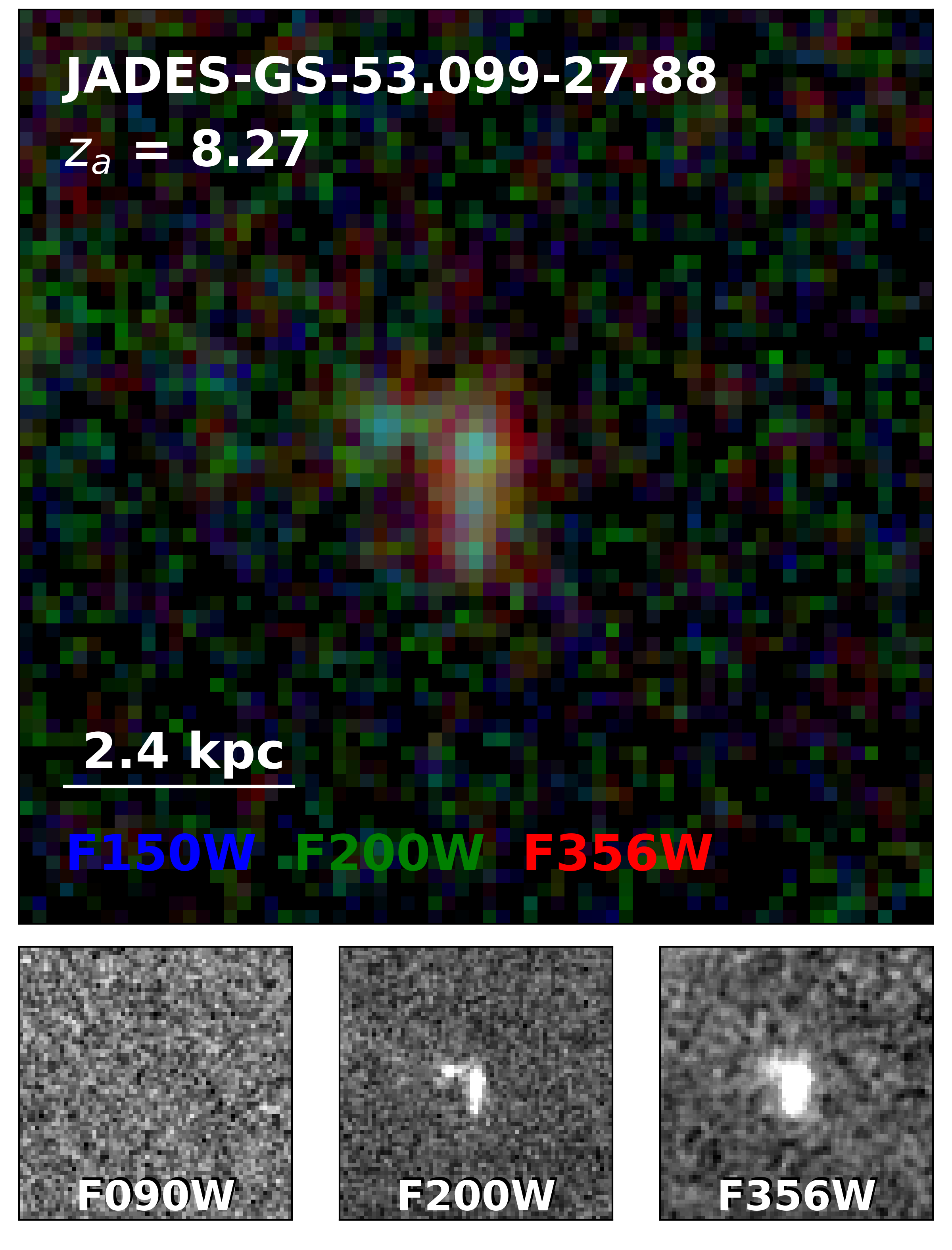}}\ 
{\includegraphics[width=0.3\textwidth]{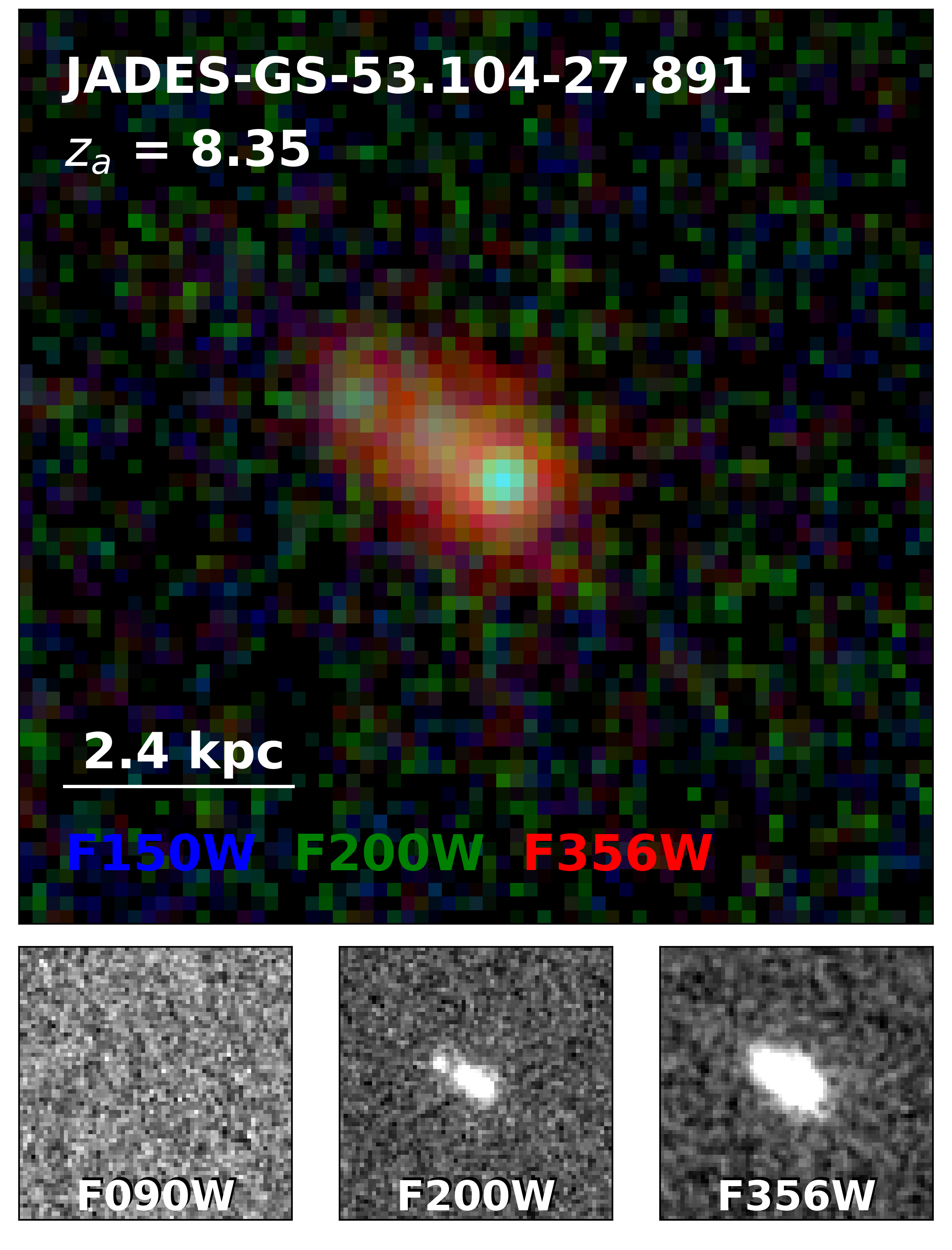}}\ 
{\includegraphics[width=0.3\textwidth]{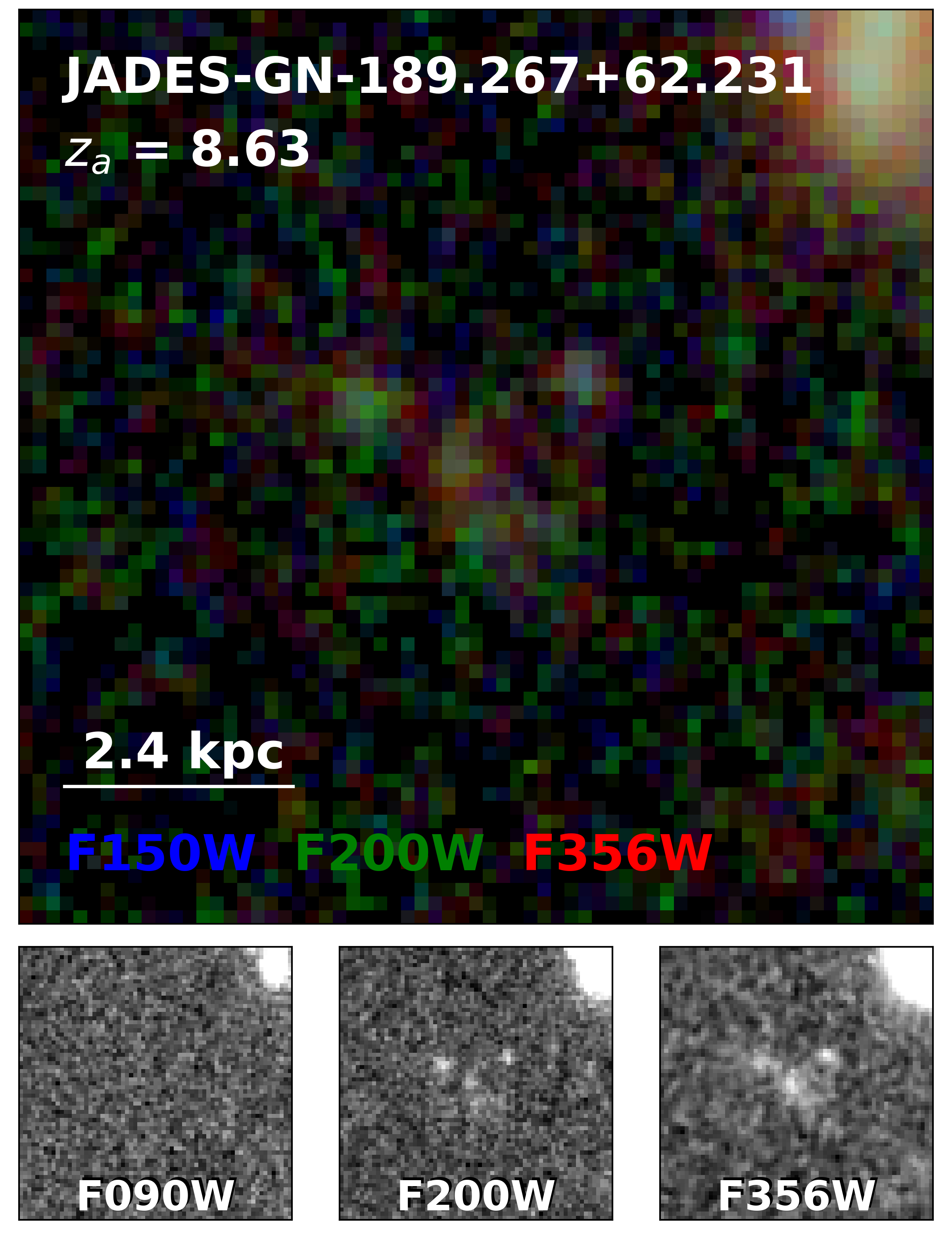}}\  
{\includegraphics[width=0.3\textwidth]{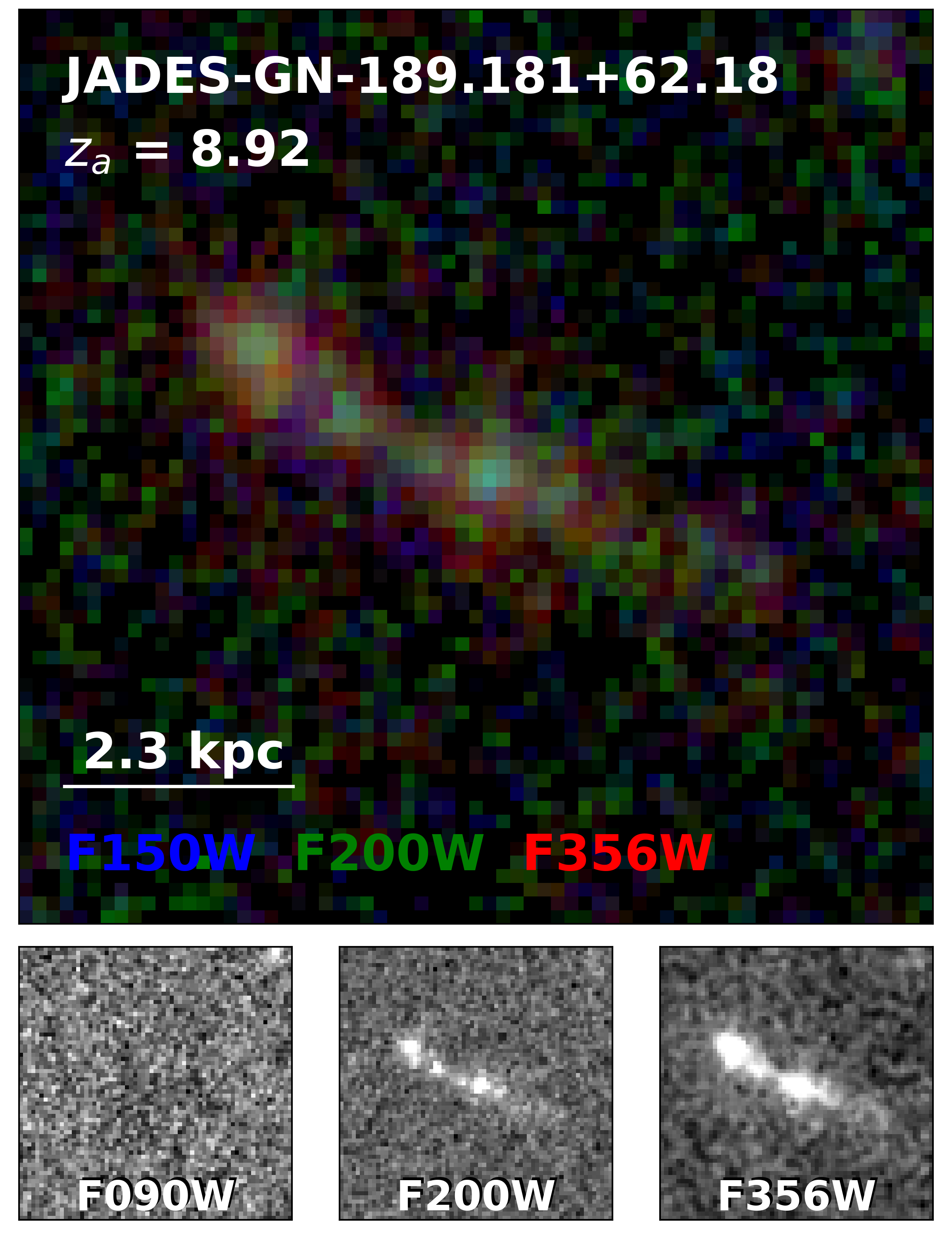}}\  
{\includegraphics[width=0.3\textwidth]{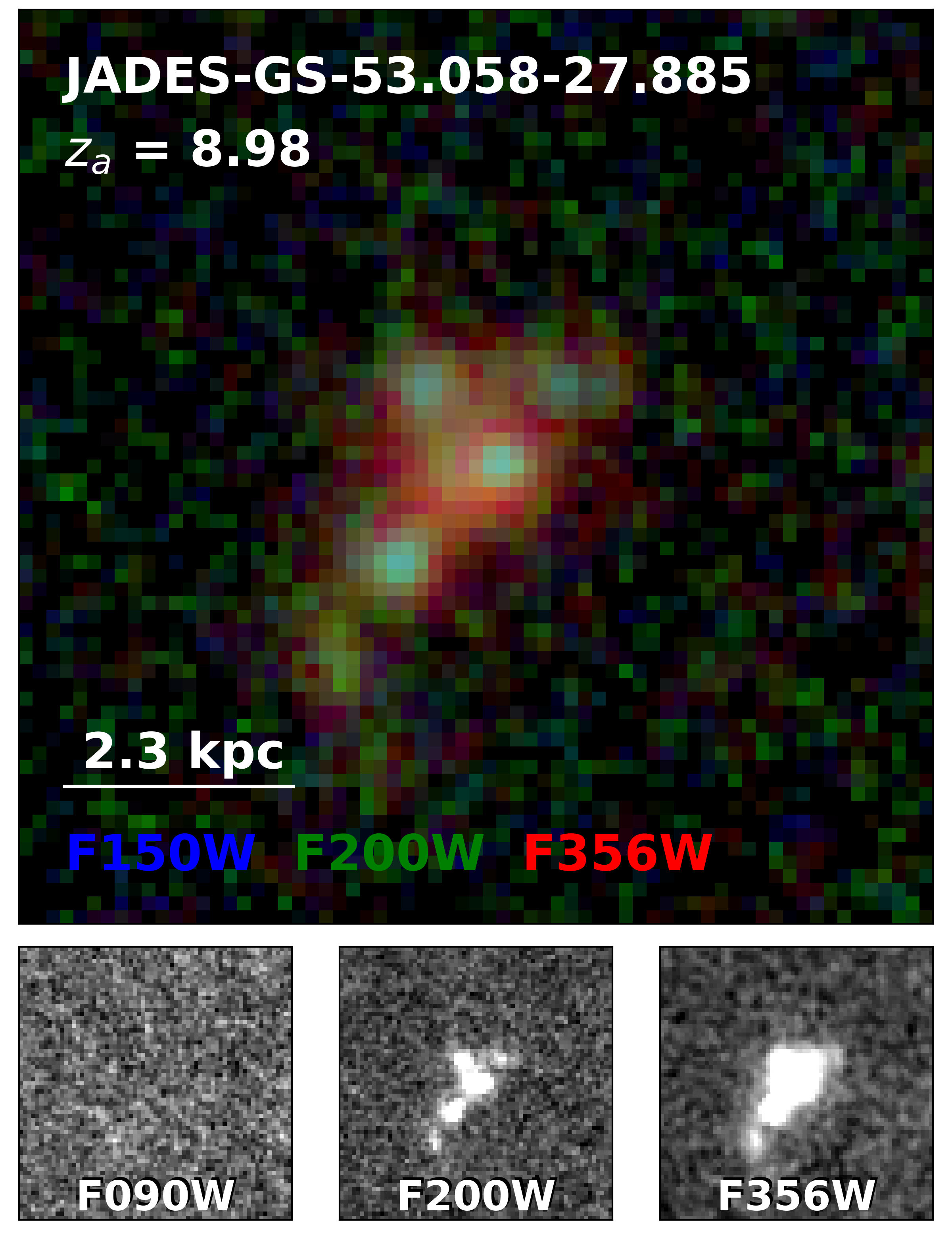}} 
\caption{
Color thumbnails for a selection of nine $z_a = 8 - 9$ resolved galaxies with multiple components. Each thumbnail is $2^{\prime\prime}$ on a side, and we include a size bar showing $0.5^{\prime\prime}$ for each object. The color image is composed of F356W, F200W, and F090W as red, green, and blue respectively. We also show images in those filters for each object separately to demonstrate the dropout nature of these objects in the F090W filter. 
}
\label{fig:resolved_z_8_10}
\end{figure*}

There are a number of sources in this photometric redshift range with extended morphologies, often seen in the JADES data as multiple clumps observed in the images at shorter wavelengths. In Figure \ref{fig:resolved_z_8_10}, we show a subsample of nine resolved galaxies with $z_a = 8 - 9$. For each object we show the F090W, F115W, and F356W thumbnails, along with a color image combining these three filters. Each thumbnail is $2^{\prime\prime}$ on a side, showcasing the resolved sizes of some of these targets. At $z = 8 - 10$, $1^{\prime\prime}$ corresponds to $4.6 - 4.9$ kpc, and we provide a scale bar of $0.5^{\prime\prime}$ in each panel. At these redshifts, F090W is to the blue of the Lyman-$\alpha$ break, so the galaxies should not appear in this filter, F200W spans the rest-frame UV and F356W spans the rest-frame optical continuum. We are then seeing UV-bright star-forming clumps in the F200W filter, and rest-frame $\sim 4000$\AA\, stellar continuum in F356W. We show two sources, JADES-GS-53.1571-27.83708 (top row, left column), and JADES-GS-53.08738-27.86033 (top row, middle column), which have spectroscopic redshifts from FRESCO at $z_{spec} = 7.67$ and  $z_{spec} = 7.96$ respetively, as indicated below the photometric redshifts in the color panel. 

These nine sources show multiple irregular morphologies, and many are elongated. JADES-GN-189.18051+62.18047 is an especially complex system at $z_a = 8.92$ with four or five clumps that span almost 7 kpc at this photometric redshift, similar to the ``chain of five'' F150W dropout system presented in \citet{yan2023}. Five of the extended sources we highlight were previously presented in the literature: JADES-GS-53.1571-27.83708, JADES-GS-53.08738-27.86033, JADES-GS-53.08174-27.89883, JADES-GS-53.1459-27.82279 and JADES-GS-53.10393-27.89059 \citep{mclure2013, bouwens2015, finkelstein2015, harikane2016, bouwens2021}. Given the depth and resolution of NIRCam, we can see new details for these sources from what was observed in the HST ACS and WFC3 observations, such as the nearly $\sim 0.8^{\prime\prime}$-long haze to the northeast for JADES-GS-53.10393-27.89059, which corresponds to about 4 kpc at the candidate photometric redshift. 

\subsection{$z_{phot} = 10 - 12$ Candidates}\label{sec:z_10_12}

\begin{figure*}
\centering
Example $z = 10 - 12$ Candidates \par\medskip
{\includegraphics[width=0.49\textwidth]{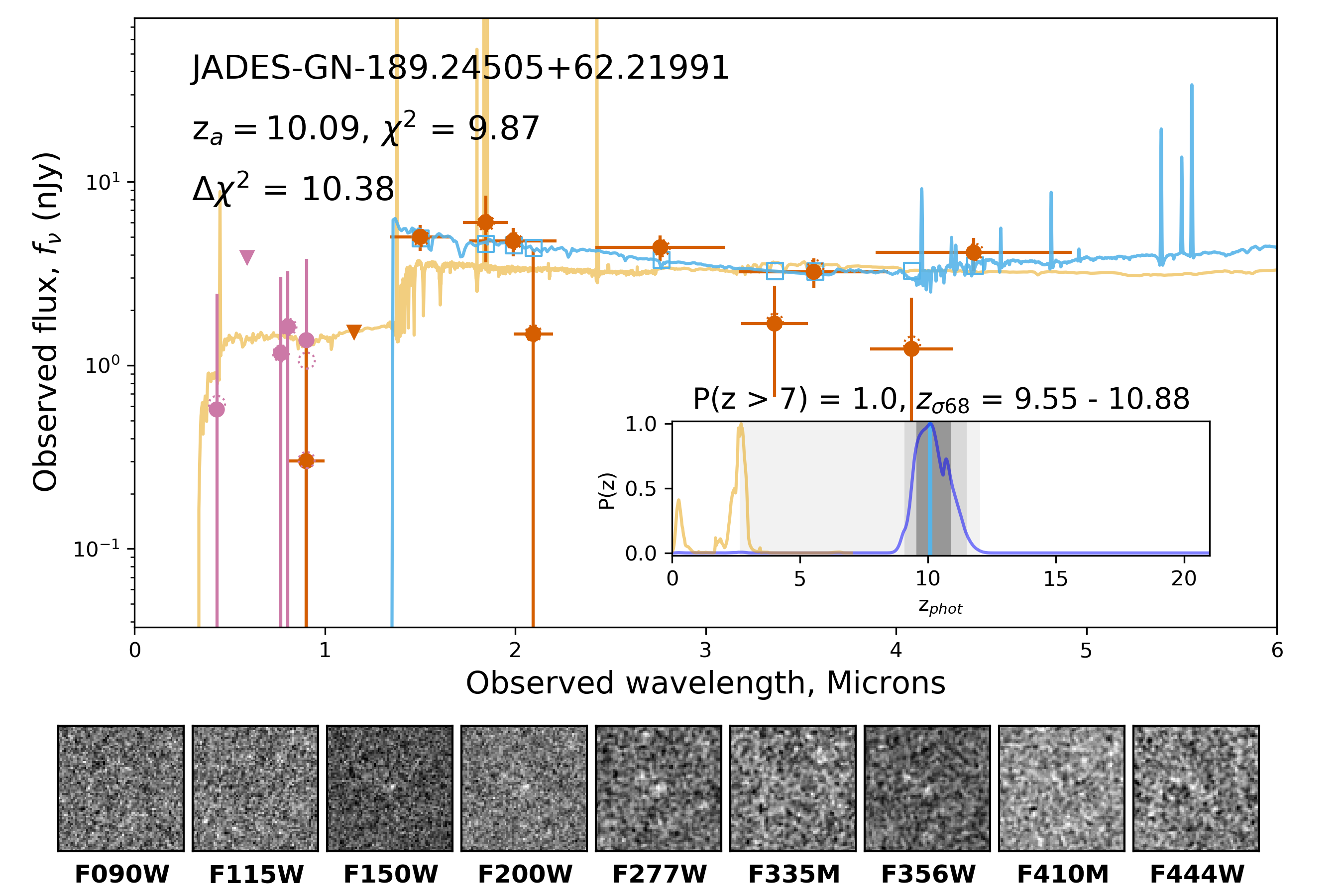}}\  
{\includegraphics[width=0.49\textwidth]{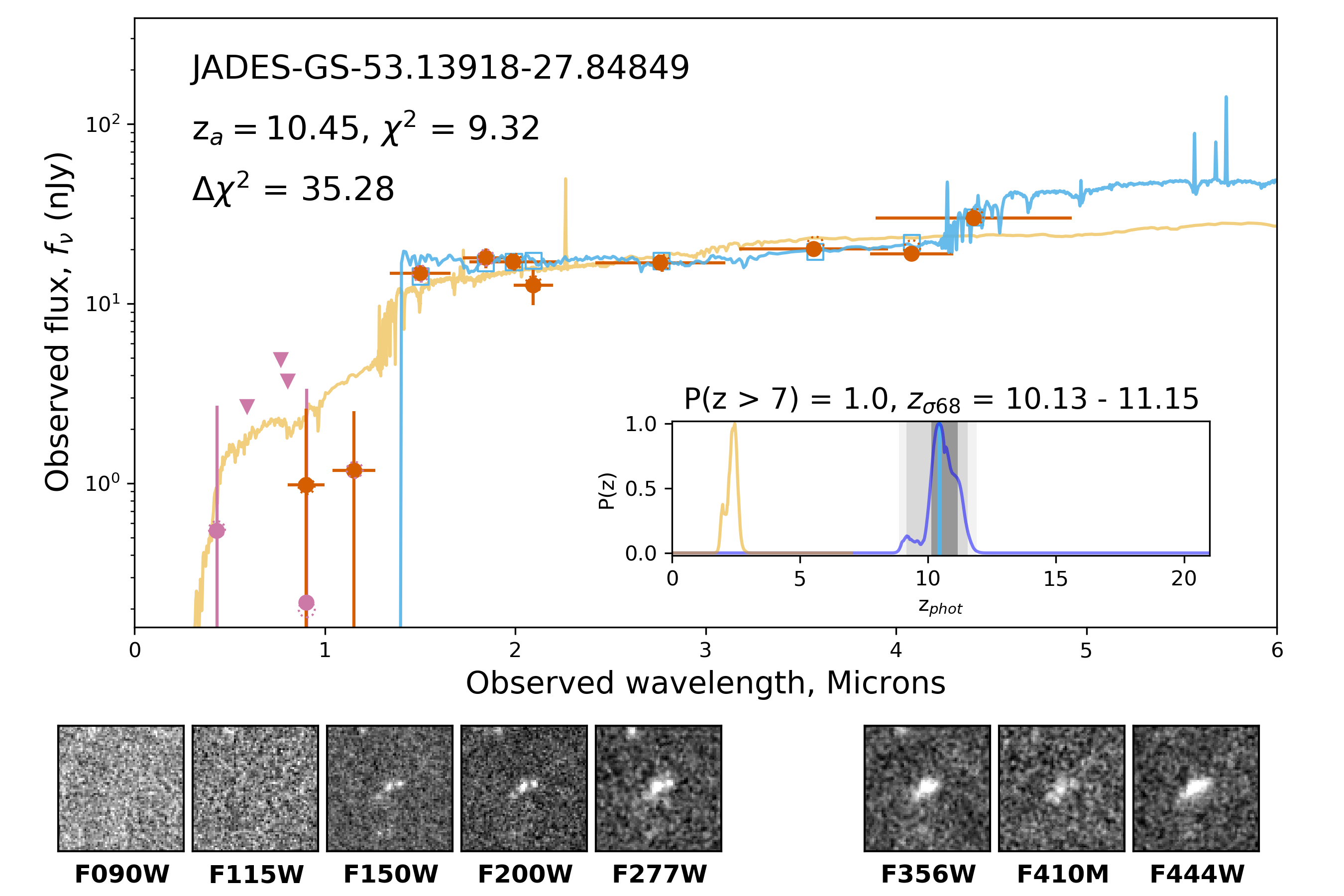}}\ 
{\includegraphics[width=0.49\textwidth]{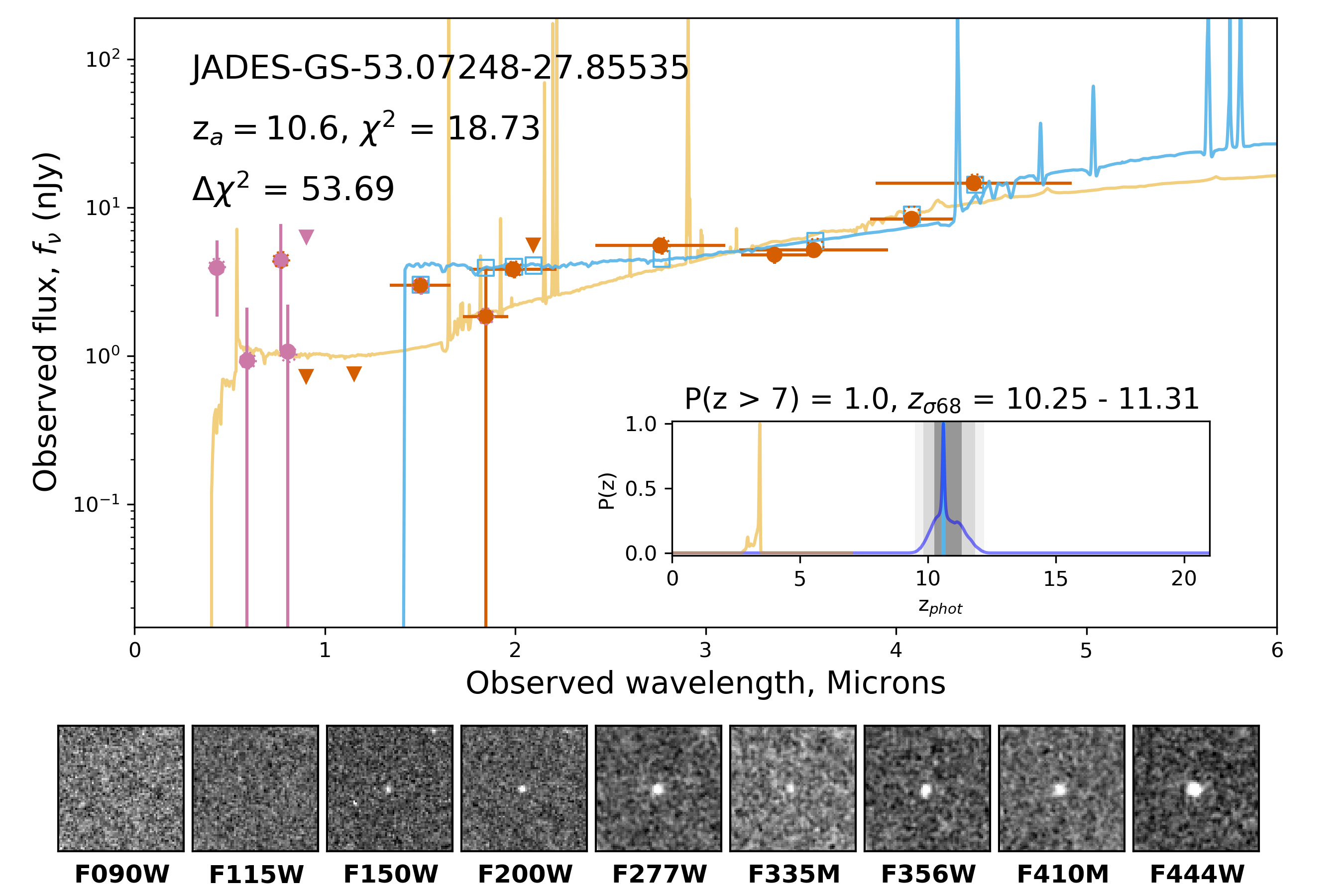}}\  
{\includegraphics[width=0.49\textwidth]{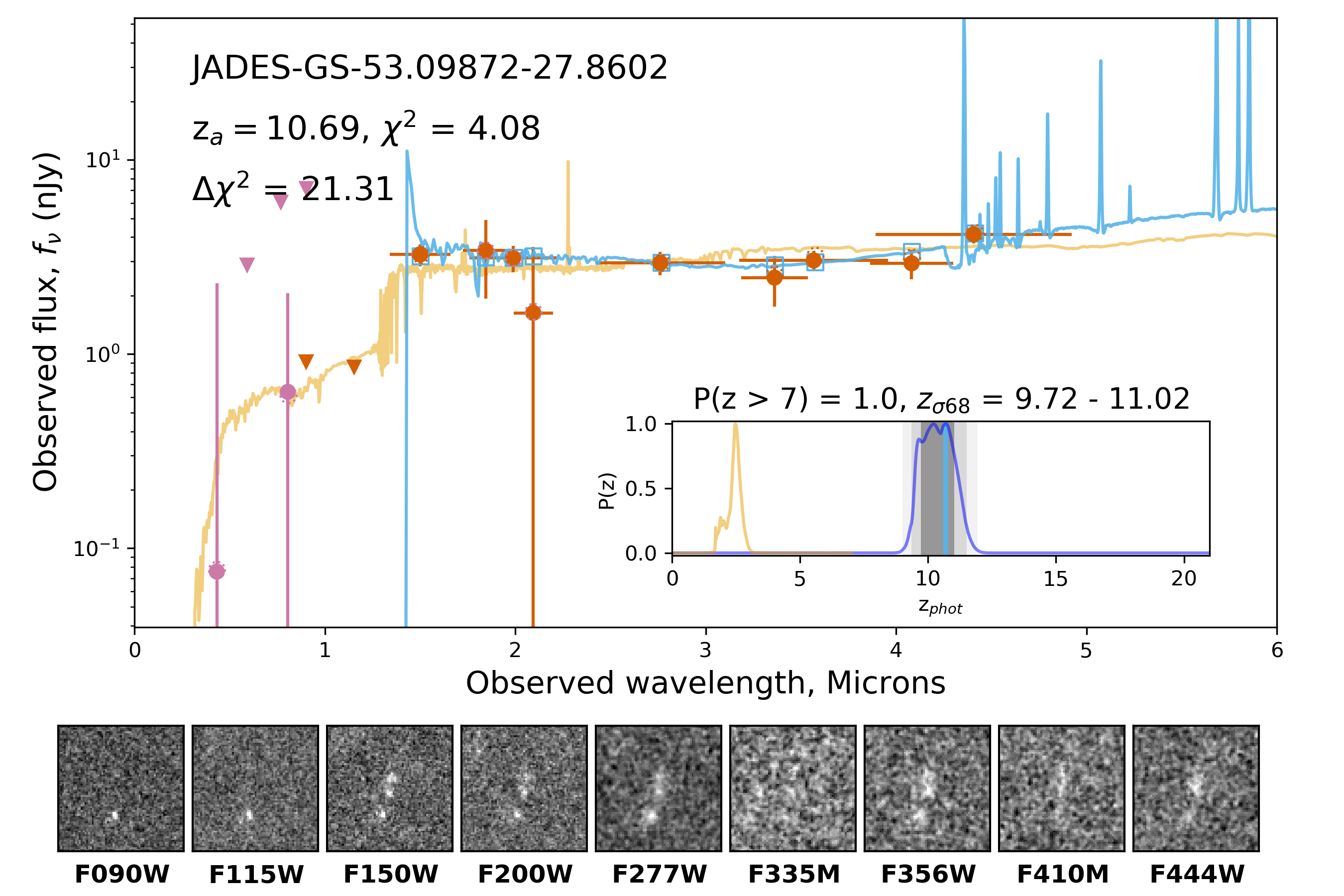}}\   
{\includegraphics[width=0.49\textwidth]{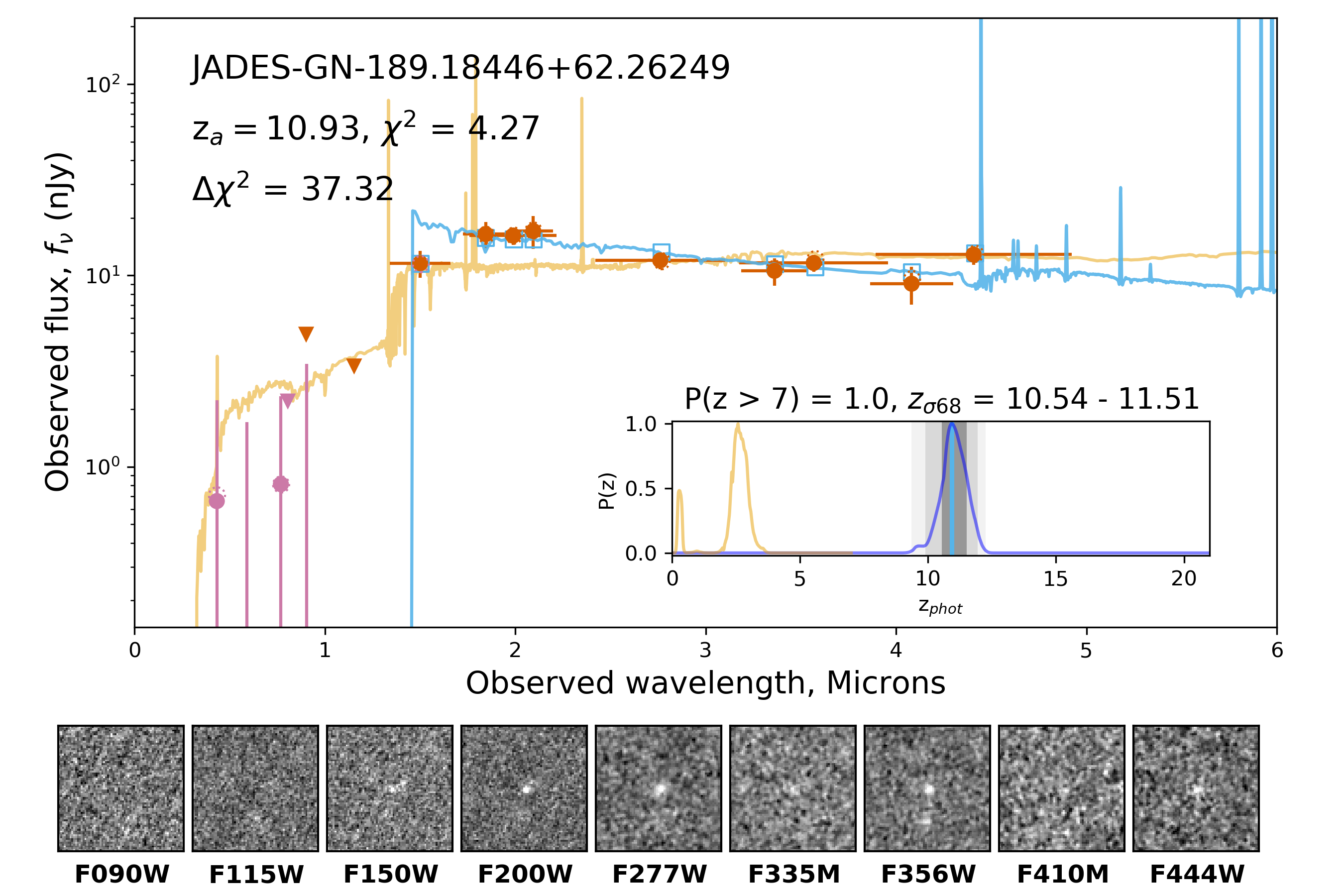}}\   
{\includegraphics[width=0.49\textwidth]{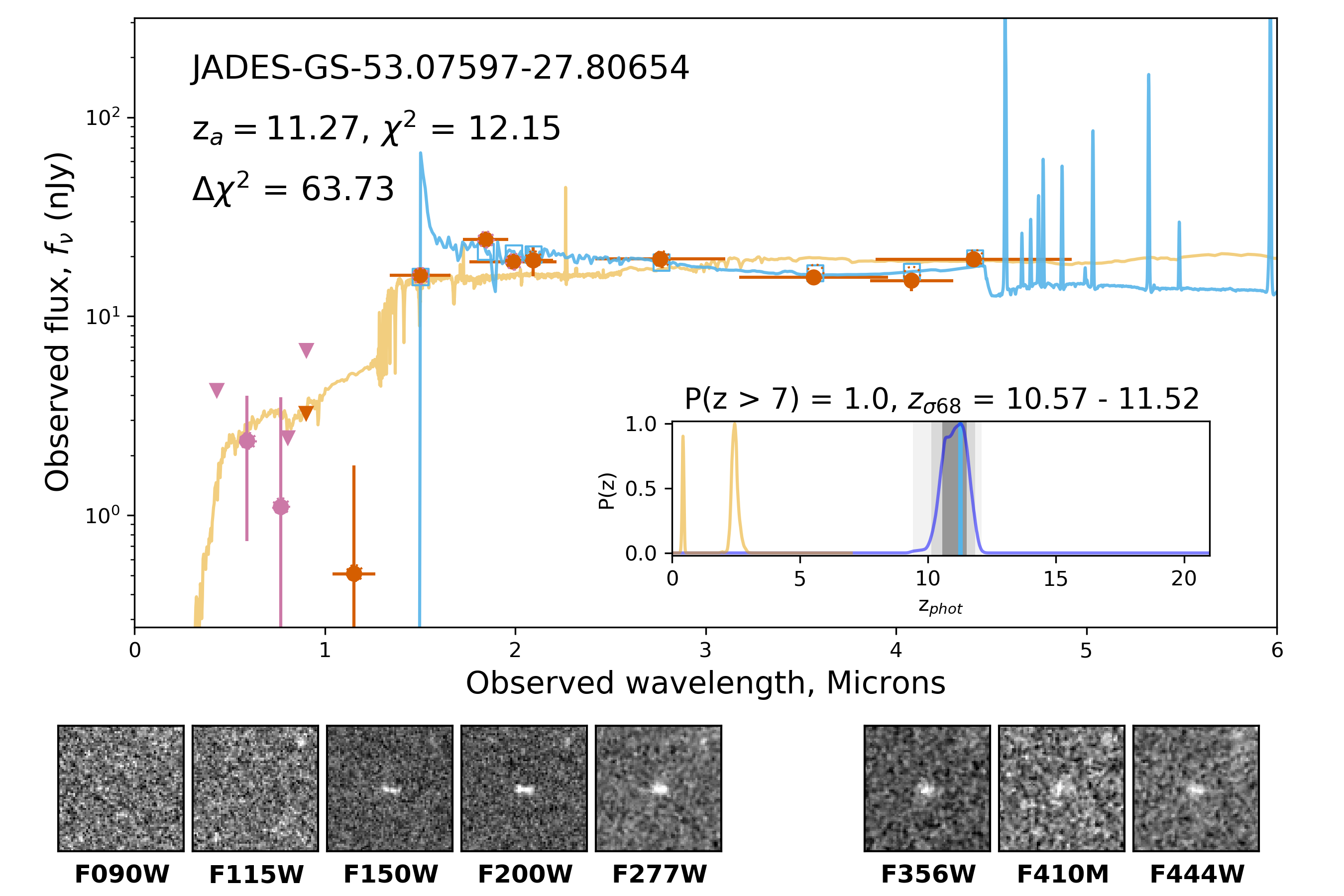}}\   
{\includegraphics[width=0.49\textwidth]{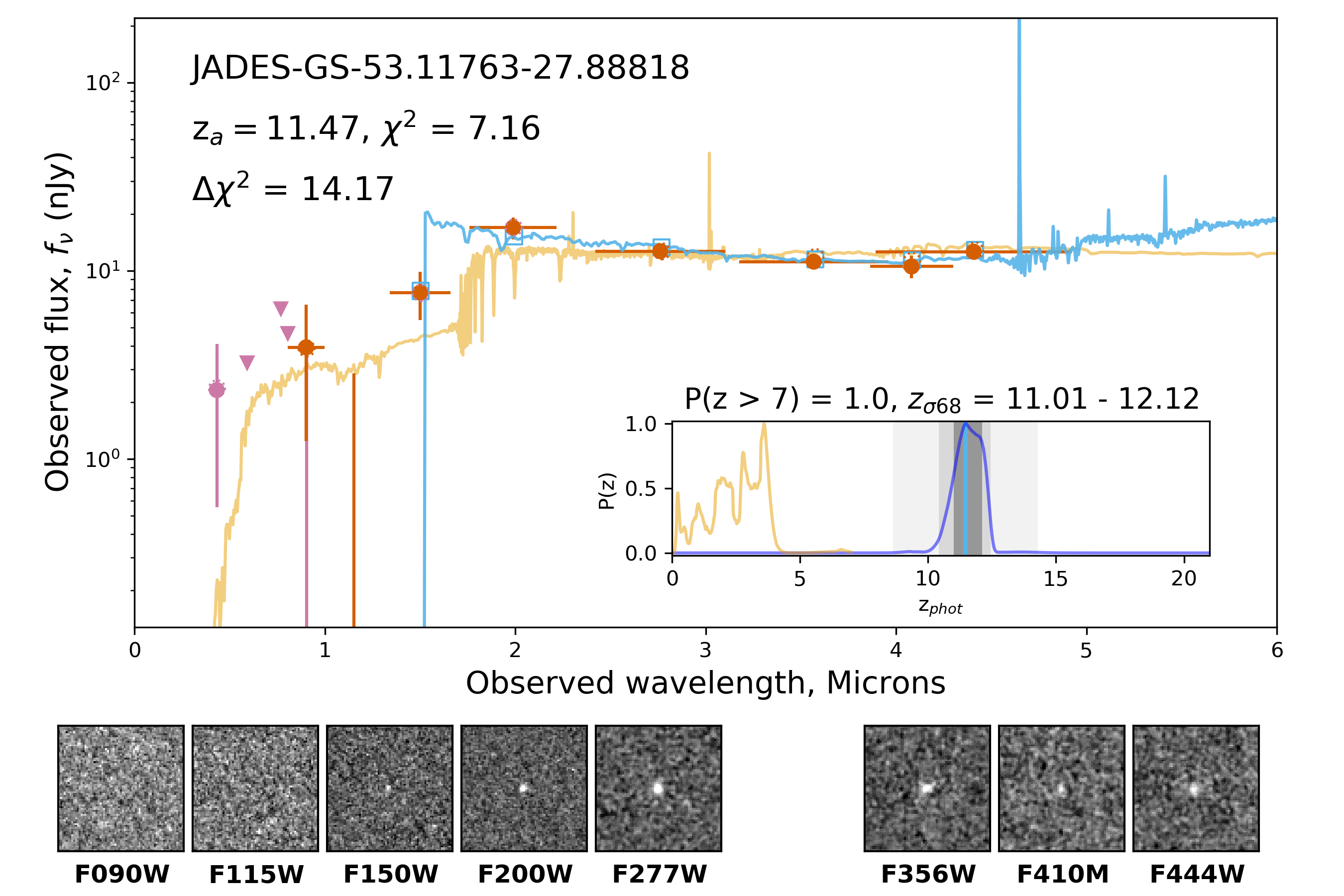}}\ 
{\includegraphics[width=0.49\textwidth]{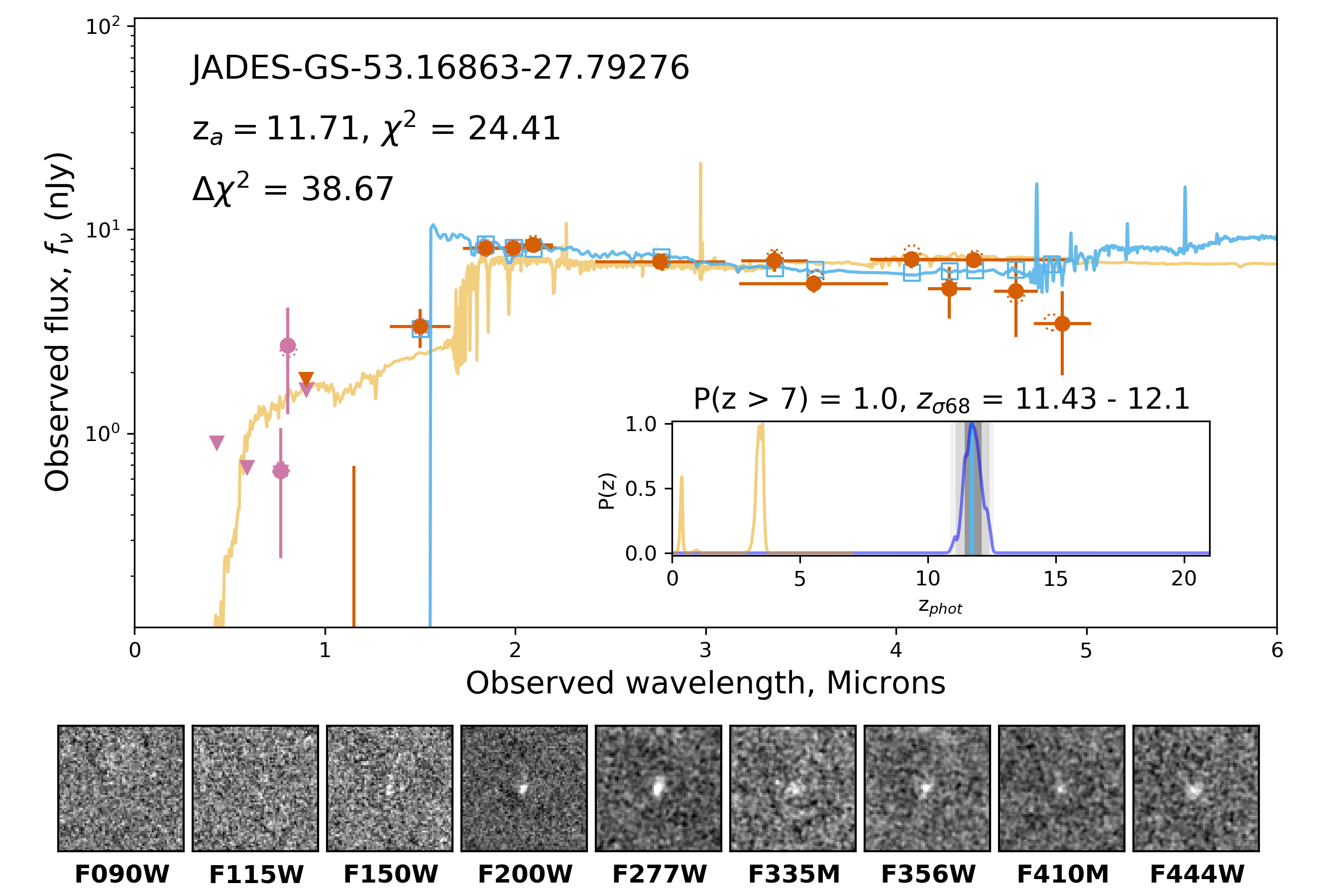}}  

\caption{
Example SEDs for eight $\Delta\chi^2 > 4$ candidate GOODS-S and GOODS-N galaxies at $z_a = 10 - 12$. In each panel, the colors, lines, and symbols are as in Figure \ref{example_SED_fit}.}
\label{fig:z_10_12_example_SEDs}
\end{figure*}

We find a total of 137 galaxies and candidate galaxies at at $z_{phot} = 10 - 12$: 92 in GOODS-S and 45 in GOODS-N. We show the {\tt EAZY} SED fits and the JADES thumbnails for eight example candidates in this photometric redshift range in Figure \ref{fig:z_10_12_example_SEDs}. 

In the GOODS-S region, this redshift range includes two of the spectroscopically-confirmed galaxies from \citet{robertson2022} and \citet{curtislake2022}, JADES-GS-z10-0 ($z_{spec} = 10.38^{+0.07}_{-0.06}$) and JADES-GS-z11-0 ($z_{spec} = 11.58^{+0.05}_{-0.05}$). The {\tt EAZY} photometric redshifts for these targets are $z_a = 10.84$ for JADES-GS-z10-0, and $z_a = 12.31$ for JADES-GS-z11-0. Both photometric redshifts are higher than the measured spectroscopic redshift, but considering the $P(z)$ uncertainty in both measurements, the measurements are within $2\sigma$ of the true values. Indeed, the $\Delta\chi^2$ between the minimum value corresponding to $z_a$ and the value at $z_{spec}$ is 10.25 for JADES-GS-z10-0 and 1.75 for JADES-GS-z11-0. In \citet{robertson2022}, the authors estimate photometric redshifts for these sources using the Bayesian stellar population synthesis fitting code \emph{Prospector} \citep{johnson2021} and recover $P(z)$ surfaces that are similarly offset to higher values than the spectroscopic redshifts. In the GOODS-N region, we find the brightest object overall in our sample (a Kron F277W aperture magnitude of 25.73 AB), GN-z11, discussed at length in \citet{tacchella2023} and spectroscopically confirmed to lie at $z = 10.603 \pm 0.001$ in \citet{bunker2023}. In our {\tt EAZY} fit, we estimate $z_a = 11.0$, which is within 2$\sigma$ of the spectroscopic redshift, but again, higher than the spectroscopic redshift. We further explore this difference in Section \ref{sec:speczs}. 

In \citet{tacchella2023}, the authors identify nine galaxies within 10 comoving Mpc ($212^{\prime\prime}$) of GN-z11 that have photometric redshifts between $z_a = 10 - 11$. Six of these sources are included in our $\Delta\chi^2 > 4$ sample \citep[JADES-GN\-+189.1162\-+62.22007, JADES-GN\-+189.07603\-+62.2207, JADES-GN\-+189.12549+62.2382, JADES-GN\-+189.08667\-+62.2395, JADES-GN\-+189.05971\-+62.2457, and JADES-GN\-+189.05166\-+62.2507, given as 465, 544, 4418, 4811, 6862, and 8597 in][]{tacchella2023},  while the other three sources \citep[JADES-GN-189.07355+62.2375, JADES-GN-189.19975+62.2703, and JADES-GN-189.05413+62.2179 given as 4155, 13543 and 62240 in][]{tacchella2023} are not in our final sample as these sources did not satisfy the requirement of having a flux SNR $> 5$ in at least two bands to the red of the potential Lyman-$\alpha$ break, or in the case of JADES-GN-189.07355+62.2375, this source has $z_a < 8$ with the updated photometry in this study. 

We want to highlight three galaxies seen in Figure \ref{fig:z_10_12_example_SEDs} because of their extended, somewhat complex morphologies. JADES-GS-53.13918-27.84849 ($z_a = 10.45$, first row, right column), an F115W dropout, has three components and spans $0.5^{\prime\prime}$, which is 2 kpc at this photometric redshift. We observe an increase in the F444W flux over what is seen at 3 - 4$\mu$m, which could either be a result of [OII]$\lambda$3727 emission at this redshift or evidence of a Balmer break. The F115W dropout JADES-GS-53.09872-27.8602 ($z_a = 10.69$, second row, right column) is the southern clump of two morphologically distinct components separated by $0.3^{\prime\prime}$ (1.2 kpc at this photometric redshift) in the rest-frame UV, which becomes less distinct at longer wavelengths. The northern clump, JADES-GS-53.09871-27.86016 ($z_a = 9.59$), is also in our sample, but the {\tt EAZY} fit prefers a lower photometric redshift which is consistent to within $1\sigma$. Finally, JADES-GS-53.07597-27.80654 ($z_a = 11.27$, third row, right column) consists of two, bright, connected clumps separated by $0.2^{\prime\prime}$ (580 pc at this photometric redshift). The sources are detected as separate clumps in the relatively shallower FRESCO F182M and F210M data as well. These sources could be interacting seed galaxies or star-forming clumps in the very early universe. 

\subsection{$z_{phot} > 12$ Candidates}\label{sec:z_gt_12}

\begin{figure*}
\centering
$z > 12$ Candidates, Part I \par\medskip
{\includegraphics[width=0.49\textwidth]{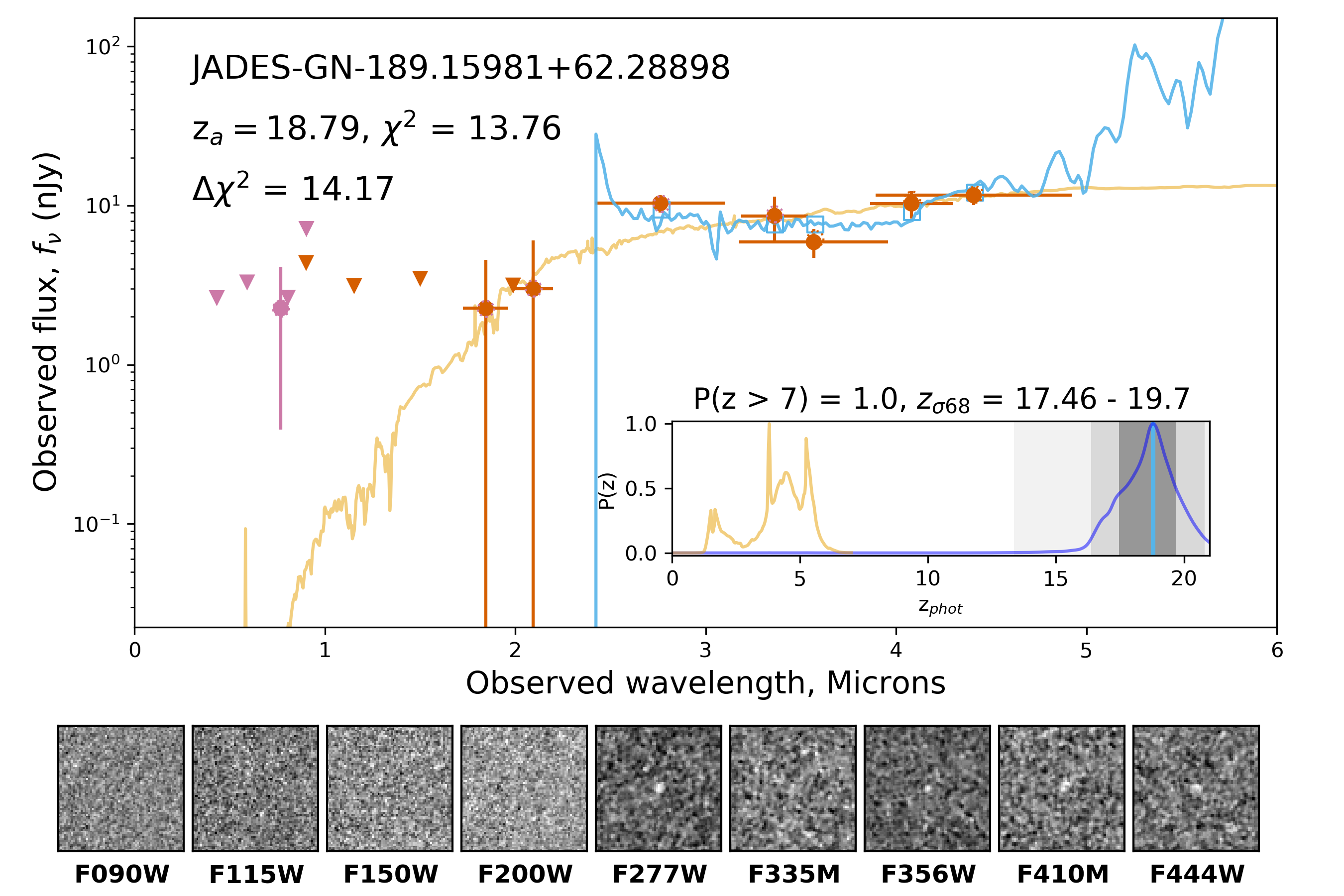}}  
{\includegraphics[width=0.49\textwidth]{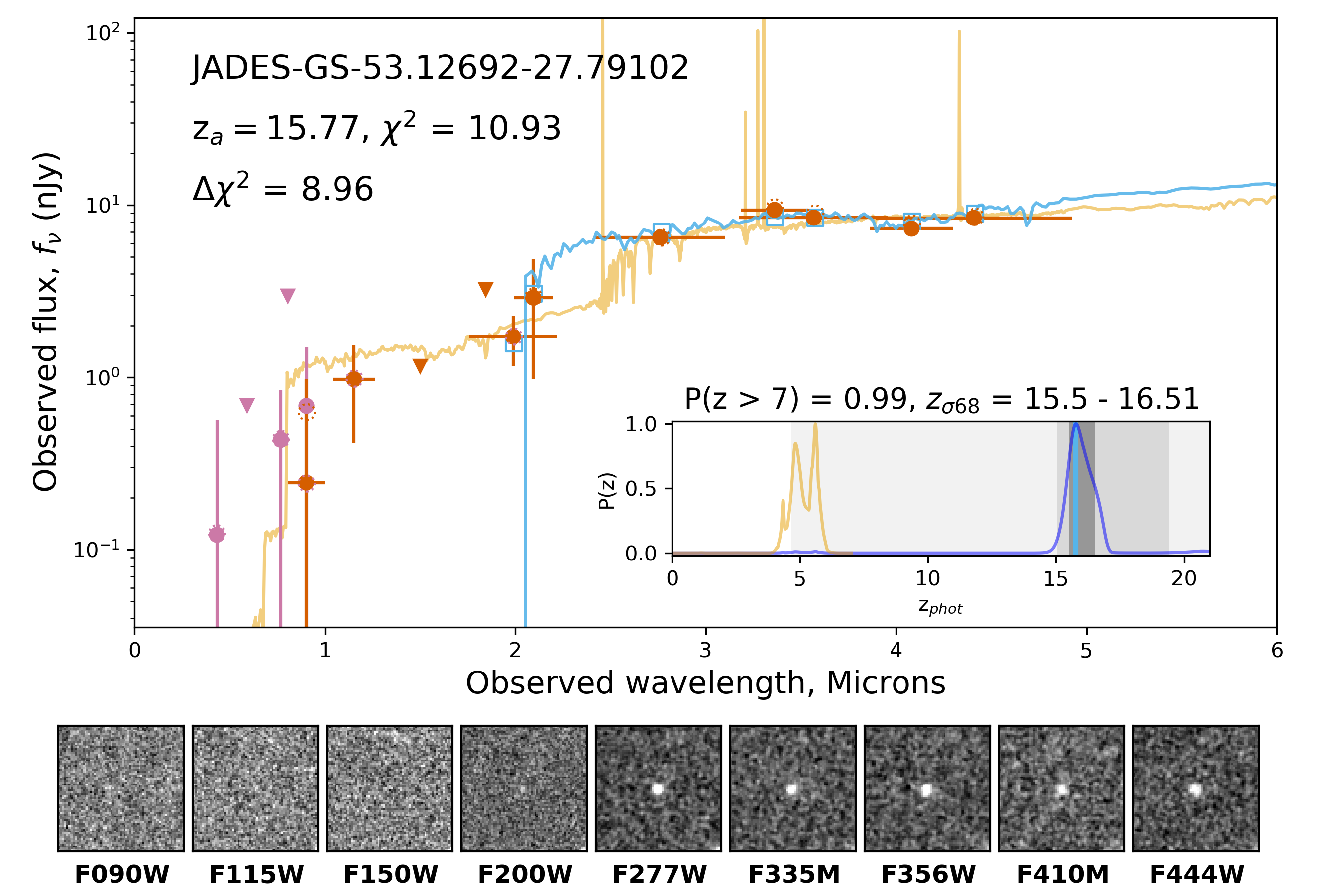}}\   
{\includegraphics[width=0.49\textwidth]{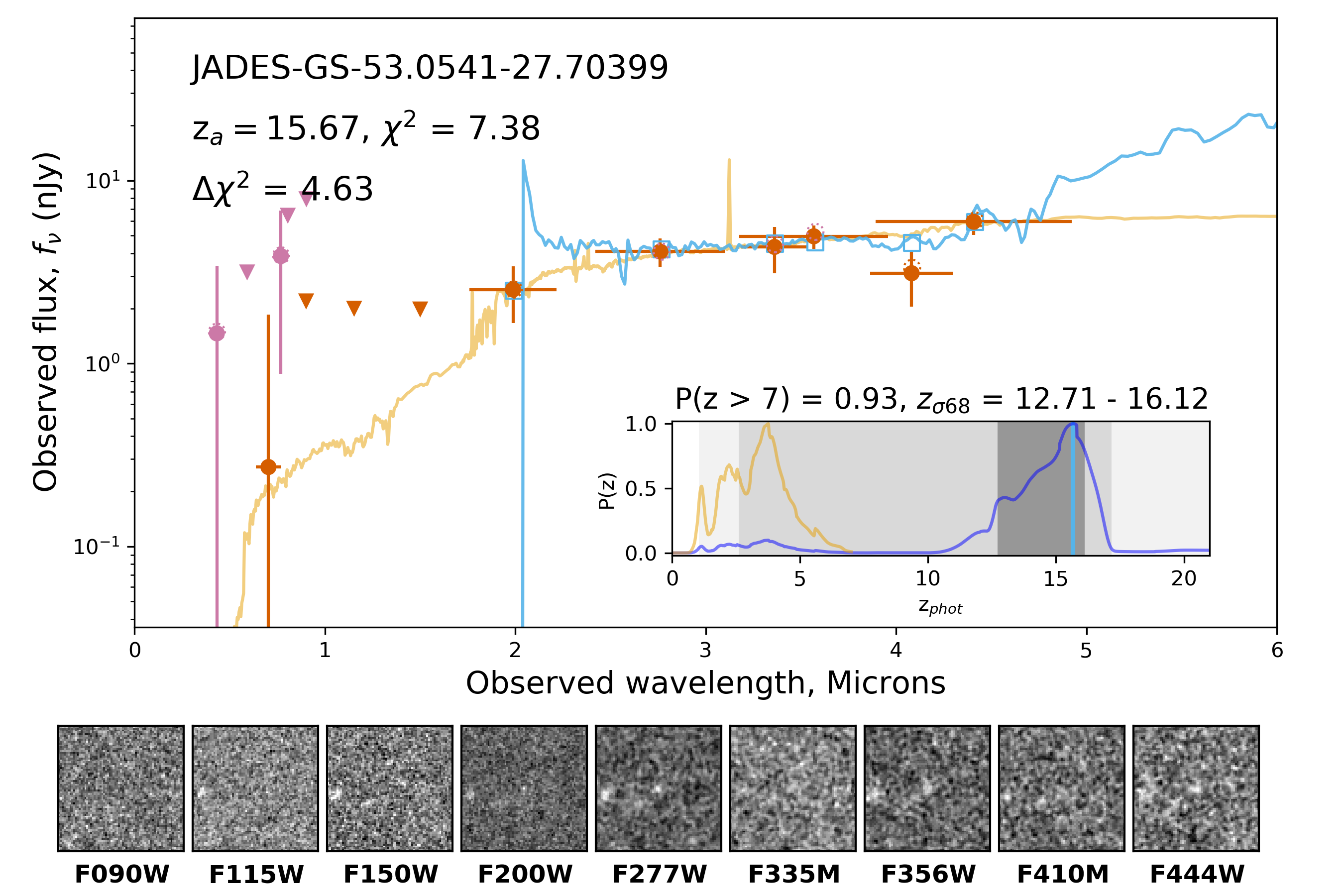}}\  
{\includegraphics[width=0.49\textwidth]{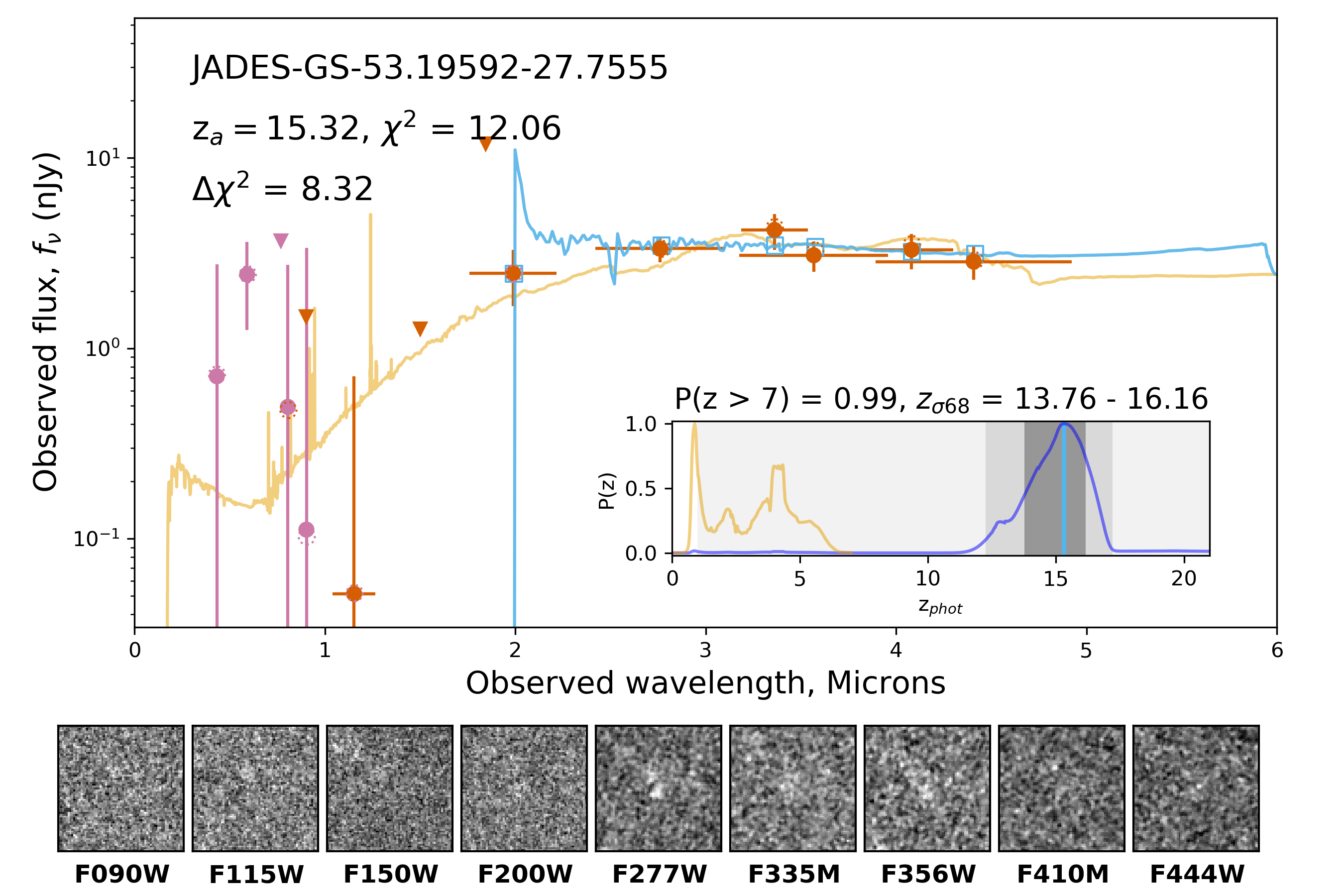}}\   
{\includegraphics[width=0.49\textwidth]{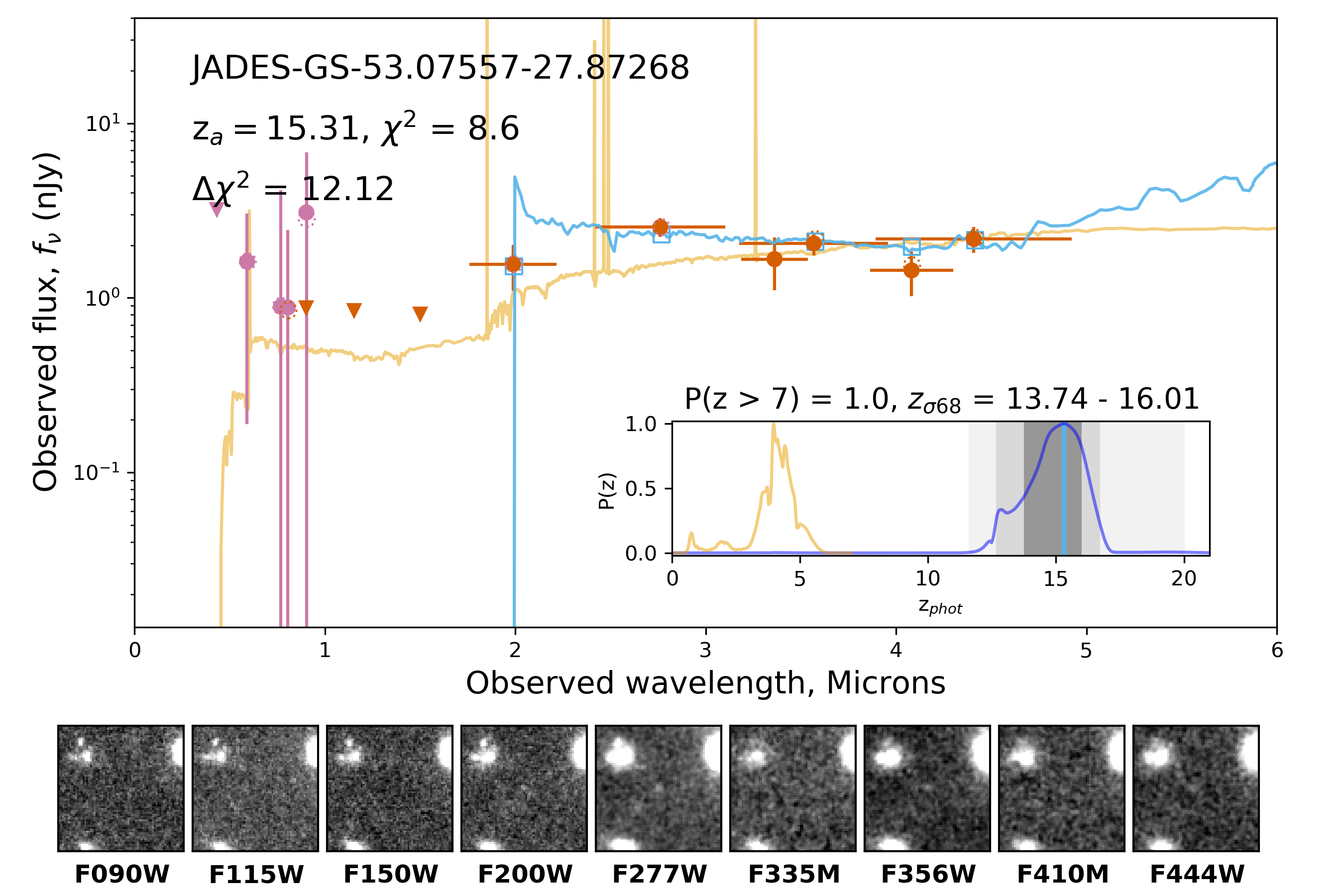}}\ 
{\includegraphics[width=0.49\textwidth]{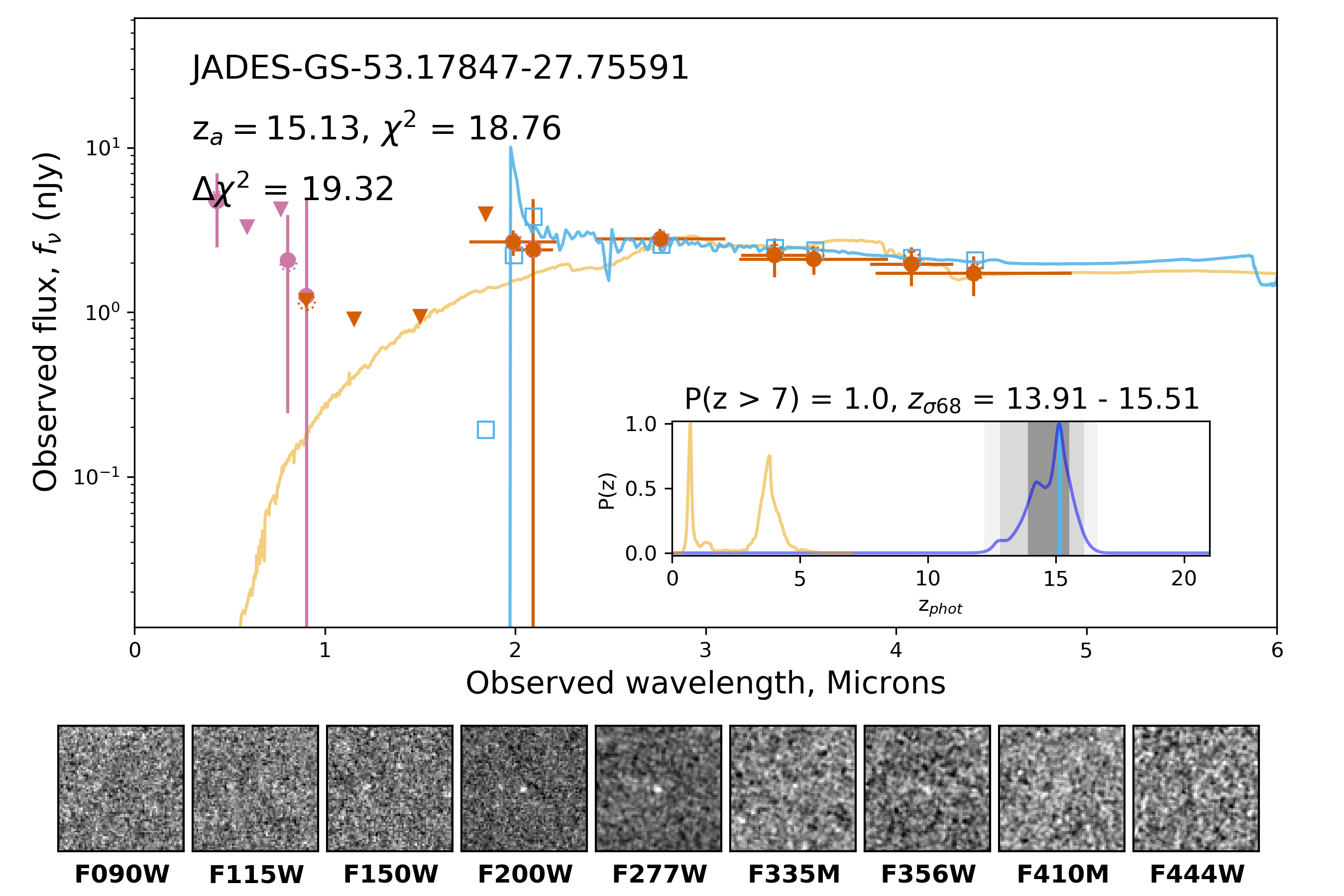}}\  
{\includegraphics[width=0.49\textwidth]{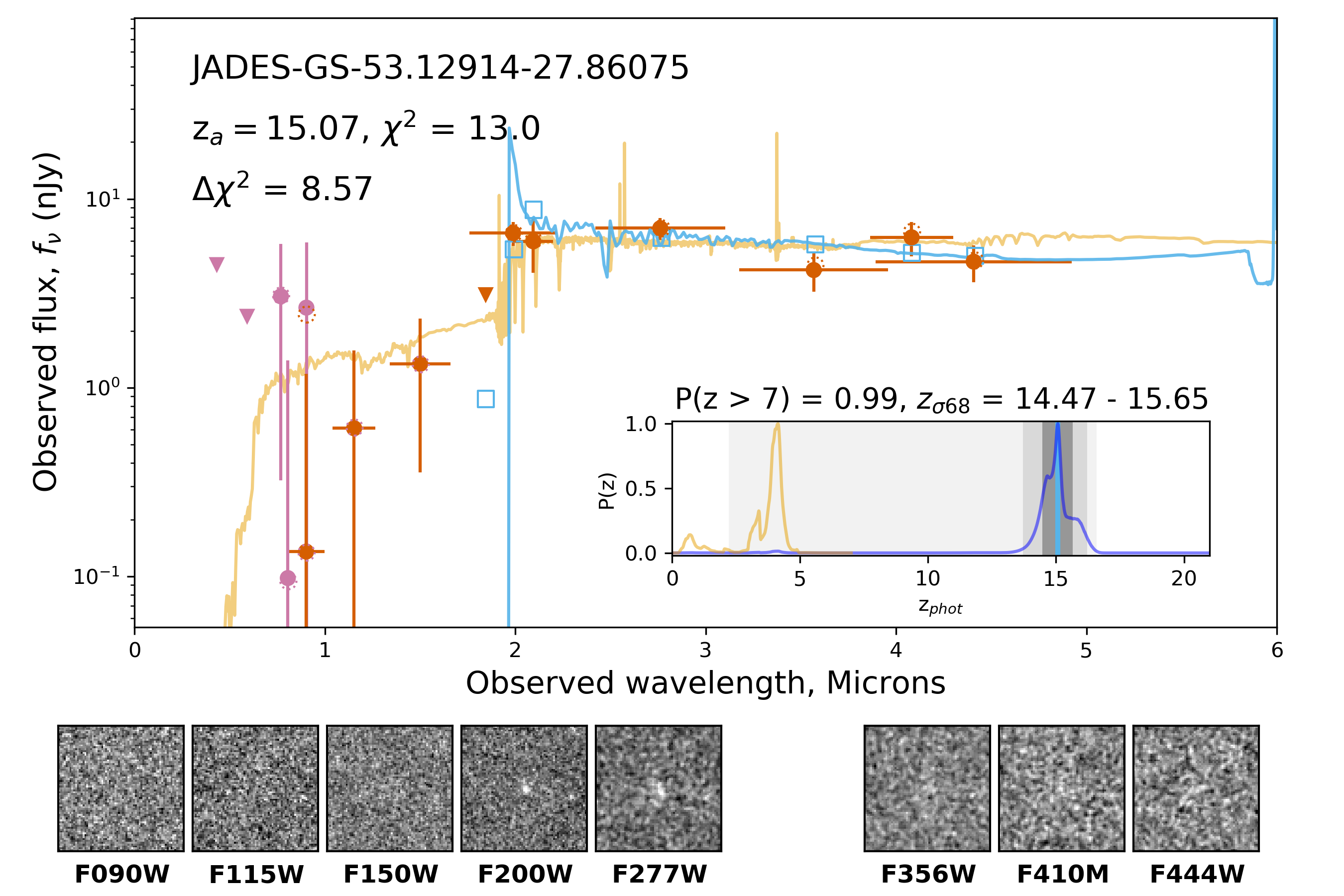}}  
{\includegraphics[width=0.49\textwidth]{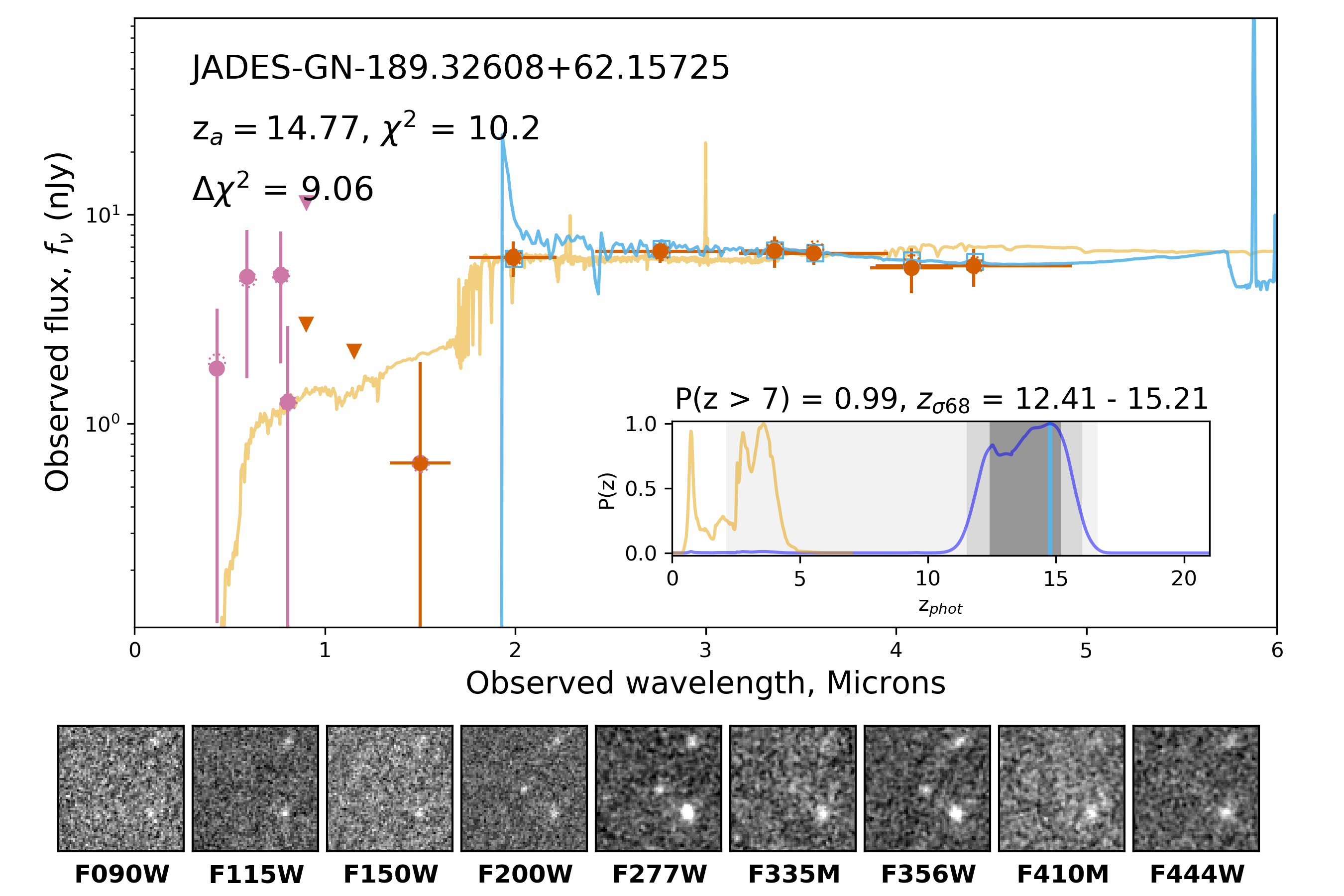}}\ 
\caption{
Example SEDs for eight candidate GOODS-S and GOODS-N galaxies at $z_a > 12$. The remainder of the objects are in Figures \ref{fig:z_gt_12_example_SEDs_pt2}, \ref{fig:z_gt_12_example_SEDs_pt3}, and \ref{fig:z_gt_12_example_SEDs_pt4} in Appendix \ref{sec:additional_tables}. In each panel, the colors, lines, and symbols are as in Figure \ref{example_SED_fit}.}
\label{fig:z_gt_12_example_SEDs_pt1}
\end{figure*}

We find 33 galaxies and  candidate galaxies across both the JADES GOODS-S (23 sources) and GOODS-N (10 sources) footprints at $z > 12$. We show their SEDs and thumbnails for eight examples in Figure \ref{fig:z_gt_12_example_SEDs_pt1}, and we show the remaining in Figures \ref{fig:z_gt_12_example_SEDs_pt2}, \ref{fig:z_gt_12_example_SEDs_pt3}, and \ref{fig:z_gt_12_example_SEDs_pt4} in Appendix \ref{sec:additional_tables}. For objects at these redshifts, the Lyman-$\alpha$ break falls in the F150W filter at $z = 12$, in between the F150W and F200W filters at $z = 13.2$, and in between the F200W and F277W filters at $z = 17.7$. The objects in our $z > 12$ sample, then, are a mixture of solid F150W dropouts and more tentative galaxies that show evidence for faint F200W flux associated with the Lyman-$\alpha$ break lying in that filter. 

Our sample in this redshift range includes the other two high-redshift spectroscopically-confirmed galaxies from \citet{robertson2022} and \citet{curtislake2022}, JADES-GS-z12-0 ($z_{spec} = 12.63^{+0.24}_{-0.08}$) and JADES-GS-z13-0 ($z_{spec} = 13.20^{+0.04}_{-0.07}$). We estimate {\tt EAZY} photometric redshifts for these targets of $z_a = 12.46$ for JADES-GS-z12-0, and $z_a = 13.41$ for JADES-GS-z13-0. While both photometric redshifts are quite uncertain due to the width of the bands used to probe the Lyman-$\alpha$ break, the range of uncertainties based on the {\tt EAZY} $\sigma_{68}$ redshifts are consistent with the spectroscopic redshifts. 

Because of the importance of these galaxies towards understanding galaxy formation in the very early universe, we will discuss the candidate galaxies in this redshift range individually, in order of decreasing photometric redshift. In our descriptions. We include brief discussions of two of the spectroscopically confirmed galaxies from \citet{robertson2022} and \citet{curtislake2022}, JADES-GS-z12-0 and JADES-GS-z13-0, but we refer the reader to these papers for more detailed discussions of these sources. 

\textbf{JADES-GN-189.15981+62.28898} ($z_a = 18.79$) This F200W dropout, the highest-redshift candidate in our sample, is clearly detected in multiple LW filters. There is no detection in the F200W filter, and we calculate a dropout color assuming a 2$\sigma$ upper-limit on the F200W flux of $m_{F200W} - m_{F277W} > 1.29$. While this source lies in the relatively shallower GOODS-N NW portion of the survey, the large $\Delta \chi^2$ provides strong evidence for this source being at high redshift. 

\textbf{JADES-GS-53.12692-27.79102} ($z_a = 15.77$) This is one of the more intriguing objects in our sample, as it is a relatively bright ($m_{F277W, Kron} = 29.37$) F150W dropout detected at greater than $16 \sigma$ in all of the detection bands. While there may be F115W flux observed in the thumbnail, it is only at SNR = 1.76. Caution should be exercised in adopting the derived redshift for this source as a result, since this object's fluxes are consistent with it being at $z = 5$. 

\textbf{JADES-GS-53.0541-27.70399} ($z_a = 15.67$) This F150W dropout has quite large photometric redshift uncertainties, but the $\sigma_{68}$ range is still consistent with it being at $z > 12$. The source $1^{\prime\prime}$ to the east is a potential F090W dropout with $z_a = 8.24$, but we measure $P(z < 7) = 0.68$ from the {\tt EAZY} fit, so it does not appear in our sample. 

\textbf{JADES-GS-53.19592-27.7555} ($z_a = 15.32$) This slightly extended F150W dropout has SNR $>5$ in three filters: F277W, F356W, and F444W. It is also over $3^{\prime\prime}$ away from any bright sources. Because of the non-detection in F150W, we estimate that $m_{F150W} - m_{F200W} > 0.74$ given a 2$\sigma$ upper limit on the observed F150W flux. 

\textbf{JADES-GS-53.07557-27.87268} ($z_a = 15.31$) This is one of the faintest $z > 12$ sources ($m_{F277W, Kron} = 30.04$), although it is observed at SNR~$> 5$ in three filters: F277W, F356W, and F444W. In the thumbnail we show how this candidate is surrounded by other, brighter sources. The sources to the northwest and southeast are both at $z_a \sim 1.0$, while the source with multiple components to the northeast is an F435W dropout at $z_a = 3.74$. 

\textbf{JADES-GS-53.17847-27.75591} ($z_a = 15.13$) This very compact F150W dropout is quite faint ($m_{F277W, Kron} = 29.79$), and is relatively isolated, with the nearest bright galaxy being almost $2^{\prime\prime}$ to the west. The lack of significant detections in the bands to the blue of the proposed Lyman-$\alpha$ break ($m_{F150W} - m_{F200W} > 1.14$) provides strong evidence of this source's photometric redshift. 

\textbf{JADES-GS-53.12914-27.86075} ($z_a = 15.07$) This F150W dropout is strongly detected (SNR $> 7$) in F200W and F277W, with $m_{F150W} - m_{F200W} = 1.73$.  It is detected at SNR $> 3$ in F210M, but not in F182M.

\textbf{JADES-GN-189.32608+62.15725} ($z_a = 14.77$) This is an F150W dropout $0.5^{\prime\prime}$ northeast of an F850LP dropout galaxy at $z_a = 5.2$, which is a slightly higher redshift than the potential secondary minimum in the $P(z)$ surface for this source. We measure $m_{F150W} - m_{F200W} = 2.46$, and find that this source is still at $z > 12$ within the $\sigma_{68}$ range on the photometric redshift.

\textbf{JADES-GS-53.02212-27.85724} ($z_a = 14.59$) This slightly diffuse F150W dropout has SNR $> 5$ in all of the bands where it is detected, and we measure $m_{F150W} - m_{F200W} = 2.91$. It is near the western edge of the JADES medium mosaic, and is $2.5^{\prime\prime}$ southeast of the star GOODS J033205.16-275124.2. 

\textbf{JADES-GN-189.23606+62.16313} ($z_a = 14.47$) This F150W dropout is faint ($m_{F277W, Kron} = 29.40$), but has a $7\sigma$ detection in F277W, and a $6\sigma$ detection with F356W. We measure a very red dropout color $m_{F150W} - m_{F200W} = 4.28$.

\textbf{JADES-GS-53.10763-27.86014} ($z_a = 14.44$) This is a faint, diffuse F150W dropout that is within $1.5^{\prime\prime}$ of a larger galaxy at $z_a = 0.9$. While there is evidence of F115W flux in the thumbnail, it is only at a SNR = 1.64.

\textbf{JADES-GS-53.07427-27.88592} ($z_a = 14.36$) This F150W dropout is not detected at SNR $< 0.8$ in each of the bands shortward of the potential Lyman-$\alpha$ break, and we measure $m_{F150W} - m_{F200W} = 4.05$. It is $1^{\prime\prime}$ away from a F435W dropout at $z_a = 4.31$, and could be associated with that source, as the secondary $P(z)$ peak indicates.

\textbf{JADES-GN-189.16733+62.31026} ($z_a = 14.33$) This  F150W dropout is very bright in F277W ($m_{F277W, Kron} = 27.28$), pushing it above the distribution at these redshifts as seen in Figure \ref{mag_vs_za}. It is within $0.5^{\prime\prime}$ of an F435W dropout at $z_a = 4.34$. As a result, the potential Lyman-$\alpha$ break for this object could be a Balmer break if these two sources are associated at similar redshifts. 

\textbf{JADES-GS-53.11127-27.8978} ($z_a = 14.22$) This source is an F150W dropout solid SNR $> 5$ detections in the LW JADES filters. The SW fluxes for this objects may be impacted by detector artifacts which are seen to the northwest and southeast of the source. 

\textbf{JADES-GN-189.24454+62.23731} ($z_a = 14.0$) This F150W dropout is only detected with $>5\sigma$ in F277W (SNR = 7.32) and F356W (SNR = 7.08), and we measure $m_{F150W} - m_{F200W} > 0.93$. 

\textbf{JADES-GS-53.06475-27.89024} ($z_a = 14.0$) This F150W dropout ($m_{F150W} - m_{F200W} > 2.27$) is detected in the F277W filter at 19.9$\sigma$, is found in the exceptionally deep GOODS-S JADES 1210 Parallel. In this region, there are no medium band observations from either JEMS or FRESCO for this source, and we do not see any detection in any of the WFC3 or ACS bands. This source a quite promising high-redshift candidate, with a $\Delta \chi^2 \sim 65$. 

\textbf{JADES-GS-53.14673-27.77901} ($z_a = 13.68$) This F150W dropout is quite well detected in multiple bands, including F182M, but it has a fairly broad P(z) surface, although the $\sigma_{68}$ values are consistent with $z > 12$ solutions. At $z_{phot} \sim 13 - 14$, the fits are more unconstrained due to the widths of the F150W and F200W photometric bands and the gap between them. 

\textbf{JADES-GS-53.14988-27.7765}, ($z_a = 13.41$) This source, also known as JADES-GS-z13-0, was spectroscopically confirmed to be at $z_{\mathrm{spec}} = 13.20$ in \citet{curtislake2022}, and the NIRCam photometry and morphology for the source was discussed in \citet{robertson2022}. The source is fairly bright ($m_{F200W, Kron} = 28.81$) with a strong observed Lyman-$\alpha$ break, and the $\sigma$68 range on the photometric redshift ($z_{\sigma68} = 12.92 - 14.06$) is in agreement with the observed spectroscopic redshift. 

\textbf{JADES-GN-189.27873+62.2112} ($z_a = 13.12$) This source has F182M and F210M fluxes boosted by flux from a diffraction spike. This source has an F150W detection at $2.4 \sigma$, potentially demonstrating that it is at a slightly lower redshift, as indicated by the large $P(z)$ distribution. 

\textbf{JADES-GN-189.11004+62.23638} ($z_a = 13.12$) The F182M and F210M fluxes for this F150W dropout ($m_{F150W} - m_{F200W} = 2.69$) are also boosted by a diffraction spike from a nearby star. There appears to be F115W flux at $2.14 \sigma$, but this is shifted to the northeast of the primary source  by $0.3^{\prime\prime}$ seen in F200W and F277W. 
 
\textbf{JADES-GS-53.06928-27.71539} ($z_a = 12.69$) This is a faint ($m_{F277W, Kron} = 29.73$) F150W dropout with a very red dropout color $m_{F150W} - m_{F200W} = 4.97$.

\textbf{JADES-GN-189.33638+62.16733} ($z_a = 12.53$) While this F150W dropout is faint ($m_{F277W, Kron} = 30.01$), the fit indicates a blue UV slope and the object is detected in multiple filters at $>5 \sigma$, with $\Delta \chi^2 = 12.94$. 

\textbf{JADES-GS-53.18129-27.81043} ($z_a = 12.52$) This very faint ($m_{F277W, Kron} = 30.11$) F150W dropout has solid 8$\sigma$ detections in F200W and F277W, and can be seen in the F356W thumbnail at 4$\sigma$. 

\textbf{JADES-GS-53.08468-27.86666} ($z_a = 12.48$) This F150W dropout ($m_{F150W} - m_{F200W} = 1.96$) has a slightly redder potential UV slope, and this may be a low-redshift dusty interloper.

\textbf{JADES-GS-53.16635-27.82156} ($z_a = 12.46$) This galaxy, also known as JADES-GS-z12-0, was originally spectroscopically confirmed to lie at $z_{\mathrm{spec}} = 12.63$ in \citet{curtislake2022}, and the NIRCam photometry and morphology for the source was discussed in \citet{robertson2022}. Recent, deeper NIRSpec observations for this source showed \ion{C}{3}]$\lambda\lambda$1907,1909 line emission, indicating a spectroscopic redshift of $z_{\mathrm{spec}} = 12.479$ \citep{deugenio2023}. This source is a bright ($m_{F277W, Kron} = 28.64$) F150W dropout ($m_{F150W} - m_{F200W} = 2.01$) with a 26$\sigma$ detection in F277W, and is observed at the $4-6\sigma$ level in the relatively shallow F182M and F210M filters. 

\textbf{JADES-GS-53.02868-27.89301} ($z_a = 12.39$) This source is an F150W dropout with strong detections (SNR $> 5$) in each filter to the red of the potential Lyman-$\alpha$ break. 

\textbf{JADES-GS-53.10469-27.86187} ($z_a = 12.27$) This source is an F150W dropout with strong detections in F200W and F277W. The fluxes at F115W and F150W are observed at 1.55$\sigma$ and 1.39$\sigma$ significance, respectively. 

\textbf{JADES-GN-189.27641+62.20724} ($z_a = 12.19$) This source is well detected in F200W and F277W (SNR $> 10$ in both filters), but is quite faint at longer wavelengths. There is a source $1.5^{\prime\prime}$ the southeast of the target at $z_{spec} = 2.44$ \citep{reddy2006}, so we caution that the observed Lyman-$\alpha$ break for the high-redshift candidate may be a Balmer break at 1.5$\mu$m. 

\textbf{JADES-GN-189.09217+62.25544} ($z_a = 12.16$) This is a bright ($m_{F277W, Kron} = 28.53$) F150W dropout. The F150W detection is at a SNR $= 2.03$, while the F115W flux is only measured at the 1.55$\sigma$ level. 

\textbf{JADES-GS-53.19051-27.74982} ($z_a = 12.08$) This is a bright ($m_{F277W, Kron} = 28.72$) F115W dropout with multiple filters with $>10\sigma$ detections. 

\textbf{JADES-GS-53.14283-27.80804} ($z_a = 12.06$) This object is an F150W dropout that is primarily seen in F200W (SNR = 5.89) and F277W (SNR = 6.72). The fit very strongly favors the high-redshift solution, and it does not seem be associated with the nearby galaxy to the east, an F814W dropout at $z_a = 5.98$ with [OIII]$\lambda$5007 potentially boosting the F335M flux. 

\textbf{JADES-GS-53.18936-27.76741} ($z_a = 12.05$) This source is a very faint, slightly diffuse F150W dropout. While the $\Delta \chi^2$ is still in favor of the $z > 8$ fit, the lower-redshift solution would help explain the boosted F210M flux as potentially arising from an [OIII] emission line, although the F210M flux is only significant at 2.6$\sigma$. 

We caution that photometric redshifts at $z > 12$ are quite uncertain, and that our sources are observed down to very faint magnitudes, and thus require deep spectroscopic follow-up to confirm. In many cases we also raise the possibility that the source is potentially associated with a nearby galaxy at lower redshift. For the GOODS-S sources, continued observations extending both the size of the JADES Medium region and the depth of JADES Deep planned for Cycle 2 as part of JADES will help to provide evidence as to whether these sources are truly at high-redshift or not.

\subsection{$z > 8$ Candidates with $\Delta\chi^2 < 4$}\label{sec:deltachisqlt4}

In the previous sections we explored those objects for which the {\tt EAZY} fit strongly favors a high-redshift solution. The fits to these sources at their proposed photometric redshifts indicate strong Lyman-$\alpha$ breaks and more robust upper limits on the photometric fluxes blueward of the break. For cases where the observed HST/ACS or short-wavelength JWST/NIRCam fluxes have higher uncertainties (for fainter objects or those objects in shallower parts of the GOODS-S or GOODS-N footprint), fits at $z < 7$ are less strongly disfavored, leading to values of $\Delta\chi^2 < 4$. 

We selected candidate $z > 8$ galaxies in our sample that satisfy our criteria outlined in Section \ref{sec:highzselection}, but where $\Delta\chi^2 < 4$. While the bulk of the output {\tt EAZY} $P(z)$ indicates that the galaxy is at high-redshift ($P(z > 7) > 0.7$), the minimum $\chi^2$ for the $z < 7$ solution is more similar to the overall minimum $\chi^2$ at $z > 8$. In this section we explore these targets, as they represent a non-insignificant number of candidates. 

Following the initial selection of these objects, they were visually inspected following the same routine as for the $\Delta\chi^2 > 4$ objects, where an object was removed from the sample if the majority of reviewers flagged it for rejection. Our final sample consists of 163 candidates in GOODS-S and 64 candidates in GOODS-N, for a total of 227 objects. These objects are also plotted with lighter symbols in Figure \ref{fig:GS_GN_footprint}. While these sources span the full redshift range of the $\Delta\chi^2 > 4$ sample, the median F277W Kron magnitude is 29.47 for the GOODS-S objects and 29.11 for the GOODS-N objects, fainter than the median F277W magnitudes of the $\Delta\chi^2 > 4$ sources (29.25 for GOODS-S and 28.62 for GOODS-N). This is expected as $\Delta\chi^2$ is strongly dependent on the observed flux uncertainties for each source.  

\begin{figure*}
\centering
Example $\Delta\chi^2 < 4$ Candidates\par\medskip
{\includegraphics[width=0.49\textwidth]{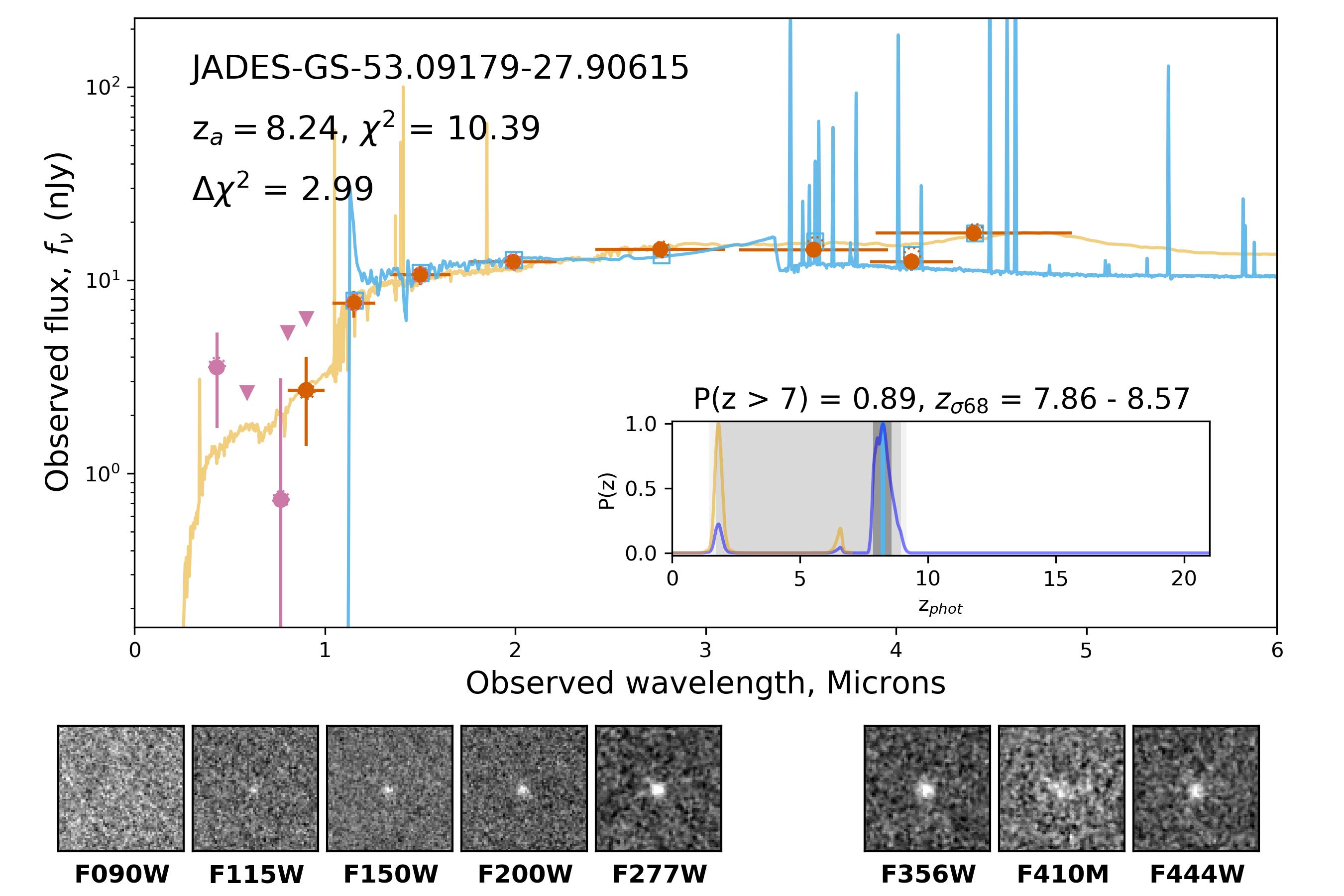}}\ 
{\includegraphics[width=0.49\textwidth]{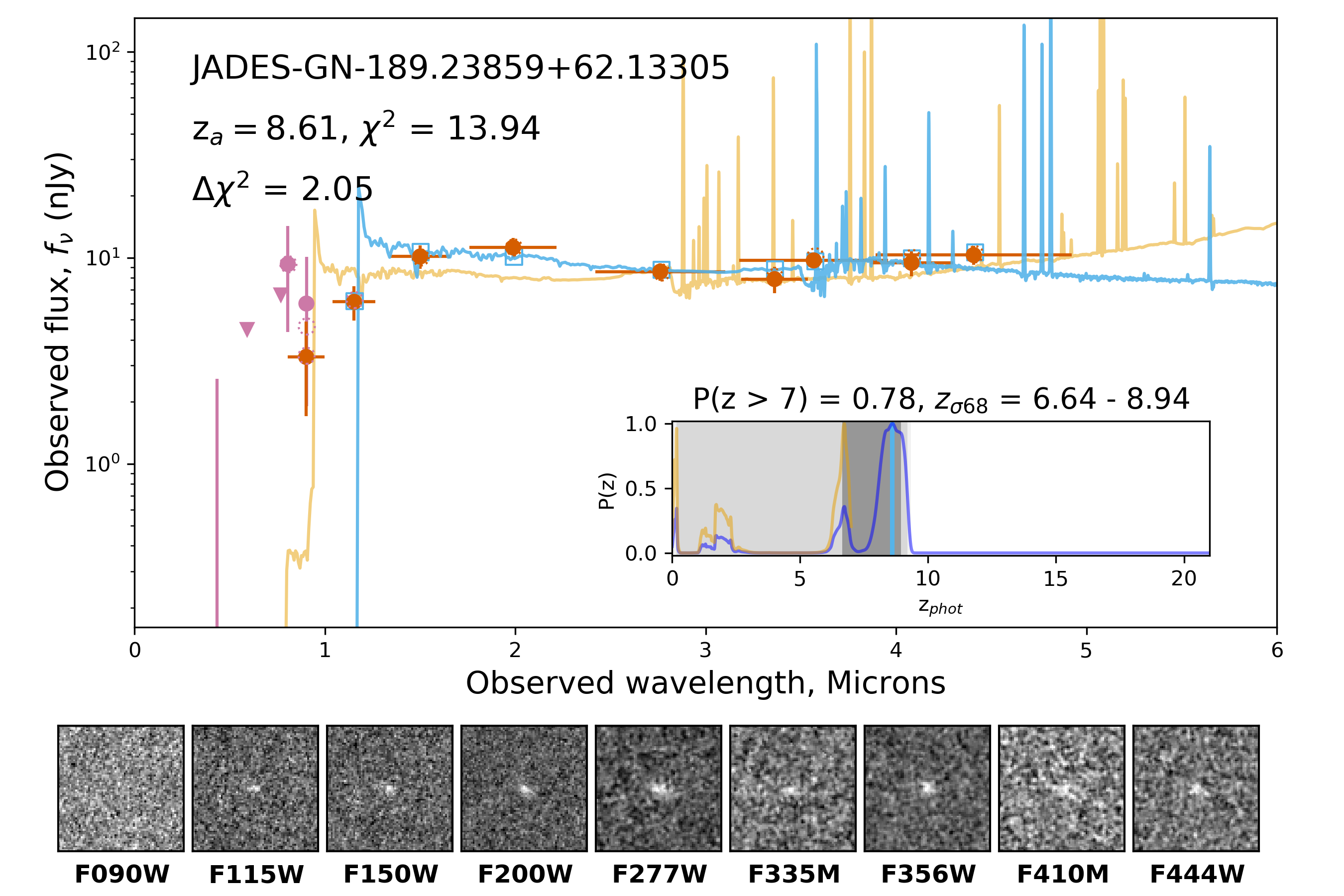}}\ 
{\includegraphics[width=0.49\textwidth]{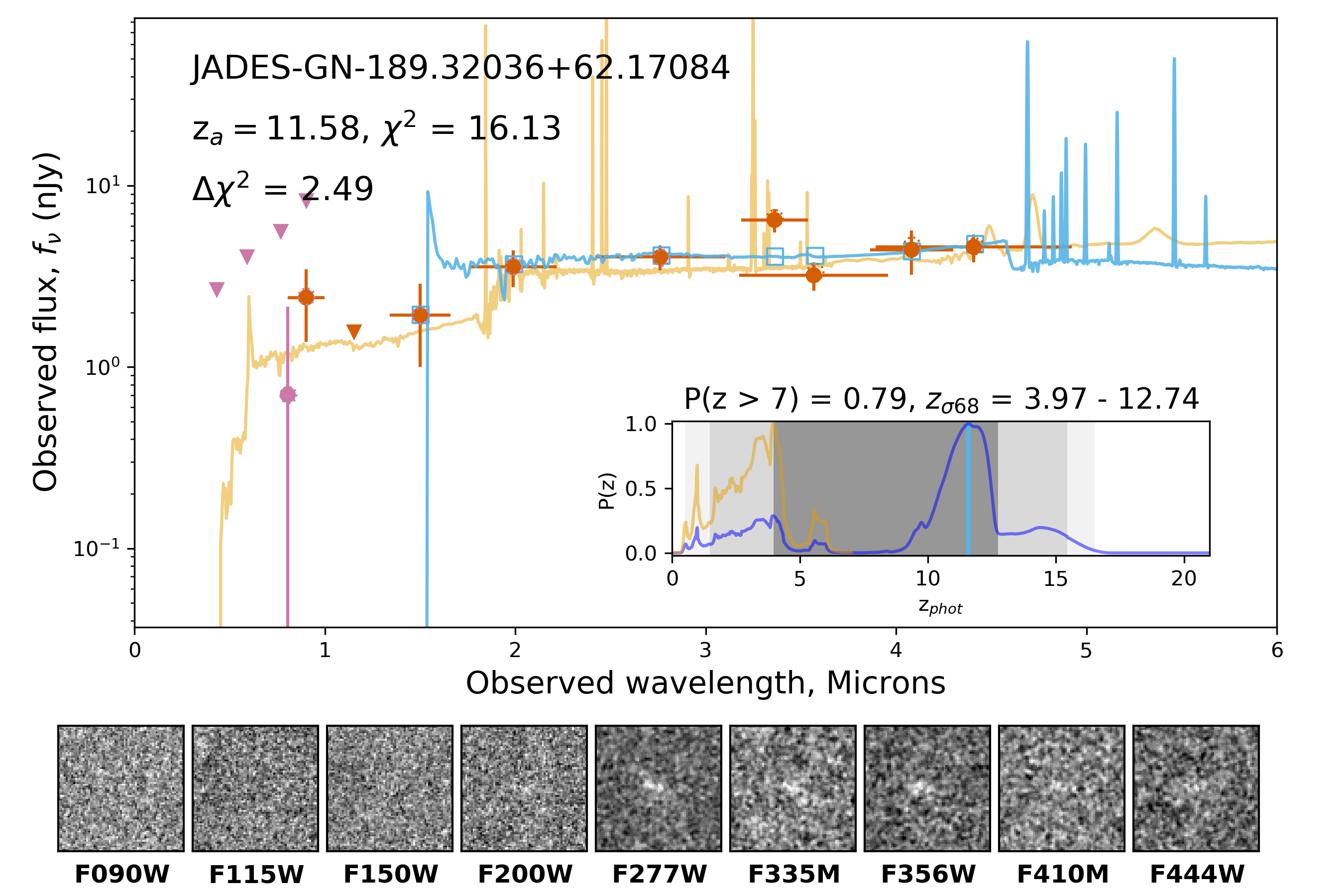}}\ 
{\includegraphics[width=0.49\textwidth]{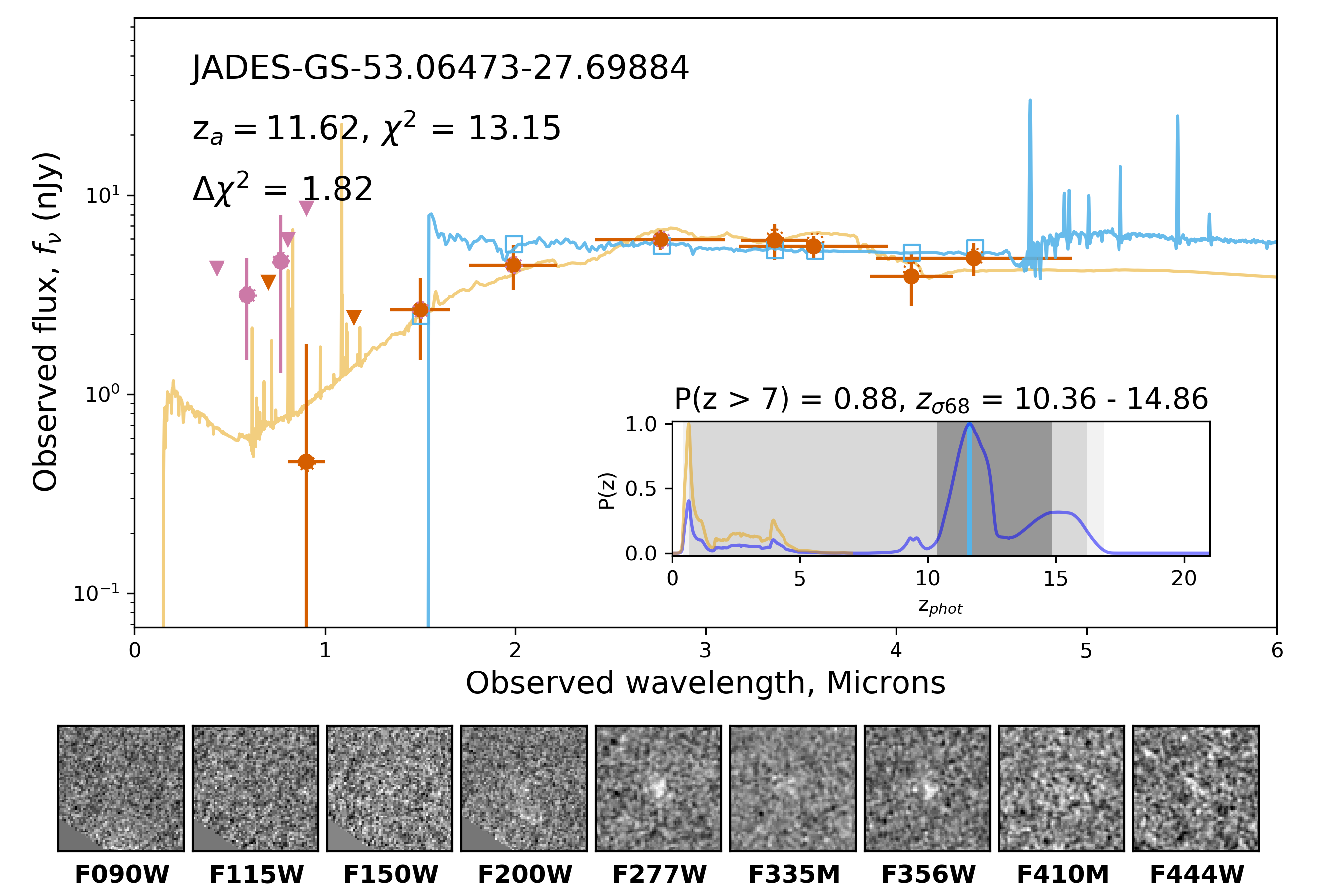}}\ 
{\includegraphics[width=0.49\textwidth]{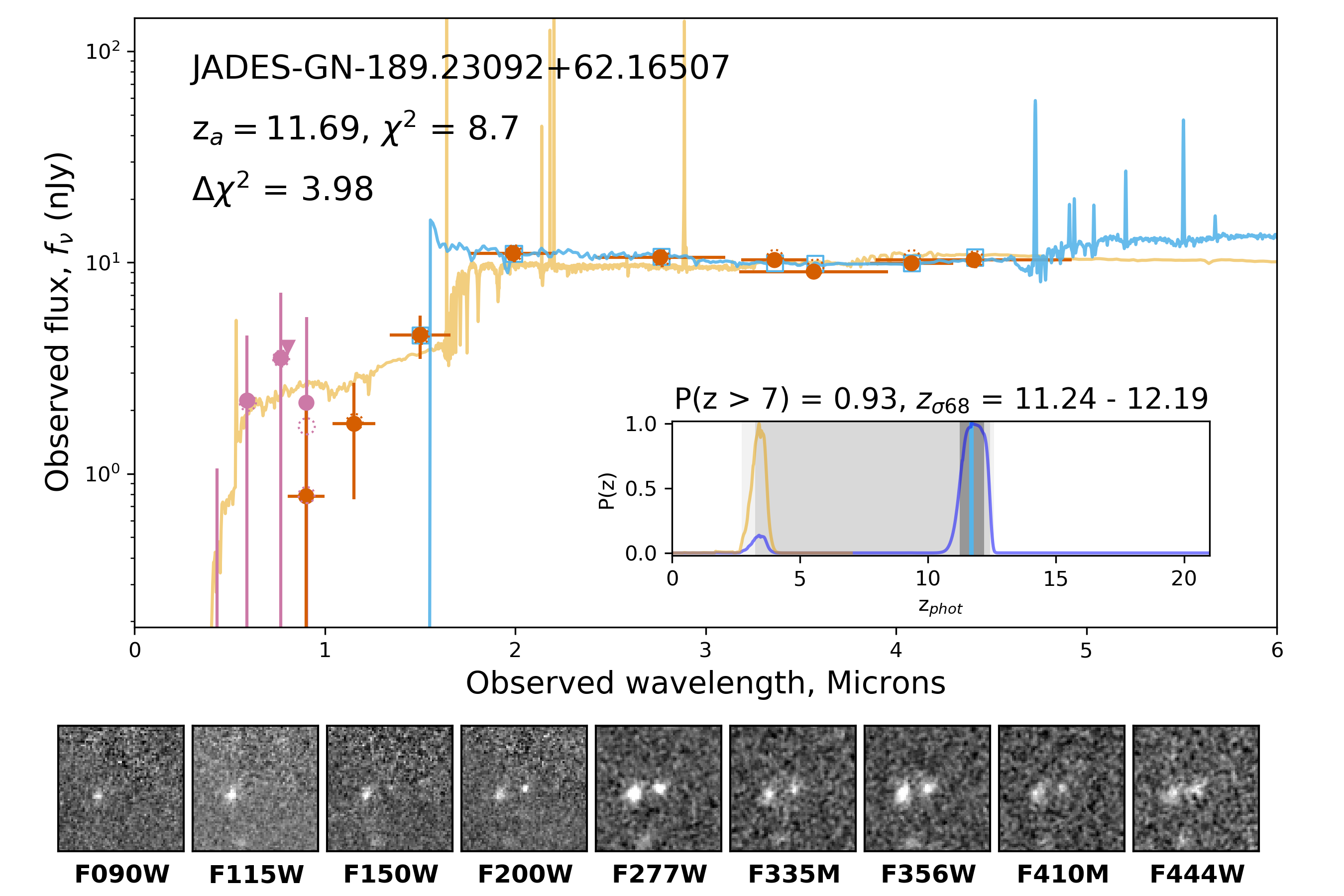}}\ 
{\includegraphics[width=0.49\textwidth]{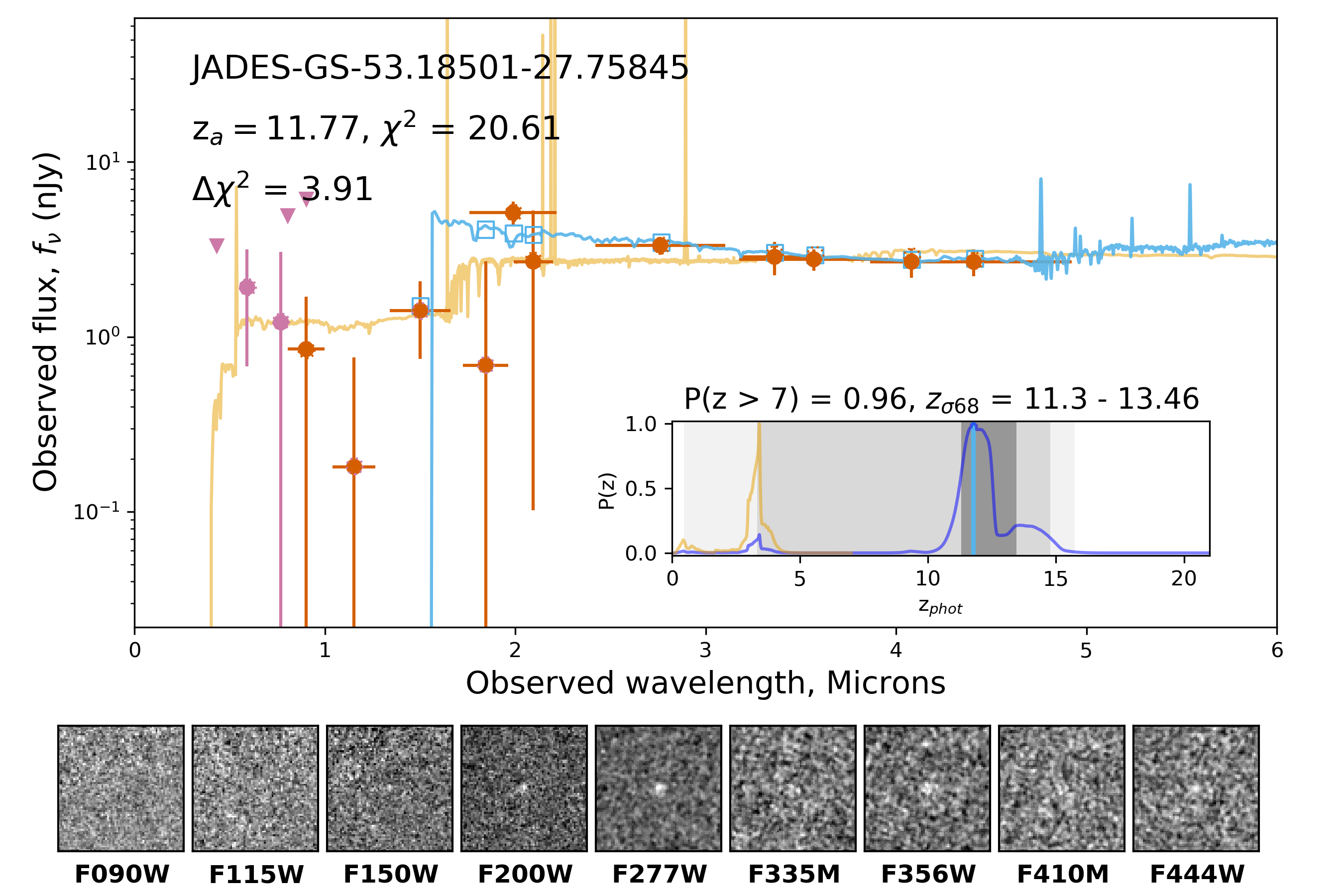}}\ 
{\includegraphics[width=0.49\textwidth]{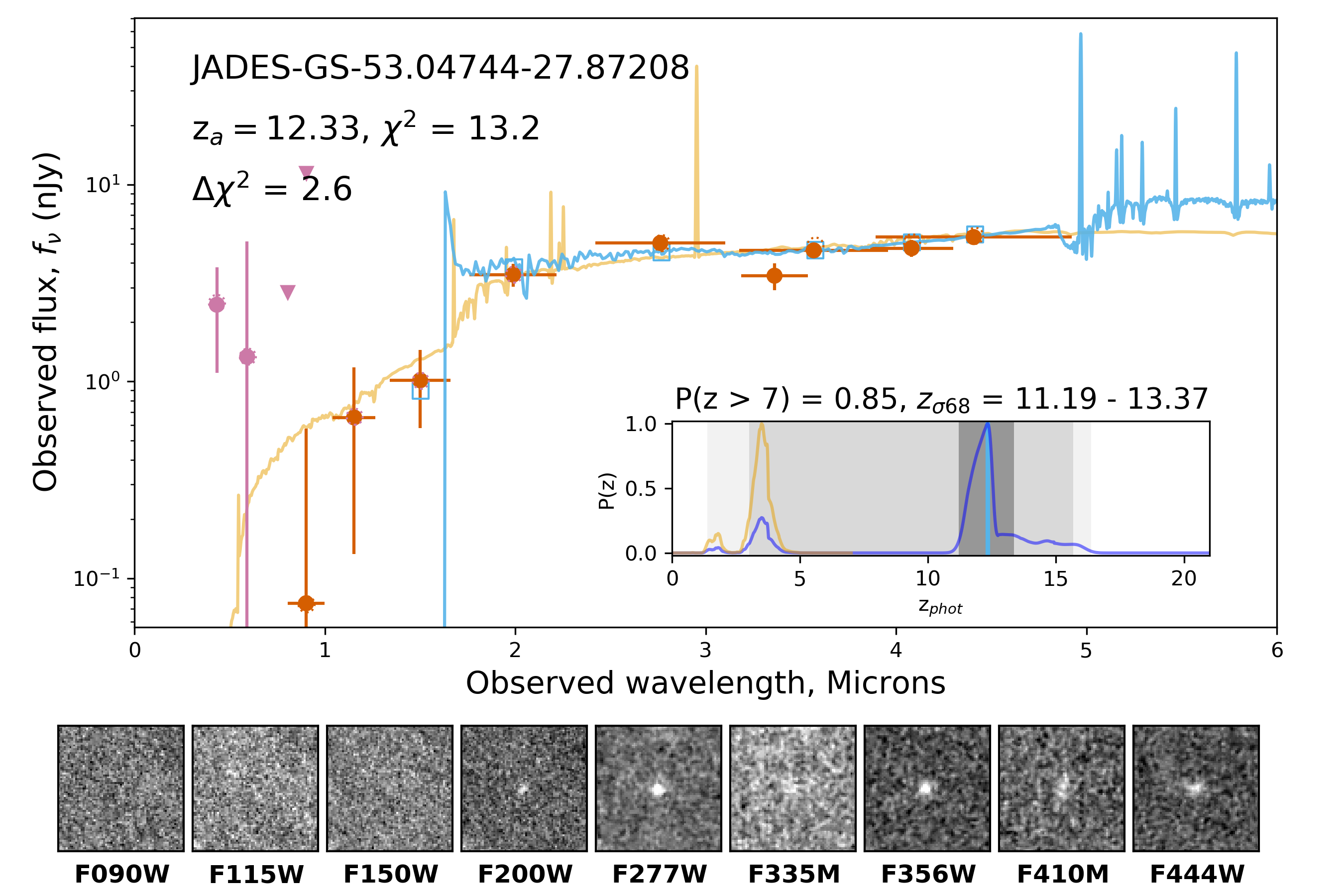}}\ 
{\includegraphics[width=0.49\textwidth]{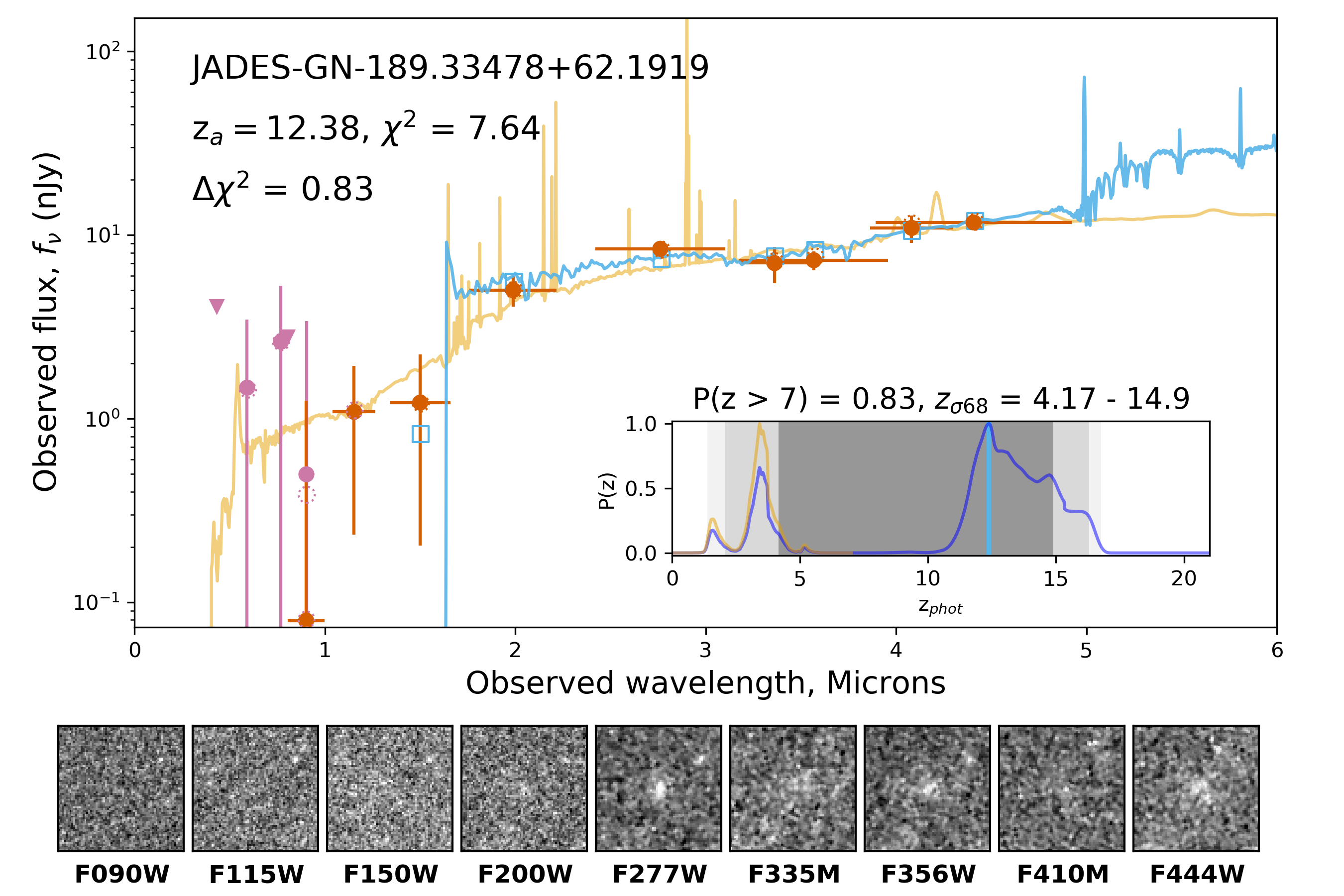}} 
\caption{
Example SEDs for eight $\Delta\chi^2 < 4$ candidate GOODS-S and GOODS-N galaxies. In each panel, the colors, lines, and symbols are as in Figure \ref{example_SED_fit}.}
\label{fig:chisq_lt_4_example_SEDs}
\end{figure*}

In Figure \ref{fig:chisq_lt_4_example_SEDs} we highlight some targets from both GOODS-S and GOODS-N with $\Delta\chi^2 < 4$, demonstrating the variety of targets in this subcategory. The median F090W flux (measured in a $0.2^{\prime\prime}$ diameter aperture) for the full sample of GOODS-S and GOODS-N $z > 8$ candidates is -0.02 nJy (the distribution is consistent with 0 nJy), while the median F090W flux for the combined sample of $\Delta\chi^2 < 4$ targets is 0.39 nJy. Targets like JADES-GS-53.04744-27.87208 ($z_a = 12.33$) and JADES-GN-189.33478+62.1919 ($z_a = 12.38$) have faint F115W and F150W flux measurements (with high uncertainties) consistent with dusty $z \sim 4$ solutions with strong emission lines, similar to CEERS-93316 observed in \citet{haro2023}. Many of these objects are also limited by the lack of deep HST/ACS data; JADES-GS-53.04744-27.87208 only has coverage with F435W and F606W due to its position in the southwest of the JADES GOODS-S footprint.

We include these sources and their fluxes to aid in the selection of high-redshift galaxies in future deep JWST surveys with different filter selection and observational depths. While a larger number of these objects may be lower redshift interlopers masquerading as $z > 8$ galaxies, this sample may serve as a pool of additional sources to be placed on multi-object slit-masks in follow-up spectroscopic campaigns to confirm source redshifts. In addition, these objects are helpful with calibrating template sets as they have colors that can be fit with models at low and high redshift.

\subsection{$z > 8$ Candidates Proximate to Brighter Sources}\label{sec:flag_bn_2_objects}

In addition to exploring the sources with $\Delta\chi^2 < 4$, we also looked at objects (at all $\Delta\chi^2$ values) that were near bright sources, being within $0.3^{\prime\prime}$ or the bounding box of a source with ten times greater brightness. There is an increased probability that proximate sources are at similar redshifts, and so we can compare the $\chi^2$ distribution of the brighter galaxy to that of the fainter high-redshift candidate. At faint magnitudes, Balmer breaks and strong line emission can be lead to sources being mistaken for higher redshift objects, which be be seen by looking at the $\chi^2$ minima for these sources. In addition, being so close to a bright source can potentially introduce flux into the circular aperture photometry and change the observed colors of the candidate galaxy and the shape of the SED. We went through the same visual classification procedure for these sources as we did for the full sample, and ended up with 41 candidates (30 have $\Delta\chi^2 > 4$) in GOODS-S and 17 candidates (14 have $\Delta\chi^2 > 4$) in GOODS-N. These sources have a median F277W Kron magnitude of $m_{AB} = 28.61$ for those in GOODS-S, and $m_{AB} = 27.50$ for those in GOODS-N, and range in redshift between $z = 8.0 - 16.7$. 

We want to specifically highlight some of the higher-redshift candidates from this subsample. Notably, there are three galaxies at $z_a > 12$: JADES-GS-53.08016-27.87131 ($z_a = 16.74$), JADES-GS-53.09671-27.86848 ($z_a = 12.03$), and JADES-GN-189.23121+62.1538 ($z_a = 12.16$). JADES-GS-53.08016-27.87131 has $\Delta\chi^2 < 4$, although this source is the most interesting due to its photometric redshift. This source has a very faint F200W detection ($\mathrm{SNR} > 4$ in an $0.2^{\prime\prime}$ diameter aperture), but is relatively bright at longer wavelengths (with an F277W AB magnitude of 29.20). This source lies $3^{\prime\prime} - 4^{\prime\prime}$ north of a pair of interacting galaxies at $z_{spec} = 1.1$ \citep{bonzini2012, momcheva2016}, and flux from the outskirts of these galaxies may be contributing to the aperture photometry for this object, leading to an artificially red UV slope. We caution that this source may be a stellar cluster associated with this pair or the dusty galaxies that are to the north of its position.

There are five objects in this subsample that have spectroscopic redshifts from FRESCO: JADES-GS-53.12001-27.85645 ($z_a = 8.33$, $z_{spec} = 7.652$), JADES-GS-53.13341-27.83909 ($z_a = 8.18$, $z_{spec} = 8.217$), JADES-GS-53.07688-27.86967 ($z_a = 8.57$, $z_{spec} = 8.270$), JADES-GS-53.10107-27.86511 ($z_a = 8.49$, $z_{spec} = 8.195$), and JADES-GN-189.27457+62.21053 ($z_a = 8.03$, $z_{spec} = 8.015$). These sources have [OIII]$\lambda$5007 line detections from FRESCO, demonstrating that while this class of sources may be associated with their nearby brighter neighbors, there are possible high-redshift galaxies among them. 

\subsection{Stellar Contamination}\label{sec:browndwarfs}

One primary source of contamination for high-redshift galaxy samples are low-mass Milky Way stars and brown dwarfs which at low temperatures can have near-IR colors similar to high-redshift galaxies, and many studies have explored the selection of these sources from within extragalactic surveys \citep{ryan2005, caballero2008, wilkins2014, finkelstein2015, ryanreid2016, hainline2020}. Candidate brown dwarfs have been observed in extragalactic surveys, such as GLASS \citep{nonino2023}. To explore whether our sample contains objects with a high probability of being a possible brown dwarf, we looked at the sizes of the targets in our sample and their fits to stellar models and observed brown dwarf SEDs.

We fit the targets in our sample using the {\tt jades-pipeline} profile fitting software, which utilizes the python {\tt lenstronomy} package \citep{birrer2018, birrer2021}. We fit each source, as well as the other nearby sources within $2^{\prime\prime}$ and up to two magnitudes fainter than the primary galaxy as Sersic profile. Objects that are fainter or farther from the source are masked instead of fit. We use the final residuals to determine the goodness of fit for each source. We measured the sizes using the NIRCam F444W mosaic, as brown dwarfs are bright and unresolved at 4 $\mu$m \citep{meisner2020}. To determine whether an object was unresolved, we looked at those where the observed half-light radius for each source was smaller than the NIRCam long-wavelength channel pixels size ($0.063^{\prime\prime}$). We note that the maximum half-light radius measured using the same procedure on a sample of stars and brown dwarfs in GOODS-S and GOODS-N was $0.02^{\prime\prime}$, but adopted a larger limit to broaden our search. These sources were identified using both photometric fits to theoretical brown-dwarf models and identification of sources with proper motions compared to HST observations, and will be described further in Hainline et al. (in prep).

We fit the NIRCam photometry of our $z > 8$ candidates using both the SONORA cloud-free brown dwarf models from \citet{marley2018} as well as a sample of observed brown-dwarf observations from the SpeX Prism Spectral Library\footnote{Compiled by Adam Burgasser and found online at \url{http://pono.ucsd.edu/~adam/browndwarfs/spexprism/}.}. As the SpeX spectra, in general, are only observed to 2.5$\mu$m, we took a group of objects across the temperature range that were detected in the Wide Field Infrared Survey Explorer (WISE) allWISE catalog \citep{cutri2014} and used their photometry at 3.4 and 4.6 $\mu$m to create an extrapolated spectra out to 5$\mu$m, which we used to estimate NIRCam photometry, following \citet{finkelstein2023}. We supplemented these with empirical NIRCAM SEDs of M dwarfs, obtained from a selection of extremely compact objects in F115W-F200W color magnitude space, consistent with stellar evolutionary models and JWST observations of globular clusters \citep[][B. Johnson priv. comm]{weisz2023}. The full set of model photometry were then fit to the observed NIRCam $0.2^{\prime\prime}$ diameter aperture photometry for the $z > 8$ candidate galaxies using a $\chi^2$ minimization approach. We compared the resulting $\chi^2$ minima for the stellar fits to those from the {\tt EAZY} galaxy templates, and if an object had a $\Delta\chi^2 < 4$ between the galaxy model fit and the stellar fit, it was flagged as a brown dwarf candidate. 

We find 303 objects in our $z > 8$ sample with $\Delta\chi^2 > 4$ that are unresolved (42\% of the full $z > 8$, $\Delta\chi^2 > 4$ sample), with a half-light radius less than $0.063^{\prime\prime}$, while only six objects across both fields have fits to stellar models with {\tt EAZY} $\chi^2_{\mathrm{min}}$ within $\Delta\chi^2 < 4$ (two of these sources have lower $\chi^2$ values with the brown dwarf fits). We flag the sources in the online table if they satisfy either of these requirements. Of these objects, only two sources are both unresolved and have stellar fits within $\Delta\chi^2 < 4$: JADES-GS-53.0353-27.87776 ($z_a = 10.82$) and JADES-GN-189.19772+62.25697 ($z_a = 8.61$). The latter source, which is detected with HST WFC3/IR, was identified as the Y-dropout candidate GNDY-6474515254 in \citet{bouwens2015}. While this object has evidence for being a brown dwarf, FRESCO identified both [OIII]$\lambda\lambda$5007,4959 emission lines ($z_{spec} = 8.28$), ruling out the brown dwarf hypothesis.

There are additional brown dwarf candidates in the $z > 8$ candidate galaxies with $\Delta\chi^2 < 4$ identified in \S \ref{sec:deltachisqlt4}. We find 83 objects with an F444W half-light radius less than $0.063^{\prime\prime}$, and 17 objects with stellar fits within $\Delta\chi^2 < 4$ (6 of these sources have lower $\chi^2$ values with the brown dwarf fits). We caution that because of the larger flux uncertainties for these objects, it is more likely that models would fit these data with comparable $\chi^2$ values, but we include flags in the online table in these cases. Only five of the sources in this subsample are unresolved with comparable brown dwarf fits to the EAZY fits: JADES-GS\-+53.02588-27.87203 ($z_a = 8.84$), JADES-GS\-+53.12444-27.81363 ($z_a = 8.33$), JADES-GS\-+53.07645-27.84677 ($z_a = 8.64$), JADES-GN\-+189.16606+62.31433 ($z_a = 8.6$), and JADES-GN\-+189.07787+62.23302 ($z_a = 8.1$). These sources, on visual inspection, do not appear to be strong brown dwarf candidates due to them being quite faint, which would indicate potentially unphysical distances compared to models of the halo brown dwarf population \citep{ryan2005}. While there are 19 unresolved sources in the sample that are proximate to brighter objects (as in Section \ref{sec:flag_bn_2_objects}), none of these have stellar fits within $\Delta\chi^2 < 4$ of the EAZY fits. 

\subsection{$z > 8$ Candidates in the Literature}\label{sec:literature}

As the GOODS-S and GOODS-N fields have been observed across a wide wavelength range and to deep observational flux limits, a number of the sources in our sample have been previously presented in the literature. As described in \citet{robertson2022}, both JADES-GS-z10-0 (JADES-GS-53.15883-27.7735) and JADES-GS-z11-0 (JADES-GS-53.16476-27.77463) were previously identified in \citet{bouwens2011}, as UDFj-38116243 and UDFj-39546284, respectively. Both of these galaxies are in our $z > 8$ sample, as JADES-GS-53.15883-27.7735 and JADES-GS-53.16476-27.77463. Similarly, we also previously discussed GN-z11, first identified in \citet{bouwens2010}, and later further explored in \citet{oesch2014,oesch2016}, which is present in our sample as JADES-GN-189.10605+62.24205. 

In \citet{bouwens2023}, the authors use the publicly available JEMS data to search for $z > 8$ candidates in GOODS-S and construct a sample of ten sources. Nine of the ten sources appear in our sample (their source XDFY-2376346017, which they measure at $z_{EAZY} = 8.3^{+0.2}_{-0.2}$, is at $z_a = 7.89$ in our fits, and we additionally measure a FRESCO $z_{spec} = 7.975$ for this source), and of those, eight sources were previously known and are in our sample. The remaining two sources were not previously known, and also appear in our sample: JADES-GS-53.13918-27.78273 ($z_a = 10.49$) and JADES-GS-53.16863-27.79276 ($z_a = 11.71$). \citet{donnan2023} perform a similar search, and find two additional candidates that fall into our sample: JADES-GS-53.17551-27.78064 ($z_a = 9.66$), and JADES-GS-53.12166-27.83012 ($z_a = 9.42$) (they also independently recover JADES-GS-53.16863-27.79276). The photometric redshifts presented in \citet{bouwens2023} for JADES-GS-53.13918-27.78273 (XDFH-2334046578 in their sample, $z_{EAZY} = 11.8^{+0.4}_{-0.5}$) and JADES-GS-53.16863-27.79276 (XDFJ-2404647339 in their sample, $z_{EAZY} = 11.4^{+0.4}_{-0.5}$) are broadly similar to our values, but we measure a much lower redshift for the former due to the availability of the F150W flux from JADES. Similarly, \citet{donnan2023}, estimate similar photometric redshifts to what we find for JADES-GS-53.17551-27.78064 (UDF-21003 in their sample, $z_{phot} =  9.79^{+0.15}_{-0.13}$) and JADES-GS-53.16863-27.79276 (UDF 16748 in their sample, $z_{phot} = 11.77^{+0.29}_{-0.44}$), but they claim a much higher redshift for JADES-GS-53.12166-27.83012 (UDF-3216 in their sample, $z_{phot} = 12.56^{+0.64}_{-0.66}$), which is inconsistent with the measured F150W flux. We note that this latter candidate appears in our catalog of sources proximate to brighter objects, although with $\Delta\chi^2 = 4.32$. 

For the full sample of $\Delta \chi^2 > 4$ candidates at $z > 8$, we additionally cross-matched their sky positions against GOODS-S and GOODS-N high-redshift catalogs in the literature, including \citet{bunker2009, bunker2010, yan2010, bouwens2011, lorenzoni2011, ellis2013, lorenzoni2013, mclure2013, oesch2013, schenker2013, oesch2014, bouwens2015, finkelstein2015, bouwens2016, harikane2016} and \citet{bouwens2021}. Because our JADES mosaics were aligned using the GAIA reference frame, we had to carefully visually match against each sample, which have different reference frames. In Table \ref{tab:literature_names}, we list the targets that were matched to sources previously discussed in the literature, the photometric redshift for these sources, and we include the references for each object. 

We find 47 objects across the full $\Delta \chi^2 > 4$ catalog have been discussed previously in the literature, 42 in GOODS-S and 5 in GOODS-N. As previously mentioned seven are at $z_a > 10$, three are at $z_a = 9 - 10$, and the remaining 37 are at $z_a = 8 - 9$.

\begin{deluxetable*}{l l c}
\tabletypesize{\footnotesize}
\tablecolumns{3}
\tablewidth{0pt}
\tablecaption{$\Delta \chi^2 > 4$ Catalog Sources in the Literature. \label{tab:literature_names}}
\tablehead{
\colhead{JADES ID} & \colhead{{\tt EAZY} $z_a$} & \colhead{Reference(s)}}
\startdata
		JADES-GS-53.15751-27.76677 & $8.00$ & 1, 2, 3, 4, 5, 6, 8, 10, 12, 13, 16  \\
		JADES-GS-53.16415-27.78452 & $8.02$ & 8, 10  \\
		JADES-GS-53.13563-27.79185 & $8.02$ & 12, 16  \\
		JADES-GN-189.27457+62.21053$^a$ & $8.03$ & 12  \\
		JADES-GS-53.08174-27.89883 & $8.04$ & 15 \\
		JADES-GS-53.13849-27.85854 & $8.04$ & 12, 16  \\
		JADES-GS-53.148-27.79571 & $8.04$ & 4, 10, 12, 15, 16  \\
		JADES-GS-53.06029-27.86353 & $8.04$ & 12, 13, 16  \\
		JADES-GS-53.17727-27.78011 & $8.08$ & 8, 12, 15, 16 \\
		JADES-GS-53.13675-27.83746 & $8.13$ & 13 \\
		JADES-GS-53.08745-27.81492 & $8.17$ & 7, 12, 16 \\
		JADES-GS-53.06035-27.86355 & $8.17$ & 12, 13, 16 \\
		JADES-GS-53.07052-27.86725 & $8.21$ & 12, 16 \\
		JADES-GS-53.05924-27.8353 & $8.22$ & 12, 16 \\
		JADES-GS-53.1459-27.82279 & $8.23$ & 8 \\
		JADES-GS-53.13569-27.83884 & $8.24$ & 13  \\
		JADES-GN-189.2032+62.24245 & $8.28$ & 12  \\
		JADES-GS-53.14585-27.82274 & $8.28$ & 8  \\
		JADES-GS-53.10393-27.89059 & $8.35$ & 12, 16  \\
		JADES-GS-53.20988-27.77928 & $8.36$ & 12, 16 \\ 
		JADES-GN-189.09186+62.25744 & $8.38$ & 12 \\
		JADES-GS-53.1571-27.83708 & $8.39$ & 12, 15, 16 \\
		JADES-GS-53.08738-27.86033 & $8.46$ & 8, 12, 13, 16  \\
		JADES-GS-53.10224-27.85925 & $8.46$ & 12 \\
		JADES-GS-53.16447-27.80218 & $8.50$ & 4, 6, 8, 9, 13, 16, 19 \\
		JADES-GS-53.0865-27.8592 & $8.50$ & 12, 16  \\
		JADES-GS-53.08741-27.8604 & $8.51$ & 8, 12, 13, 16  \\
		JADES-GS-53.15891-27.76508 & $8.52$ & 1, 2, 3, 4, 5, 6, 8, 9, 10, 12, 13, 16, 19  \\
		JADES-GS-53.08932-27.8727 & $8.53$ & 8, 13  \\
		JADES-GS-53.1777-27.78478 & $8.53$ & 6, 8, 9, 19 \\
		JADES-GS-53.07581-27.87938 & $8.55$ & 8, 12, 13, 16 \\
		JADES-GS-53.15784-27.76271 & $8.57$ & 12, 16 \\
		JADES-GN-189.19772+62.25697 & $8.61$ & 12, 13, 16 \\
		JADES-GS-53.16767-27.80017 & $8.64$ & 9, 12, 14, 16, 19  \\
		JADES-GS-53.16337-27.77569 & $8.65$ & 6, 8, 9, 10, 12, 16, 19 \\
		JADES-GS-53.15342-27.77844 & $8.81$ & 10 \\
		JADES-GN-189.2114+62.1703 & $8.92$ & 12, 16 \\
		JADES-GS-53.13363-27.84499 & $9.36$ & 14, 16 \\
		JADES-GS-53.12166-27.83012$^a$ & $9.42$ & 20 \\
		JADES-GS-53.17551-27.78064 & $9.66$ & 20 \\
		JADES-GS-53.13918-27.78273 & $10.49$ & 19 \\
		JADES-GS-53.15883-27.7735$^b$ & $10.84$ & 6, 9, 12, 14, 16, 17, 18, 19, 20 \\
		JADES-GN-189.10604+62.24204$^c$ & $11.00$ & 11, 12, 14, 16 \\
		JADES-GS-53.16863-27.79276 & $11.71$ & 19, 20 \\
		JADES-GS-53.16476-27.77463$^d$ & $12.31$ &  6, 8, 17, 18, 19, 20 \\
		JADES-GS-53.16635-27.82156$^e$ & $12.46$ & 17, 18, 21\\  
		JADES-GS-53.14988-27.7765$^f$ & $13.41$ & 17, 18  
\enddata
\tablecomments{a: Proximate to a brighter source, as described in \S \ref{sec:flag_bn_2_objects}, b: JADES-GS-z10-0, c: GN-z11, d: JADES-GS-z11-0, e: JADES-GS-z12-0, f: JADES-GS-z13-0, References: 1: \citet{bunker2009}, 2: \citet{bunker2010}, 3: \citet{yan2010}, 4: \citet{bouwens2011}, 5: \citet{lorenzoni2011}, 6: \citet{ellis2013}, 7: \citet{lorenzoni2013}, 8: \citet{mclure2013}, 9: \citet{oesch2013},  10: \citet{schenker2013},  11: \citet{oesch2014},  12: \citet{bouwens2015},  13: \citet{finkelstein2015},  14: \citet{bouwens2016},  15: \citet{harikane2016},  16: \citet{bouwens2021},  17: \citet{curtislake2022}, 18: \citet{robertson2022},  19: \citet{bouwens2023},  20: \citet{donnan2023}, 20: \citet{deugenio2023}.}
\end{deluxetable*}

\subsection{Alternate {\tt EAZY} Template Fitting Results}\label{sec:alternatetemplates}

In \citet{larson2022}, the authors present a series of theoretical galaxy templates\footnote{\url{https://ceers.github.io/LarsonSEDTemplates}} designed to be used with {\tt EAZY} to better model the bluer UV slopes expected for very high redshift galaxies. To create these templates, the authors first used {\tt EAZY} to calculate photometric redshifts for mock galaxies in the CEERS Simulated Data Product V32 catalog using the {\tt EAZY} ``tweak\_fsps\_QSF\_v12\_v3'' templates, which were derived from \texttt{fsps}. At this point, the authors created an additional set of templates that better matched the simulated $m_{F200W} - m_{F277W}$ colors for the $z > 8$ galaxies in their sample using the binary stellar evolution models BPASS \citep{eldridge2017} with nebular emission derived from the spectral synthesis code CLOUDY \citep{ferland2017}. These templates resulted in significantly better photometric redshift estimations for the mock galaxies with the CEERS filter set. 

To explore how our choice of {\tt EAZY} templates affects our final $z > 8$ sample, we fit the photometry for all of the objects recovered across GOODS-N and GOODS-S with {\tt EAZY} with the recommended template set from \citet{larson2022} for fitting $z > 8$ galaxies: tweak\_fsps\_QSF\_v12\_v3 along with the BPASS-only ``Set 1'' and the ``BPASS + CLOUDY -- NO LyA'' ``Set 4'' templates. We ran {\tt EAZY} in an otherwise identical manner, including the template error function used, but we utilized the same photometric offsets as provided in Table \ref{tab:photometric_offsets}. 

The resulting photometric redshifts for our primary sample of $z > 8$ sources provide no significant differences or noticeably improved photometric redshift fits: only 4\% have $|z_{Larson} - z_a| / (1 + z_a) > 0.15$ (23 sources in GOODS-S and 5 sources in in GOODS-N). More importantly, if we look at those sources where $z_{Larson} < 8$, only 2.5\% (14 sources in GOODS-S and 4 sources in GOODS-N) have significantly different photometric redshifts. In the majority of these cases, the lower-redshift solution offered by the \citet{larson2022} templates is at the same secondary $\chi^2$ minimum seen for our own template fits, and the validity of the fit is strongly dependent on the observed F090W or F115W fluxes. 

The sources in our sample with $\Delta\chi^2 < 4$ fits are less robust, as has previously been discussed, and show more discrepancy between the fits with our {\tt EAZY} templates and those from \citet{larson2022}. Here, 37\% have $|z_{Larson} - z_a| / (1 + z_a) > 0.15$ (67 sources in GOODS-S and 17 sources in in GOODS-N). 64 of these GOODS-S sources and 15 of the GOODS-N sources (for a total fraction of 35\% of the $\Delta\chi^2 < 4$ objects) have $z_{Larson} < 8$. 

In addition, we derived a sample of $z_a > 8$ sources from fits with the \citet{larson2022} templates after applying the same SNR, $P(z > 7)$ and $\Delta\chi^2$ cuts as described in \S\ref{sec:highzselection}. We compared the resulting candidates with those from the original template set, and after visual inspection found a total of 10 additional $z > 8$ candidates (7 in GOODS-S and 3 in GOODS-N) which we list in Table \ref{tab:larson_template_sources}. Of those sources, 5 are at $z_a = 6 - 8$ and 2 have $P(z < 7) < 0.7$ in our own {\tt EAZY} fits. The remaining three objects are quite faint, but should be considered alongside the main sample. We conclude that our results would not be significantly improved by using the \citet{larson2022} templates. 

\subsection{Spectroscopic Redshifts}\label{sec:speczs}

In total, we have spectroscopic redshifts for 42 objects in our sample. As discussed previously, five of the high-redshift galaxies have been spectroscopically confirmed to lie at $z > 10$: JADES-GS-z10-0, JADES-GS-z11-0, JADES-GS-z12-0, JADES-GS-z13-0 \citep{curtislake2022, deugenio2023}, and GN-z11 \citep{bunker2023}. In this section, we discuss the other objects in our sample with spectroscopic confirmation from both JWST NIRSpec and JWST NIRCam grism spectroscopy from FRESCO. We also compare the photometric redshifts to the spectroscopic redshifts and discuss the observed offset between the two values. Four additional GOODS-S sources have NIRspec spectroscopic redshifts at $z > 8$. Besides GN-z11, there are no GOODS-N NIRSpec spectroscopic redshifts for our sample. An additional 28 sources in our sample have FRESCO spectroscopic redshifts, with 19 objects in GOODS-S and 8 in GOODS-N. As described in Section \ref{sec:flag_bn_2_objects}, there are 4 additional GOODS-S sources, and 1 GOODS-N object with FRESCO $z_{spec}$ that are proximate to other bright sources. The photometric redshifts for these sources were derived from either single [OIII]$\lambda5007$ line detections or, in some brighter cases, multiple line detections. Fifteen of the sources in our sample of $z > 8$ candidates have FRESCO spectroscopic redshifts at $z < 8$ (13 in GOODS-S, and 1 in GOODS-N) in all cases at $z_{spec} > 7.6$. We chose to include these objects as they satisfy our {\tt EAZY} selection criteria. 

\begin{figure*}
  \centering
  {\includegraphics[width=0.49\textwidth]{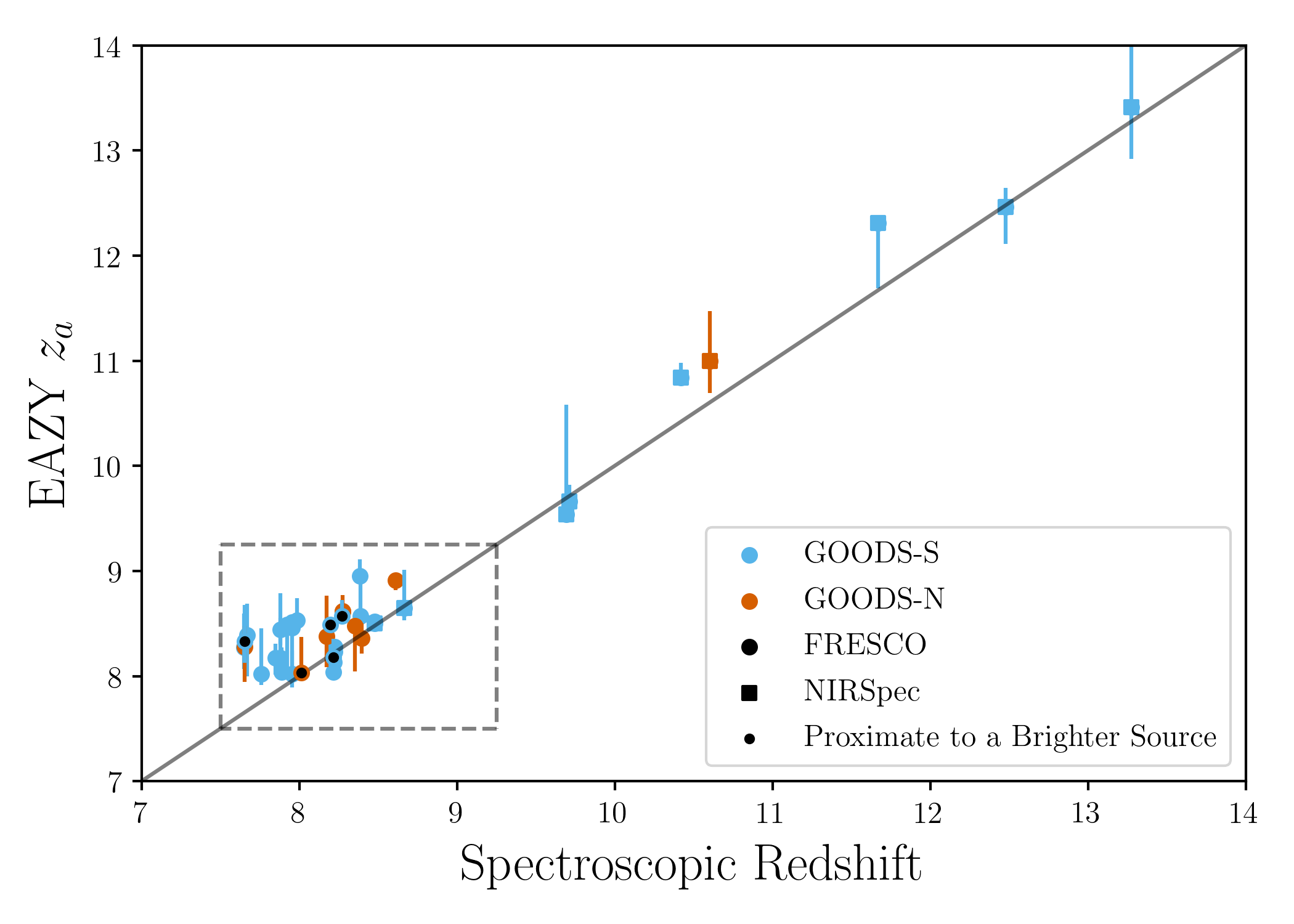}}\ 
  {\includegraphics[width=0.49\textwidth]{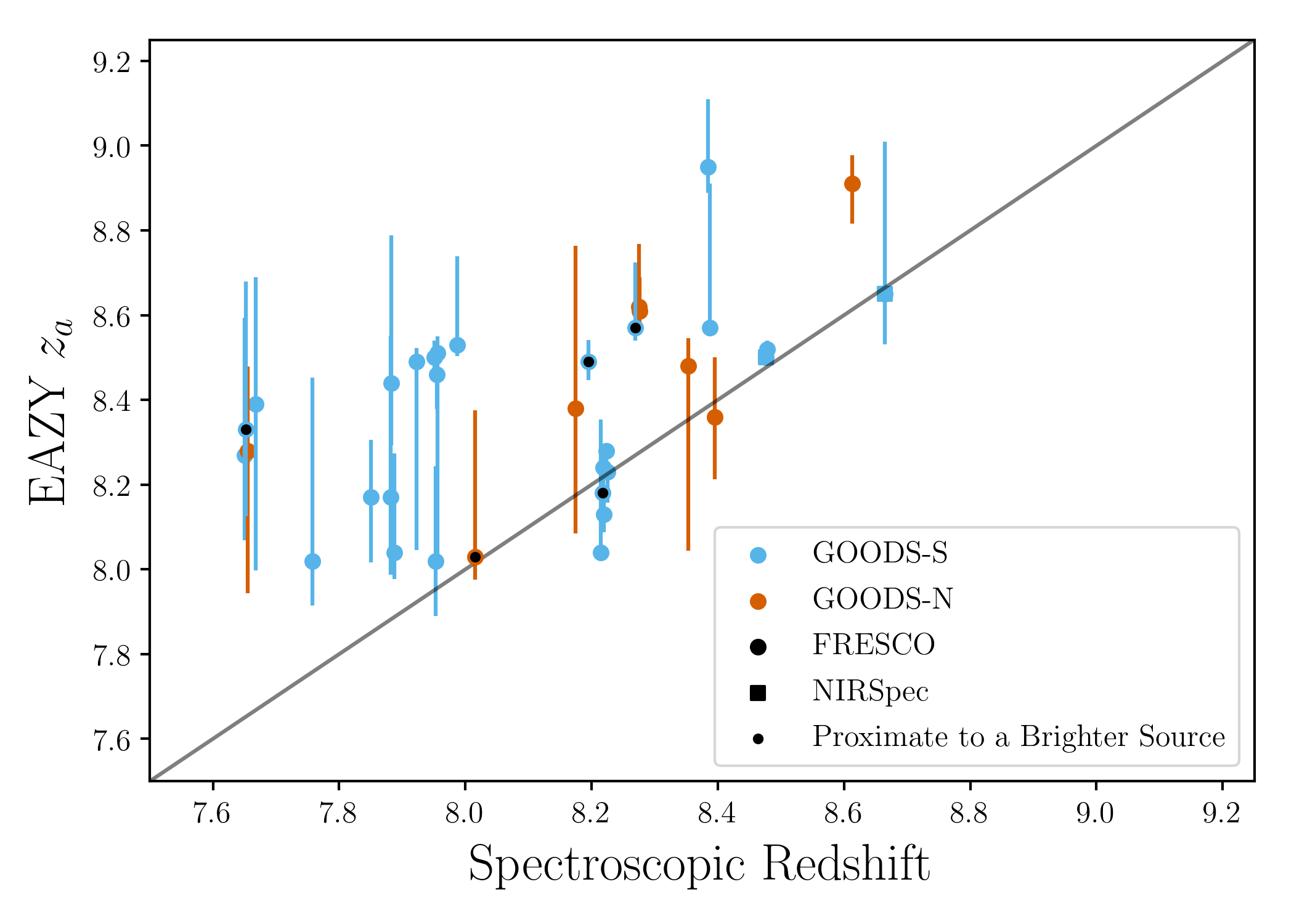}}\ 
  \caption{(Left) Spectroscopic Redshifts for objects in our sample measured from both NIRSpec (square symbols) and FRESCO (circular symbols) spectroscopy, as compared to the {\tt EAZY} photometric redshifts. We highlight those four sources proximate to brighter sources, as discussed in Section \ref{sec:flag_bn_2_objects}, with black filled circles. (Right) A zoom in on the dashed region at $z = 7 - 9$ in the left panel. We find that the estimated photometric redshifts overpredict the spectroscopic redshifts ($\langle \Delta z = z_a - z_{spec} \rangle= 0.26$), potentially due to the differences between our adopted templates and high-redshift galaxies. We explored removing Lyman-$\alpha$ emission in our template set, and this lowers the offset to $\langle \Delta z = z_{spec} - z_a \rangle= 0.19$.}
  \label{fig:spec_z_vs_phot_z}
\end{figure*}

In Figure \ref{fig:spec_z_vs_phot_z} we show the spectroscopic redshifts of the objects in our sample against their photometric redshift. There are no catastrophic outliers, defined here as those objects where $|z_{spec} - z_a| / (1 + z_{spec}) > 0.15$. As discussed previously with individual objects, the photometric redshifts have a systematic offset such that {\tt EAZY} is slightly overpredicting the distances to these galaxies ($\langle \Delta z = z_a - z_{spec} \rangle= 0.26$). To estimate the scatter on the relationship, we also calculated the normalized mean absolute deviation $\sigma_{NMAD}$, defined as:

\begin{equation}
\sigma_{\mathrm{NMAD}} = 1.48 \times \mathrm{median}\left(\left| \frac{\delta z - \mathrm{median}(\delta z)}{1 + z_{spec}} \right| \right) 
\end{equation}

where $\delta z = z_{spec} - z_{phot}$. For all of our sources with spectroscopic redshifts, $\sigma_{NMAD} = 0.05$. Understanding the source of this offset is quite important given the usage of photometric redshifts in deriving statistical parameters like the UV luminosity function. By constraining the {\tt EAZY} fit for each of these sources to be at the spectroscopic redshift, we find that the primary reason for these higher-redshift fits is due to the flux of the filter that spans the Lyman-$\alpha$ break. In the fits where redshift is constrained to be at $z_{spec}$, the observed fluxes in the band at the Lyman-$\alpha$ break are overestimated in the template fits. While this effect maybe be due to photometric scatter upwards in those bands, it is more likely due to the templates themselves. In \citet{haro2023b}, the authors present $z_{spec} = 8 - 10$ objects with NIRSpec spectroscopy which show a larger offset ($\langle \Delta z = z_a - z_{spec} \rangle = 0.46 \pm 0.11$) to higher photometric redshifts, and these authors also hypothesize that this might be a result of potential differences between the observed high-redshift galaxy SEDs and the templates used to model high-redshift galaxies. 

One potential source of excess flux in the UV is the strength of the Lyman-$\alpha$ emission line in our templates. To explore the effect of this line, we first took our {\tt EAZY} templates and removed the Lyman-$\alpha$ contribution by cutting out the flux between 1170 to 1290 \AA\, and replacing that portion in each template with a linear fit. Using these templates without Lyman-$\alpha$, we re-fit every one of our sources with spectroscopic redshifts, and calculated a new $\sigma_{NMAD} = 0.02$, as well as a difference in the average offset $\langle \Delta z = z_a - z_{spec} \rangle = 0.19$. While this is smaller, the offset is still present, indicating that Lyman-$\alpha$ flux is not the dominant factor. One alternate possibility is that the strength of the optical emission lines at long wavelength may not be fully reflected in our limited template set, and for those $z_{spec} < 9$ sources (where the optical emission is not redshifted out of the NIRCam filters), this may have an effect of pushing fits at higher redshifts. At the redshift range of our sample, the FRESCO redshifts are calculated preferentially for those objects with strong line emission, which may not be probed by our template set. Understanding this offset may prove important for future fits to high-redshift galaxies, and it will necessitate the creation of templates derived from high-resolution NIRSpec spectra of these sources once larger samples are observed.  

\subsection{Rejected High-Redshift Candidates}\label{sec:rejectedobjects}

\begin{figure*}
\centering
Example Visually Rejected Candidates \par\medskip
{\includegraphics[width=0.49\textwidth]{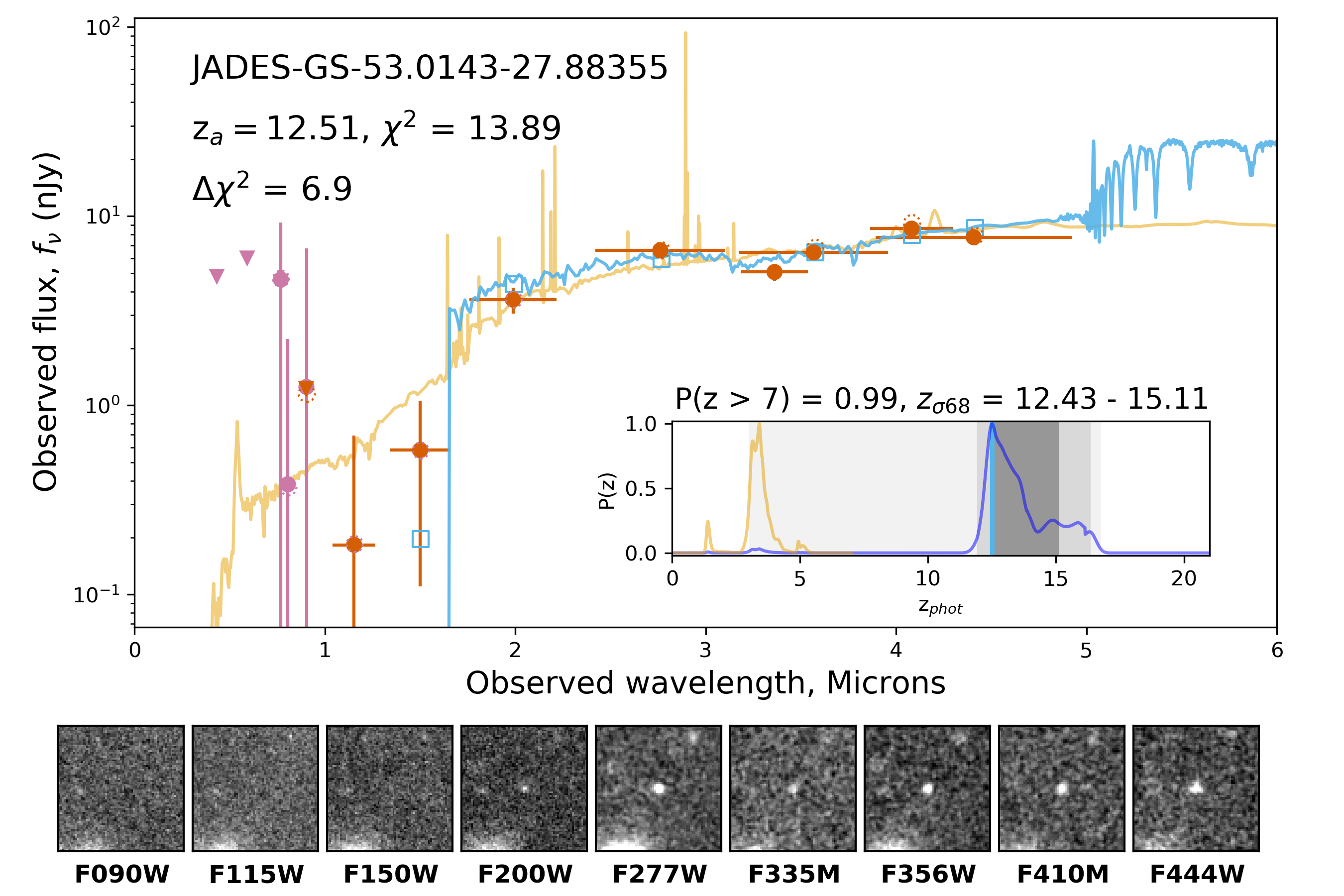}}\ 
{\includegraphics[width=0.49\textwidth]{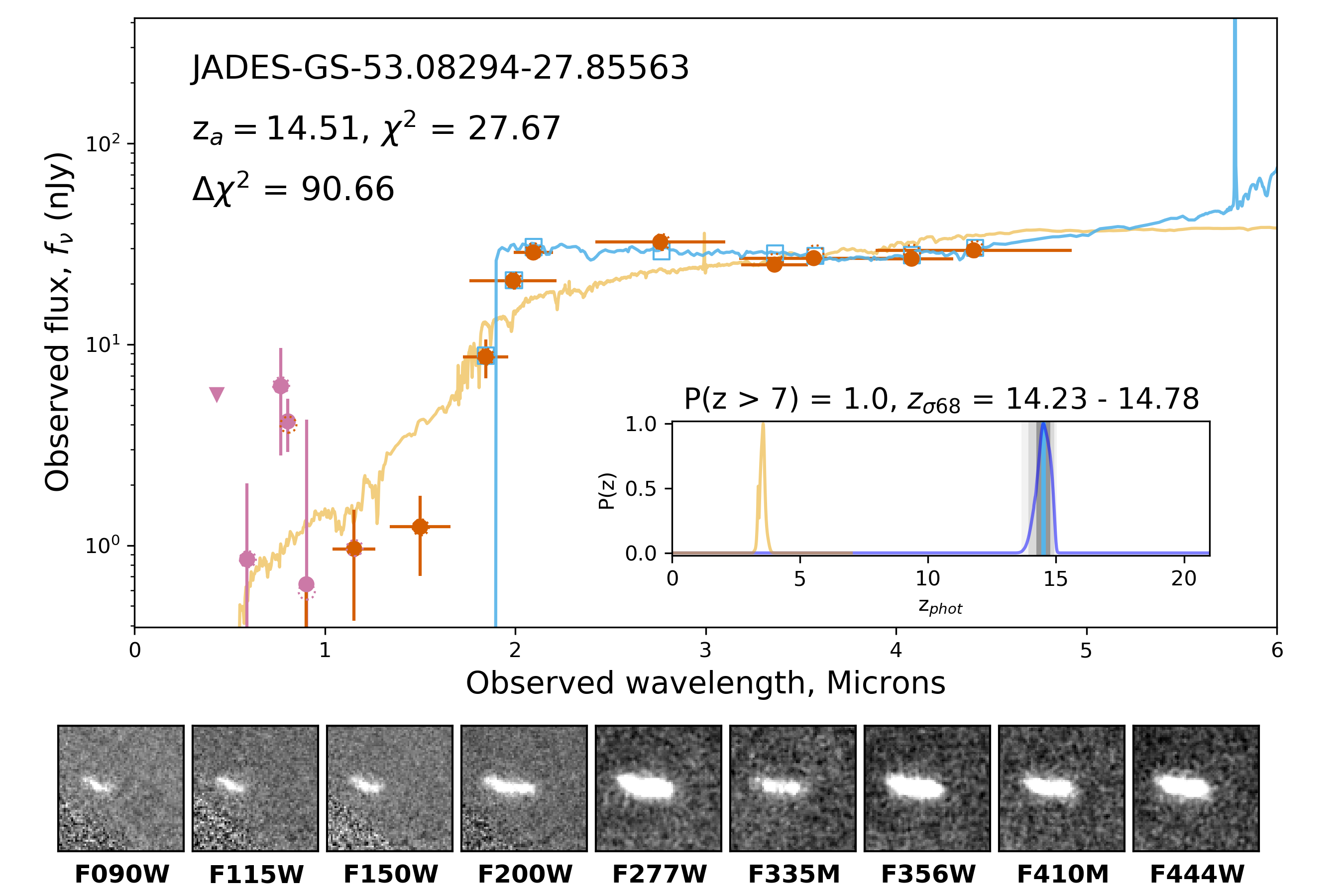}}\ 
{\includegraphics[width=0.49\textwidth]{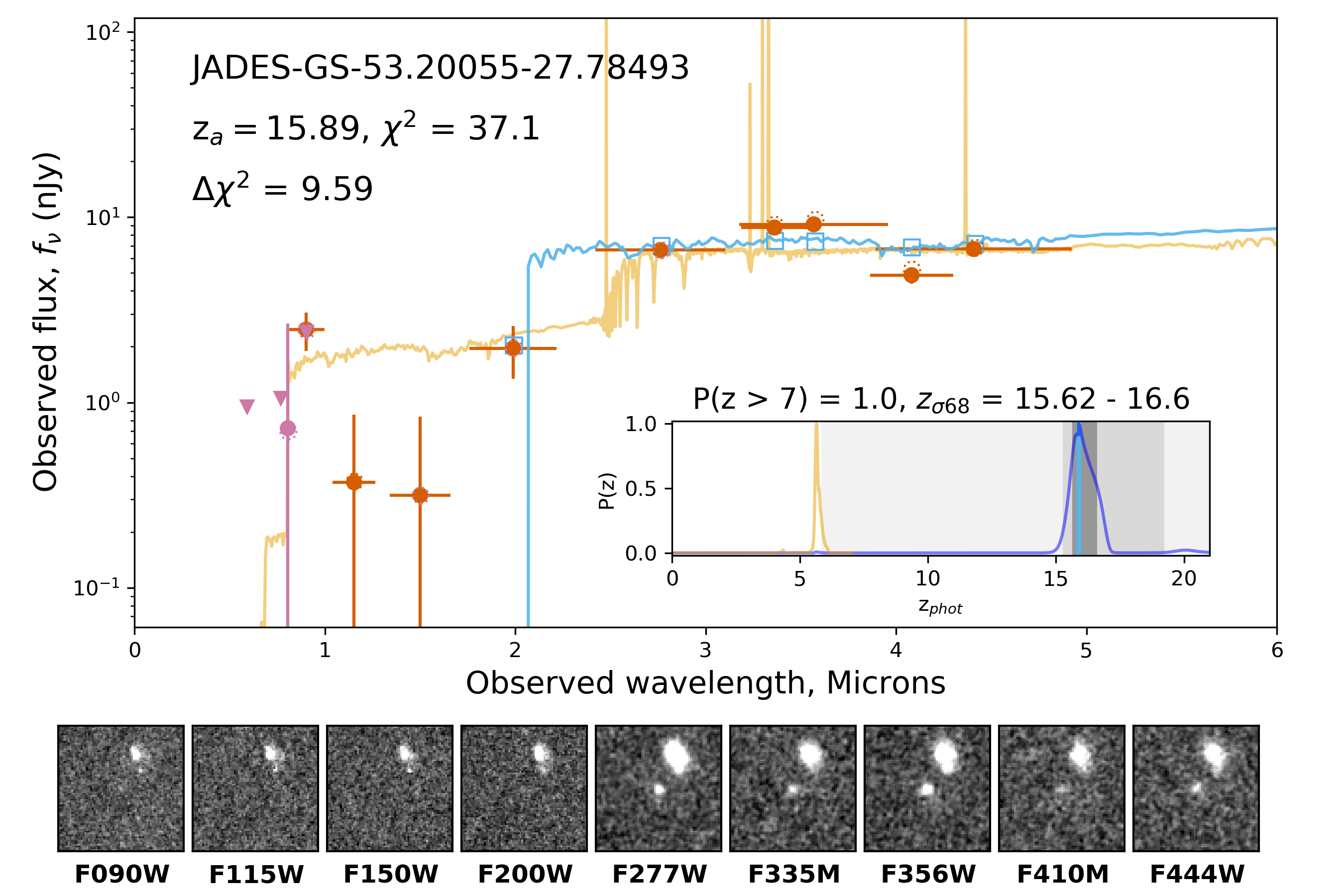}}\
{\includegraphics[width=0.49\textwidth]{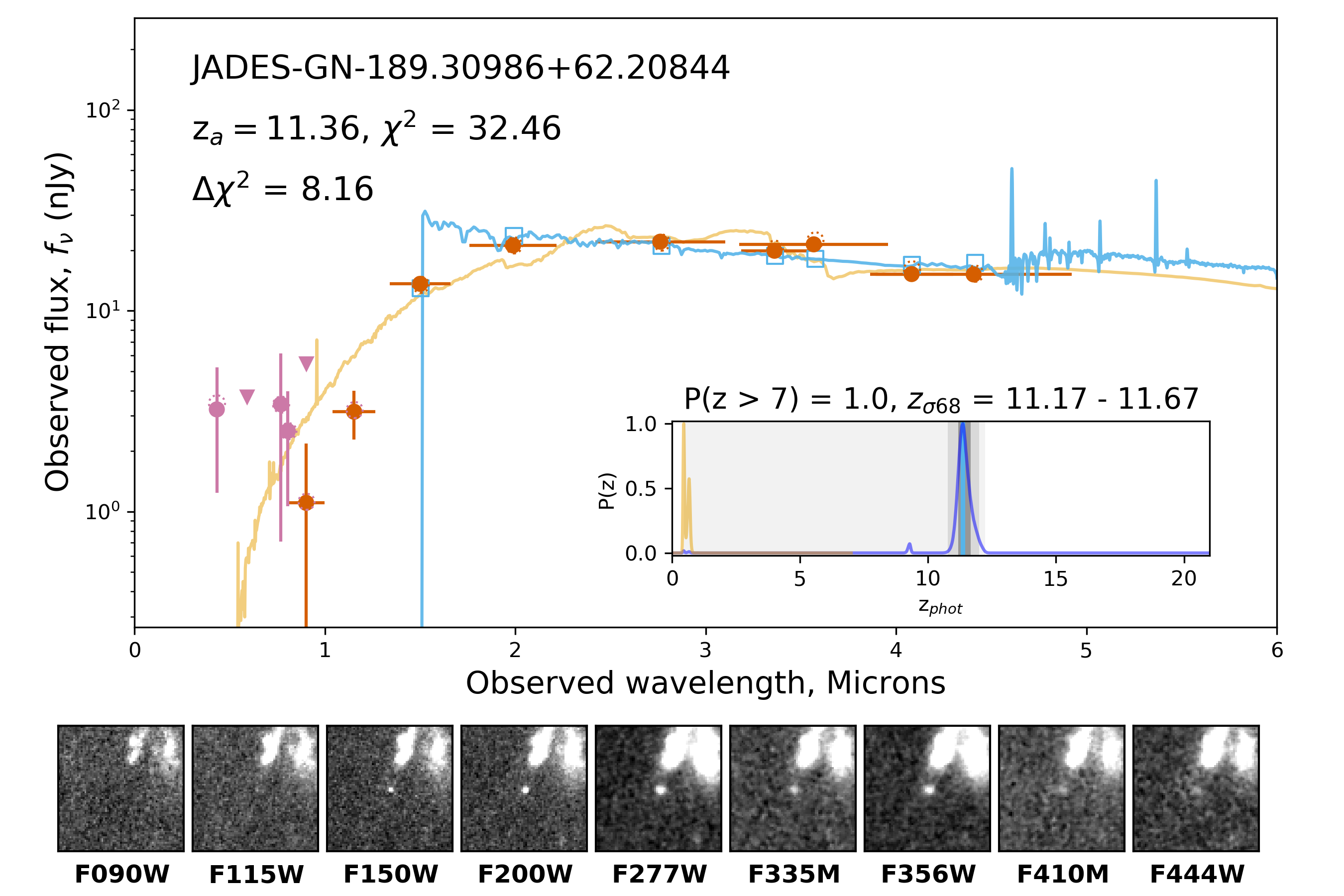}}

\caption{
Example SEDs for four rejected GOODS-S and GOODS-N galaxies. In each panel, the colors, lines, and symbols are as in Figure \ref{example_SED_fit}.}
\label{fig:rejectedsample}
\end{figure*}

Finding and characterizing high-redshift galaxies is a complex process, even given the IR filters on board JWST. In our visual inspection, we found a number of bright galaxies that we rejected from our $z > 8$ sample because of multiple reasons, and in this section, we will provide four examples as case studies to demonstrate the sorts of galaxies with colors that can mimic those of high-redshift galaxies. This analysis follows discussions in \citet{naidu2022} and \citet{zavala2023}, and seen directly with CEERS-93316, a candidate galaxy at $z_{phot} = 16.4$ which was shown to be at $z_{spec} = 4.912$ \citep{haro2023}. 

In Figure \ref{fig:rejectedsample} we provide SEDs for JADES-GS-53.0143-27.88355, JADES-GS-53.08294-27.85563, JADES-GS-53.20055-27.78493, and JADES-GN-189.30986+62.20844. Here, we highlight the solution at $z < 7$ in each, while also leaving the overall minimum $\chi^2$ solution. 

JADES-GS-53.0143-27.88355 ($m_{Kron, AB} = 29.3$), appears from the thumbnails and from the {\tt EAZY} minimum $\chi^2$ fit to be an F150W dropout at $z_a = 12.51$. However, the red UV slope indicates that perhaps this object is much dustier and at low-redshift ($z_{alt} = 3.41$), where the H$\alpha$ emission line was boosting the observed F277W flux. This UV slope could also arise from the bright source to the southeast \citep[at $z_{spec} = 0.2472$, ][]{cooper2012}. In addition there is what appears to be a flux detection in the F115W thumbnail which helps to rule out the high-redshift solution. 

JADES-GS-53.08294-27.85563 ($m_{F277W, Kron} = 26.8$) appears to be a bright F150W dropout clump immediately adjacent to another object. The SED is well fit at $z_a = 14.51$, and the fit constrained to be at $z < 7$ is significantly worse ($z_{alt} = 3.56$). The secondary source, which is detected with all five of the JADES HST/ACS bands (although only at $2\sigma$ significance for F435W), has an EAZY template redshift $z_a = 3.4$. This redshift puts the Balmer break between the NIRCam F150W and F200W filters, and we cannot rule out the possibility that this is what we are observing for JADES-GS-53.08294-27.85563. This source appeared in a sample of ``HST dark'' galaxies in \citet{williams2023b}, where they present a photometric redshift of $z_{\mathrm{phot}} = 3.38$, although they use a larger aperture for measuring photometry which may introduce flux from the nearby source. JADES-GS-53.08294-27.85563 was further imaged in six NIRCam medium-band filters as part of the JWST Cycle-2 GO observations of the JADES Origins Field \citep[JOF, PID 3215;][]{eisenstein2023b}, and a more detailed analysis of this source will be presented in a forthcoming paper from the collaboration (Robertson, B. et al. in prep.). Both sources will be observed by the NIRSpec prism in JWST GTO program 1287 (Willott, C. in prep).  

JADES-GS-53.20055-27.78493 ($m_{F277W, Kron} = 28.8$) appears to be an F200W dropout at $z_a = 15.89$ southeast of another, brighter source. While positive flux is observed at the 1 - 2 nJy level in F115W and F150W, this is at a SNR $< 0.8$ in both cases. This source was ruled out as an F200W dropout because of the detection at the 4$\sigma$ of F090W flux, which can be seen in the thumbnail. 

JADES-GN-189.30986+62.20844 ($m_{F277W, Kron} = 27.4$) is best fit at $z_a = 11.36$, placing the observed Lyman-$\alpha$ break at 1.5$\mu$m. This object is proximate to another, brighter galaxy with an EAZY fit at $z_a = 1.87$ with a complex morphology first observed as part of the GOODS survey in \citet{giavalisco2004}. This is at a lower redshift than the alternate {\tt EAZY} result for JADES-GN-189.30986+62.20844, $z_a = 2.58$, but the faint F115W detection (SNR = 3.64) demonstrates that the minimum $\chi^2$ redshift solution for this object is erroneous. 

\section{Discussion} \label{sec:discussion}

In this section we explore the selection and derived properties of this large sample of candidate high-redshift galaxies in more detail. A full description of the theoretical implications of these sources is outside the scope of this paper. The stellar mass and star formation histories for these sources will be the focus of a study by Tacchella et al. (in prep), while the full estimation of the evolution of the UV luminosity function at  $z > 8$ from the JADES sources will be presented in Whitler et al. (in prep). 

\subsection{UV Magnitudes}

\begin{figure*}
  \centering
  \includegraphics[width=\textwidth]{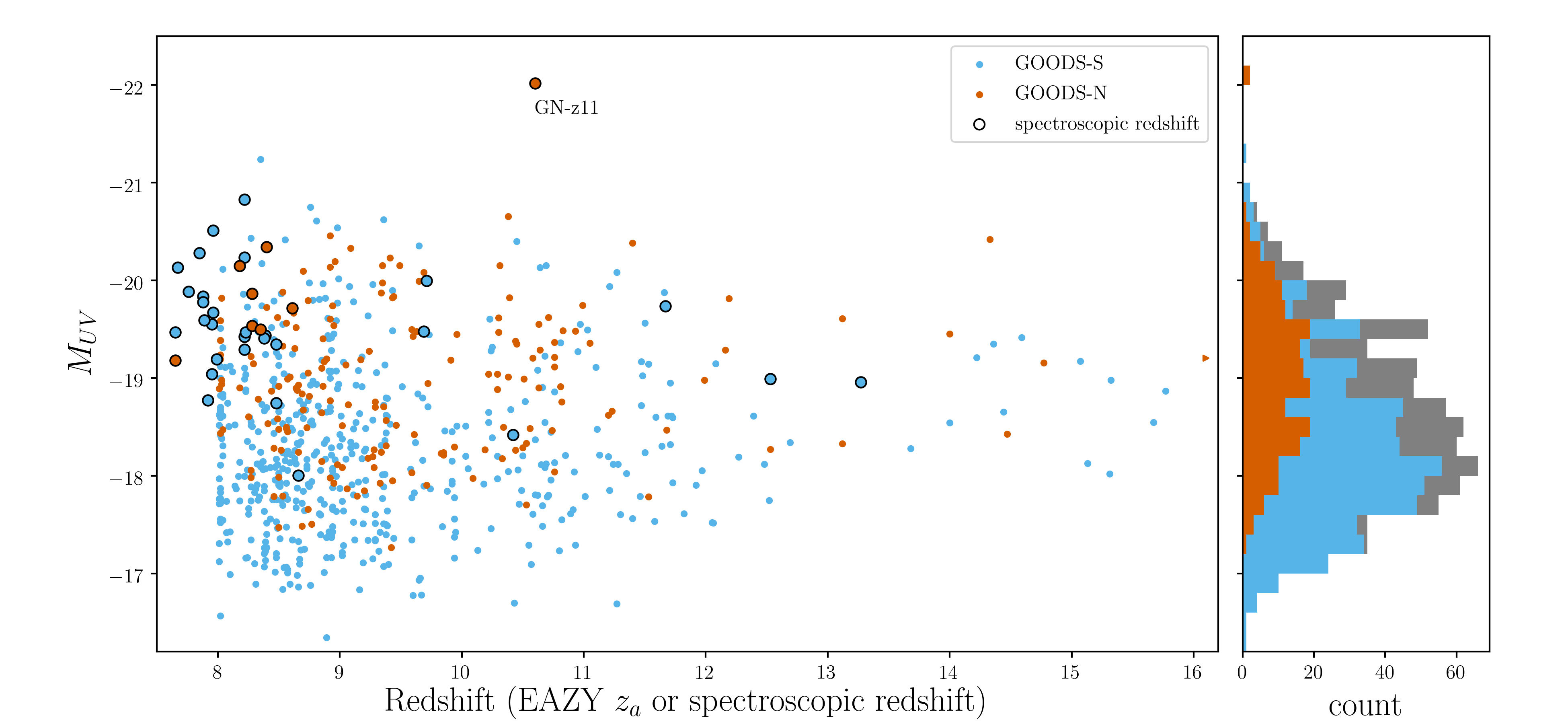}
  \caption{MUV plotted against the best-fitting {\tt EAZY} $z_a$ photometric redshift or the observed spectroscopic redshift for the $z > 8$ galaxies and candidate galaxies in the GOODS-S (blue) and GOODS-N (red) $z > 8$ samples. The points and colors are the same as Figure \ref{mag_vs_za}. On the right we show the UV magnitude distribution. GN-z11 stands out for its extreme $M_{UV}$.}
  \label{fig:MUV_vs_za}
\end{figure*}

We calculated the UV magnitudes from the {\tt EAZY} fits to explore the range of intrinsic UV brightnesses for the sample. To calculate $M_{UV}$ we started by fitting the Kron magnitude catalog fluxes forced to be at the redshifts derived from the smaller circular apertures, or, if available, the spectroscopic redshifts for each source. This was done to not bias the resulting UV magnitudes against more extended objects by encompassing more of the total flux. From here, we took the best-fitting rest-frame {\tt EAZY} template for each object and passed it through mock top-hat filter centered at $1500$\AA\, with a width of $100$\AA, and calculated the intrinsic UV magnitude based on the resulting flux. 

In Figure \ref{fig:MUV_vs_za} we show the resulting $M_{UV}$ values against the photometric and spectroscopic redshift for the sample. As can be expected, GN-z11 is by far the brightest source in the sample at $M_{UV} = -22.0$\footnote{In \citet{bunker2023}, the authors calculate $M_{UV} = -21.5$ for GN-z11 from JWST/NIRSpec spectroscopy for the source, and in \citet{tacchella2023} the authors present a value of $M_{UV} = -21.6$ from the {\tt ForcePho} fits to the NIRCam photometry to the source. Our value is higher due to the fact that we estimated $M_{UV}$ from the Kron magnitudes to the source, which includes excess flux from a ``haze''  observed near the source \citep{tacchella2023}. The Kron fluxes we use are $\sim1.5$ times brigher than what are presented in \citet{tacchella2023}, and if we scale our fluxes by this amount our value changes to $M_{UV} = -21.58$ for GN-z11.}. Excitingly, we find 227 objects in our sample with $M_{UV} > -18$, and 16 objects (all in GOODS-S) with $M_{UV} > -17$, entirely at $z_a < 11.5$. These UV-faint high-redshift galaxy candidates demonstrate the extraordinary depth of the JADES survey. In addition, these results stand in contrast to the decline in the number counts of HST-observed galaxies discussed in \citet{oesch2018} and \citet{bouwens2019}, and help to confirm results from other JWST surveys \citep{finkelstein2023, harikane2023, perezgonzalez2023}.

\subsection{Dropout Colors}
As discussed in the introduction, traditionally, high-redshift samples are assembled by targeting Lyman-$\alpha$ dropout galaxies in color space. In \citet{hainline2020}, the authors used the JAGUAR mock catalog \citep{williams2018} to explore the NIRCam colors of simulated dropout samples, and demonstrated the tradeoff between sample completeness and accuracy for high-redshift dropout galaxies. Because of the utility of dropout selection, we sought to explore how successful this technique alone would be at finding the JADES $z > 8$ candidate galaxies. We utilized a uniform two-color selection scheme to target F090W, F115W, and F150W dropouts within our primary $z > 8$ sample, where in each case the color limit for the filters that targeted the Lyman-$\alpha$ break was $m_1 - m_2 > 1.0$, while the color limit for the filters that targeted the rest-frame UV was $m_2 - m_3 < 0.5$. 

\begin{figure*}
\centering

{\includegraphics[width=0.9\textwidth]{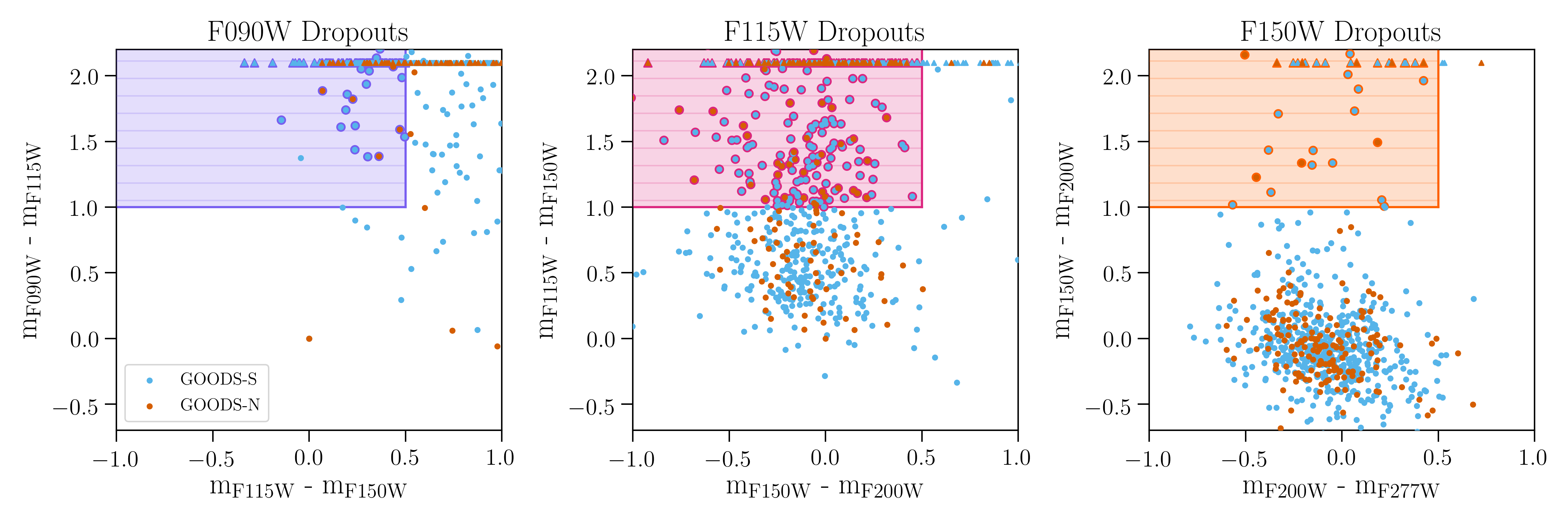}}\ 
{\includegraphics[width=0.9\textwidth]{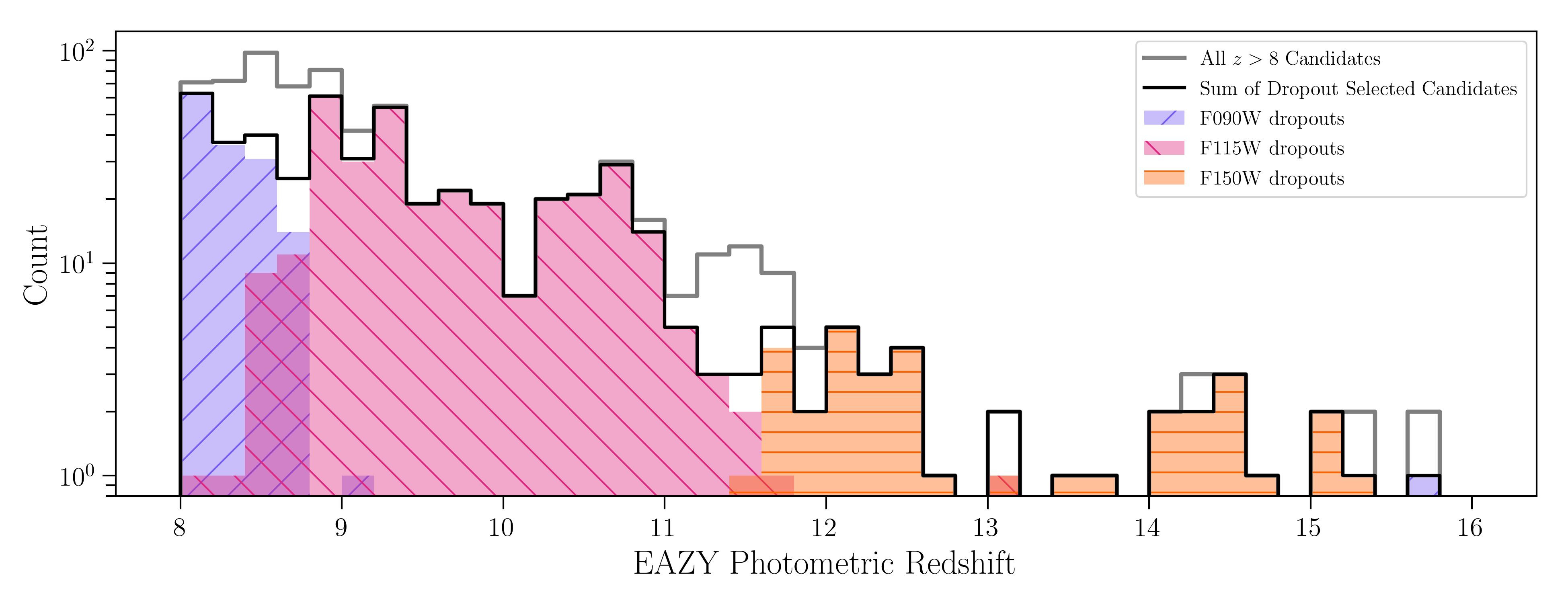}} 

\caption{
(Top Row) F090W, F115W, and F150W color selection diagrams for the GOODS-S and GOODS-N $z > 8$ sample. We show all of the sources in each panel, with upward pointing arrows plotted for those whose colors place them off the top of each figure. The colored region indicates the simple dropout selection in each panel, and we indicate which sources are selected with thick lines around the symbol. (Bottom Row) Photometric redshift distribution for the $z > 8$ sample plotted with a thick grey line, with the photometric redshifts for each color-selected samples overplotted in colored regions. In black, we plot the sum of all three of the dropout distributions. While the F090W, F115W, and F150W dropouts broadly map to photometric redshifts of $z = 8 - 8.5$, $z = 8.5 - 11.5$, and $z = 11.5 - 15$ respectively, there are a large number of sources in our sample that would not be selected via color selection, as seen by comparing the black and grey histograms.}
\label{fig:color_selection}
\end{figure*}

In Figure \ref{fig:color_selection}, we show the F090W, F115W, and F150W color selection in the top row plots, targeting the entire $z > 8$ sample in each panel. In the bottom panel we show a photometric redshift histogram for the sources in the sample with a thick grey line and, in the shaded regions the F090W, F115W, and F150W dropout sample distributions. We sum these distributions and plot that with a thick black line. The lone F090W dropout at $z_a > 15$ is JADES-GS-53.12692-27.79102, which we discuss in Section \ref{sec:z_gt_12} and plot in the upper-right panel of Figure \ref{fig:z_gt_12_example_SEDs_pt1}. We find that $71\%$ of the $z > 8$ sample would be selected as dropouts with these color criteria, while 171 GOODS-S and 37 GOODS-N objects in the sample are not selected by any scheme, which are are predominantly at $z \sim 8.5$ and $z \sim 11.5$, as seen from the bottom panel of the figure. These candidates have colors just outside of the selected color space, where $z \sim 8.5$, while $m_{F090W} - m_{F115W} > 1.0$,  $m_{F115W} - m_{F150W} = 0.5 - 1.0$. A similar effect is seen for the F115W and F150W dropouts at $z \sim 11.5$. This effect could be mitigated by expanding the selection criteria, but this is at the risk of including significantly more lower-redshift interlopers \citep{hainline2020}.

\begin{figure}
  \centering
  \includegraphics[width=\linewidth]{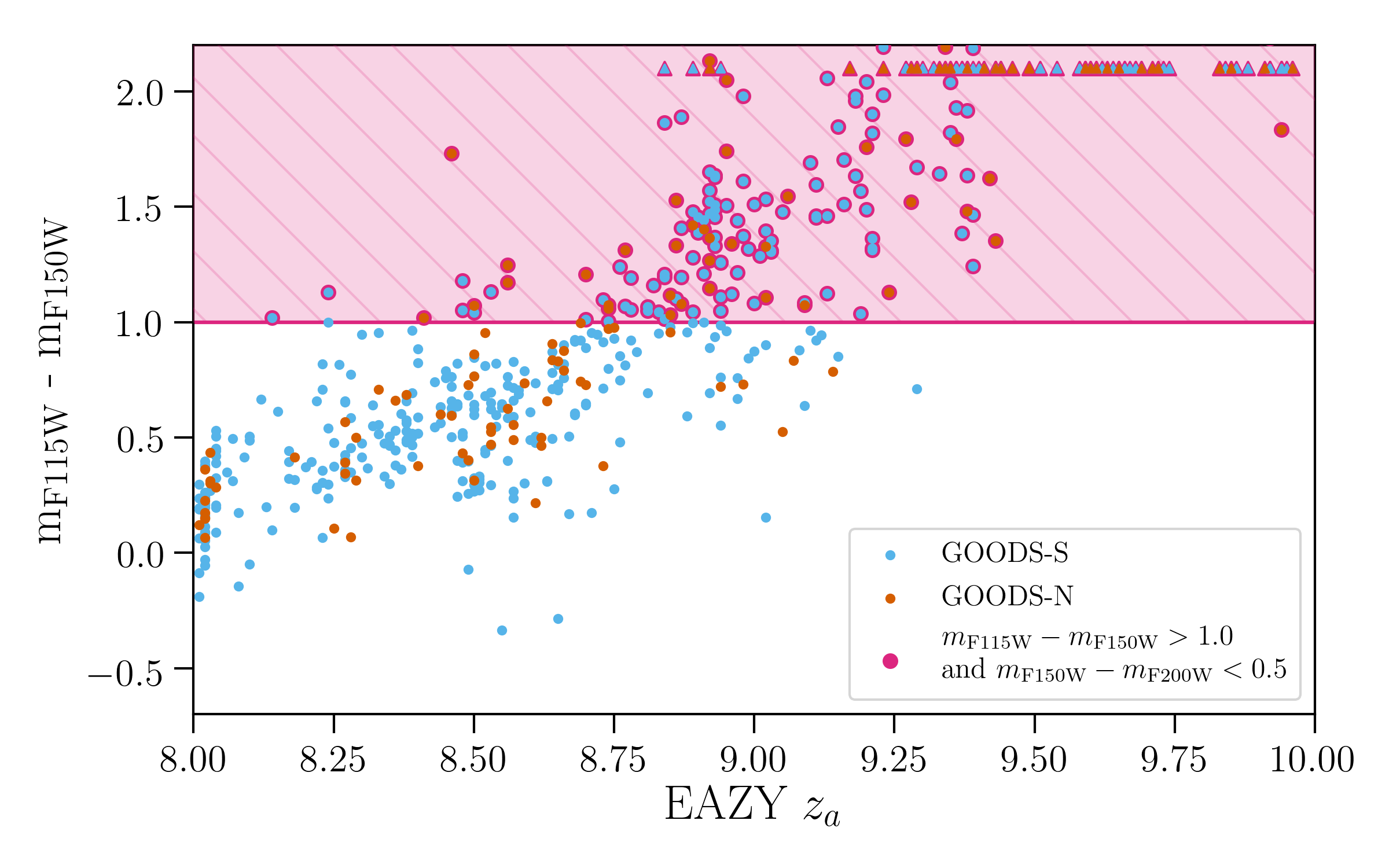}
  \caption{$m_{F115W} - m_{F150W}$ color, as measured using $0.2^{\prime\prime}$ diameter aperture photometry, plotted against the best-fitting {\tt EAZY} $z_a$ photometric redshift for the GOODS-S (blue) and GOODS-N (red) samples. As photometric redshift increases, the effect of the Lyman-$\alpha$ break can be seen in the redder $m_{F115W} - m_{F115W}$ color. We shade the region and highlight those objects selected by a $m_{F115W} - m_{F150W} > 1.0$ and $m_{F150W} - m_{F200W} < 0.5$ color cut with a thick outline around the point.}
  \label{f150w_m_f115w_color_vs_z}
\end{figure}

Another way of looking at color selection is by directly plotting the dropout color against the {\tt EAZY} photometric redshift. At $z_{phot} = 8$, the Lyman-$\alpha$ break is at $\sim 1.1 \mu$m, which is on the blue edge of the NIRCam F115W band, and by $z_{phot} = 10$, the Lyman-$\alpha$ break should sit between the F115W and F150W filters, so for the objects at increasing photometric redshifts in this range, the F115W SNR will vary as the galaxy's rest-frame UV emission drops out of this band. In Figure \ref{f150w_m_f115w_color_vs_z}, we plot the $m_{F115W} - m_{F150W}$ color against the {\tt EAZY} $z_a$ value for the GOODS-N and GOODS-S objects at $z_{phot} = 8 - 10$. As expected, the $m_{F115W} - m_{F150W}$ color increases in this redshift range. We find that 95\% of the candidate high-redshift galaxies selected as F115W dropouts by our cuts have $z_a > 8.75$, while 16\% of the candidates at $z_a = 8.75 - 10.0$ in our sample would still fall outside of this simple color cut. 

\subsection{Using $\Delta\chi^2$ to Discern Between High- and Low-Redshift Template-fitting Solutions}

Fitting a galaxy's SED with templates or stellar population synthesis models enables a measurement of the probability of a galaxy being at a range of photometric redshifts. In this study, we have used the difference in $\chi^2$ values between the best-fit model and the model constrained to be at $z < 7$ as a our metric of accuracy. The exact $\Delta \chi^2$ value we measure for each object is dependent on the template set used, as well as the flux uncertainties and, in our case, the template error function and photometric offsets used. As a result, as is the case for any continuous value of merit, choosing a specific cut is a tradeoff between sample accuracy and completeness. 

In \citet{harikane2023}, the authors discuss that $\Delta \chi^2 > 4$, the value we adopt in this current work \citep[following ][]{bowler2020,donnan2023,finkelstein2023,harikane2022} is not sufficient for properly removing low-redshift interlopers, through injecting and recovering mock galaxies in the CEERS extragalactic data. Instead, these authors recommend the stricter cut of $\Delta \chi^2 > 9$. Because we have a larger number of observed photometric filter in the JADES data, choosing a low $\Delta \chi^2$ limit may be resulting in the inclusion of more potential interlopers, which has lead to our releasing output catalogs which include all of the sources we visually inspected regardless of the chosen $\Delta \chi^2$ cut. 

If we do instead look only at those objects in our sample with $\Delta \chi^2 > 9$, our primary sample is reduced to 483 candidates (358 in GOODS-S and 125 in GOODS-N), or 67\% of the 717 $\Delta \chi^2 > 4$ sources (67\% in GOODS-S and 69\% in GOODS-N). This subsample selected with a stricter cut has a similar redshift distribution to our full sample (19 of the 33 candidates at $z > 12$ would still be included), but the sources have brighter F277W magnitudes, as would be expected. The median F277W magnitude for the $\Delta \chi^2 > 4$ sample is 29.11, while the median F277W magnitude for the $\Delta \chi^2 > 9$ sample is 28.96. It should be noted that every source in our sample with a spectroscopic redshift has $\Delta \chi^2 > 13$. Pushing the cut to even stricter values, we find that 45\% of the original sample has $\Delta \chi^2 > 15$ and 36\% of the original sample has $\Delta \chi^2 > 20$.

\subsection{Candidate Galaxies with Red Long-Wavelength Slopes} 

In our visual inspection of the galaxy candidates, we find a number of high-redshift candidates with very red long-wavelength slopes, following the discovery of similar sources at $z_{phot} = 5 - 9$ in \citet{furtak2022}, \citet{barro2023}, \citet{labbe2023}, \citet{endsley2022} \citet{leung2023}, \citet{akins2023}, and \citet{williams2023b}. These objects are often very bright and unresolved in F444W, and in many cases are comparatively faint at shorter wavelengths. To systematically search for these sources in our full sample, we selected those objects that have $m_{F277W} - m_{F444W} > 1.3$ and $m_{200W} - m_{F356W} > 0.0$. These color limits ensure that the observed red long-wavelength slope is not due to an emission line boosting the F444W flux, and return sources similar to those presented in the literature. 

\begin{figure*}
\centering
Example Red Long-Wavelength Slope Candidates \par\medskip
{\includegraphics[width=0.49\textwidth]{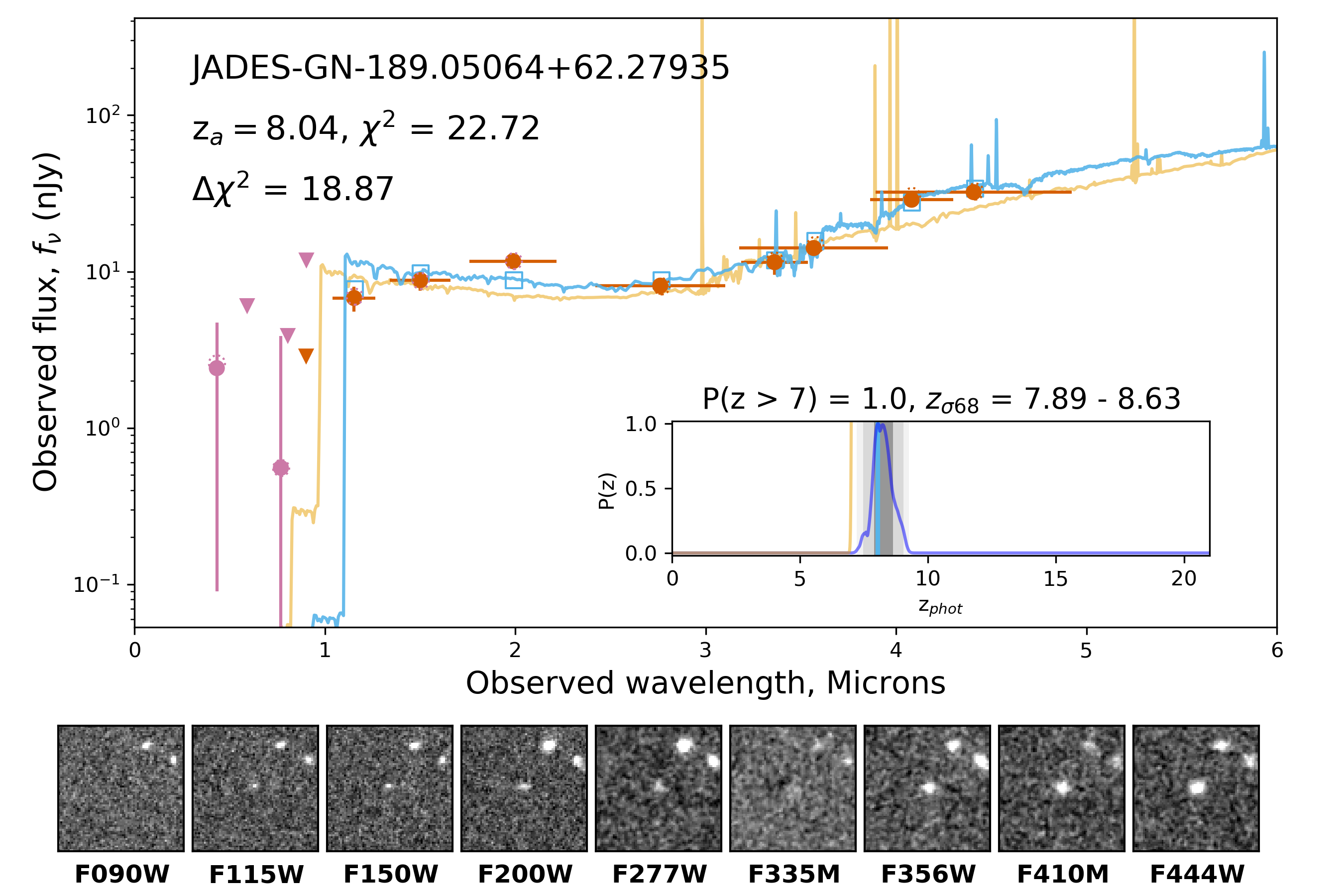}}\
{\includegraphics[width=0.49\textwidth]{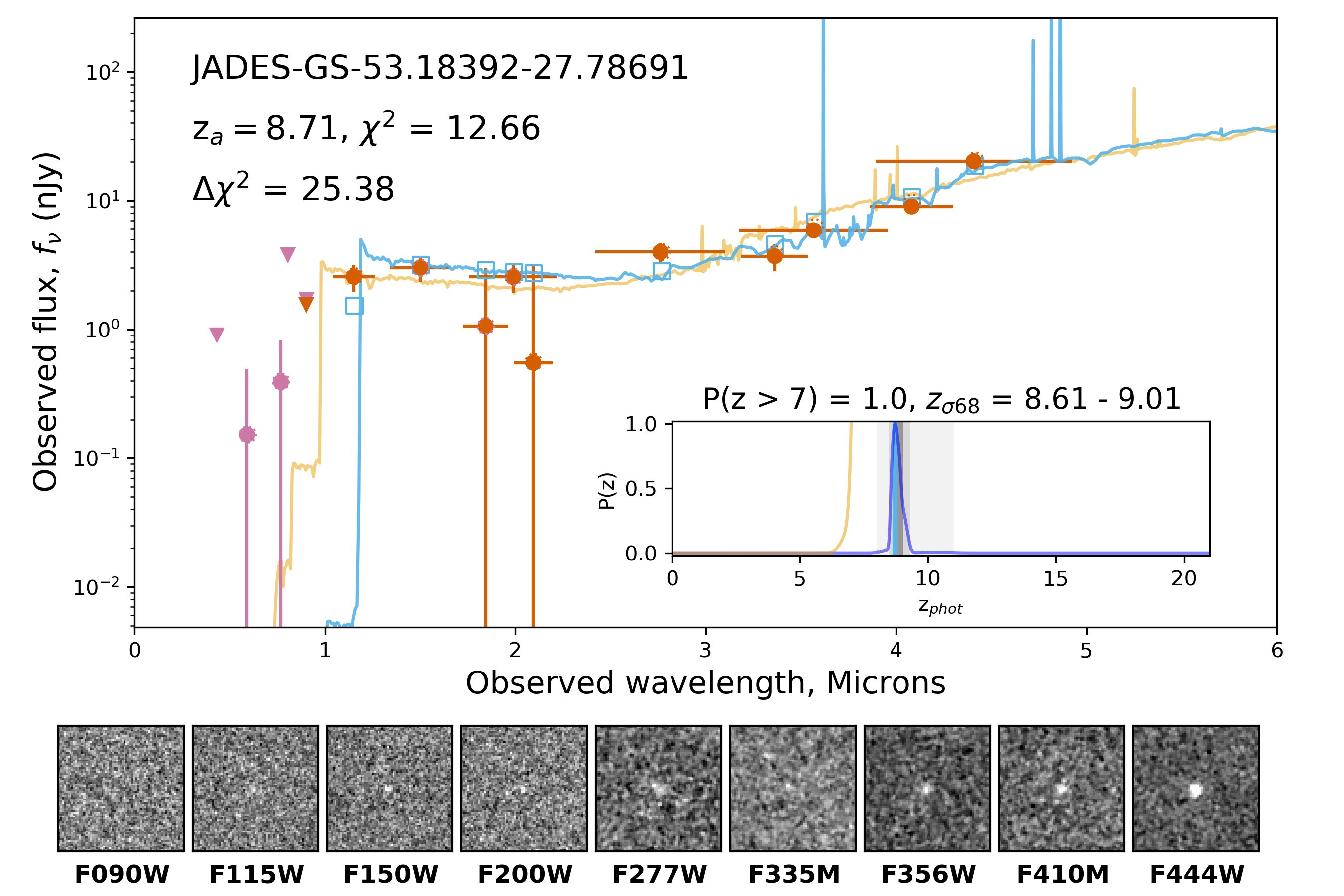}}\
{\includegraphics[width=0.49\textwidth]{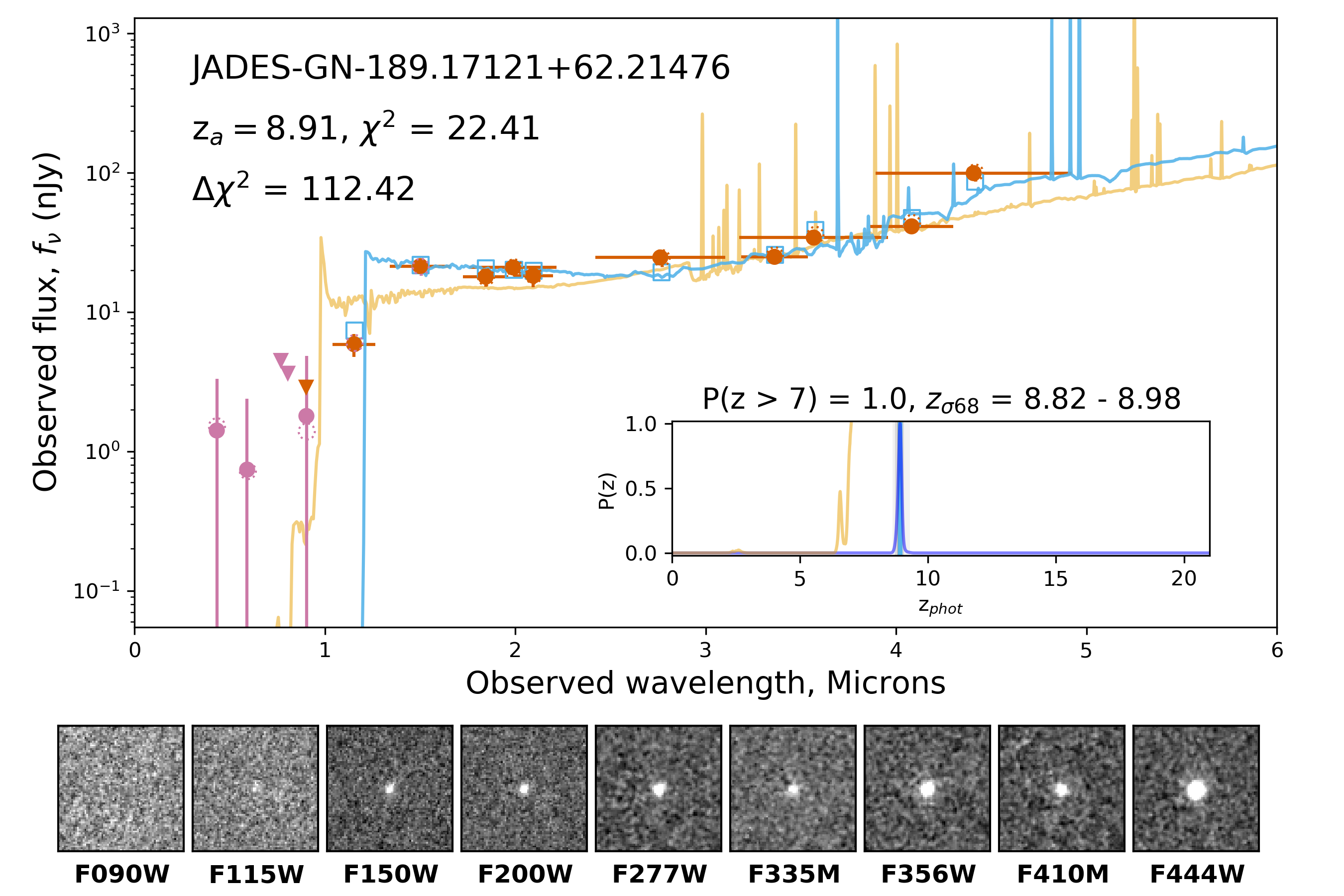}}\ 
{\includegraphics[width=0.49\textwidth]{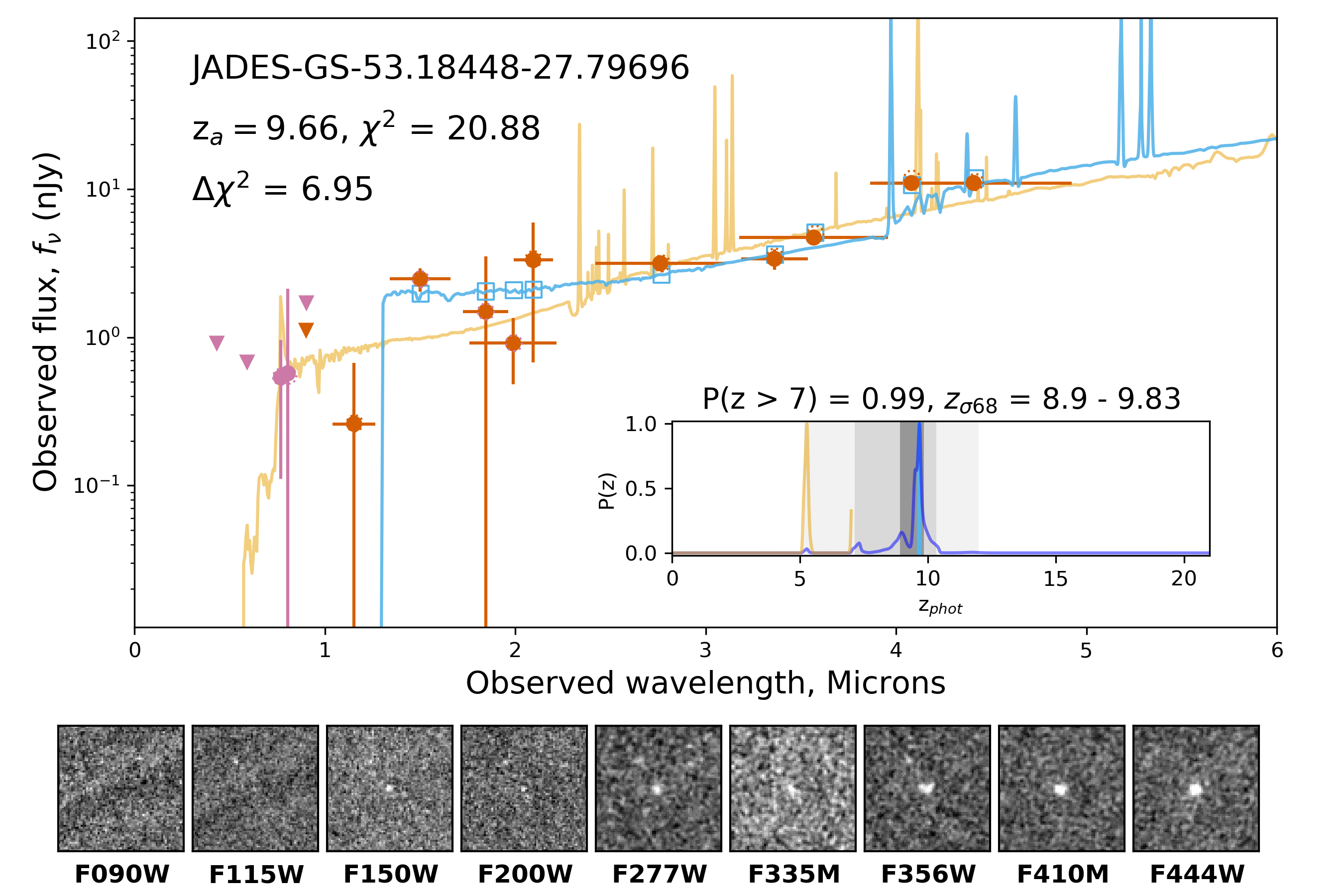}}\ 
{\includegraphics[width=0.49\textwidth]{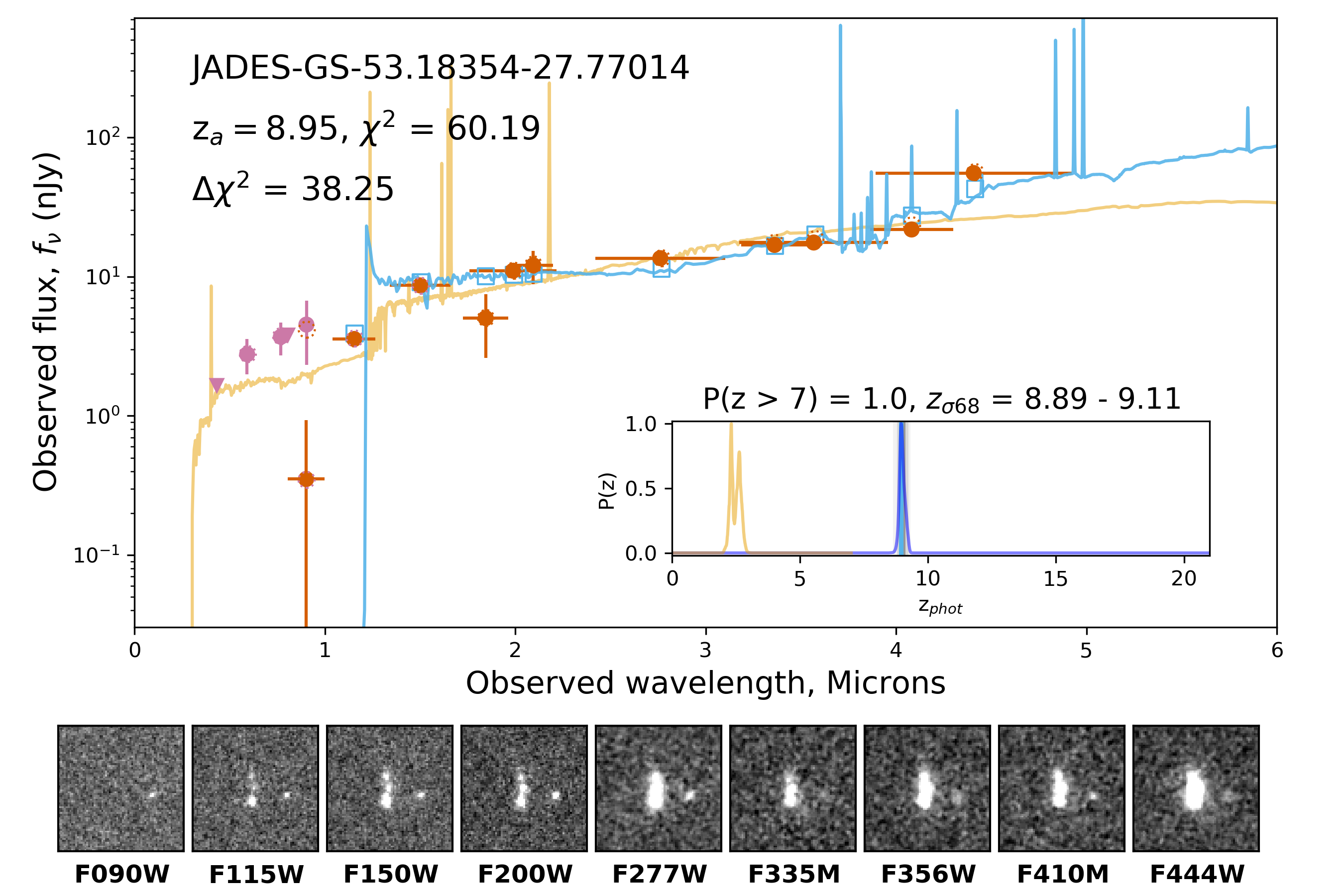}}\ 
{\includegraphics[width=0.49\textwidth]{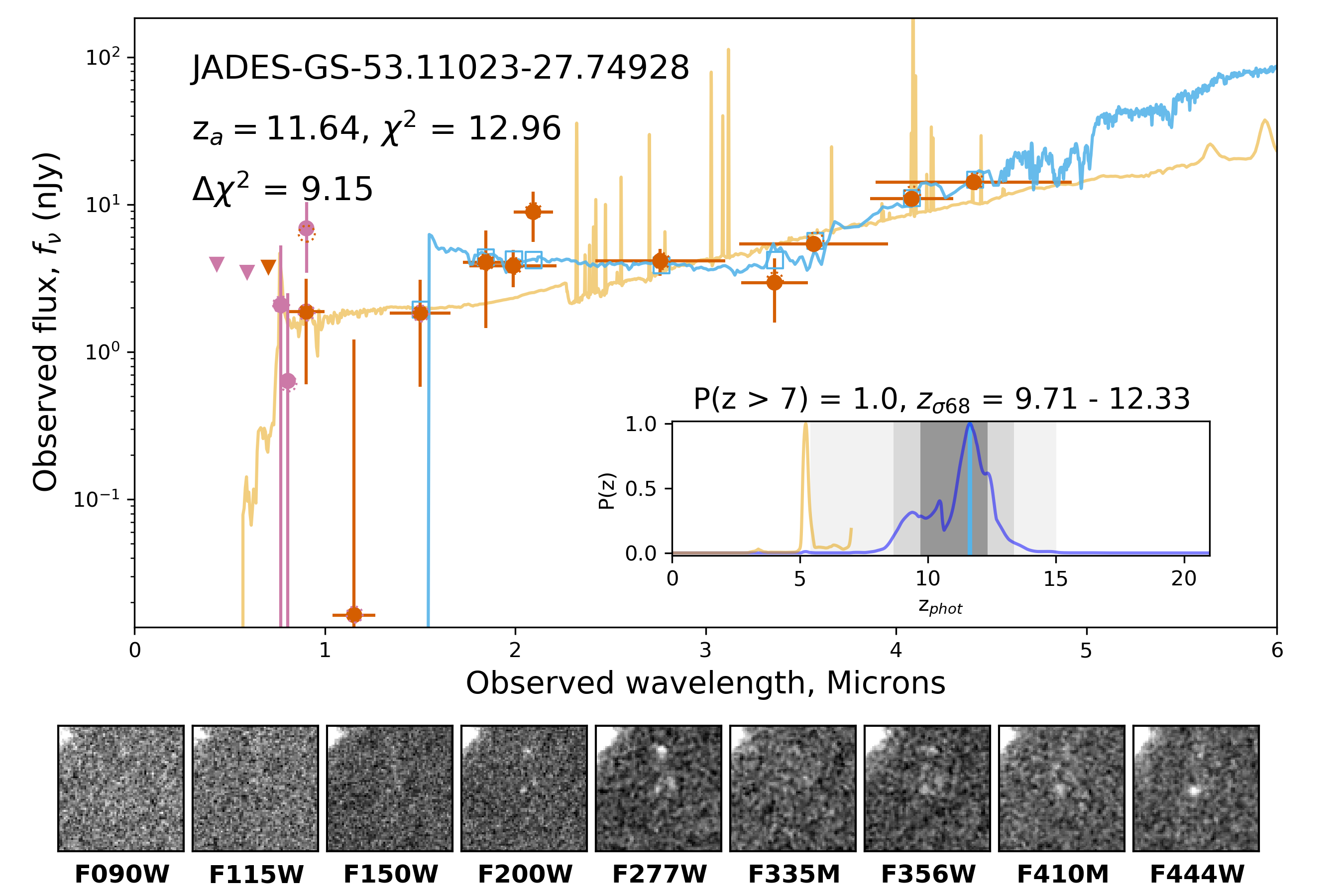}}
\caption{
Example SEDs for six candidate galaxies with red long-wavelength slopes. In each panel, the colors, lines, and symbols are as in Figure \ref{example_SED_fit}.}
\label{fig:red_slope_objects}
\end{figure*}

For our sample, these cuts select 12 objects  (9 in GOODS-S and 3 in GOODS-N). Of those sources, 11 are at $z_a = 8 - 10$, while one source is at $z_a = 11.64$. We provide the IDs, $z_a$ values, F444W magnitudes (measured in an $0.2^{\prime\prime}$ aperture), and colors for these sources in Table \ref{tab:red_slope_objects}, and we show six of these sources in Figure \ref{fig:red_slope_objects}. Outside of the highest-redshift source, JADES-GS-53.11023-27.74928, these sources have fairly tight lower limits on their redshift due to both the lack of flux observed in the F090W band and the red slope not being easily reproduced at low redshift. The {\tt EAZY} templates used in the present analysis are able to fit the observed SEDs as high-redshift sources, as is demonstrated with the blue lines in Figure \ref{fig:red_slope_objects}. However, JADES-GS-53.11023-27.74928 is very faint ($\sim 1$ nJy) at wavelengths shorter than 2$\mu$m, making a photometric redshift estimate difficult. JADES-GS-53.18354-27.77014, a source with a FRESCO spectroscopic redshift ($z_{spec} = 8.38$) is extended with three visible clumps spanning $0.6^{\prime\prime}$ (2.9 kpc at $z_a = 8.38$), of which the central knot has a very observed red UV through optical slope. 

\begin{deluxetable*}{l c c c c}
\tabletypesize{\footnotesize}
\tablecolumns{5}
\tablewidth{0pt}
\tablecaption{Sources with Red Long-Wavelength Slopes \label{tab:red_slope_objects}}
\tablehead{
\colhead{JADES ID} &  \colhead{{\tt EAZY} $z_a$} & \colhead{$m_{F444W}$}  &  \colhead{$m_{F277W} - m_{F444W}$} & \colhead{$m_{F200W} - m_{F356W}$} }
\startdata
		JADES-GN-189.05064+62.27935 & $8.04$ & 27.63 & 1.495 & 0.209 \\
		JADES-GS-53.04601-27.85399 & $8.3$ & 26.44 & 1.493 & 0.602 \\
		JADES-GS-53.19904-27.77207 & $8.31$ & 29.14 & 1.313 & 0.182 \\
		JADES-GS-53.19211-27.75252 & $8.53^{a}$ & 26.79 & 1.354 & 0.608 \\
		JADES-GN-189.18036+62.28851 & $8.69$ & 28.42 & 1.51 & 1.021 \\
		JADES-GS-53.18392-27.78691 & $8.71$ & 28.14 & 1.76 & 0.907 \\
		JADES-GS-53.1387-27.79248 & $8.87$ & 28.4 & 1.88 & 0.475 \\
		JADES-GN-189.17121+62.21476 & $8.91^{b}$ & 26.41 & 1.514 & 0.535 \\
		JADES-GS-53.18354-27.77014 & $8.95^{c}$ & 27.04 & 1.529 & 0.506 \\
		JADES-GS-53.18087-27.80577 & $9.33$ & 29.42 & 1.414 & 0.185 \\
		JADES-GS-53.18448-27.79696 & $9.66$ & 28.8 & 1.355 & 1.783 \\
		JADES-GS-53.11023-27.74928 & $11.64$ & 28.52 & 1.339 & 0.372
\enddata
\tablecomments{a: $z_{spec} = 7.99$, b: $z_{spec} = 8.62$, c: $z_{spec} = 8.38$}
\end{deluxetable*}

The origin of these sources is not obvious. One possible cause of such a red slope is the presence of a dust-obscured accretion disk from supermassive black hole growth in these objects, as discussed in \citet{furtak2022}, \citet{barro2023} and \citet{akins2023}. This would be of interest given the lack of ultra-high-redshift active galaxies currently known, and the short timescales by which these supermassive black holes could have grown in the early universe. Another alternative is that these sources could have strong optical line emission that boosts the long-wavelength flux, similar to what is presented in \citet{endsley2022}. In this work, the authors describe how galaxy models with young stellar populations or supermassive black hole growth can replicate the photometry for a sample of sources selected from JWST CEERS. An alternate view is offered in \citet{labbe2023}, who argue that sources like these are instead very massive, and the red long-wavelength slope is indicative of an evolved population, although this interpretation is in contrast to theoretical models of galaxy growth \citep{prada2023}. A continued exploration of the stellar properties of JADES sources at $z = 7 - 9$ with red long-wavelength slopes is discussed in Endsley et al. (in prep). However, until a number of these sources are followed-up with deep spectroscopy, their nature will remain elusive. 

\section{Conclusions} \label{sec:conclusions}

In this paper, we have assembled a sample of 717 galaxies and candidate galaxies at $z > 8$ selected from the 125 sq. arcmin JWST JADES observations of GOODS-N and GOODS-S. We combined these data with publicly available medium-band observations from JEMS and FRESCO, and describe our data reduction and photometric extraction. Our primary results are listed below:

\begin{itemize}
  \item Using the template-fitting code {\tt EAZY}, we calculated photometric redshifts for the JADES sources, and selected $z > 8$ candidates based on source SNR, the resulting probability of the galaxy being at $z > 7$, $P(z > 7)$, and the difference in $\chi^2$ between the best-fit at $z > 8$ and the fit at $z < 7$. The final sample was visually inspected by seven of the authors, and contains 182 objects in GOODS-N and 535 objects in GOODS-S, consistent with the areas and observational depths in the different portions of the JADES survey.
  \item The photometric redshifts of these sources extend to $z \sim 18$, with an F277W Kron magnitude range of $25 - 31$ (AB). The brightest source in our sample is the previously studied galaxy GN-z11 ($m_{F277W, Kron} = 25.73$). We find 33 galaxy candidates at $z_a > 12$, with the highest-redshift candidate being JADES-GN-189.15981+62.28898 with a photometric redshift of $z_a = 18.79$.  
 \item We find a number galaxies and galaxy candidates at $z = 8 - 12$ that are visually extended across many kpc and consist of multiple UV-bright clumps with underlying diffuse optical emission, potentially demonstrating very early massive galaxy growth. 
  \item Forty-two of the sources in our sample have spectroscopic redshift measurements. Each spectroscopic redshift agrees with the photometric redshift for the source within ${|z_{spec} - z_a| / (1 + z_{spec}) < 0.15}$. We find an average offset between the calculated photometric redshifts and the spectroscopic redshifts of ${\langle \Delta z = z_{spec} - z_a \rangle= 0.26}$, lower than the results seen with other high-redshift samples in the literature. We speculate that the offset may be due to differences between the templates used to fit these objects and the observed galaxy SEDs, which will be mitigated as more accurate templates are created using high-redshift galaxy spectra from JWST/NIRSpec.  
  \item To explore whether any of the sources are consistent with being low-mass stars, we fit our sources with brown dwarf models and measure whether the objects are unresolved. The galaxy templates fit the photometry with better accuracy than the brown dwarf templates for the vast majority of cases.
  \item We demonstrate that while traditional color selection would find most of the sources in our sample, at specific redshift ranges there are a number of sources that fall outside of typical color selection criteria.
  \item These results are robust to the exact {\tt EAZY} templates used; the vast majority of sources found in our sample have similar redshifts when fit using the independently derived templates from \citet{larson2022}. 
  \item Our sample includes a number of intriguing sources with red long-wavelength slopes, potentially from dust heated by a growing supermassive black hole at $z > 8$. This red slope could also be due to an abundance of strong optical line emission from young stellar populations. 
\end{itemize}

Taken together, these sources represent an exciting and robust sample for follow-up studies of the early universe. The detailed stellar populations, as well as the resulting evolution of the mass and luminosity functions for the $z > 8$ JADES galaxies will be found in forthcoming studies from the JADES collaboration members. We also look forward to JADES Cycle 2 observations which will push to fainter observed fluxes. In addition, many of these sources will be observed with JADES NIRSpec MSA spectroscopy to both confirm their redshifts and to explore their ionization and metallicity properties. JWST has only just opened the door to the early universe, and the years to come promise to be the most scientifically fruitful in the history of extragalactic science.   

$ $\\

We want to thank the anonymous referee for their comments and suggestions which significantly improved this paper. This work is based on observations made with the NASA/ESA/CSA James Webb Space Telescope. The data were obtained from the Mikulski Archive for Space  Telescopes at the Space Telescope Science Institute, which is operated by the Association of Universities for Research in Astronomy, Inc., under NASA contract NAS5-03127 for JWST. These observations are associated with PID 1063, 1345, 1180, 1181, 1210, 1286, 1963, 1837, 1895, and 2738. Additionally, this work made use of the lux supercomputer at UC Santa Cruz which is funded by NSF MRI grant AST1828315, as well as the High Performance Computing (HPC) resources at the University of Arizona which is funded by the Office of Research Discovery and Innovation (ORDI), Chief Information Officer (CIO), and University Information Technology Services (UITS). We acknowledge support from the NIRCam Science Team contract to the University of Arizona, NAS5-02015. 

DJE is supported as a Simons Investigator.  E.C.L acknowledges support of an STFC Webb Fellowship (ST/W001438/1). S.C acknowledges support by European Union’s HE ERC Starting Grant No. 101040227 - WINGS. AJB, AJC, JC, IEBW, AS, \& GCJ acknowledge funding from the ``FirstGalaxies'' Advanced Grant from the European Research Council (ERC) under the European Union’s Horizon 2020 research and innovation programme (Grant agreement No. 789056). JW, WB, FDE, LS, TJL, and RM acknowledges support by the Science and Technology Facilities Council (STFC) ERC Advanced Grant 695671, ``QUENCH''. JW also acknowledges support from the Foundation MERAC. RM also acknowledges funding from a research professorship from the Royal Society. The research of CCW is supported by NOIRLab, which is managed by the Association of Universities for Research in Astronomy (AURA) under a cooperative agreement with the National Science Foundation. REH acknowledges support from the National Science Foundation Graduate Research Fellowship Program under Grant No. DGE-1746060. LW acknowledges support from the National Science Foundation Graduate Research Fellowship under Grant No. DGE-2137419. Funding for this research was provided by the Johns Hopkins University, Institute for Data Intensive Engineering and Science (IDIES). This research is supported in part by the Australian Research Council Centre of Excellence for All Sky Astrophysics in 3 Dimensions (ASTRO 3D), through project number CE170100013. DP acknowledges support by the Huo Family Foundation through a P.C. Ho PhD Studentship. The Cosmic Dawn Center (DAWN) is funded by the Danish National Research Foundation under grant no.140.


\appendix

\section{Exploring the Origin of Photometric Redshift Gaps At $z\sim10$ and $z\sim13$} \label{sec:PhotoZGaps}

\begin{figure}
\centering
{\includegraphics[width=0.48\textwidth]{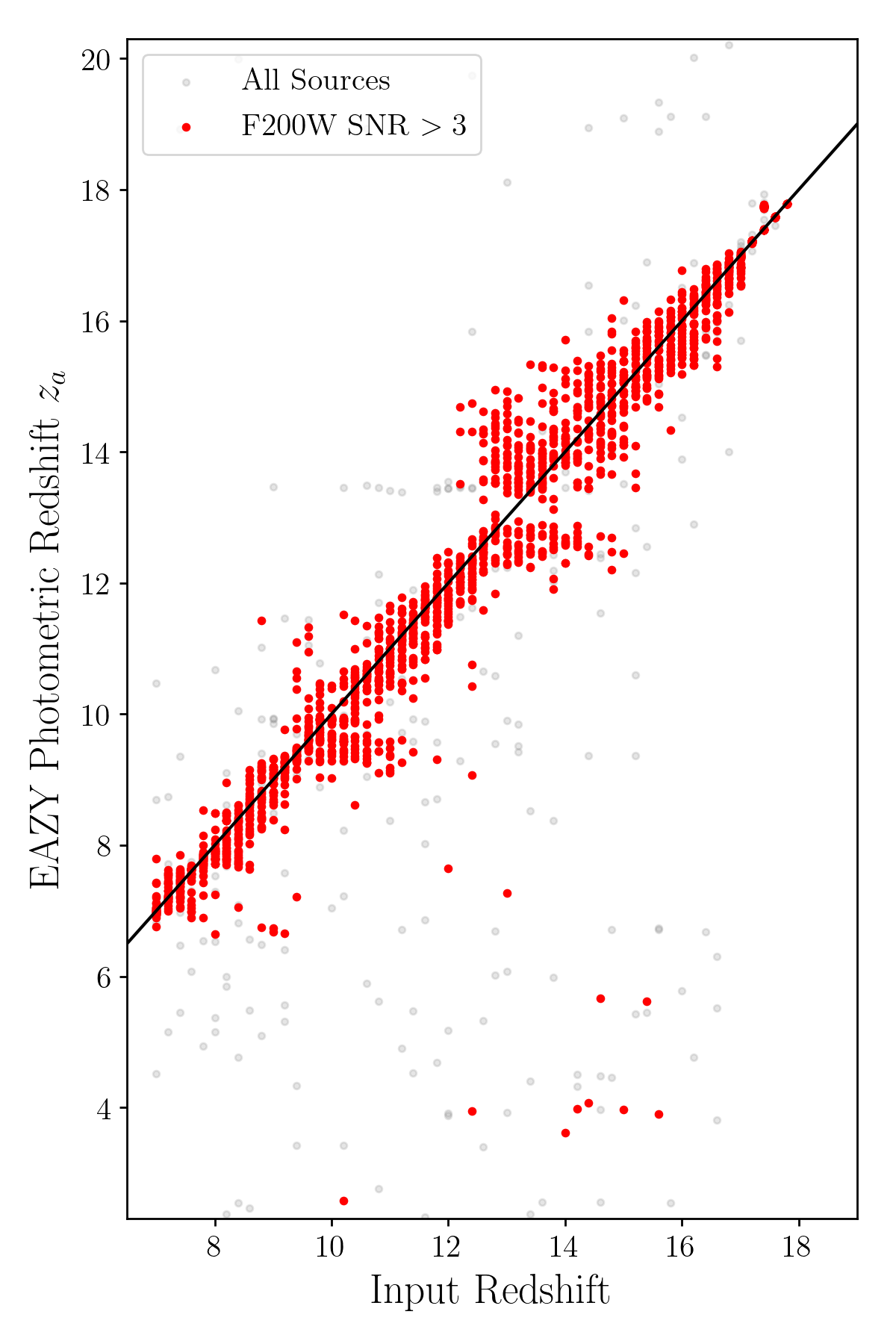}}\
\caption{
Photometric redshifts plotted against input redshifts for a simulated {\tt EAZY} SED placed at a grid of uniformly spaced redshifts between $z = 7 - 18$ with $\Delta z = 0.2$, and at F200W SNR values between 0.5 and 20. We plot all of the resulting photometric redshifts in grey, and those with F200W SNR $> 3$ in red. We see a pile-up of sources at photometric redshifts of $z \sim 10$ and $z \sim 13$.}
\label{fig:appendix_photoz_vs_specz}
\end{figure}

In Figure \ref{mag_vs_za}, we plot the photometric redshift distribution of the $z > 8$ candidates, and note in Section \ref{sec:results} that there are relative gaps in the distribution at $z \sim 10$ and $z \sim 13$. As mentioned in this Section, these gaps arise due to the usage of wide NIRCam F090W, F115W, and F150W filters to estimate redshifts. To help explore this effect, we calculated photometric redshifts using the same {\tt EAZY} fitting procedure described in the text but for a simulated SED placed at known redshifts. We started with the best-fit SED from {\tt EAZY} for a bright source in our sample and artificially redshifted this SED to between $z = 7 - 18$ with $\Delta z = 0.2$. We calculated the photometry for the SED at these redshifts with our JADES filter set, and added Gaussian noise in agreement with the values presented in Table \ref{tab:five_sigma_limits} for the JADES Deep footprint, with a uniform grid of F200W SNR values between 0.5 and 20 and $\Delta \mathrm{SNR} = 0.5$. We then fit the noisy photometry for these artificial sources using the same {\tt EAZY} templates and procedure as done on the full sample, and we plot the resulting photometric redshifts against the input redshifts in Figure \ref{fig:appendix_photoz_vs_specz}. In Figure \ref{fig:appendix_photoz_distribution}, we plot the photometric redshift distribution from the {\tt EAZY} fits. In each Figure, we plot sources at all SNR values in light grey, and those sources with F200W SNR $> 3$ in red. 

We observe the same gaps in the photometric redshifts for these simulated source fits as are seen for the true galaxies in Figure \ref{mag_vs_za}. The gaps are more easily visible in the simulated plots due to the uniform distribution of the input redshifts. There is a pile-up of sources at redshifts just lower than each observed gap, comprised of objects at higher simulated spectroscopic redshift, but where the Lyman-$\alpha$ break falls between two adjacent filters such that photometrically they are not distinguishable from a galaxy at slightly lower redshift.

\begin{figure}
\centering
{\includegraphics[width=0.48\textwidth]{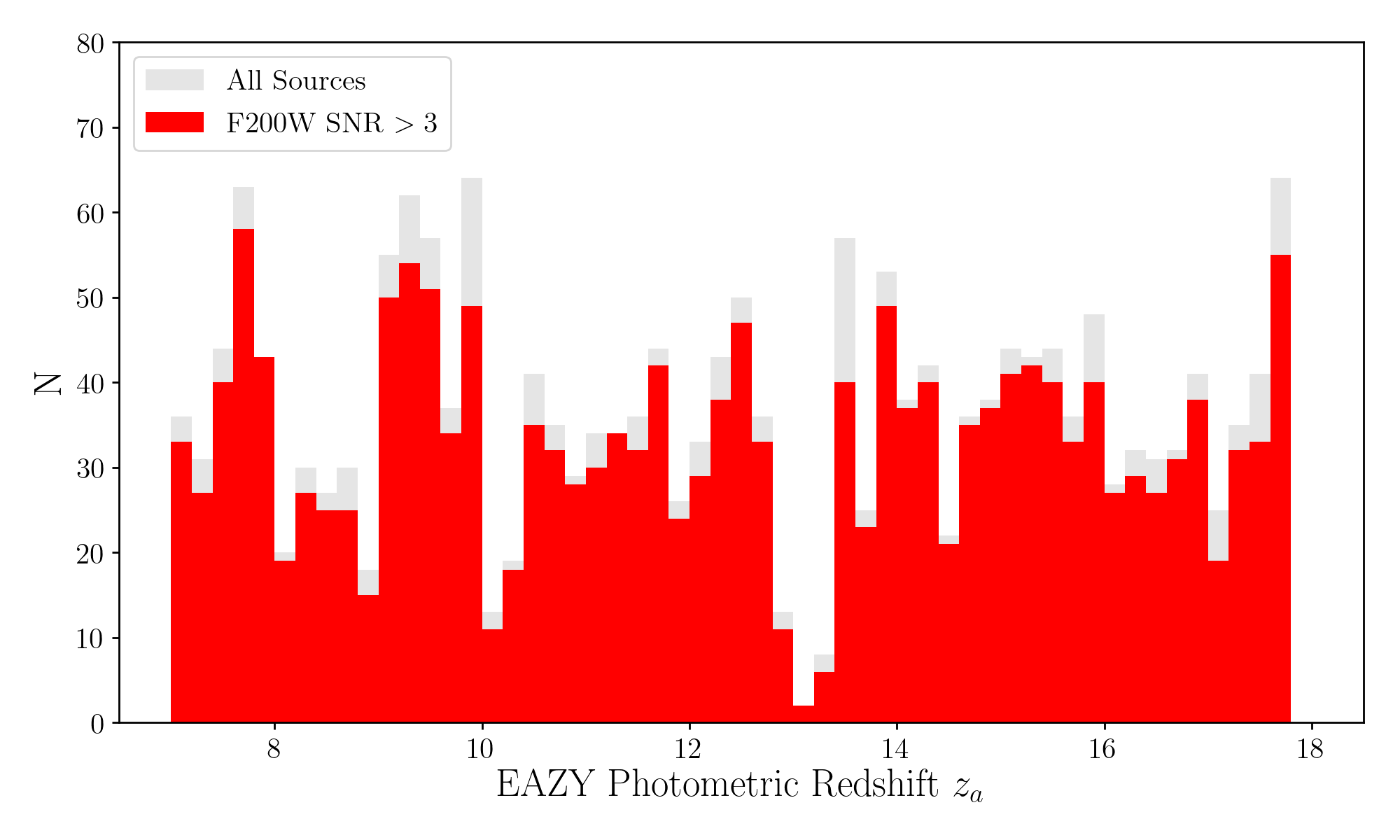}}\
\caption{
Distribution of photometric redshifts from the results shown in Figure \ref{fig:appendix_photoz_vs_specz}. We plot the distribution of all of the sources with grey bars, and we plot the distribution of sources with F200W SNR $> 3$ with red bars. The gaps we observe at $z \sim 10$ and $z \sim 13$ in Figure \ref{mag_vs_za} are more easily visible here for the simulated galaxies.}
\label{fig:appendix_photoz_distribution}
\end{figure}

\section{Additional Tables and Figures} \label{sec:additional_tables}

\begin{figure*}
\centering
$z > 12$ Candidates, Part II \par\medskip
{\includegraphics[width=0.49\textwidth]{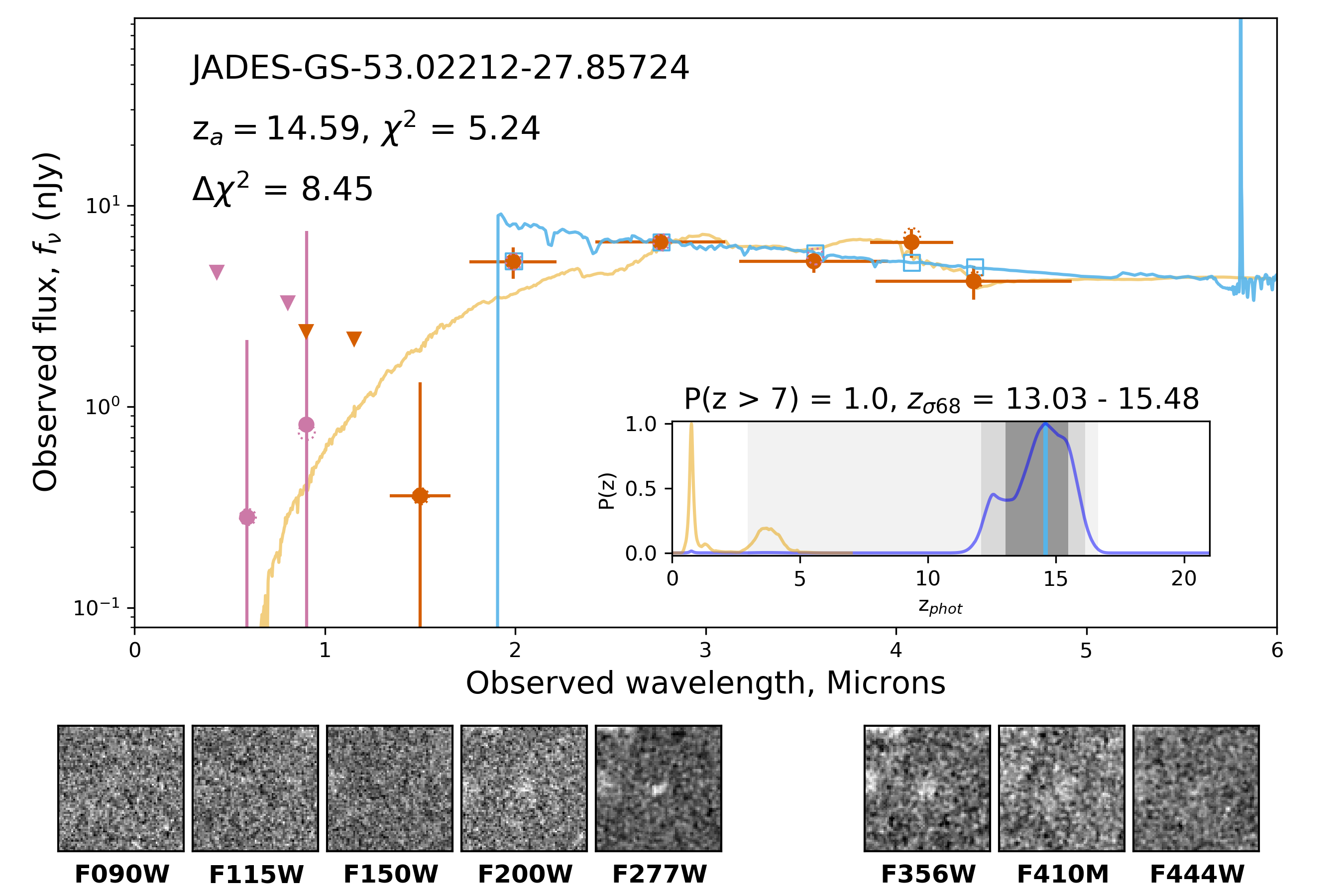}}\
{\includegraphics[width=0.49\textwidth]{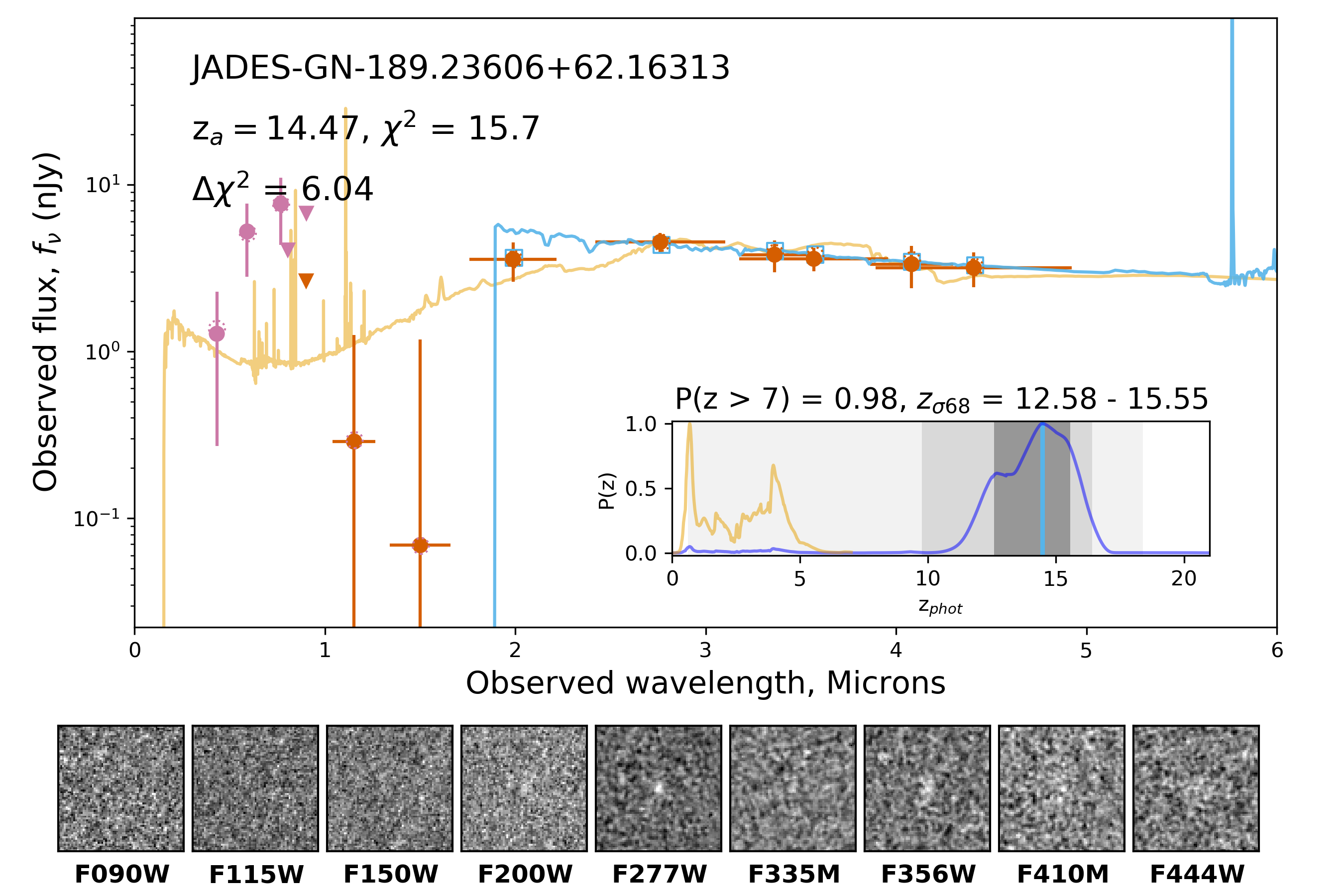}}\  
{\includegraphics[width=0.49\textwidth]{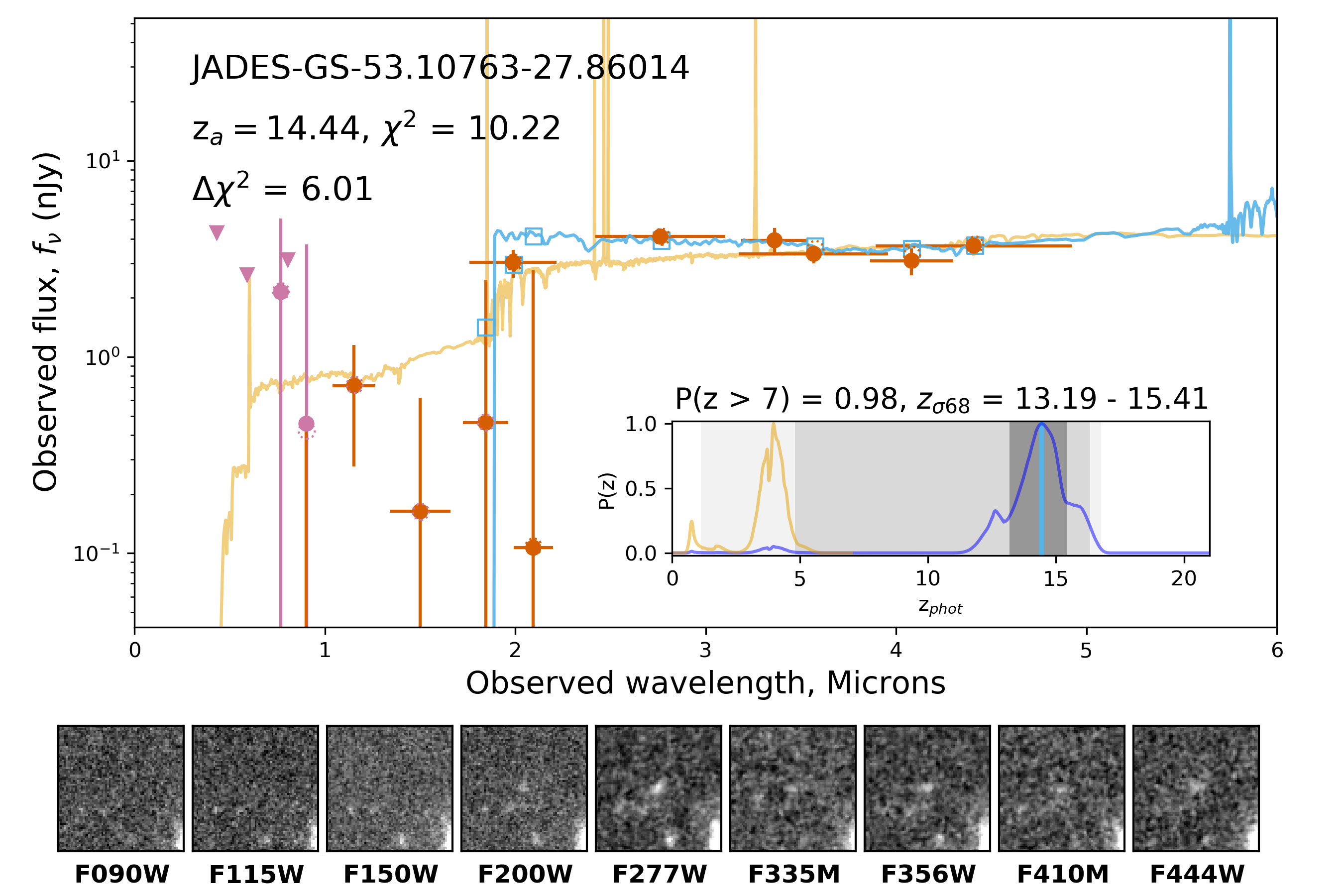}}\  
{\includegraphics[width=0.49\textwidth]{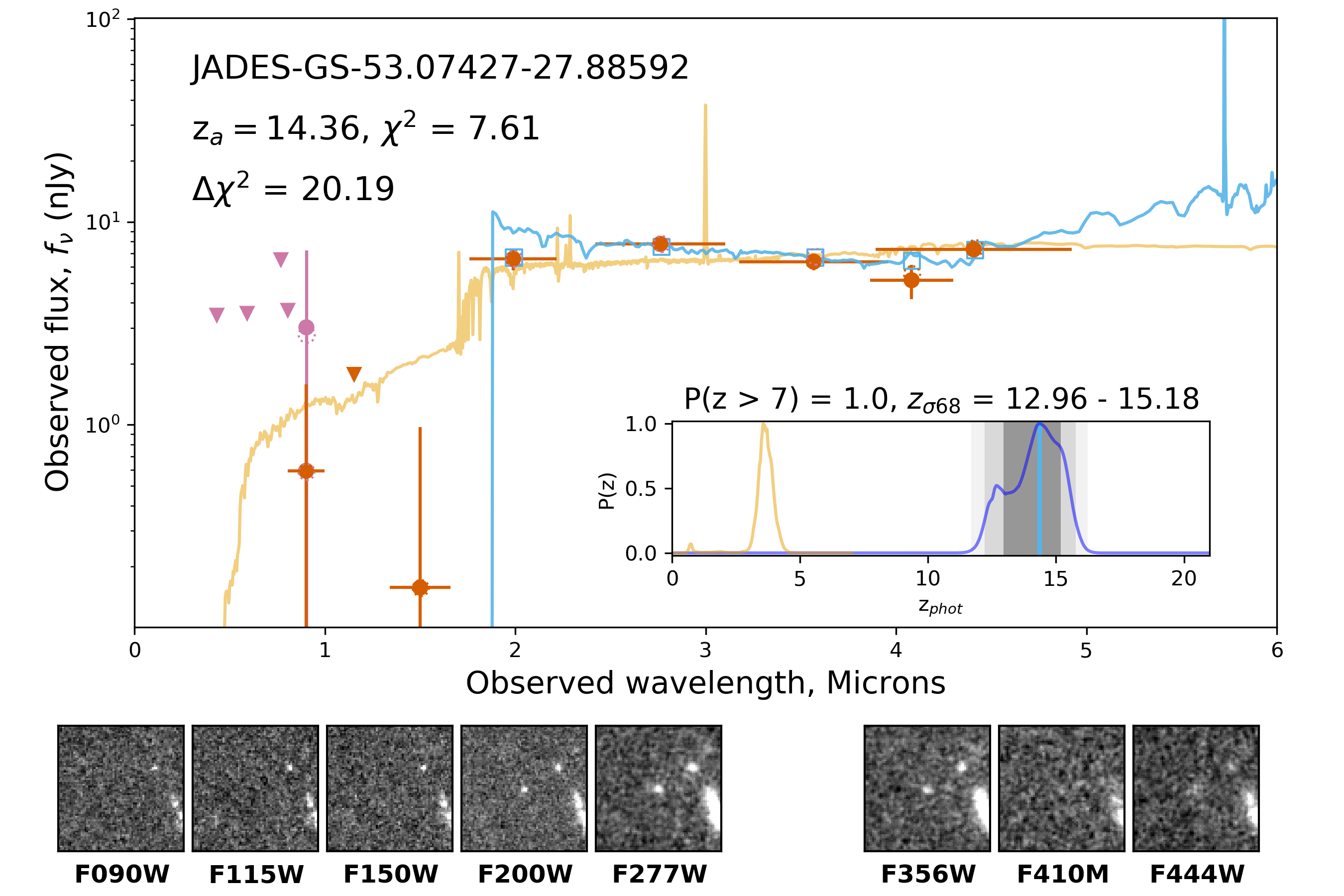}}\ 
{\includegraphics[width=0.49\textwidth]{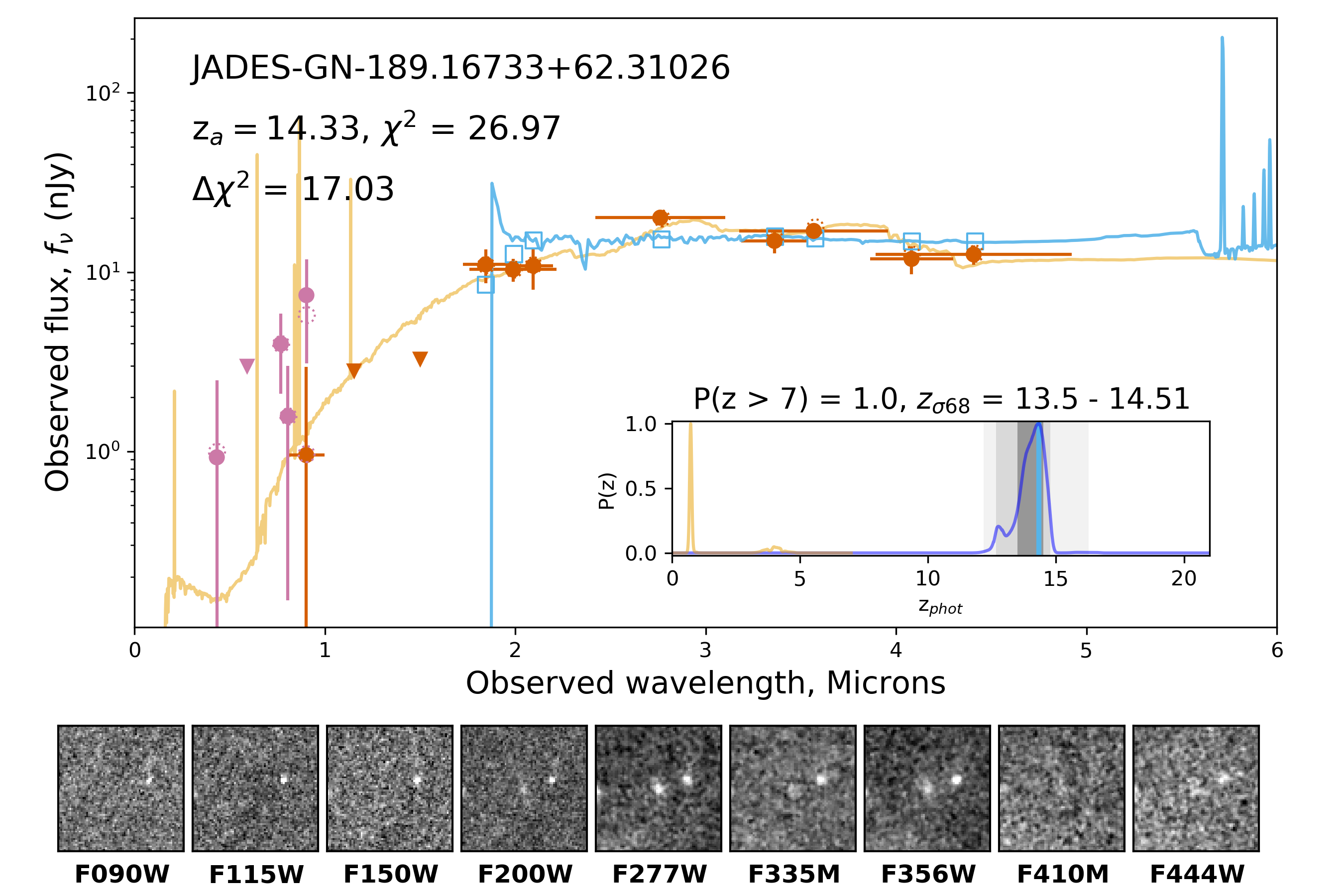}}\ 
{\includegraphics[width=0.49\textwidth]{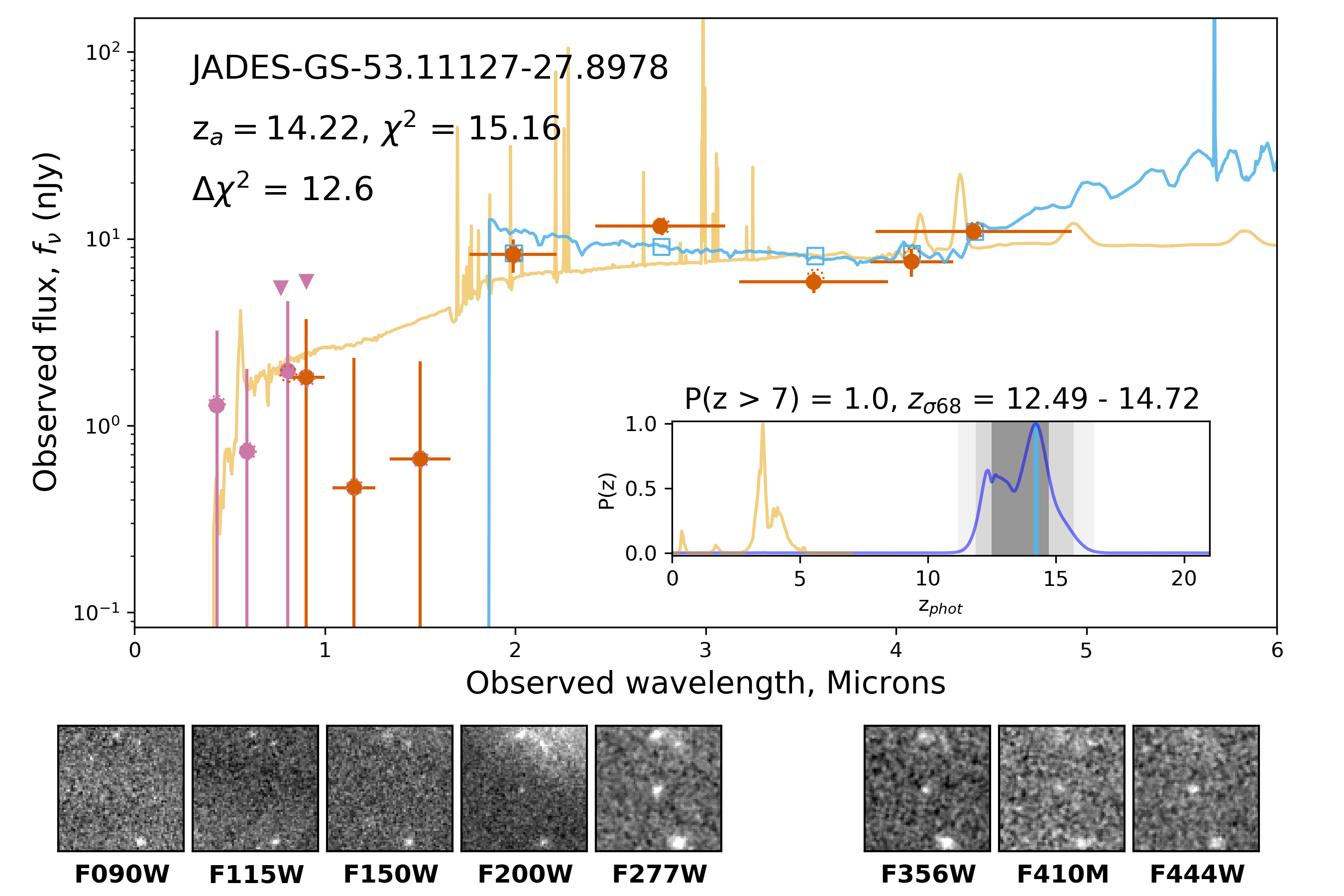}}\  
{\includegraphics[width=0.49\textwidth]{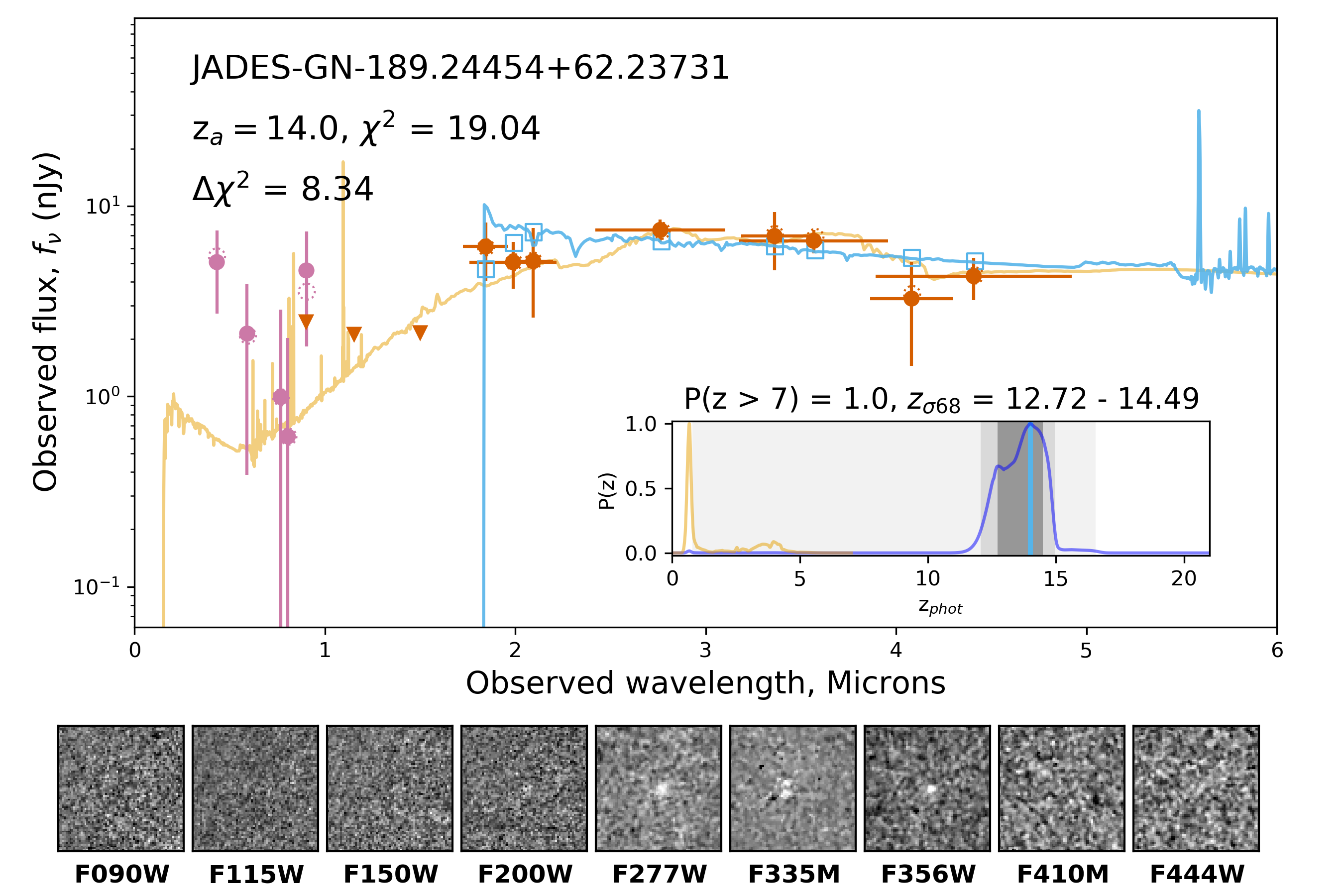}} 
{\includegraphics[width=0.49\textwidth]{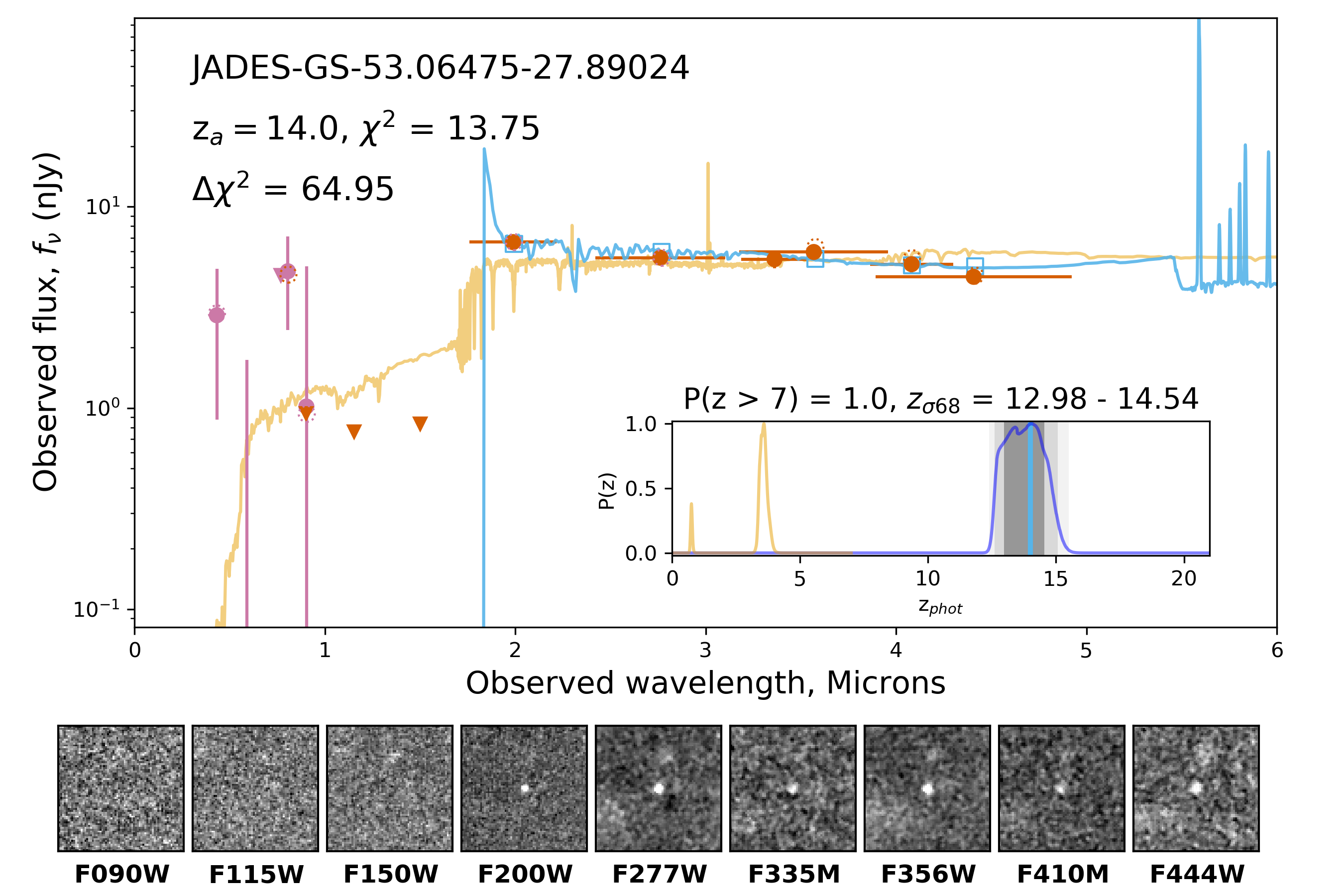}}\  
\caption{
Continuation of Figure \ref{fig:z_gt_12_example_SEDs_pt1}. In each panel, the colors, lines, and symbols are as in Figure \ref{example_SED_fit}.}
\label{fig:z_gt_12_example_SEDs_pt2}
\end{figure*}

\begin{figure*}
\centering
$z > 12$ Candidates, Part III \par\medskip
{\includegraphics[width=0.49\textwidth]{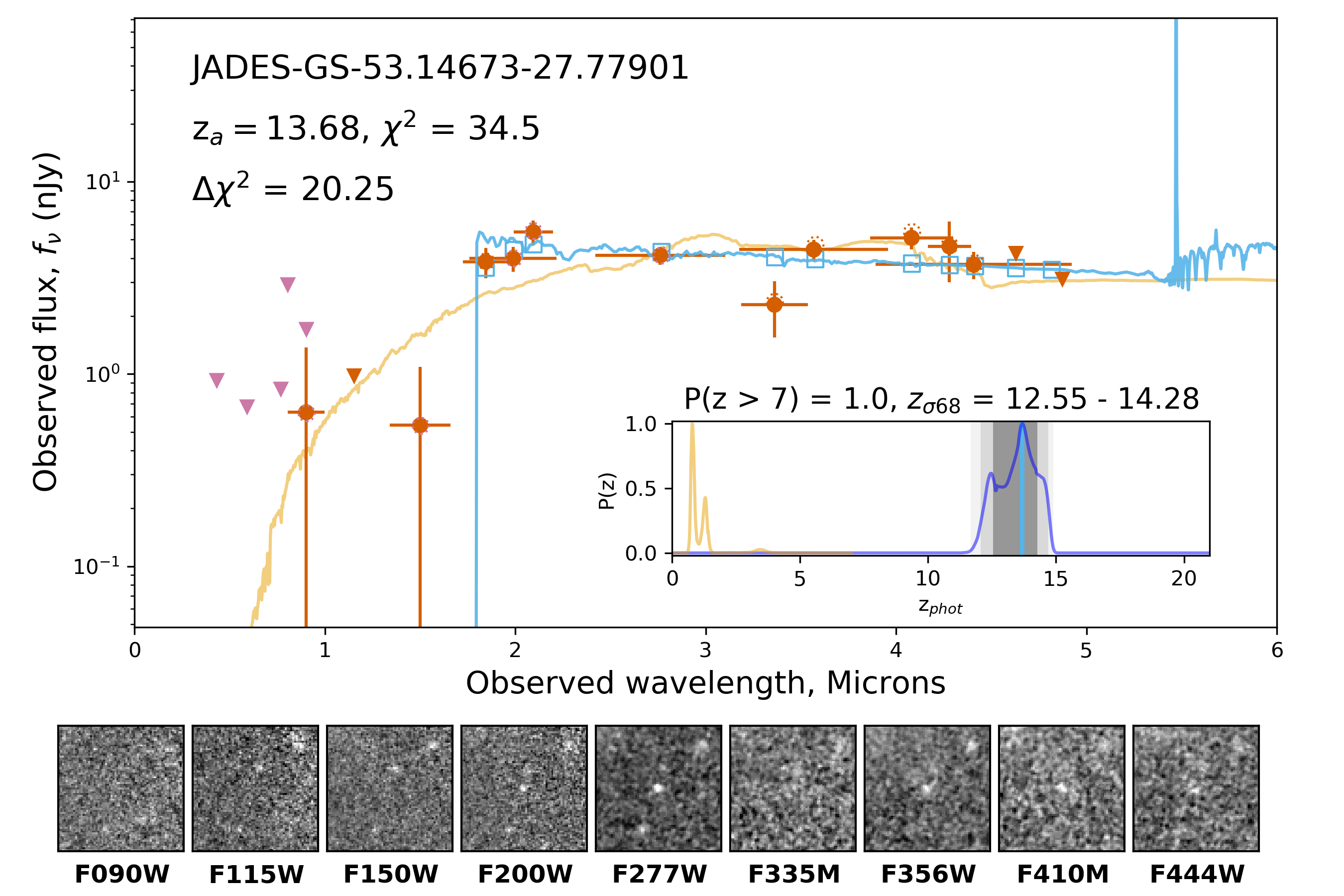}}\  
{\includegraphics[width=0.49\textwidth]{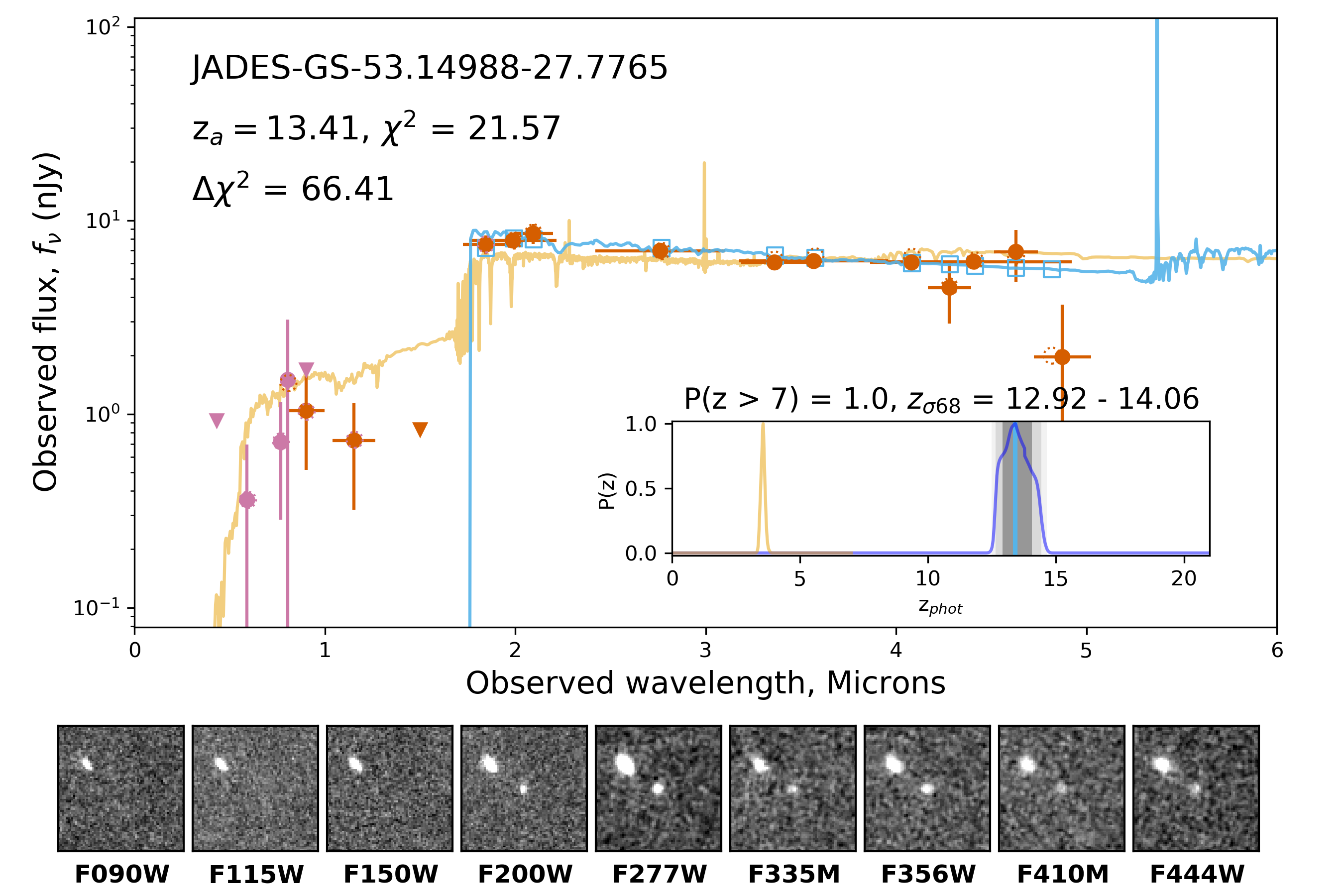}}\  
{\includegraphics[width=0.49\textwidth]{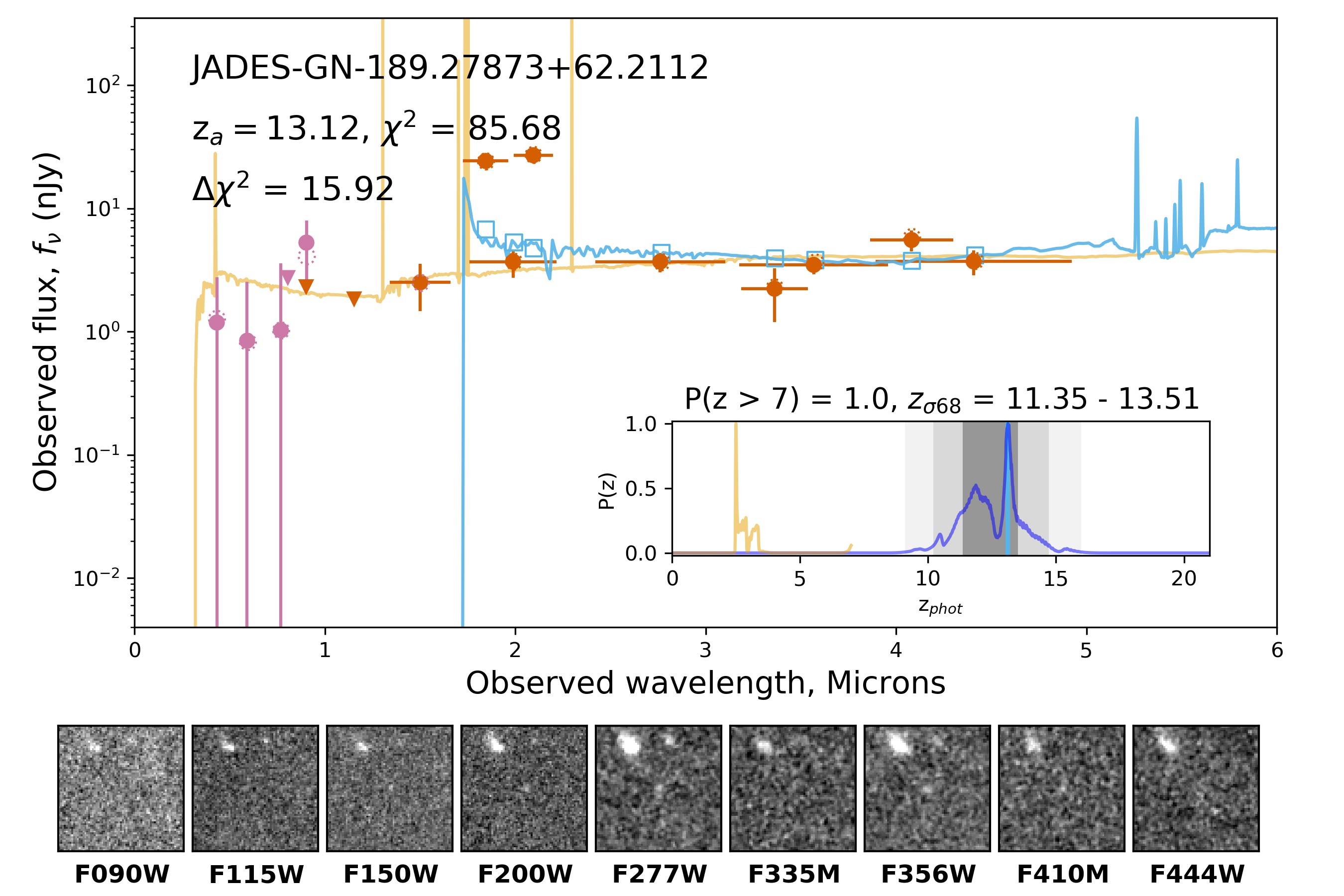}}\   
{\includegraphics[width=0.49\textwidth]{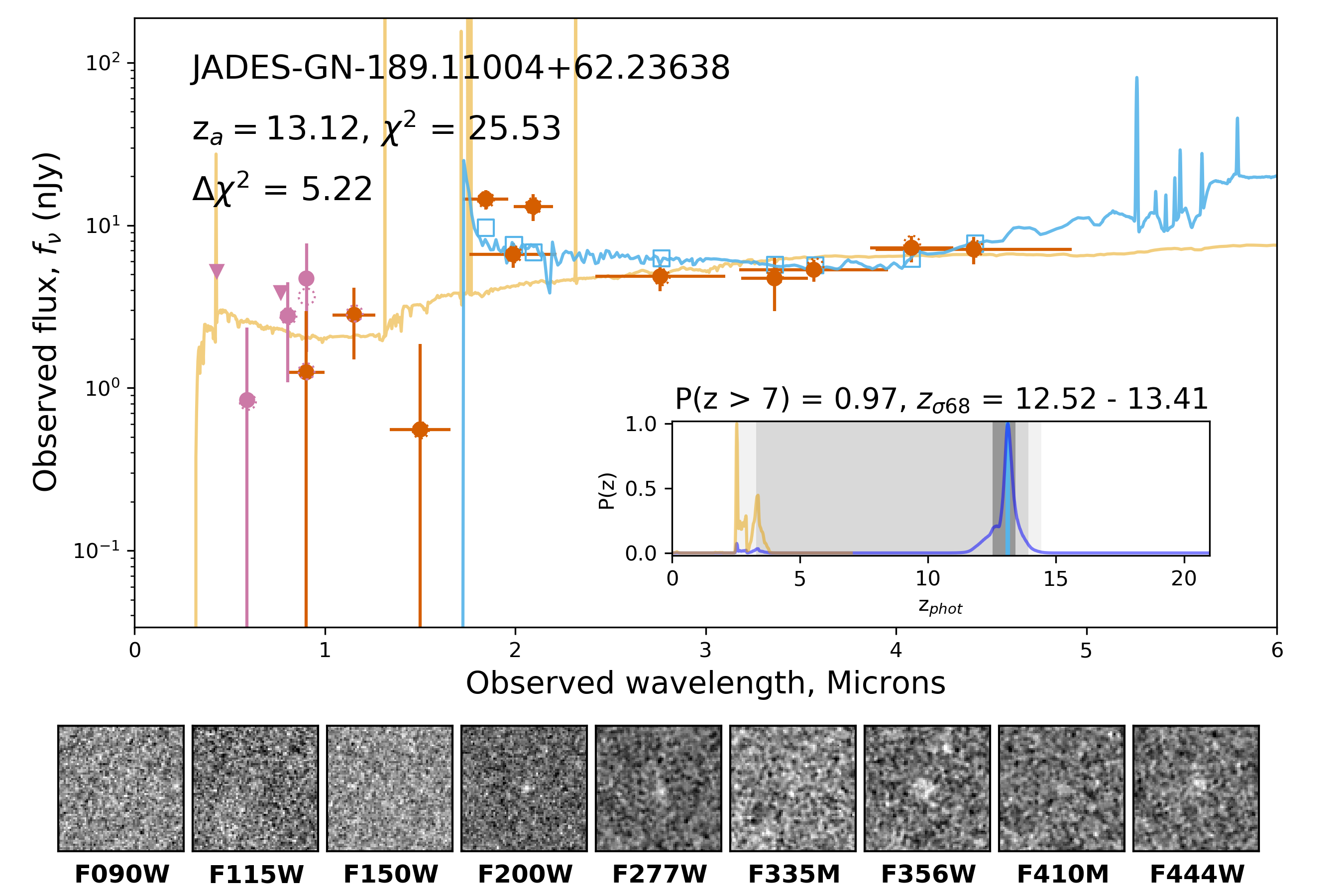}}\   
{\includegraphics[width=0.49\textwidth]{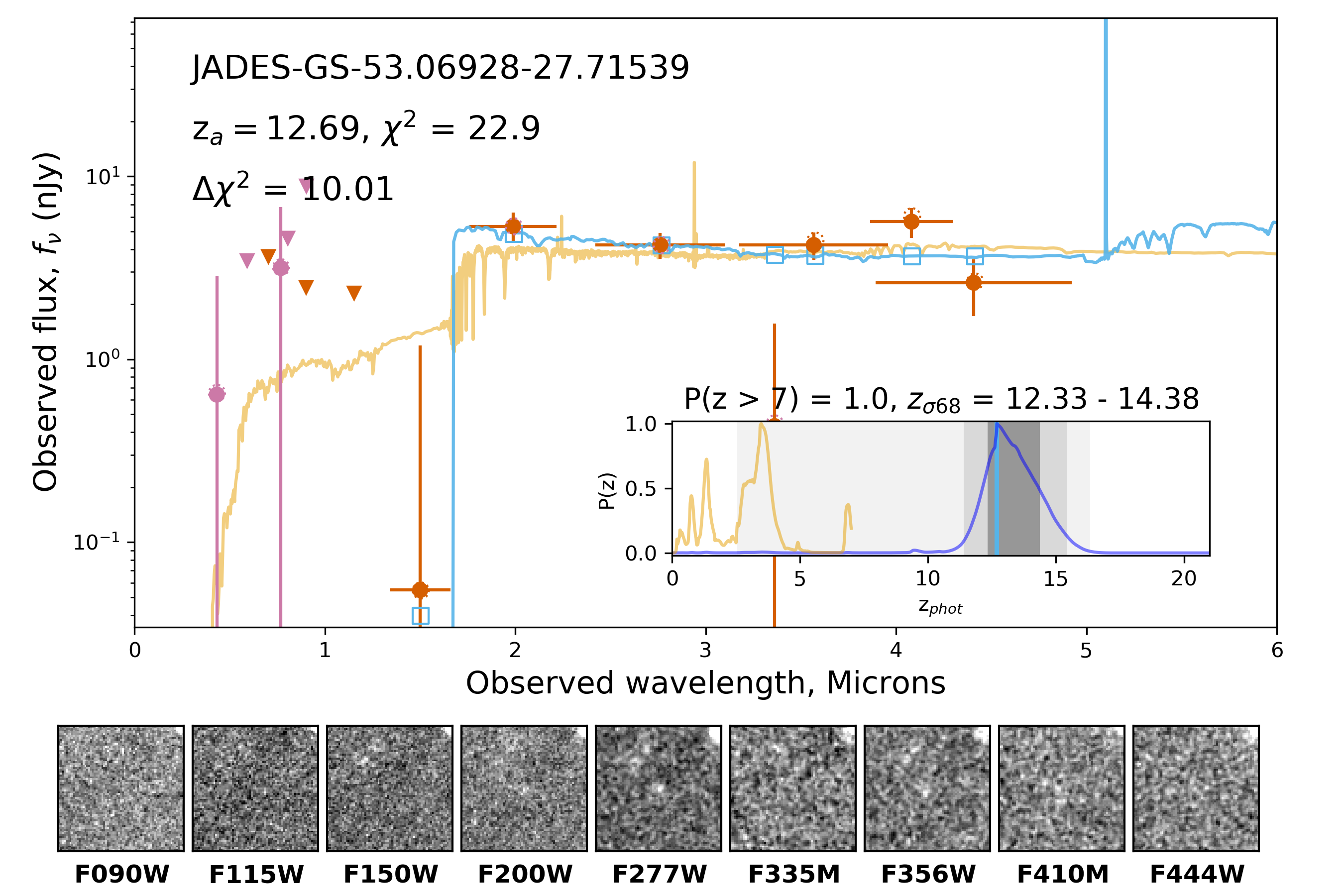}}\   
{\includegraphics[width=0.49\textwidth]{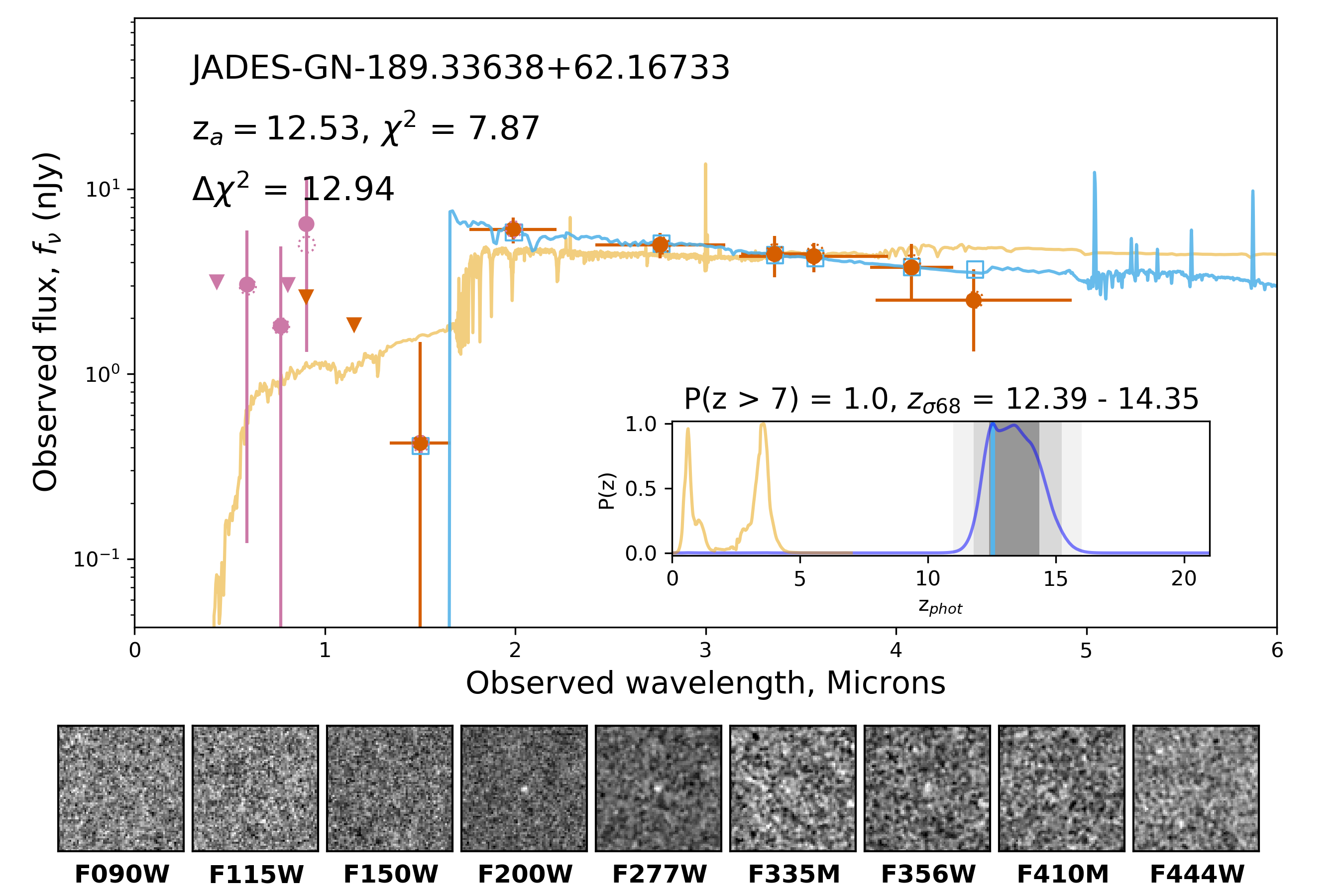}}\   
{\includegraphics[width=0.49\textwidth]{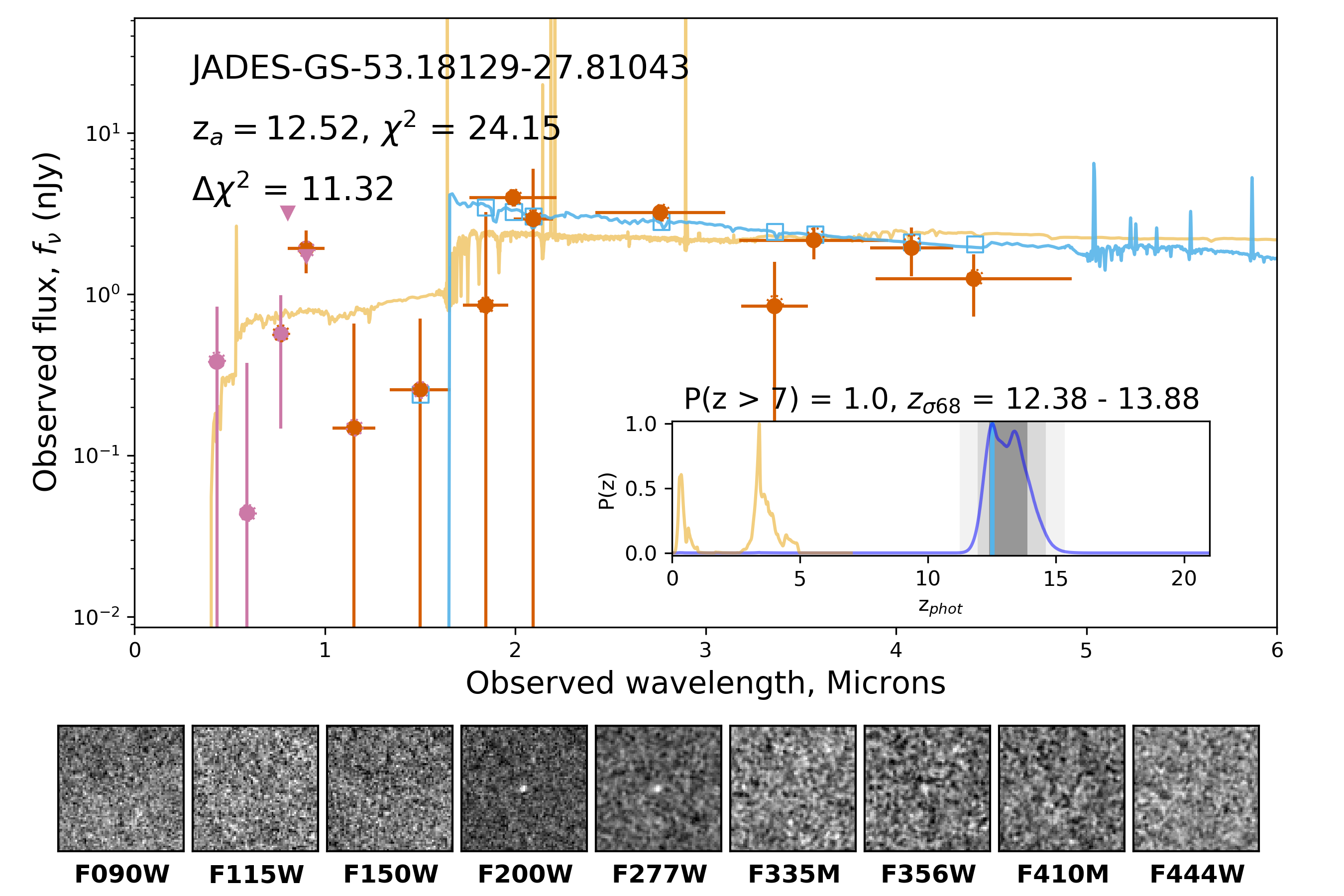}}\  
{\includegraphics[width=0.49\textwidth]{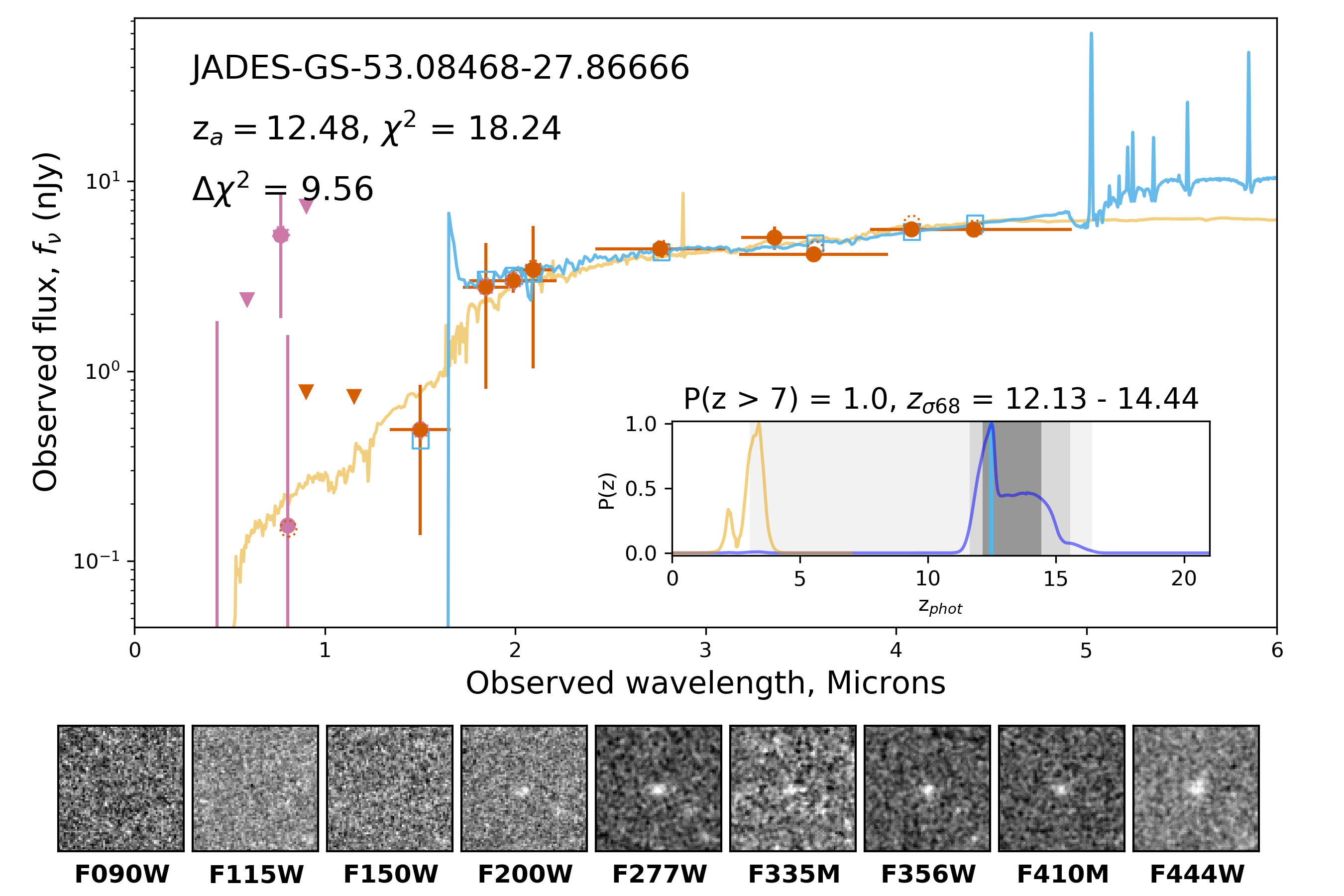}}\
\caption{
Continuation of Figure \ref{fig:z_gt_12_example_SEDs_pt2}. In each panel, the colors, lines, and symbols are as in Figure \ref{example_SED_fit}.}
\label{fig:z_gt_12_example_SEDs_pt3}
\end{figure*}

\begin{figure*}
\centering
$z > 12$ Candidates, Part IV \par\medskip
{\includegraphics[width=0.49\textwidth]{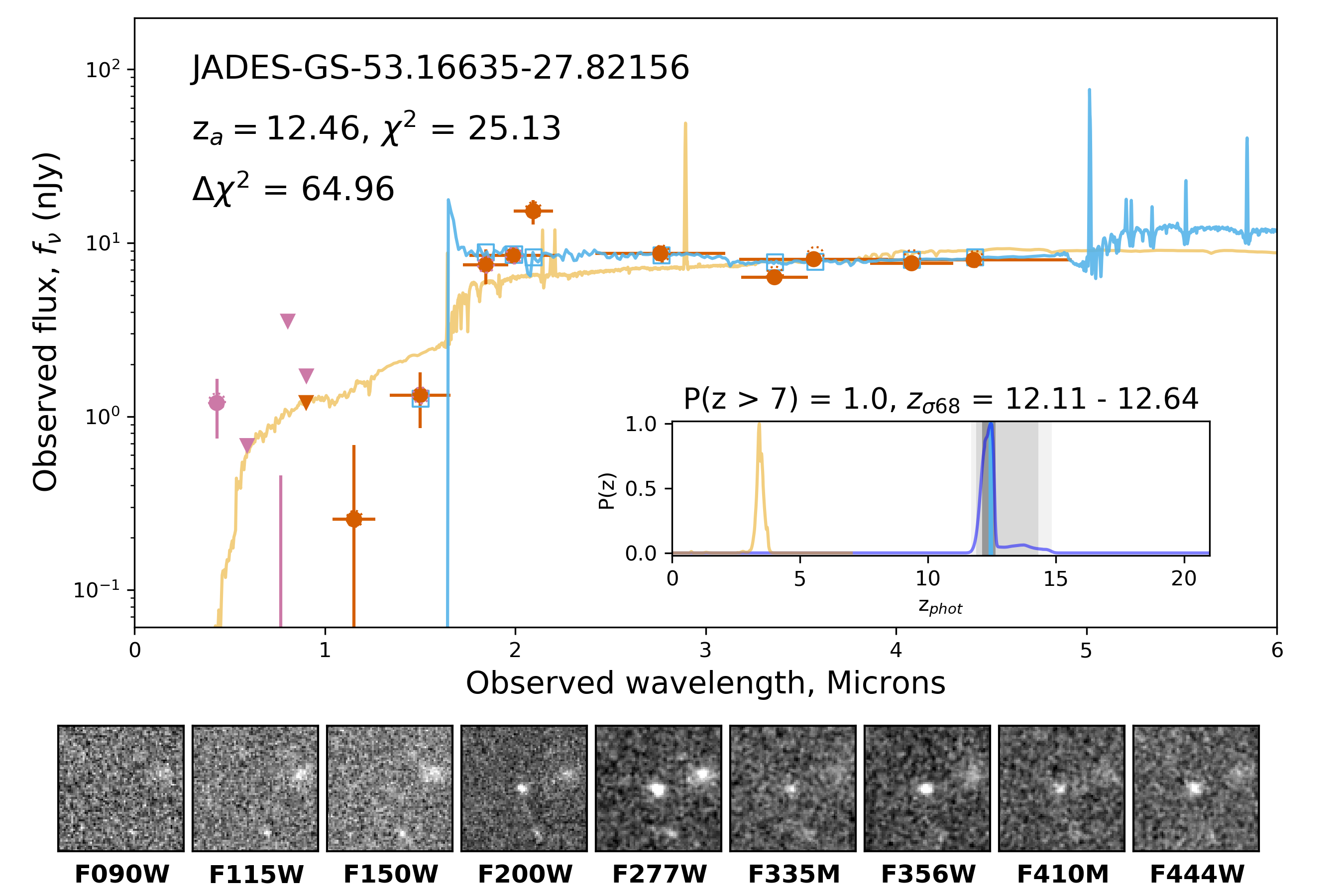}}\ 
{\includegraphics[width=0.49\textwidth]{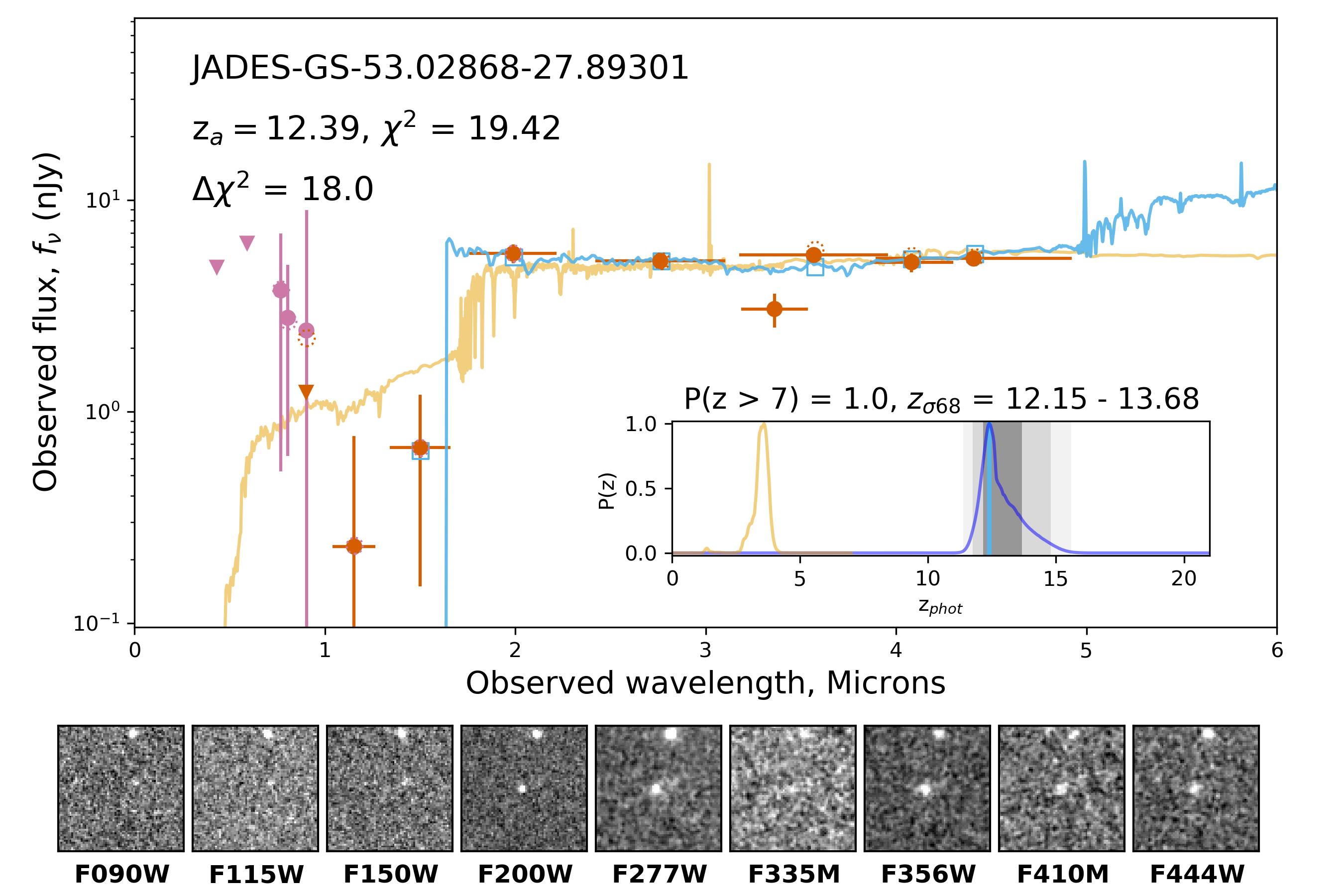}}\ 
{\includegraphics[width=0.49\textwidth]{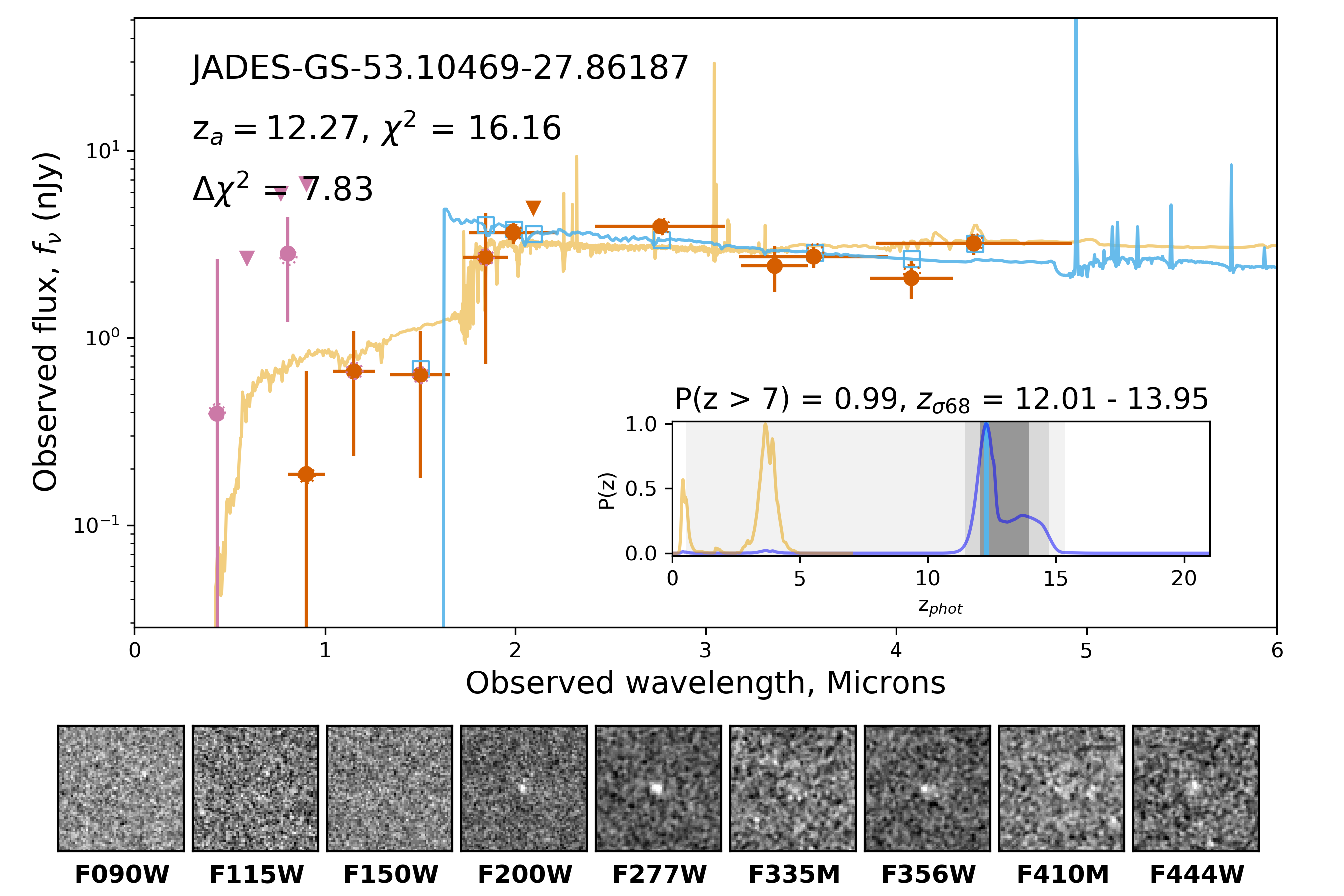}}\ 
{\includegraphics[width=0.49\textwidth]{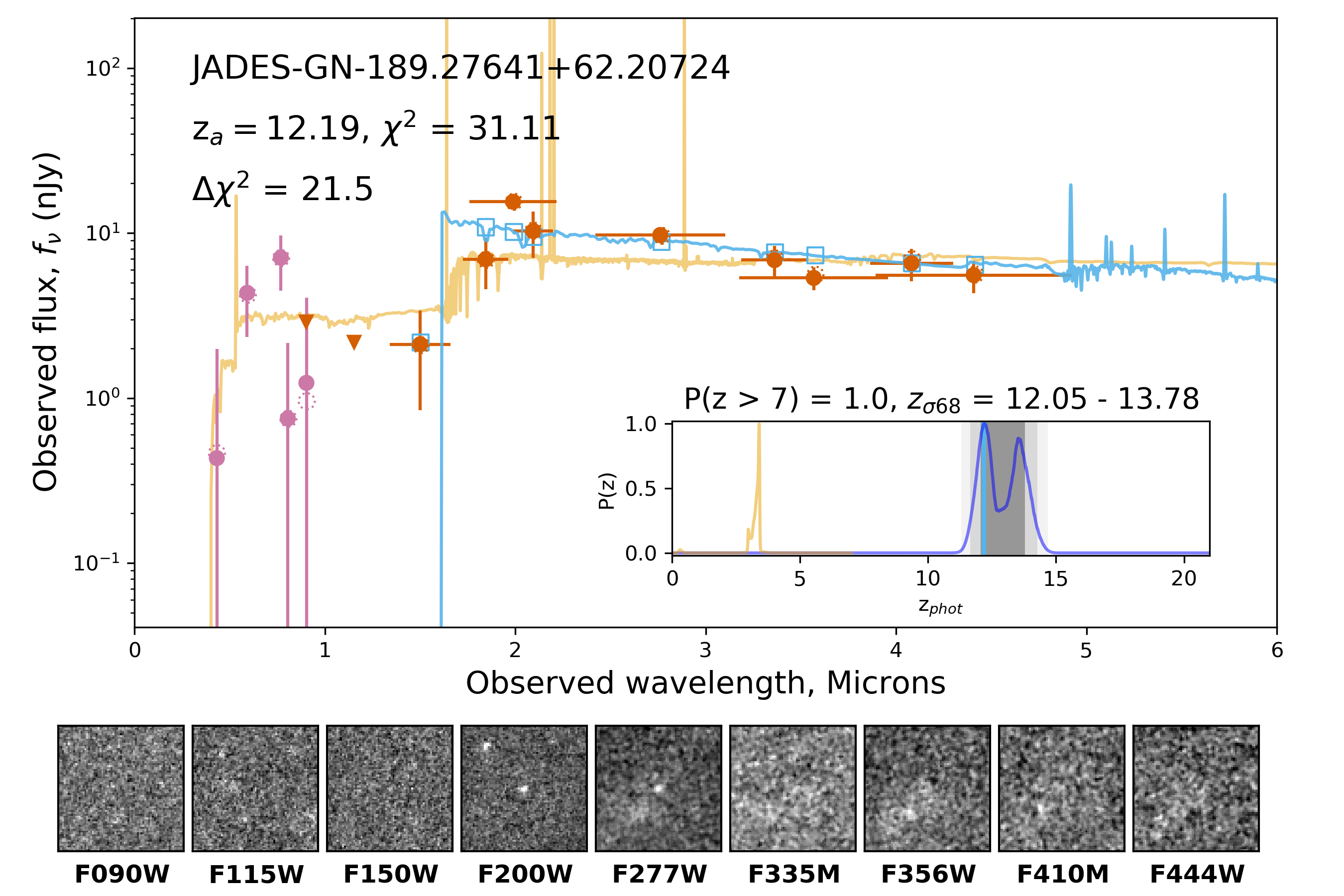}}\   
{\includegraphics[width=0.49\textwidth]{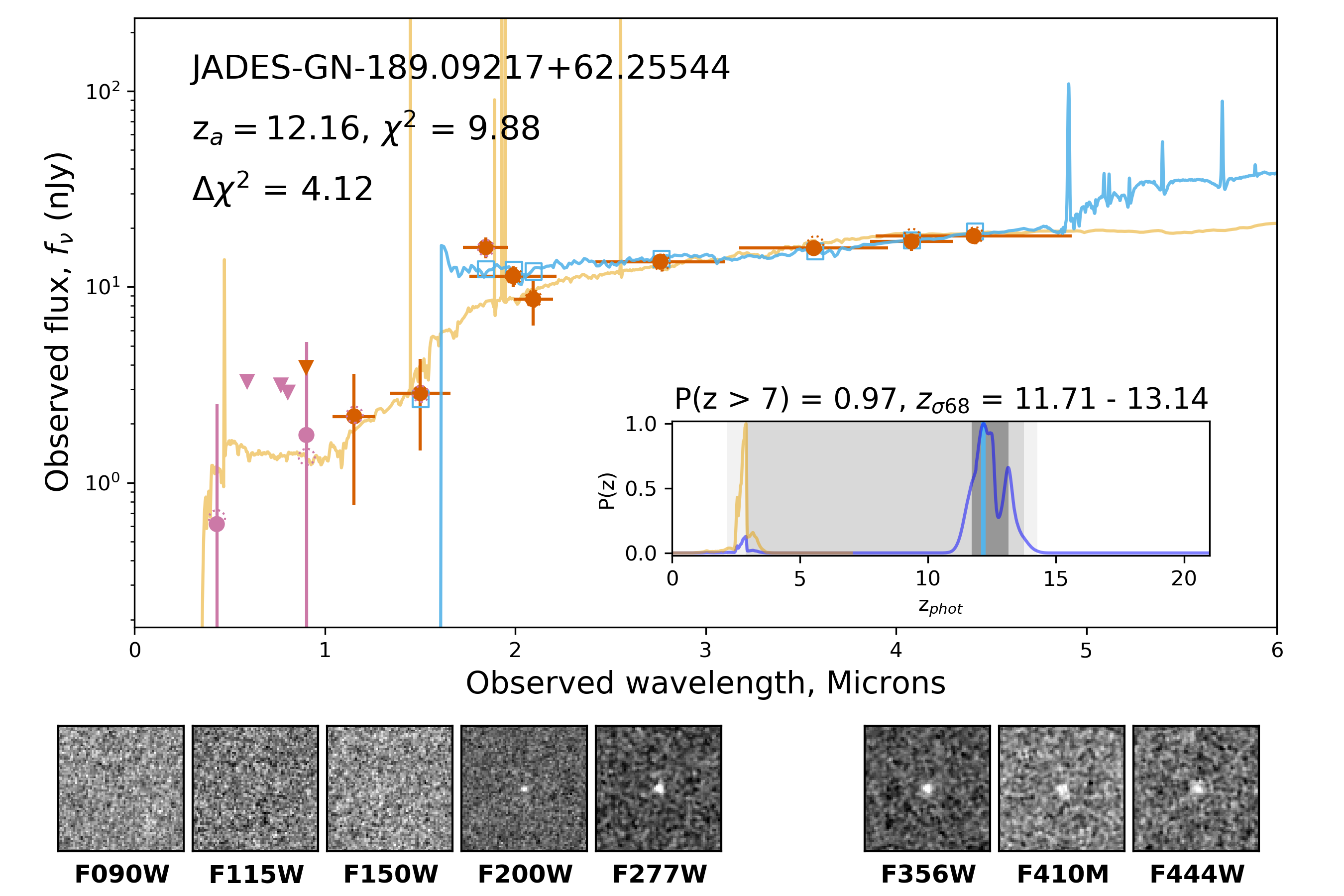}}\   
{\includegraphics[width=0.49\textwidth]{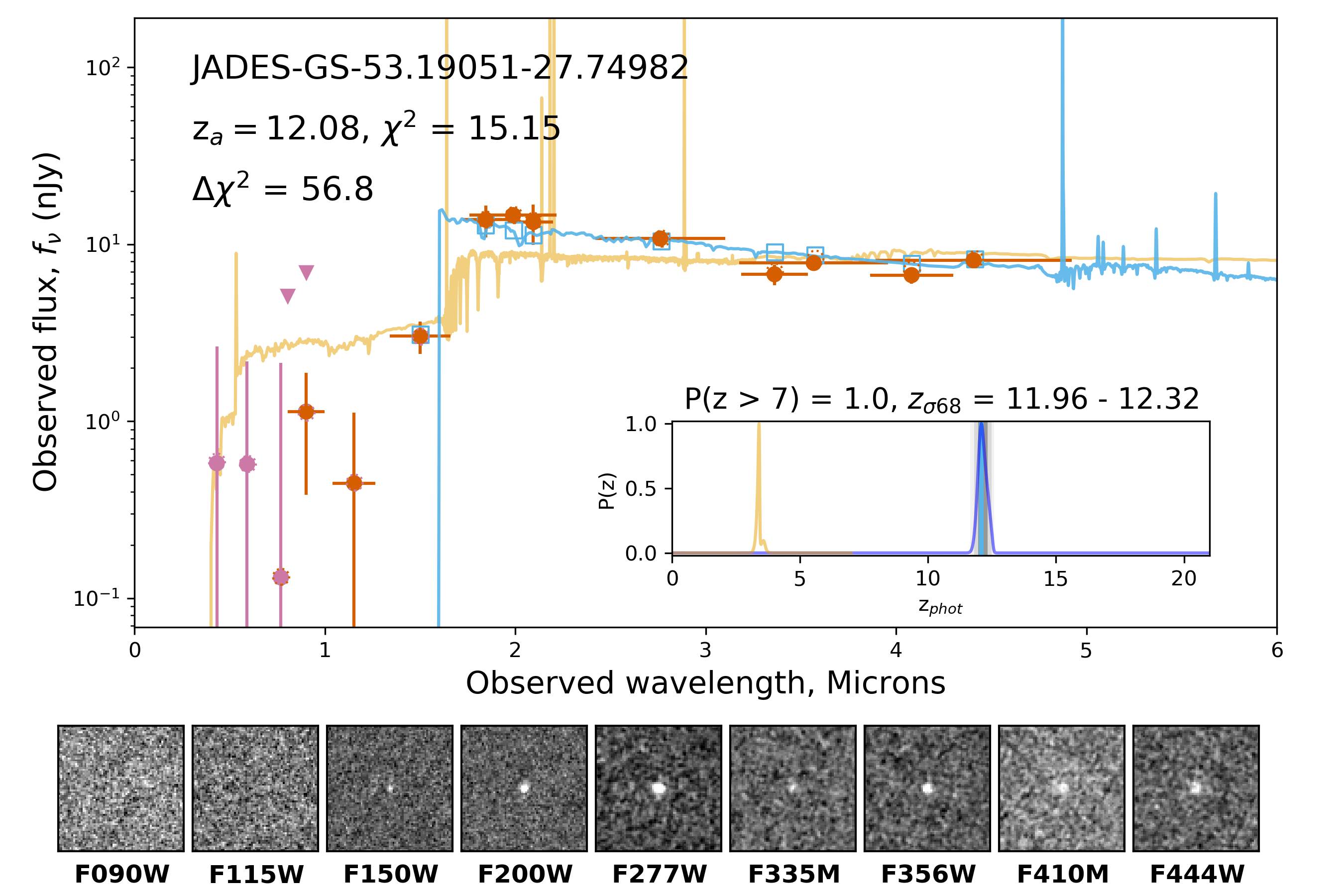}}\   
{\includegraphics[width=0.49\textwidth]{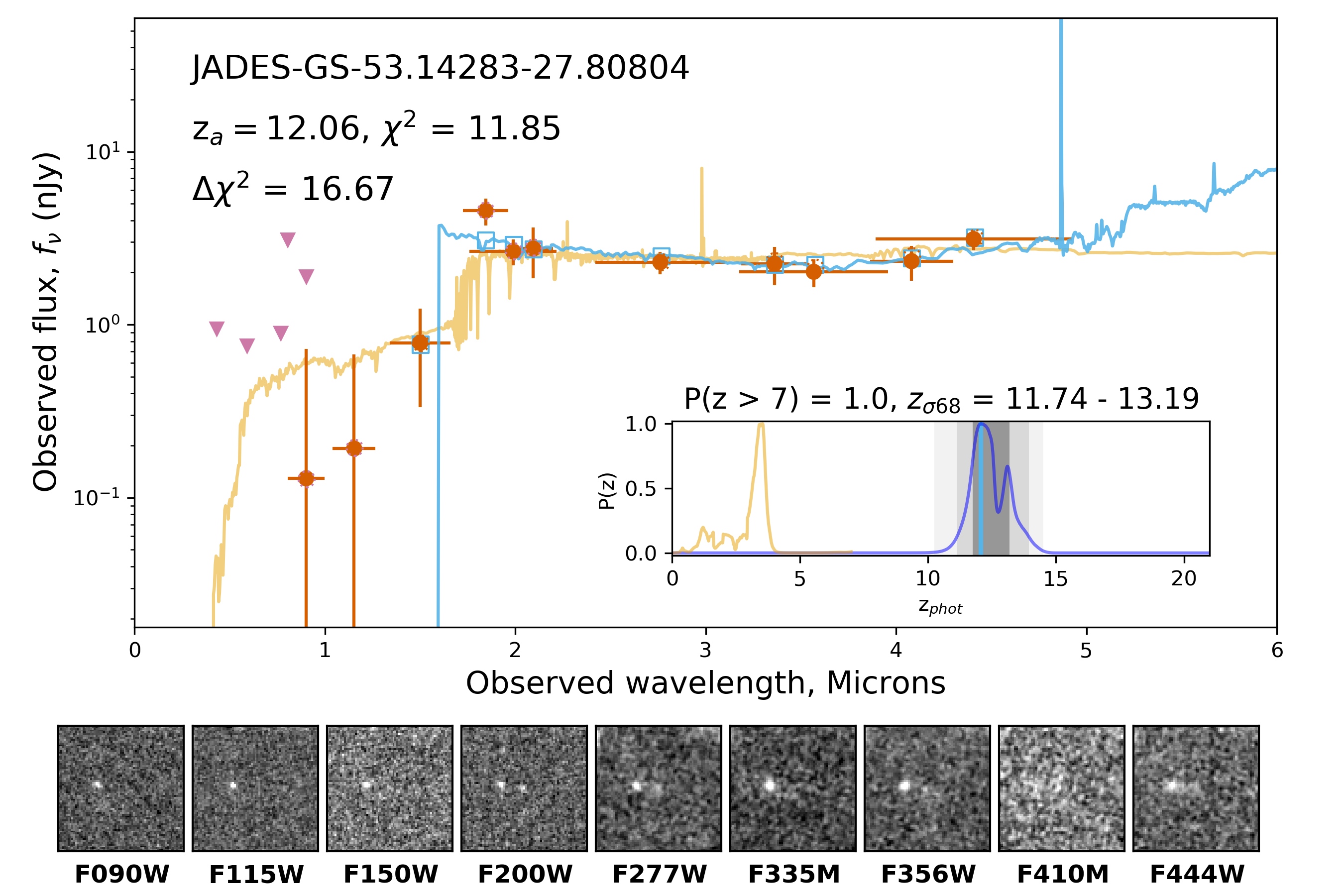}}\  
{\includegraphics[width=0.49\textwidth]{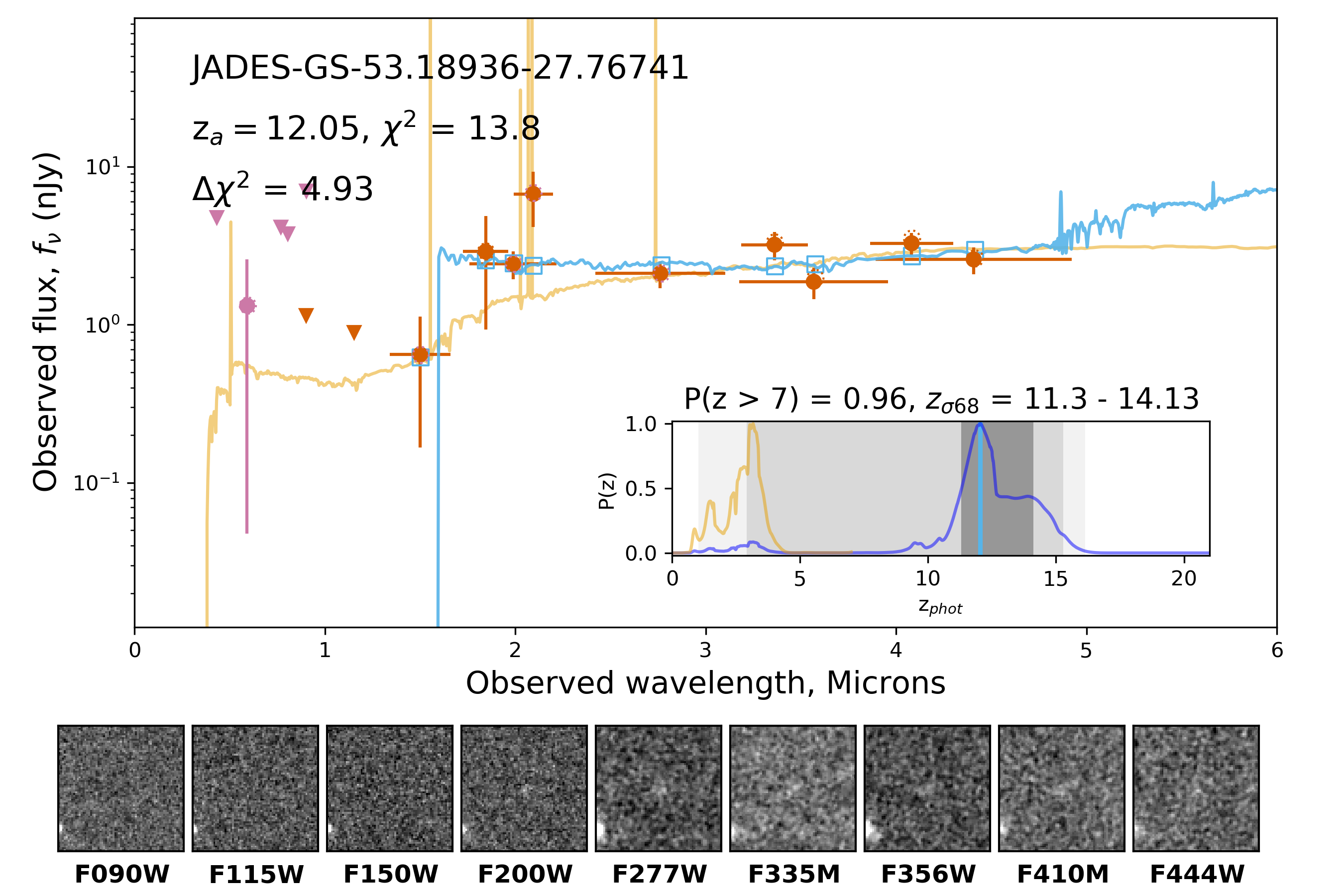}}\ 

\caption{
Continuation of Figure \ref{fig:z_gt_12_example_SEDs_pt3}. In each panel, the colors, lines, and symbols are as in Figure \ref{example_SED_fit}.}
\label{fig:z_gt_12_example_SEDs_pt4}
\end{figure*}

\begin{deluxetable*}{l | c c c c c c | c c c c c c}
\tabletypesize{\footnotesize}
\tablecolumns{13}
\tablewidth{0pt}
\tablecaption{Additional $z > 8$ candidates from {\tt EAZY} \citet{larson2022} Template Fits  \label{tab:larson_template_sources}}
\tablehead{
\multicolumn{1}{c}{} & \multicolumn{6}{c}{Photometric Redshifts (\citet{larson2022} Templates)} & \multicolumn{6}{c}{Photometric Redshifts (This Study)}\\
\colhead{JADES ID} &  \colhead{{\tt EAZY} $z_a$} & \colhead{$\chi^2_{min}$} & \colhead{$z_{\sigma68, low}$} & \colhead{$z_{\sigma68, high}$} & \colhead{$P(z > 7)$} & \colhead{$\Delta\chi^2$} & \colhead{{\tt EAZY} $z_a$} & \colhead{$\chi^2_{min}$} & \colhead{$z_{\sigma68, low}$} & \colhead{$z_{\sigma68, high}$} & \colhead{$P(z > 7)$} & \colhead{$\Delta\chi^2$}}
\startdata
JADES-GS-53.05706-27.81652  & 8.92  & 13.72 & 8.54 & 10.01  & 0.984 &  6.190 & 1.89 & 15.14 & 2.14 & 10.25 & 0.793 &  0.0 \\
JADES-GS-53.1153-27.80992    & 8.52  & 6.16   & 7.42 &  8.99  & 0.935 &  4.108  & 7.39 & 6.45   & 7.36 & 9.55 & 0.915 &  3.226 \\
JADES-GS-53.13383-27.82825  & 8.02  & 13.13 & 7.58 &  8.09  & 1.000 & 16.976 & 7.89 & 10.72 & 7.72 &  7.98 & 1.000 &  32.386 \\
JADES-GS-53.14036-27.79026  & 8.61  & 20.72 & 8.28 &  8.77  & 0.979 &  5.167  & 6.99 & 23.51 & 7.01 &  8.69 & 0.859 &  0.0 \\
JADES-GS-53.14712-27.77639  & 8.14  & 18.44 & 8.04 &  8.30  & 0.876 &  5.791  & 8.30 & 20.71 & 6.06 &  8.44 & 0.691 &  2.222 \\
JADES-GS-53.14992-27.88179  & 8.94  & 26.30 & 8.14 &  8.95  & 0.919 &  4.484  & 8.96 & 25.67 & 2.42 &  8.98 & 0.678 &  2.254 \\
JADES-GS-53.18389-27.82345  & 8.20  & 15.47 & 7.90 &  8.38  & 1.000 & 20.058 & 1.83 & 20.18 & 1.84 &  8.28 & 0.578 &  0.0 \\
JADES-GN-189.07044+62.29257 & 8.34  & 23.40 & 7.50 &  8.45  & 1.000 & 12.288 & 7.20 & 15.42 & 7.17 &  8.22 & 1.000 &  12.651 \\ 
JADES-GN-189.26946+62.19909 & 8.11  & 13.93 & 7.60 &  8.29  & 0.997 &  7.348  & 7.84 & 15.67 & 7.40 &  8.12 & 0.995 &  5.804 \\
JADES-GN-189.29444+62.14231 & 8.79  & 16.89 & 8.16 &  9.00  & 0.984 &  7.340  & 1.86 & 18.56 & 1.85 &  8.79 & 0.499 &  0.0
\enddata
\end{deluxetable*}

\vspace{5mm}
\facilities{JWST(NIRCam, NIRSpec), HST(ACS)}
\software{{\tt astropy} \citep{astropy2013,astropy2018}, {\tt matplotlib} \citep{matplotlib2007}, {\tt numpy} \citep{numpy2020}, {\tt scipy} \citep{scipy2020}, {\tt Photutils}  \citep{photutils2023}, {\tt lenstronomy} \citep{birrer2018, birrer2021}, {\tt EAZY} \citep{brammer2008}, {\tt fsps} \citep{conroy2010}. 
          }

\bibliography{main}{}
\bibliographystyle{aasjournal}

\end{document}